\documentclass[reqno]{amsart}
\usepackage{mathabx,mathtools,amsmath,amssymb,enumitem,amsthm,nccmath}
\usepackage{hyperref}
\usepackage{dsfont}
\usepackage{suetterl} 
\usepackage{pgfplots}	
\usepackage{caption}
\usepackage{subcaption}		
\usepackage{braket}

\usepackage[T1]{fontenc}
\usepackage{currvita}

\usepackage{csquotes} 

\usepackage{simpler-wick} 	
\newlength{\wdth}

\usepgfplotslibrary{fillbetween}
\usetikzlibrary{patterns}

\usetikzlibrary{arrows}

\allowdisplaybreaks		

\def\C{{\mathbb C}}
\def\N{{\mathbb N}}	
\def\Q{{\mathbb Q}}				
\def\R{{\mathbb R}}				 
\def\Z{{\mathbb Z}}				
\def\T{{\mathbb T}}

\def\bk{{\bf k}}

\def\bx{{\bf x}}
\def\bp{\mathbf{p}}
\def\bk {\mathbf{k}}				
\def\bq {\mathbf{q}}				
\def\bs{\mathbf{s}}					
\def\br{\mathbf{r}}

\def\bS{\mathbf{S}}
\def\bP{\mathbf{P}}
\def\bX{\mathbf{X}}

\def\cA{{\mathcal A}}

\def\cD{{\mathcal D}}
\def\cE{{\mathcal E}}
\def\cF{{\mathcal F}}
\def\cH{{\mathcal H}}
\def\cI{{\mathcal I}}						
\def\cK{{\mathcal K}}					

\def\cM{{\mathcal M}}
\def\cN{{\mathcal N}}
\def\cP{{\mathcal P}}
\def\cQ{{\mathcal Q}}

\def\cS{{\mathcal S}}
\def\cU{{\mathcal U}}
\def\cV{{\mathcal V}} 
\def\cW{{\mathcal W}}

\renewcommand{\Re}{\mathrm{Re}}
\renewcommand{\Im}{\mathrm{Im}}			
\def\complement{{c}}									
\newcommand{\tr}{\operatorname{Tr}}   

\def\1{{\bf 1}}
\def\alg{{\mathfrak A}}
\def\Bra{\Big\langle}
\def\Ket{\Big\rangle}

\newcommand{\scp}[2]{\langle#1,#2\rangle}
\def\jb#1{\langle#1\rangle} 
\newcommand{\floor}[1]{\left\lfloor #1 \right\rfloor}

\def\eqnn{\begin{eqnarray*}}			
	\def\eeqnn{\end{eqnarray*}}			
\def\eqn{\begin{eqnarray}}				  
	\def\eeqn{\end{eqnarray}}
\def\prf{\begin{proof}}
	\def\endprf{\end{proof}} 

\makeatletter
\def\namedlabel#1#2{\begingroup					
	#2%
	\def\@currentlabel{#2}%
	\phantomsection\label{#1}\endgroup
}
\makeatother

\theoremstyle{plain}
\newtheorem{theorem}{Theorem}[section]		
\newtheorem{definition}[theorem]{Definition}				
\newtheorem{proposition}[theorem]{Proposition}		   

\newtheorem{lemma}[theorem]{Lemma}
\newtheorem{corollary}[theorem]{Corollary}
\newtheorem{remark}[theorem]{Remark}
\newtheorem*{remark*}{Remark}

\numberwithin{equation}{section}

\def\cHcorz{\cH_{cor}^{\phi_0}}
\def\cHquart{\cH_{quart}}

\def\cHmfz{\cH_{HFB}^{\phi_0}}
\def\cHmft{\cH_{HFB}}
\def\cVmf{\cV_{HFB}}

\def\cHcubz{\cH_{cub}^{\phi_0}}
\def\cHcubt{\cH_{cub}}

\def\nc#1{\|#1\|_{2(\floor{\frac\reg2}+1),c}}
\def\nd#1{\|#1\|_{d}}
\def\wnc#1{\|#1\|_{2(\floor{\frac\reg2}+1),w,c}}		 
\def\wnd#1{\|#1\|_{w,d}}													
\def\fc{\nc{f_0}}																	
\def\fd{\nd{f_0}}
\def\vc{\wnc{\hat{v}}}
\def\vd{\wnd{\hat{v}}}
\def\An#1{\|#1\|_{A}}
\def\tind#1{\|#1\|_{2\cap\infty,d}}
\def\Jtn{\tind{J}}
\def\Jin{\|J\|_{\ell^\infty(\lattice)}}
\def\Jcn{\nc{J}}

\def\chif{\mathds{1}_{A=f[J] }}					
\def\chig{\mathds{1}_{A=g[J] }}				 
\DeclareMathOperator{\osg}{sign}		 
\def\fbar{\widetilde{f_0}}
\def\hb{\widetilde{h}}
\def\kro{\delta_{\lattice}}

\def\veps{\vep}

\newcommand{\Rem}{\operatorname{Rem}}   
\def\rr{\operatorname{err}}				 	  
\def\eeb{{\rr}_{Bog}}								 
\def\eed{{\rr}_{disc}}								 
\def\rrb{\rr^{(Bog)}}

\def\rrbogc{\rr^{(Bog,\con)}}
\def\rrbogb{\rr^{(Bog,\cB)}}

\def\rrf{\rr^{(fd)}}
\def\rrec{\rr^{(ec)}}
\def\rrfec{\rr^{(fec)}}
\def\rreff{\rr^{(cen)}}
\def\rrd{\rr^{(dis)}}

\def\vep{{\varepsilon}}
\def\ups{{\tau}}							   

\def\dEcub{{\Delta_{cub}E}}						
\def\dOcub{{\Delta_{cub}\Omega}}		
\def\dJcub{{\Delta_{cub}J}}						 

\def\reg{r}
\def\dO{m_\Omega}					  
\def\rate{\delta}

\def\vol{|\Lambda|}
\def\lattice{{\Lambda^*}}
\def\nb{\mathcal{N}_b}
\def\vac{\Omega_0}

\def\gc{\bar{\lambda}}

\def\bbg{\operatorname{col}}
\def\bbgd{\bbg_d}							
\def\bbgc{\bbg_c}							

\def\con{\operatorname{Con}}	
\def\cCd{\con_d}							 
\def\cCc{\con_c}							 

\def\bbf{\operatorname{bol}}
\def\cB{\Col}
\def\Col{\operatorname{Bol}}				
\def\cBd{\Col_d}										 
\def\cBc{\Col_c}
\def\BBF{\Col}
\def\cBfec{\cB_{fec}}

\def\ac{\mathrm{abs}}

\def\acd{\ac_{cub,d}}
\def\acc{\ac_{cub,c}}

\def\acqd{\ac_{quart,d}}
\def\acqc{\ac_{quart,c}}

\makeatother

\begin{document} 
	
	\title[quantum fluctuations around BEC]
	{On the emergence of quantum Boltzmann fluctuation dynamics near a Bose-Einstein Condensate}
	
	\author{Thomas Chen}
	\address[T. Chen]{Department of Mathematics, University of Texas at Austin, Austin TX 78712, USA}
	\email{tc@math.utexas.edu} 
	\author{Michael Hott}
	\address[M. Hott]{School of Mathematics, University of Minnesota Twin Cities, Minneapolis MN 55455, USA}
	\email{mhott@umn.edu}

	\begin{abstract}
		In this work, we study the quantum fluctuation dynamics in a Bose gas on a torus $\Lambda=(L\mathbb{T})^3$ that exhibits Bose-Einstein condensation, beyond the leading order Hartree-Fock-Bogoliubov (HFB) theory. Given a Bose-Einstein condensate (BEC) with density $N\gg 1$ surrounded by thermal fluctuations with density $1$, we assume that the system dynamics is generated by a Hamiltonian with mean-field scaling. We derive a quantum Boltzmann type dynamics from a second-order Duhamel expansion upon subtracting both the BEC dynamics and the HFB dynamics, with rigorous error control. Given a quasifree initial state, we determine the time evolution of the centered correlation functions $\langle a\rangle$, $\langle aa\rangle-\langle a\rangle^2$, $\langle a^+a\rangle-|\langle a\rangle|^2$ at mesoscopic time scales $t\sim\lambda^{-2}$, where $0<\lambda\ll1$ is the coupling constant determining the HFB interaction, and $a$, $a^+$ denote annihilation and creation operators. While the BEC and the HFB fluctuations both evolve at a microscopic time scale $t\sim1$, the Boltzmann dynamics is much slower, by a factor $\lambda^2$. For large but finite $N$, we consider both the case of fixed system size $L\sim1$, and the case $L\sim \lambda^{-2-}$. In the case $L\sim1$, we show that the Boltzmann collision operator contains subleading terms that can become dominant, depending on time-dependent coefficients assuming particular values in $\mathbb{Q}$; this phenomenon is reminiscent of the Talbot effect. For the case $L\sim \lambda^{-2-}$, we prove that the collision operator is well approximated by the expression predicted in the literature. In either of those cases, we have $\lambda\sim \Big(\frac{\log \log N}{\log N}\Big)^{\alpha}$, for different values of $\alpha>0$. 
	\end{abstract}

	\maketitle

	\tableofcontents

   \section{Introduction}
   
	\subsection{Quantum dynamics and Boltzmann equations}
	
	The main question we set out to answer in this work is:
	\begin{displayquote}
	Is there a scaling regime in an interacting quantum field theory, for which the emergence of collisional processes described by a Boltzmann equation can be rigorously established?
	\end{displayquote}
	\subsubsection{Emergence of a quantum Boltzmann equation} In analogy to Maxwell's and Boltzmann's theory of collisions in classical systems, Nordheim \cite{nordheim} in 1928 was the first to propose a Boltzmann equation for Bose and Fermi gases given by 
	\eqn
		\lefteqn{\partial_t f (p)=Q_4(f):=}\nonumber\\
			&& \int d\bp_4 \, \delta(p_1+p_2-p_3-p_4) \delta(E(p_1)+E(p_2)-E(p_3)-E(p_4))\nonumber\\ &&|\cM_{22}(\bp_4)|^2(\delta(p-p_1)+\delta(p-p_2)-\delta(p-p_3)-\delta(p-p_4)) \nonumber\\
			&& \big((1\pm f(p_1))(1\pm f(p_2))f(p_3)f(p_4)-f(p_1)f(p_2)(1\pm f(p_3))(1\pm f(p_4))\big) \, . \label{BUU}
	\eeqn 
	Here, $f$ denotes the particle density in the spatially homogeneous case; '$+$' refers to the bosonic, and '$-$' refers to the fermionic equation, and $\bp_4=(p_1,p_2,p_3,p_4)$. In addition, $E(p)=\frac12|p|^2$ denotes the free dispersion, and $\cM_{22}$ is the (microscopic) scattering cross section for $2\leftrightarrow2$ processes describing two thermal fluctuation scattering off of each other. As Nordheim already argues, the distribution of the outgoing particles needs to be taken into account, resulting in a quartic collision operator, in contrast to classical particles that are described by a quadratic collision operator. It is shown in \cite{nordheim} that the equilibrium is given by the Bose-Einstein, respectively the Fermi-Dirac statistics, and that an $H$-theorem holds true. In 1933, Uehling and Uhlenbeck \cite{uhue} studied the linearization about the equilibrium, in order to determine the associated hydrodynamics, and to compute the heat conductivity and the viscosity coefficient.
	\par Subsequently, physicists have given formal derivations of the above quantum Boltzmann equation, using diagrammatic techniques from quantum field theory, see, e.g., \cite{AGD,kaba,vanleeuwen}. This has given rise to interesting fundamental effective theories, such as the {\it Kadanoff-Baym equations}, see, e.g., \cite{hokali,shewespriwa}. We also mention the important contributions by Bogoliubov and collaborators \cite{bogboltz} and Prigogine and collaborators \cite{prig}. 
	\par The first mathematically rigorous works on the derivation of the classical Boltzmann equation, a billiards model for a classical gas, go back to Cercignani \cite{cerc} in 1972 and Lanford \cite{lanford} in 1975, where they studied the Grad-limit \cite{grad} of a hard-sphere model. These works were later revisited and completed through works by Uchiyama \cite{uc}, Cercignani-Illner-Pulvirenti \cite{cerillpul}, Spohn \cite{spohn81bo}, Cercignani-Gerasimenko-Petrina \cite{cegepe}, and Gallagher--Saint-Raymond--Texier \cite{gasrte}.
	\par In 1983, Hugenholtz \cite{hu} considered the commutator perturbation expansion with respect to the weak coupling constant $\lambda$. Implementing the kinetic time scale $t=T\lambda^{-2}$ used by van Hove \cite{vanhove}, it was shown that, in the translation-invariant case, terms of order $O(\lambda)$ vanish as $\lambda\to0$, and that terms of order $O(\lambda^2)$ are proportional to $T$. Using a selection rule, it is conjectured in \cite{hu} that only two-point correlations of higher orders in $\lambda$ survive, motivating the assumption of {\it quasifreeness}. Quasifreeness, as we will see below, is, in a sense, a quantum analogue to the 'Sto\ss zahlansatz', also known as 'molecular chaos'. Hence, Hugenholtz argues, at leading order, the Boltzmann equation should arise for the evolution of the two-point function. Ho and Landau \cite{hola} later proved that, to second order in $\lambda$, this holds true. 
	\par In 2004, Erd\"os, Salmhofer, and Yau \cite{ESY} extended the results by Hugenholtz, and by Ho and Landau, by introducing the concept of {\em restricted quasi-freeness}, i.e., quasi-freeness only up to six- or eight-point correlations. Assuming propagation of restricted quasi-freeness, they showed that a (time-dependent) Boltzmann equation arises from the second-order Duhamel expansion, under certain assumptions. Around the same time, Benedetto, Castella, Esposito and Pulvirenti, began a series of works that used Bogoliubov-Born-Green-Kirkwood-Yvon (BBGKY) hierarchies to derive a quadratic Boltzmann equation with a quantum collision kernel for a system obtained by iterating Duhamel's formula, ignoring the tail, and truncating the obtained hierarchy in the small-coupling limit \cite{BCEP04,BCEP08} respectively in the low-density regime \cite{BCEP06}. In \cite{BCEP05}, they went on to show that the contributions to second order in the coupling recover \eqref{BUU} to leading order, evaluated at initial time with $f_0$ instead of $f_t$.  Further works studying the BBGKY hierarchy in the context of the quantum Boltzmann equation include works by Gerasimenko and collaborators \cite{gera2011,gera2012,getsv}, see also references therein. A recent review in this direction can be found in \cite{gera2021}. For works using a second quantization approach, we refer to Spohn and collaborators \cite{fulusp,LuSp1,Spohn94,Spohn06,Spohn07}, see also references therein. 
	\par In 2015, X. Chen and Y. Guo \cite{xchguo} showed that, if the marginals of the BBGKY hierarchy converge in the weak-coupling limit in a strong sense, and if the $W^{4,1}$-regularity per particle remains uniformly bounded for some positive time, then the limiting hierarchy is that associated to the quadratic Boltzmann equation with a quantum collision kernel, instead of the Boltzmann equation derived in \cite{ESY}. 
	\par In a different line of work addressing the weakly disordered Anderson model, scattering of electrons at impurities in a lattice have been investigated extensively. For works studying the derivation of the linear Boltzmann equation in this context, we refer to \cite{chen05,chro,chsa,EY00,hikiolsc,holawi,Spohn77}. For a more recent treatment, we refer to \cite{griffin}. 
	\par Lukkarinen-Spohn \cite{LuSp1,LuSp2}  showed that the nonlinear Schr\"odinger equation (NLS) with random initial data leads to the wave kinetic equation: They present this system as a simplified model to gain insights into the emergence of Boltzmann-type dynamics in a quantum Bose gas. The study of wave-turbulence in the context of the long-time behavior of the NLS is a very active area of research, see, e.g., \cite{amcoge,bugehash,coge19,coge20,deha19,deha21,esveTurbo,geiotr,kive,rost}.
	\par The derivation of Boltzmann equations from a system of classical interacting particles is an extraordinarily active research field, see \cite{ammipa,ampa,bogasr,bogasrs20,bogasrs21,bogasrs22,cegepe,cerillpul,den,gasrte,king,lanford,papusi,pulvi,pusi,pusasi,sisp,sp,uc} and references therein. The methods  differ vastly from the quantum field theoretic approach developed in the work at hand. For works studying well-posedness and other analytical properties of the classical Boltzmann equation, we refer to \cite{akopian,akga,amuxy2011,amuxy2012,alvi,alonso,alga,amgapata,arke,bobylev1984,bogapan,beto1985,cerillpul,chdepa,chdepa2,delgapa,desgol,deswenn,diperlio,duan,glassey,goudon,grstr, guo,hano,hanoyu,ilshi,kashi,miper,toscani1986,taalgapa,toscani1988,ukai,villani1998,villanibook,wennberg} and references therein.

	\subsubsection{Well-posedness\label{sec-wellposed}} The first well-posedness results for \eqref{BUU} go back to Dolbeaut \cite{dolbeaut} and Lions \cite{lions94} for the fermionic case. X. Lu and collaborators have made significant progress on the fermionic Boltzmann-Uhlenbeck-Uehling (BUU) equation, see, e.g., \cite{lu01fer,lu08fer,luwe}. For a recent work on the fermionic problem, we refer to \cite{zli}.
	\par A unified treatment of bosons and fermions can be found in \cite{zhlu02I,zhlu02II}, and more recently \cite{ouwu}. In addition, generalized statistics such as anyons \cite{arno15an}, and Haldane statistics \cite{ar10,arno19ha} have also been included in this line of study. 
	\par Works studying the relativistic quantum Boltzmann equation include \cite{bajayu,esmiva03,esmiva} and references therein. 
	\par The quantum Landau equation, which can be viewed as a limit of the Boltzmann equation accounting for long-range interactions, has been studied in \cite{albadelo,albalo,ba,bale,lemou,liu}. Recently, He-Lu-Pulvirenti \cite{helupu} showed that it can be obtained as a weak semi-classical limit from the quantum Boltzmann equation.
	\par In 1924, Bose \cite{bose}, and, independently in 1925, Einstein \cite{einstein} predicted that, below a critical temperature, the ground state becomes gradually more populated, forming a macroscopic state called the {\it Bose-Einstein condensate} (BEC). In 1995, this phenomenon was independently experimentally verified by groups around Cornell and Wiemann \cite{corwie} and Ketterle \cite{ketterle}. Both groups were awarded the 2001 Physics Nobel Prize.
	\par In the mathematically rigorous PDE literature, the bosonic problem has been investigated first by X. Lu \cite{lu00bos}. The long-time behavior for radial initial data was studied, obtaining global existence, local stability, conservation of energy, and  estimates on moment production. Moreover, at low temperatures, it is shown in \cite{lu00bos} that a solution concentrates at $p=0$ for large time, and that, at high temperatures, the solution converges weakly to the Bose-Einstein distribution. Escobedo-Mischler-Valle \cite{esmiva} showed that bosonic entropy maximizers are given by 
	\eqn
		f_{eq}(p) \, = \, \frac1{e^{\beta(E(p)-\mu)}-1} \, + \, m_0\delta(p) \, ,
	\eeqn 
	where $m_0\cdot \mu=0$, see, e.g., \cite{esve2014}. The BEC density $m_0$, the chemical potential $\mu$ and the inverse temperature $\beta$ are uniquely determined by the moments $\int dp f_{eq}(p)$ and $\int dp E(p) f_{eq}(p)$, see, e.g., \cite{lu04bos} and references therein. One has that $m_0$ vanishes above a critical temperature $T_c$, and is non-zero below $T_c$.
	
	\par Well-posedness results have been formulated to account for solutions that form a Dirac mass at temperatures below $T_c$, and that stay bounded for temperatures above $T_c$. We refer to \cite{ar10,cai,esve2014,esve2015,lu04bos,lu05bos,lu13bos,lu14bos,lu16bos,luzh} for the isotropic and space-homogeneous case, to \cite{arno15an,arno17,roy,ouwu,zhlu02I,zhlu02II} for the space-dependent case, and to \cite{brei,cailu,lilu,ouwu} for the anisotropic case. Further works studying the blow-up behavior related to condensation include \cite{bave,esve,Spohn10}.

	\subsection{Collisions of Fluctuations about a Bose-Einstein Condensate}
	
	\par Pioneering works by Kirkpatrick and Dorfmann \cite{kidr83,kidr85} and Eckern \cite{eck} started analyzing the interplay of the BEC with thermal excitation cloud surrounding the condensate. They formally obtained a Boltzmann equation of the form
	\eqn
		\partial_t f^{(ex)}(p) = n_c Q_3(f^{(ex)}) \, + \, Q_4(f^{(ex)}) \, ,
	\eeqn 
	where $Q_4$ is given by \eqref{BUU} with $E$ replaced by the Bogoliubov dispersion $\Omega=\sqrt{E(E+2n_c\gc)}$, and 
	\eqn
		Q_3(f) & := & \int d\bp_3 \, \delta(p_1+p_2-p_3) \delta(\Omega(p_1)+\Omega(p_2)-\Omega(p_3))\nonumber\\ &&|\cM_{21}(\bp_3)|^2(\delta(p-p_1)+\delta(p-p_2)-\delta(p-p_3)) \nonumber\\
		&& \big((1+f(p_1))(1+f(p_2))f(p_3)-f(p_1)f(p_2)(1+ f(p_3))\big) \, . \label{BUU3}
	\eeqn 
	Here $n_c$ denotes the BEC density, $f^{(ex)}$ the density of thermal fluctuation particles, $\gc$ is the coupling strength of the hard-sphere pair interaction $\gc\delta$, $\cM_{21}$ is the cross section for $2\leftrightarrow1$ processes describing collisions of 2 thermal particles, where one is either being absorbed into or emitted from the BEC, and $\bp_3=(p_1,p_2,p_3)$. Zaremba, Niguni and Griffin \cite{ZNG}, see also \cite{GNZ}, later extended their approach to include the dynamics of the condensate. In the {\it Hartree-Fock-Bogoliubov-Popov} approximation, and for a translation-invariant initial state, they formally argue that the condensate wave function $\Phi$ satisfies 
	\eqn
		i\partial_t \hat{\Phi}(p) & = & (h_{Har} \, + \, 2\gc f^{(ex)}(p)-iQ_3(f^{(ex)}))\hat{\Phi}(p) \nonumber\\
		&& + \, \gc g^{(ex)} (p)\overline{\hat{\Phi}}(-p) \, , \label{HFBP}
	\eeqn 
	and it is linked to the density via $n_c(t,x)=|\Phi_t(x)|^2$. Here $h_{Har}$ denotes the Hartree Hamiltonian in momentum representation, see Section \ref{sec-hartree} below, and $g^{(ex)}$ denotes the rate of pair absorption into the condensate, which they discarded as a lower-order contribution. One of the motivations to study the coupled system between the condensate and the thermal cloud is to understand the nucleation process of the BEC, see, e.g., \cite{bizast,gazo97,jopo,lalapori,pobrm,setka}. For a review, we also refer to \cite{progada,retr}.
	\par Observe that, for large values of the condensate density $n_c$, we expect that $Q_4$ is of subleading order.
	\par For mathematical works studying the system describing the two-component gas consisting of the condensate and the excitation cloud, we refer to \cite{algatr,arno15,coes,crtr,espeva,ngtr,potr,potr20,potr21,sotr,sotrKI}.

	\subsection{Definition of the mathematical model}
	As a starting point for our analysis, we choose a single-species Bose gas at positive temperature trapped in a periodic box $\Lambda=(L\T)^3$ of linear length $L$. We assume that the gas consists of two phases:
	\begin{enumerate}
		\item A Bose-Einstein condensate (BEC) with density $N$,
		\item thermal fluctuations with density $\sim 1$.
	\end{enumerate}
	We note that a significant part of this paper addresses the case of a fixed volume, where we may think of $N$ as the number of bosons when $L=1$. On the other hand, we will also consider the limit of large volume, where $N>0$ will denote the number of bosons per unit volume (that is, the density); for convenience, we are not changing the notation.  
	We assume mean-field interactions for which the kinetic energy of the condensate and the total (pair) interaction potential among particles are balanced, and analyze the interplay between the condensate dynamics and the dynamics of the fluctuation particles.  
	
	\subsubsection{Definition of the model}
	Let $\cF=\C\oplus\bigoplus_{n\geq1}(L^2(\Lambda))^{\otimes_{sym}n}$ denote the bosonic Fock space, equipped with the inner product
	\begin{equation} \label{def-fock-prod}
		\scp{\Psi}{\Phi}_{\cF} \, = \, \sum_{n\in\N_0}\scp{\Psi_n}{\Phi_n}_{L^2(\R^{3n})}
	\end{equation}
	for all $\Psi=(\Psi_n)_{n\in\N_0}$, $\Phi=(\Phi_n)_{n\in\N_0}\in\cF$.
	For $n\in\N$, $f\in L^2(\R^3)$, $\Psi \in (L^2(\Lambda))^{\otimes_{sym}n}$, and $\bx_{n-1}\in\R^{3(n-1)}$, we define
	\begin{equation}\label{def-annihilation}
		(a(f)\Psi)(\bx_{n-1})  \, := \, \sqrt{n}\int_\Lambda dx \,  \overline{f}(x)\Psi(x,\bx_{n-1}) \, .
	\end{equation}
	For any $n\in \N$, let $\cS_n$ denote the permutation group of $\{1,\ldots,n\}$. For $n\in\N_0$, $f\in L^2(\R^3)$, $\Psi \in (L^2(\Lambda))^{\otimes_{sym}n}$, and $\bx_{n+1}=(x_1,\ldots,x_{n+1})\in\R^{3(n+1)}$, we define 
	\begin{equation} \label{def-creation}
		(a^+(f)\Psi)(\bx_{n+1}) \, := \,  \frac{\sqrt{n+1}}{(n+1)!} \sum_{\pi \in \cS_{n+1}} f(x_{\pi(1)})\Psi(x_{\pi(2)},\ldots,x_{\pi(n+1)}) \, .
	\end{equation}
	Then we have that for all $f\in L^2(\Lambda)$, $\Phi,\Psi\in \cF_{fin}$
	\begin{equation}\label{eq-adjoint}
		\scp{\Phi}{a(f)\Psi}_{\cF} \, = \, \scp{a^+(f)\Phi}{\Psi}_{\cF} \, .
	\end{equation}
	In addition, $a$ and $a^+$ satisfy the canonical commutation relations (CCR) for any $f,g\in L^2(\Lambda)$
	\begin{equation}\label{eq-smeared-out-CCR}
		[a(f),a^+(g)] \, = \, \scp{f}{g} \quad , \quad  [a(f),a(g)] \, = \, [a^+(f),a^+(g)] \, = \, 0 \, .
	\end{equation}
	Moreover, we
	introduce the operator-valued distributions $a_x$, $a^+_y$ by requiring
	\begin{align}
		a(f) \, & = \, \int_\Lambda dx \, \overline{f}(x) a_x \, , \\
		a^+(g) \, & = \, \int_\Lambda dx \, f(x) a^+_x
	\end{align}
	for all $f,g\in L^2(\Lambda)$. 
	These satisfy the CCR
	\eqn 
	[a_x,a_y^+]=\delta_\Lambda(x-y) 
	\quad,\quad
	[a_x,a_y]=[a_x^+,a_y^+]= 0 \, .
	\eeqn 
	We call $\vac:=(1,0,0,\ldots)\in\cF$ the \emph{Fock vacuum}. Then we have that $a_x \vac =0$ for all $x\in\Lambda$. In addition, for $\Psi\in\cF_{fin}$, $x\in\Lambda$, $n\in\N$, and $\bx_n\in\R^{3n}$, we have that
	\begin{equation}
		(a_x\Psi)^{(n)}(\bx_n) \, = \, \sqrt{n+1}\Psi^{(n+1)}(x,\bx_n) \, .
	\end{equation}
	We introduce the number operator  
	\begin{equation} \label{def-nb}
		\nb \, := \, \int_{\Lambda} \, dx \, a_x^+a_x \, ,
	\end{equation}
	which satisfies
	\begin{equation}
		(\nb \Psi)^{(n)} \, = \, n\Psi^{(n)} \,,
	\end{equation}
	see, e.g., \cite{BPS2}. 
	\par
	The Hamiltonian of the system studied in this paper has the following form. Let $v$ be a sufficiently regular pair potential, see section \ref{sec-assumptions}, and $\lambda>0$ a small coupling constant. Given $N>0$, we define
	\eqn\label{def-hamiltonian} 
	\cH_N \, := \,   \frac12\int_{\Lambda} \, dx \, a_x^+(-\Delta_x)a_x +\frac\lambda{2N}\int_{\Lambda^2} dx \, dy \, v(x-y) \, a_x^+ a_y^+ a_ya_x \, .
	\eeqn

	\par Our aim is to study the evolution associated with the Hamiltonian $\cH_N$ given in \eqref{def-hamiltonian}. The condensate is accounted for by a coherent state $\cW[\sqrt{N\vol}\phi_0]\vac$, where
	\eqn
		\cW[f] \, := \,\exp(a^+(f)-a(f)) \, = \, \exp\big(\int_{\Lambda}dx \, (f(x)a_x^+-\overline{f}(x)a_x)\big)\label{def-weyl}
	\eeqn 
	denotes the Weyl operator, and $\phi_0$ is normalized, $\|\phi_0\|_{L^2(\Lambda)}=1$. By the Baker-Campbell-Hausdorff formula, 
	\eqn
		\cW[f]\Omega \, = \, e^{-\frac12\|f\|_{L^2(\Lambda)}} \big( \frac{f^{\otimes n}}{\sqrt{n!}}\big)_{n\in\N_0} \, , 	
	\eeqn 
	see, e.g., \cite{BPS2}. That is, in each fixed particle sector, the wave function is the product state determined by a single wave function $f$.
	\par We assume that, at initial time $t=0$, the system is described by a state
	\eqn
	 \nu_0(A) \, := \, \frac{ \tr\big( e^{-\cK} A \big)}{\tr(e^{-\cK})}
	 \eeqn 
	for all observables $A\in\alg$, where $\alg$ denotes the Weyl algebra generated by $\cW[f]$, where $f\in \cS(\Lambda)$ is a Schwartz function, see \cite{BR2}, section 5.2.3. In addition, the Baker-Campbell-Hausdorff formula yields
	\eqn
		\cW^*[f]a_x\cW[f] \, = \, a_x \, + \, f(x) \, . \label{eq-a-weyl-trafo}
	\eeqn 
	
	\begin{definition}[Quasifree state]
		Let $\nu$ be a state and 
		\eqn
			\nu^{(cen)}(A):=\nu(\cW[\nu(a)]A\cW^*[\nu(a)])
		\eeqn 
		denote its centering. We say $\nu$ is {\normalfont quasi-free} iff 
		\begin{equation} \label{eq-wick-thm}
			\begin{cases}
				\nu^{(cen)}(a^{\#_1} a^{\#_2}\ldots a^{\#_{2n}})  &= 
				\wick{
					\c1{a}^{\#_1} \c2{a}^{\#_2} \c2{.}\c2{.}\c1{.} \c2{a}^{\#_{2n}}
				}\\
			& \quad +\mbox{all\ pair\ contractions}\\ 
				\nu^{(cen)}(a^{\#_1} a^{\#_2}\ldots a^{\#_{2n-1}}) &=  0 
			\end{cases} \, ,			
		\end{equation}
		where $\wick{\c a^{\#_1} \c a^{\#_2}}:=\nu^{(cen)}(a^{\#_1}a^{\#_2})$. \eqref{eq-wick-thm} is referred to as {\normalfont Wick's Theorem}.
	\end{definition}
	
	\begin{definition}[Number conserving state]
		A state $\nu$ is called {\normalfont number conserving} iff $\nu([A,\nb]) = 0$ for every observable $A\in\alg$.
	\end{definition}
	
	We assume that the initial state $\nu_0$ is number conserving, quasifree, and translation-invariant. In particular, we assume that the translation-invariant generator $\cK$ is given by 
	\eqn\label{def-K}
	\cK \, := \, \int_{\lattice} \, dp \, K(p) a_p^+a_p \, ,
	\eeqn
	where $K(p)\geq\kappa_0$ for some $\kappa_0>0$. Observe that, by being number conserving, $\nu_0$ is already centered.
	
	\par The state describing the two-phase Bose gas is then given by
	\eqn \label{def-rho0}
		\rho_0(A) \, := \, \frac{1}{\tr(e^{-\cK})} \tr\Big( \, \cW[\sqrt{N\vol}\phi_0]e^{-\cK}\cW^*[\sqrt{N\vol}\phi_0] A  \, \Big)
	\eeqn 
	for all $A\in\alg$. The initial value problem (IVP) associated with the Hamiltonian $\cH_N$ and the initial state $\rho_0$ is then given by the von Neumann equation
	\eqn\label{eq-schroedinger-non-rel}
		i\partial_t \rho_t(A) \, = \, \rho_t([A,\cH_N]) 
	\eeqn
	for all observables $A\in\alg$. Below, we impose assumptions on $v$ ensuring that $\cH_N$ is self-adjoint and that it induces a unitary evolution $e^{-it\cH_N}$. Then, the solution of \eqref{eq-schroedinger-non-rel} is given by 
	\eqn
		\rho_t(A) \, = \, \rho_0(e^{it\cH_N}Ae^{-it\cH_N}) \, .
	\eeqn 
	By making specific choices for $A$, we will study effective equations of key correlation functions characterized by the Bose gas.
	
	\subsection{Leading order condensate dynamics: Hartree equation\label{sec-hartree}}
	We expect the leading order dynamics of \eqref{eq-schroedinger-non-rel} to be described by the leading order condensate dynamics, as the BEC describes the bulk component of the Bose gas. Indeed, for instance, \cite{GM1,GM2,GM3,grmama,rosc}, based on Hepp's method and using coherent states, have shown, that, in a precise sense, $e^{-it\cH_N}W[\sqrt{N\vol}\phi_0]\vac$ is well approximated by $W[\sqrt{N\vol}\phi_t]\vac$ for $N\gg1$, with approximation errors $o_N(1)$, where $\phi_t$ satisfies the Hartree equation
	\eqn \label{eq-hartree}
		i\partial_t\phi_t \, = \,  -\frac12 \Delta\phi_t  + \lambda \vol (v*|\phi_t|^2)\phi_t \, .
	\eeqn 
	The volume factor in the nonlinear interaction term accounts for our assumption that the $L^2$-mass of the condensate is 
	\eqn
	\|\sqrt{N\vol} \phi_t\|_{L^2(\Lambda)}^2 \, = \, N\vol \, .
	\eeqn 
	In the case $v=\delta$, \eqref{eq-hartree} yields the nonlinear Schr\"odinger equation (NLS). Analogous statements have been proved for different choices of $v$ with alternative approaches involving the corresponding BBGKY hierarchy, see, e.g., \cite{AGT,CHPS-1,CPBBGKY,chtal,xchho0,xchho1,xchho2,xchho3,xchho4,xchho5,esy1,esy2,esy3,esy4,EY,GreSohSta-2012,heso,hotaxie,kiscst,klma,Liard,Spohn81}, and other approaches, see \cite{alon,AFP,ALR,AN,AH,AHH,BGM,bossmann,bote,dese2020,dese2021,deseyn,frknpi,frknsc,frtsya,GV,hainzl,hott,jelepi,kiess,KP,Lee1,Lee,lenaro,lenasc,narose,lise,lisesoyn,liseyn,liseyn2,Luehr,miol,misc,pickl1,pickl2,rou,shen,Sohinger,sohinger2020}. For more background on the derivation of Hartree theory, we refer to \cite{BPS2,golse,GS,lewin,lisesoyn,N,schlein}.
	
	\subsubsection*{Stationary, translation-invariant condensate}
	For simplicity, we choose to consider a stationary and translation-invariant solution of \eqref{eq-hartree}. Due to the normalization constraint $\|\phi_0\|_2=1$, we have
	\eqn\label{eq-phi0-set}
	\phi_t=\phi_0=\vol^{-\frac12} \; \in \R_+ \, .
	\eeqn
	Substituting this into \eqref{eq-hartree} yields
	\eqn
		0 \, = \, \lambda\vol |\phi_0|^2\phi_0 \int_\Lambda dx \, v(x)  =\frac{\lambda}{\sqrt{\vol}} \int_\Lambda dx \, v(x) \, . \label{eq-v-int-ass-0}
	\eeqn 	
	In particular, we assume
	\eqn
		\int_\Lambda dx \, v(x) \, = \, 0 \, , \label{eq-int-v-0}
	\eeqn 	
	with additional regularity properties introduced below. Henceforth, we assume that $\phi$ is stationary, translation-invariant and satisfies \eqref{eq-phi0-set}.
	
	\subsection{Leading order fluctuation dynamics: HFB equations}
	
	We next turn to the leading order corrections of the full dynamics past the leading order BEC Hartree dynamics. For this purpose, we consider the fluctuation dynamics described by $W^*[\sqrt{N\vol}\phi_0]e^{-it\cH_N}W[\sqrt{N\vol}\phi_0]$. We show in Lemma \ref{lem-FD-1}
	\eqn
		W^*[\sqrt{N\vol}\phi_0]\cH_NW[\sqrt{N\vol}\phi_0] \, = \, \cHmft \, + \, \cHcubt \, + \, \cHquart \, , \label{eq-fluct-ham}
	\eeqn 
	where
	\eqn
		\cHmft &:=& \int_{\Lambda^2} dx \, dy \,  \Big(a_x^+(-\frac12\delta(x-y)\Delta_x+\lambda v(x-y))a_y \nonumber\\
		&& + \, \frac{\lambda v(x-y)}2 (a_x^+a_y^++a_xa_y)\Big) \, , \label{def-hmf}\\
		\cHcubt &:=& \frac{\lambda}{\sqrt N}
		\int_{\Lambda^2} dx \, dy \, v(x-y)  a_x^+(a_y + a_y^+) a_y \, , \label{def-hcubz}\\
		\cHquart &:=& \frac{\lambda}{2N }
		\int_{\Lambda^2} dx \, dy \, v(x-y)  a_x^+ a_y^+ a_y a_x \, , \label{def-hquartz}
	\eeqn 
	noting that \eqref{eq-phi0-set} has been used to obtain these expressions. In particular, \eqref{eq-fluct-ham} implies that the fluctuation dynamics is determined by the unitary operator
	\eqn
		\widetilde\cU_N(t) \, :=\, W^*[\sqrt{N\vol}\phi_0]e^{-it\cH_N}W[\sqrt{N\vol}\phi_0] \, ,
	\eeqn 
	where 
	\eqn 
		 i\partial_t \widetilde\cU_N(t) \, = \, (\cHmft \, + \, \cHcubt \, + \, \cHquart)\widetilde\cU_N(t) \, .
	\eeqn 
	\par In the unitary evolution relative to the Hartree dynamics, the dynamics of thermal bosons is determined by two types of processes: 
	\begin{enumerate}
		\item Emission and absorption of thermal bosons from and into the BEC, respectively.
		\item Collisions between thermal bosons.
	\end{enumerate}	
	In particular, the Hamiltonian $\cHmft + \cHcubt + \cHquart$ describing this relative dynamics is not number conserving, as opposed to the original Hamiltonian $\cH_N$. Observe that conjugation by the Weyl operator $\cW[\sqrt{N\vol}\phi_0]$ 'subtracts' the condensate dynamics, thereby revealing the relative dynamics. As a consequence, our focus will be on the IVP
	\begin{equation} \label{eq-rel-BEC}
		\begin{cases}
			i\partial_t \rho_t^{(rel. BEC)}(A) &=\rho_t^{(rel. BEC)}([A,\cHmft + \cHcubt + \cHquart]) \, , \\
			\rho_0^{(rel. BEC)} (A)&= \frac{1}{\tr(e^{-\cK})} \tr\big( \, e^{-\cK} A  \, \big) \, .
		\end{cases}
	\end{equation}
	Notice that the initial state $\rho_0^{(rel. BEC)}$ is chosen to be particle number conserving, in contrast to $\rho_0$.
	\par Observe that $\cHmft=O(1)$, while $\cHcubt=O(\frac{\lambda}{\sqrt{N}})$ and $\cHquart=O(\frac{\lambda}N)$ are of lower order as $N\gg1$. We thus expect the leading order fluctuation dynamics to be determined by the Hartree-Fock-Bogoliubov (HFB) dynamics described by the Hamiltonian $\cHmft$. Bogoliubov \cite{bogoliubov} observed that a Hamiltonian of the form of $\cHmft$ can be diagonalized in terms of rotated creation and annihilation operators $b_x= a(u_x)+a^+(v_x)$. Based on this idea, and considering the more general case for a non-stationary, non-translation invariant condensate wave function $\phi_t$, there are many works analyzing the emergence of the HFB dynamics, including \cite{ABS,BBCFS,BBCFS18,bocca,BBCS,bocesc,BPaPS,BPePS,bopese,brcasc,brscsc,cena,chong2018,chong2021,chgrmazh,chzh,defrpipi,GM1,GM2,GM3,kuz1,kuz2,lenaseso,MPP,NN1,NN2,NN3,NN4,NNRT,namsa,namsei,pepiso} and references therein. See also \cite{lewin,N} for more details.
	\par Let $(\cVmf(t))_{t\in\R}$ denote the unitary group associated with the generator $\cHmft$, see \eqref{def-HFB} below. Then the density of HFB fluctuation particles, we have that
	\eqn
		\cVmf^*(t) \nb \cVmf(t) \, \lesssim \, \nb+\vol \, ,
	\eeqn 
	see Remark \ref{rem-HFB-density} for details. In particular, the HFB fluctuation density has the order of magnitude
	\eqn
		\frac1{\vol}\rho_0^{(rel. BEC)}\big(\cVmf^*(t) \nb \cVmf(t)\big) \lesssim 1 \, ,
	\eeqn 
	compared to the BEC density $N$, see also Lemma \ref{lem-num-mom}. 
	\par As we verify in Lemma \ref{cor-errorest}, the HFB evolution captures oscillations between absorption into and emission from the BEC with frequency 
	\eqn
		\Omega \, = \, \sqrt{E(E+2\lambda\hat{v})} \, , \label{def-bog}
	\eeqn  
	which is the Bogoliubov dispersion relation. Here $\hat{v}$ is the Fourier transform of $v$, see definition \eqref{def-fourier}, and $E(p)=|p|^2/2$ denotes the free kinetic energy. We will assume $\hat{v}\geq0$ to be non-negative. The Bogoliubov dispersion corresponds to the propagation of acoustic excitations, see, e.g., \cite{defrpipi,sei} for more details.

	\subsection{Fourier transform\label{sec-FT}}
	
	Before moving on to the next order corrections, we fix our conventions for the Fourier transform. Let the Fourier transform be given by
	\eqn \label{def-fourier}
	\hat{f}(p) \, := \, \int_\Lambda \, dx\, e^{ip\cdot x} f(x) \, ,
	\eeqn 
	for all $p\in\lattice=(\frac{2\pi}L\Z)^3$, where $\Lambda=(L\T)^3$. We denote by 
	\eqn
	a_p & := & a(e^{-ip\cdot(\cdot)}) \, = \, \int_\Lambda \, dx\, e^{ip\cdot x}a_x \, , \\
	a_p^+ &:=& a^+(e^{-ip\cdot(\cdot)}) \, = \, \int_\Lambda \, dx\, e^{-ip\cdot x}a_x^+ 
	\eeqn 
	the Fourier transforms of the operator-valued distributions $a_x$, $a_y^+$. These satisfy the discrete CCR
	\eqn 
	[a_p,a_q^+] &=& \vol\delta_{p,q} \, =: \, \kro(p-q) \, , \label{def-kro} \\
	[a_p,a_q] & = & [a_p^+,a_q^+] \, = \, 0 \, .
	\eeqn 
	When the context is clear, we will omit the subscript '$\lattice$' in $\kro$.
	\par Recalling \eqref{def-nb}, the number operator is given by
	\eqn 
		\nb  \, = \,  \int_{\lattice} dp \, a_p^+  a_p \, .
	\eeqn
    Then $a_p$, $a_q^+$ satisfy the bounds
	\eqn
	\|a_p\Psi\|_{\cF} &\leq & \|e^{-ip\cdot(\cdot)}\|_{L^2(\Lambda)}\|\nb^{\frac12}\Psi\|_{\cF} = \vol^{\frac12} \|\nb^{\frac12}\Psi\|_{\cF} \, , \label{eq-ap-bd} \\
	\|a_p^+\Psi\|_{\cF} &=& \sqrt{\|a_p\Psi\|_{\cF}^2+\vol} \, \leq \,  \vol^{\frac12} \|(\nb+1)^{\frac12}\Psi\|_{\cF} \, , \label{eq-ap+-bd}
	\eeqn 
	see, e.g., \cite{BPS2}.
	\par For convenience, we will use the notation  
	\eqn\label{eq-disc-int}
	\int_{\lattice} dp \, f(p) \equiv \frac{1}{\vol}\sum_{p\in\lattice} f(p) \, . 
	\eeqn
	In contrast, we will denote the Lebesgue integral by $\int_{\R^3} \, dq\, f(q)$. Moreover, we will denote an $n$-tuple $(q_1,q_2,\ldots,q_n)$, $n\in\N$, of vectors $q_j\in\R^d$ for some $d\in\N$, as
	\eqn
	\mathbf{q}_n \, := \, (q_1,q_2,\ldots,q_n) \, .
	\eeqn 
	
	\subsection{Lower order fluctuations: Emergence of Boltzmann dynamics}
	
	In order to study corrections to the HFB dynamics, we subtract it from the dynamics \eqref{eq-rel-BEC} relative to the BEC Hartree dynamics by conjugation with $\cVmf(t)$, the unitary group induced by $\cHmft$. The resulting relative dynamics is determined by
	\begin{equation}\label{eq-schroedinger-rel}
		\begin{cases}
			i\partial_t\nu_t(A) &= \nu_t([A,\cHcubt(t)+\cHquart(t)]) \, , \\
			\nu_0(A) &= \frac{1}{\tr(e^{-\cK})} \tr\big( e^{-\cK} A \big) \, ,
		\end{cases}
	\end{equation}
	where 
	\eqn
	\cHcubt(t) & := &  \cVmf^*(t) \cHcubt \cVmf(t)\nonumber\\
	&=& \frac{\lambda}{\sqrt{N}}\int_{(\lattice)^3} d\bp_3 \, \delta(p_1+p_2-p_3)\hat{v}(p_2)\nonumber\\
	&& \Big(e^{i(\Omega(p_1)+\Omega(p_2)-\Omega(p_3))t} a_{p_1}^+a_{p_2}^+a_{p_3} + \, h.c.\Big) \, + \, O(\frac{\lambda^2}{\sqrt{N}}) \, \label{eq-hcub-approx}\\
	\cHquart(t) & := & \cVmf^*(t)\cHquart \cVmf(t) \nonumber\\
	&=& \frac\lambda{2N} \int_{(\lattice)^4} d\bp_4 \, \delta(p_1+p_2-p_3-p_4)\hat{v}(p_1-p_3)\nonumber\\
	&& e^{i(\Omega(p_1)+\Omega(p_2)-\Omega(p_3)-\Omega(p_4))t} a_{p_1}^+a_{p_2}^+a_{p_3}a_{p_4} \, + \, O(\frac{\lambda^2}N) \, , \label{eq-hquart-approx}
	\eeqn
	see Corollary \ref{cor-errorest}. For a derivation of \eqref{eq-schroedinger-rel}, we refer to Lemma \ref{lem-FD-1}.
	\par We are interested in the evolution of the density of fluctuation particles
	\eqn
		f_t(p) \, := \, \frac{\nu_t(a_p^+a_p)}\vol \, .
	\eeqn 	
	We have that
	\eqn \label{def-f0}
		f_0(p) \, = \, \frac1{e^{K(p)}-1} \, ,
	\eeqn 
	see Remark \ref{rem-f0-computation}. The case $K(p)=\beta(E(p)-\mu)$, with inverse temperature $\beta>0$, and chemical potential $\mu<0$, corresponds to the Bose-Einstein distribution of the ideal Bose gas. 
	\par In order to study $f_t$, we need to extend \eqref{eq-schroedinger-rel} to hold for a more general class of operators. We prove in Lemma \ref{lem-num-mom} a quantitative version of
	\eqn
	\nu_0(\nb^{k}) <\infty \, , 
	\eeqn 
	and in Corollary \ref{cor-mom-prop} a quantitative version of
	\eqn
		\nu_t (\nb^k) < \infty 
	\eeqn 
	for all $k\in\N$. Thus, we may extend the IVP \eqref{eq-schroedinger-rel} to observables
	\eqn
	A \, = \, a_{p_1}^{\sharp_1}a_{p_2}^{\sharp_2}\ldots a_{p_k}^{\sharp_k} \, . \label{eq-prod-form-0}
	\eeqn 
	\par In order to expand $f_t$, we follow \cite{hu,hola,ESY} and apply Duhamel's formula three times
	\eqn
		\lefteqn{f_t(p)-f_0(p)=}\nonumber\\
		&& -\frac{i}{\vol}\int_0^t ds \, \nu_0([a_p^+a_p,\cH_I(s)]) \label{eq-transport-f}\\
		&& - \frac1\vol\int_{[0,t]^2} d\bs_2 \mathds{1}_{s_1\geq s_2} \nu_0([[a_p^+a_p,\cH_I(s_1)],\cH_I(s_2)]) \label{eq-boltzmann-f} \\
		&& + \Rem_t(p) \, , \label{eq-remainder-f} 
	\eeqn 
	where we abbreviated
	\eqn
		\cH_I(t) & :=& \cHcubt(t)+\cHquart(t) \, , \\
		\Rem_t(p) &:=&  \frac{i}\vol\int_{[0,t]^3} d\bs_3 \mathds{1}_{s_1\geq s_2\geq s_3} \nu_{s_3}([[[a_p^+a_p,\cH_I(s_1)],\cH_I(s_2)],\cH_I(s_3)]) \, ,
	\eeqn  
	and $\bs_j=(s_1,\ldots,s_j)$, $j\geq2$.
	\par 
	Because $a_p^+a_p$ commutes with $e^{-\cK} $, the transport term \eqref{eq-transport-f} vanishes. 
	Due to translation-invariance and $\nu_0$ being number conserving, the transport term \eqref{eq-transport-f} vanishes. 
	\par For \eqref{eq-boltzmann-f}, observe that $\cHcubt(t)\sim \frac{\lambda}{\sqrt{N}}$ is much larger than $\cHquart(t)\sim\frac{\lambda}N$. We thus expect the main contribution in \eqref{eq-boltzmann-f} to stem from the terms involving $\cHcubt$. 
	\par It is key to our analysis that we exploit the fact that the HFB dynamics happens on a much {\it shorter} time scale than the corrections coming from $\cHcubt$ and $\cHquart$. Observing that $\lambda$ defines the coupling strength at the level of the HFB evolution, we consider the kinetic time scale defined by $t\sim\lambda^{-2}\gg1$. In order to separate the corrections in \eqref{eq-hcub-approx} and \eqref{eq-hquart-approx} from the HFB oscillations, we choose $0<\lambda\ll 1$.
	\par Using quasifreeness of $\nu_0$, we thus expect the main contributions in \eqref{eq-boltzmann-f} to be given by
	\eqn
		\lefteqn{\frac1N \int_0^T \, dS \, Q_S^{(mol)}(f_0)(p)} \\
		&& + \, \Big|-\frac{i}\vol\int_0^{T\lambda^{-2}}\, ds\,\nu_0([a_0, \cHcubt(s)])\Big|^2 \delta(p) \label{eq-boltzmann-f-con} \, ,
	\eeqn 
	where 
	\eqn
		\lefteqn{ Q^{(mol)}_S(h)(p)}\nonumber\\
		&=& \int_{(\lattice)^3} d\bp_3 \,  \frac{\sin\Big(
			\frac{T-S}{\lambda^2}\big(\Omega(p_1)+\Omega(p_2)-\Omega(p_3)\big)\Big)}{\Omega(p_1)+\Omega(p_2)-\Omega(p_3)}\delta(p_1+p_2-p_3)\nonumber\\
		&&(\hat{v}(p_1)+\hat{v}(p_2))^2\big(\delta(p-p_1)+\delta(p-p_2)-\delta(p-p_3)\big) \nonumber\\
		&&\big((1+h(p_1))(1+ h(p_2)) h(p_3) \, - \, h(p_1) h(p_2)(1+ h(p_3))\big) \label{def-qmol}
	\eeqn 
	is a mollification of the cubic Boltzmann operator given in \eqref{BUU3}. The fact that energy conservation only holds approximately up to an error of order $\lambda^2=O(t^{-1})$ is consistent with the time-energy Heisenberg uncertainty principle, see Remark \ref{rem-approx-ec} for more details. $Q^{(mol)}_S$ describes collisions between fluctuation particles, one of which is being absorbed into or emitted from the BEC. We expect that \eqref{eq-boltzmann-f-con} is of size $\frac{T^2}{N\lambda^2}$ and that it dominates the Boltzmann collision term $\frac1N \int_0^T dS\, Q^{(mol)}_S$. It turns out that the presence of \eqref{eq-boltzmann-f-con} owes to the fact that, above, we only subtracted the leading-order condensate dynamics. Thus, in order to resolve Boltzmann dynamics collisions, we need to pass to centered moments according to $a_p\to a_p-\nu_t(a_p)$. Denoting
	\eqn
		F_T(p) &:= & \frac{\nu_{T\lambda^{-2}}\big((a_p^+-\nu_{T\lambda^{-2}}(a_p^+))(a_p-\nu_{T\lambda^{-2}}(a_p))\big)}\vol\nonumber\\
		&=& \frac{\nu_{T\lambda^{-2}}(a_p^+a_p)-|\nu_{T\lambda^{-2}}(a_p)|^2}{\vol} \, , 
	\eeqn 
	and using that $\nu_0$ is number conserving, we expect -- and will indeed prove -- to have
	\eqn
		F_T(p) - F_0(p) \, = \, \frac1N \int_0^T dS \, Q^{(mol)}_S(f_0)(p) + \Rem_{T\lambda^{-2}}(p) + l.o.t. \, , \label{eq-Ft-eff}
	\eeqn 
	where "l.o.t." abbreviates "lower order terms".
	\par Before moving on, we would like to reflect on the validity of this identity.
	
	\subsubsection{Fixed, $N$-independent lattice $\lattice\cong \Z^3$ \label{intro-sec-fixed-lattice}} 
		Recall that the fluctuation particles, at leading order, propagate with the Bogoliubov dispersion $\Omega$. These acoustic waves have the phase velocity
		\eqn
			v_P(p) \, := \, \frac{\Omega(p)}{|p|} \,= \, \sqrt{\frac{E(p)}2+\lambda\hat{v}(p)} \, .
		\eeqn 
		Averaging this over all particles yields
		\eqn
			\jb{v_P}_0 \, := \, \int_{\lattice} dp \, f_0(p) v_P(p) \sim 1 \, ,
		\eeqn 
		where we assume that $f_0$ is sufficiently regular for this argument. During the time $t\sim \lambda^{-2}$, the corresponding acoustic waves propagate a distance $\sim \lambda^{-2}$. In particular, we have 
		\begin{center}
			$\lambda^{-2} \gtrsim L$ \quad $\Leftrightarrow$ \quad  acoustic waves interfere with themselves.
		\end{center}
		Thus, when $\lambda$ is small enough, lower-order terms in \eqref{eq-hcub-approx} and \eqref{eq-hquart-approx} can constructively interfere to an extent as to contribute to leading order terms of $F$ in \eqref{eq-Ft-eff}. As we will see below, the effect of these contributions is large, depending on whether certain time-dependent expressions, coming from HFB oscillations, have particular values in $\Q$. This phenomenon is slightly reminiscent of the {\em Talbot effect} \cite{ertz,karo}. The absence of this effect has been discussed in the context of the kinetic wave equation, see, e.g., \cite{bugehash,coge19,coge20,deha19,deha21}. To the best of our knowledge, this phenomenon has not previously been discussed in the literature in the context of the quantum field theoretic emergence of Boltzmann equations.

	\subsubsection{Continuum approximation $\lattice\to \R^3$} 
	In order to elucidate the link with the expression for the BUU collision operator that is widely discussed in the literature, we present a continuum limit, that is derived here. However, we emphasize that this limit applies to the kernel itself, but not to the dynamics because sharp energy conservation cannot hold for finite times, due to the Heisenberg uncertainty. 
\par As observed above, we expect that, for $L\gg \lambda^{-2}$, the self-interactions of the HFB waves to be negligible. In fact, we show that in this limit, we can approximate sums $\int_{\lattice}dp \, \equiv\frac1\vol\sum_p$ by Lebesgue-integrals $\frac1{(2\pi)^3}\int_{\R^3}$, with control of errors that include oscillatory contributions, see Lemma \ref{lem-disc}. In this case, we establish that
	\eqn
		F_T(p) - F_0(p) \, = \, \frac{T}N Q(f_0)(p) + \Rem_{T\lambda^{-2}}(p) + O(\frac1{N^{1+\delta_N}}) \, , \label{eq-Ft-con-eff}
	\eeqn 
	where
	\eqn
		\lefteqn{Q(f_0)[J]}\nonumber\\
		&:=& \frac\pi{(2\pi)^6} \int_{\R^9} \, dp_1 \, dp_2 \, dp_3 \, \delta\big(E(p_1)+E(p_2)-E(p_3)\big)\delta(p_1+p_2-p_3)\nonumber\\
		&& (\hat{v}(p_1)+\hat{v}(p_2))^2 \big(J(p_1)+J(p_2)-J(p_3)\big)\nonumber\\
		&& \big((1+f_0(p_1))(1+f_0(p_2))f_0(p_3) \, - \, f_0(p_1)f_0(p_2)(1+f_0(p_3))\big)
	\eeqn 
	is the energy-conserving cubic Boltzmann operator found in the above mentioned literature.
	
	\subsubsection{Propagation of quasifreeness}
	If we can show that $\Rem_{T\lambda^{-2}}(p)$ is, in fact, of lower order compared to $\frac1N \int_0^T dS \, Q^{(mol)}_S$, then, by rearranging \eqref{eq-Ft-eff}, we find that
	\eqn \label{eq-boltzmann-rough}
		F_T(p) - F_0(p) \, = \, \frac1N \int_0^T dS\, Q^{(mol)}_S(F)(p) \, + \, l.o.t. \, ,
	\eeqn 
	which would prove that the next-to-leading order correction to the HFB dynamics of the particle density is described by a cubic Boltzmann equation. Comparing with the analysis of collisions for classical systems, bounding $\Rem_t(p)$ in the present context includes controlling recollisions, as is necessary in the context of classical systems. For more details on the role of recollisions in the classical case, we refer, e.g., to \cite{cerillpul}.
		
	\par While drawing comparisons to the classical case, we address the role of quasifreeness. With the given choice of a number conserving and translation-invariant initial state $\nu_0$, we have that the joint distribution function of $n$ particles satisfies
	\eqn
		f_{joint}(\bp_n) & := & \frac{\nu_0(a_{p_1}^+a_{p_2}^+\ldots a_{p_n}^+a_{p_n}\ldots a_{p_2} a_{p_1})}{\vol^n} \\
		&=& f_0^{\otimes n}(\bp_n) \, + O(\vol^{-1}) \, ,
	\eeqn 
	where $\bp_n=(p_1,\ldots,p_n)$.	In particular, we have that asymptotically, for large $L\gg1$, quasifreeness implies {\it molecular chaos} in the classical sense, which refers to factorization of the joint distribution. Propagation of the factorized form is called {\it propagation of chaos} in the classical context. 
	\par In the present quantum field theoretic context, it is an important task to understand in what sense and in which scaling regime propagation of quasifreeness can be observed. An important property of the HFB evolution is that it {\it preserves} quasifreeness. In particular, if a state $\jb{\cdot}_0$ is quasifree, then so is $\jb{\cVmf^*(t)(\cdot)\cVmf(t)}_0$. This is a natural consequence of the fact that the HFB dynamics arises as a quasifree reduction of the full evolution. In the recent literature, this is explained on the basis of the Dirac-Frenkel principle, see, for instance, \cite{bachfrsi} and \cite{besoso} for more details. 
	\par Notice that the full evolution clearly does not preserve quasifreeness, as is expected for an interacting gas. In order to study propagation of quasifreeness, it is crucial to control the Duhamel term $\Rem_t(p)$ accounting for all quantum 'recollisions'. Indeed, if we expand the evolution of arbitrary expectations
	\eqn
		\nu_{T\lambda^{-2}}(a_{p_1}^{\sharp_1}a_{p_2}^{\sharp_2}\ldots a_{p_k}^{\sharp_k}) \, = \, \nu_0(a_{p_1}^{\sharp_1}a_{p_2}^{\sharp_2}\ldots a_{p_k}^{\sharp_k}) \, + \, \Rem_{T\lambda^{-2}}(\bp_k) \, , \label{eq-prop-of-qf}
	\eeqn 
	controlling $\Rem_{T\lambda^{-2}}(\bp_k)$ implies that $\nu_{T\lambda^{-2}}$ is quasifree to $k^{th}$ order. Similar to \cite{ESY}, we do not need to propagate quasifreeness to arbitrary orders to derive a Boltzmann equation; instead, adopting their notion of {\it restricted quasifreeness}, it is sufficient to show that $\nu_{T\lambda^{-2}}$ is approximately restricted quasifree up to eight-point correlation functions, see \eqref{eq-boltzmann-f}. However, we choose not to explicitly prove such a statement, and, instead, calculate the evolution of $f$ explicitly. We leave the proof of a more general result of the form \eqref{eq-prop-of-qf} to the interested reader, which will be straightforward using the tools developed in this work.

	\par It remains to understand how we can control $\Rem_t(p)$. For that, we use the fact that $\hat{v}$ is bounded to show that
	\eqn
		\cHcubt(t) &\lesssim & \frac{\lambda}{\sqrt N}(\nb+\vol)^{\frac32} \, , \label{eq-hcub-bd}\\
		\cHquart(t) &\lesssim & \frac{\lambda}N(\nb+\vol)^2 \, , \label{eq-hquart-bd}
	\eeqn 
	see Lemmata \ref{lem-hcub-est} and \ref{lem-hquart-est}. When bounding products of non number conserving operators, one needs to take into account the growth of the particle number, see Lemma \ref{lem-prod}. This is consistent with the fact that fluctuation particles are being absorbed into and emitted from the BEC. It turns out that in order to control $\Rem_t(p)$, it suffices to only consider three, and thus a fixed number of Duhamel iterations. Hence, we apply the bounds \eqref{eq-ap-bd}, \eqref{eq-ap+-bd} on $a_p^+a_p$ with the bound \eqref{eq-hcub-bd} and \eqref{eq-hquart-bd} on $\cH_I(t)$ to obtain
	\eqn
		|\Rem_{T\lambda^{-2}}(p)| \lesssim \frac{T^3}{N^{\frac32}\lambda^3} \sup_{t\in[0,T\lambda^{-2}]}\nu_t\Big((\nb+\vol)^{\frac{11}2}\big(1 \, + \, \frac{\sqrt{\nb+\vol}}{\sqrt{N}}\big)^3\Big) \, . \label{eq-rem-bd}
	\eeqn
	In order to bound $\nu_t(\nb^k)$, we employ a result by Rodnianski and Schlein \cite{rosc}, see also \cite{busasc}, to obtain that
	\eqn
		\nu_t\big( (\nb+\vol)^{\frac\ell2}\big) \, \lesssim_{\vol,\ell} \, e^{K_\ell \lambda\vol t}  \, .
	\eeqn 
	In particular, \eqref{eq-rem-bd} then yields the constraint
	\eqn
		\frac{\vol}\lambda \lesssim \frac{\log N}{\log \log N} \, ,
	\eeqn 
	in order to suppress $\Rem_{T\lambda^{-2}}(p)$ compared to the leading order Boltzmann term in the evolution of $F$, see \eqref{eq-Ft-eff}. In particular, this implies the scaling
	\begin{enumerate}
		\item $L\sim1$ fixed: $\lambda\sim\frac{\log \log N}{\log N}$ \, ,
		\item $L\sim \lambda^{-2-}$: $\lambda\sim\Big(\frac{\log \log N}{\log N}\Big)^{\frac17-}$.
	\end{enumerate}

	\subsection{Fluctuations beyond HFB}
	Now that we understand that approximate restricted quasifreeness can be propagated, we would like to comment on the fact that the fluctuation dynamics does not preserve number conservation. We have that, for a non number conserving quasifree state $\mu$, a more general version of the Wick-Theorem holds: Let $b^{\#}:=a^{\#}-\mu(a^{\#})$. Then $\mu$ is quasifree if and only if it satisfies \eqref{eq-wick-thm} with the operators $a^{\#}$ replaced by their centered counterparts $b^{\#}$. 
	\par Since quasifreeness implies that $n-$point correlation functions $\jb{a^{\#_1}\ldots a^{\#n}}$ are determined by the one- and two-point correlation functions, and since by \eqref{eq-prop-of-qf}, $\nu_{T\lambda^{-2}}$ is approximately restricted quasifree, we need to include the dynamics of
	\eqn
		\Phi_t &:=& \frac{\nu_t(a_0)}{\vol} \, , \\
		g_t(p) &:=& \frac{\nu_t(a_pa_{-p})}{\vol} 	
	\eeqn  
	in our analysis. We have that $\Phi_t$ captures the corrections to the condensate dynamics. $g_t$ is the rate of absorption into or emission from the BEC of pairs of thermal bosons.
	\par Recall that, due to translation-invariance of the condensate, $\nu_t$ is also translation-invariant, which is why we have that $\nu_t(a_p)=\Phi_t\delta(p)$, see Lemma \ref{lem-TI}. Due to $\nu_0$ being number conserving, we have that $\Phi_0=g_0=0$. Arguing as in the case of $f$ above, we are also interested in the dynamics of the centered, mesoscopic counterparts
	\eqn
		\Psi_T &:=& \frac{\nu_{T\lambda^{-2}}(a_0)}{\vol} \, ,  \\
		G_T(p) &:=& \frac{\nu_{T\lambda^{-2}}(a_pa_{-p})-\nu_{T\lambda^{-2}}(a_{p})\nu_{T\lambda^{-2}}(a_{-p})}{\vol} \, .
	\eeqn 
	We show that
	\eqn
		\Psi_T &=& \frac{-ic_1(f_0)T}{N^{\frac12}\lambda} \, + \, O(\frac1{N^{\frac12+\delta_N}\lambda}) \, , \label{eq-condensate-rough} \\
		G_T &=& (T+T^2)O(\frac{1}{N\lambda^2}) , \label{eq-absorption-rough}
	\eeqn 
	where, assuming that $\hat{v}$ is real-valued, $c_1(f_0)$ is real-valued. At leading order, we show that the dynamics of $G$ is completely determined by $F$. As a consequence of \eqref{eq-absorption-rough}, and recalling \eqref{eq-transport-f} and \eqref{eq-boltzmann-f}, we have that $G$ merely contributes lower-order corrections to the Boltzmann dynamics for $F$.

	\subsection{Scaling of $\lambda$ and $N$} We emphasize that the parameter $N\gg1$ accounts for the $L^2$ mass of the BEC per unit volume, and is unrelated to the $\sim1$ density of fluctuation particles around the BEC. Hence $1/N$ yields a small perturbation parameter in the expansion of the full dynamics, in addition to the coupling constant $0<\lambda\ll 1$ characterizing the HFB dynamics. Bo\ss mann et al. \cite{BPePS} give an entire expansion of the fluctuation dynamics in powers of $\lambda/N$, in terms of effective Hamiltonians for a given order of precision. Our analysis differs in that we choose $\lambda\ll 1$ and $N\gg 1$ suitably in order to be able to extract effective equations for the moments $F$, $G$, $\Psi$, while keeping the error sufficiently small. In particular, our time scale is $O((\log N/\log \log N)^\alpha)$, $\alpha>0$, instead of $O(1)$. In the latter case, the Boltzmann dynamics cannot be observed.
	\par In the derivation of Boltzmann equations in classical collisional systems along the lines of Lanford's approach, $O(1)$ many collisions take place during the relevant time scale (which is inversely proportional to the mean free path). In our result, we encounter a similar situation; the error of order $O(N^{-1-})$, in \eqref{eq-boltzmann-rough} dominates after $O(1)$ collisions have taken place; this can be easily seen by iterating the Duhamel formula \eqref{eq-boltzmann-rough} twice (as the $k$-th order terms in the Duhamel expansion, of size $O(N^{-k})$, account for $k$ collisions). We also note that, compliant with a kinetic scaling regime, particle velocities do not scale in our problem.
	\par Our results are limited to the parameter regime $\lambda\sim(\log \log N/\log N)^{1/7 -}$ when $L\sim\lambda^{-2-}$, similar for $\lambda\sim \log \log N/\log N$ when $L\sim1$, and $N\gg1$. This is, in part, due to technical reasons, but we do not expect the fluctuation dynamics to remain of the form \eqref{eq-boltzmann-rough}, \eqref{eq-condensate-rough}, \eqref{eq-absorption-rough} for longer time scales, even if our approach is extended to the next order of magnitude. We expect the analogous to hold for the parameter regime $\lambda\sim\frac{\log \log N}{\log N}$ when $L\sim1$.
	
	\begin{remark}
		For works studying the perturbation expansion for a fermionic gas, we refer to \cite{babrpepitz,BNPSS2020,BNPSS2021,BNPSS2021b,BPS1,besoso,chhopa1,chhopa2,chro,chsa,chlasa,chrhana,lesa1,lesa2,nase,porasasc} and references therein.
	\end{remark} 
	
	\section{Main Results}
	
	\subsection{Notation \label{sec-notation}}
	
	We introduce the rescaled $L^a(\lattice)$ norms
	\eqn
	\|f\|_{L^a(\lattice)} & := & \vol^{-\frac1a} \|f\|_{\ell^a(\lattice)} \quad\mbox{\ if\ } 1\leq a<\infty \, , \label{def-rescaled}
	\eeqn 
	and accordingly
	\eqn
	\|f\|_{a\cap\infty,d} &:=& \|f\|_{L^a(\lattice)} \, + \, \|f\|_{\ell^\infty(\lattice)} \, , \\
	\nd{f}&:=& \|f\|_{1\cap\infty,d}  \, , \label{def-nd}\\
	\|f\|_{m,c} &:=& \sum_{n=0}^{m}\Big\|\jb{|\cdot|}^nD^{m-n} f\Big\|_{L^1(\R^3)} \quad\mbox{\ if\ } m\in\N \, . \label{def-nc}
	\eeqn
	Moreover, we introduce the weight 
	\eqn
	w(p) \, := \, 1+ \frac1{|p|^2} \, . \label{def-weight}
	\eeqn 
	Whenever $wf \in L^\infty(B_R(0)\setminus\{0\})$ for some $R>0$, we define the weighted norms
	\eqn
	\wnd{f} &:=& \nd{wf} \, ,  \label{def-wnd}\\
	\|f\|_{m,w,c} &:=& \|wf\|_{m,c} \, . 
	\eeqn
	Recall that
	\eqn 
	\cHcubt(t) &=& \frac{\lambda}{\sqrt N}\int_{(\lattice)^3}  \, d\bp_3 \, \hat{v}(p_2)\delta(p_1+p_2-p_3) \nonumber\\ 
	&& 
	e^{it\cHmft}\big(a_{p_1}^+a_{p_2}^+a_{p_3}+h.c.\big)e^{-it\cHmft} \, , \label{def-hcubt} \\
	\cHquart(t) &=&\frac{\lambda}{2N}\int_{(\lattice)^3} \, d\bp_3 \, \hat{v}(p_1-p_3)\delta(p_1+p_2-p_3-p_4)\nonumber\\
	&& e^{it\cHmft}a_{p_1}^+a_{p_2}^+a_{p_3}a_{p_4} e^{-it\cHmft} \, , \label{def-hquart} 
	\eeqn 
	where
	\eqn
	\cHmft &=& \int_\lattice dp \, \big(E(p)+\lambda\hat{v}(p)\big)a_p^+a_p \, + \, \frac\lambda2
	\int_\lattice \, dp \,  \hat{v}(p)  \, \big(  a_p^+ a_{-p}^+ \, +  \, a_p a_{-p}  \big) \, , \;\;\;
	\eeqn 
	$E(p)=\frac{|p|^2}2$, and $\int_{\lattice}\equiv\frac1\vol\sum_{\lattice}$. In this context, we also recall the Bogoliubov dispersion relation
	\eqn
		\Omega(p) \, = \, \sqrt{E(p)(E(p)+2\lambda\hat{v}(p))} \, .
	\eeqn 
	\par We are interested in the evolution of the correlation functions
	\eqn
	\Psi_T &=& \frac{\nu_{T\lambda^{-2}}(a_0)}{\vol} \, , \\
	F_T(p) &= & \frac{\nu_{T\lambda^{-2}}(a_p^+a_p)-|\nu_{T\lambda^{-2}}(a_p)|^2}{\vol} \, , \\
	G_T(p) &=& \frac{\nu_{T\lambda^{-2}}(a_pa_{-p})-\nu_{T\lambda^{-2}}(a_{p})\nu_{T\lambda^{-2}}(a_{-p})}{\vol} \, .
	\eeqn 
	\par As explained in \ref{intro-sec-fixed-lattice}, there are additional dominant terms in the Boltzmann collision terms for $F$, and also $G$ in the case of $L\sim1$ fixed. Thus, we introduce the collision operators
	\eqn
	\cQ_{d,G;T,\lambda}(h)[J] & := & \frac1{\lambda^2}\int_{[0,T]^2} \mathds{1}_{S_1\geq S_2} d\bS_2 \, \bbgd(h_{\cdot\lambda^2})[J](\bS_2/\lambda^2) \, , \label{def-qdG} \\
	q^{(j)}_{d,F;\bS_2,\lambda}(H_{S_2})[J] &:=& \bbf^{(j)}(H_{\cdot \lambda^2})[J](\bS_2/\lambda^2) \, , \quad j\in\{1,2\} \, , 
	\eeqn
	and the pair absorption operator
	\eqn
	\cA_{d;T,\lambda}(h)[J] & := & \int_0^T \, dS \, \acqd(h_{\cdot\lambda^2})[J](S/\lambda^2) \nonumber\\ 
	&&+ \, \frac{1}{\lambda^2}\int_{[0,T]^2} \mathds{1}_{S_1\geq S_2} d\bS_2 \acd(h_{\cdot\lambda^2})[J](\bS_2/\lambda^2) \, , \label{def-cBgdf}
	\eeqn 
	for any test function $J$, where $\bS_2=(S_1,S_2)$. The expressions for $\bbf^{(j)}$, $j\in\{1,2\}$, $\bbgd$, $\acqd$, and $\acd$ are lengthy, and we refer the reader to Section \ref{sec-fixed-notation} for their definition. In the case $L\sim \lambda^{-2-}$, we also define the continuous counterparts $\cQ_{c,G;T,\lambda}$, $\cA_{c;T,\lambda}$ by replacing sums $\int_{\lattice}dp$ over momenta by Lebesgue integrals $\frac1{(2\pi)^3}\int_{\R^3}$. 
	\par The subscript '$d$' in the notation refers to the fact that momentum is summed over the discrete set $\lattice$; '$c$' on the other hand refers to the continuum approximation.  Henceforth, we refer to the case with $L\sim 1$ fixed as the 'discrete case', and $L\sim\lambda^{-2-}$ as the 'continuum approximation'.
	\par $\cQ_{d,G;T,\lambda}(h)$ resp.$\cQ_{c,G;T,\lambda}(h)$ is a Boltzmann collision operator for $G$, and it is, after cancellations, quadratic in $h$, and $\cA_{d;T,\lambda}(h)$ resp. $\cA_{c;T,\lambda}(h)$ corresponds to the leading order of the expected rate of absorption of a pair of fluctuation bosons into the BEC, and it is linear in $h$. 
	\par We denote the mollified cubic Boltzmann operator in the evolution of $F$ by
	\eqn
		\lefteqn{Q_{d;T-S,\lambda}(F_S)[J]}\nonumber\\
		&:=& \int_{(\lattice)^3} d\bp_3 \frac{\sin\Big(\frac{\Omega(p_1)+\Omega(p_2)-\Omega(p_3)}{\lambda^2}(T-S)\Big)}{\Omega(p_1)+\Omega(p_2)-\Omega(p_3)} \delta(p_1+p_2-p_3)\nonumber\\
		&&(\hat{v}(p_1)+\hat{v}(p_2))^2\big(J(p_1)+J(p_2)-J(p_3)\big) \nonumber\\
		&&\big((1+F_S(p_1))(1+F_S(p_2)) F_S(p_3) \,  - \, F_S(p_1) F_S(p_2)(1+F_S(p_3))\big) \, .\label{def-Qd}
	\eeqn 
	Again, in the continuum approximation, $Q_{c;T-S,\lambda}$ is defined by replacing the lattice sum $\int_{\lattice}$ by the Lebesgue integral $\frac1{(2\pi)^3}\int_{\R^3}$.
	\par In addition, $q^{(j)}_{d,F;\bS_2,\lambda}$ are higher order Boltzmann type collision terms for the equations governing $F_T$, where $j$ accounts for the order $\lambda^j$ from which they are derived. As explained above, they are of lower order when $L\sim \lambda^{-2-}$.
	
	\subsection{Assumptions\label{sec-assumptions}}
	
	We summarize all the assumptions described in the previous section. We also add the following restrictions required in our results. 
	\begin{enumerate}
		\item \eqn\label{def-nu0}
		\nu_0(A) = \frac{1}{\tr{e^{-\cK}}} \tr\Big( \, e^{-\cK} A   \Big) 
		\eeqn
		for all observables $A$, is a quasifree, translation-invariant state that is number conserving, with
		\eqn
			\cK &=& \int_{\lattice}dp \, K(p) a_p^+a_p \, , \\
			f_0(p) &=& \frac{\nu_0(a_p^+a_p)}{\vol} \, = \, \frac1{e^{K(p)}-1} \, ,
		\eeqn 
		and $K(p)\geq \kappa_0>0$.
		\item $\nu_t$ satisfies 
		\begin{equation}\label{LVN}
			\begin{cases}
				i\partial_t\nu_t(A) &= \nu_t([A,\cHcubt(t)+\cHquart(t)]) \, , \\
				\nu_0(A) &= \frac{1}{\tr(e^{-\cK})} \tr\big( e^{-\cK} A \big) 
			\end{cases}
		\end{equation}
		for all observables $A$.
		\item The Fourier transform $\hat{v}$ of $v$, see Section \ref{sec-FT}, is a non-negative, radial function.
		\item If $L\sim 1$ is fixed, assume $\vd,\fd<\infty$.
		\item If $L\sim \lambda^{-2-}$, assume $\vc,\fc<\infty$.  
	\end{enumerate}
	In either case, we assume that $\hat{v}$ satisfies 
	\eqn
		\int_\Lambda dx \, v(x) \, = \, \hat{v}(0) \, = 0 \, . \label{ass-v-0}
	\eeqn 
	Moreover, this implies that the leading order condensate wave function $\phi_0\equiv\vol^{-1/2}$ can be chosen to be a stationary, translation-invariant solution of the Hartree equation
	\eqn
		i\partial_t \phi_t \, = \, -\Delta \phi_t + \lambda\vol v*|\phi_t|^2 \phi_t \, .
	\eeqn 
	As described above, $N$ denotes the BEC density, and $\lambda>0$ is an (additional) coupling constant, defining the HFB coupling, see \eqref{def-hmf}.
	\par We are now ready to formulate our main results.

	\subsection{Statement of results} 
	
	\begin{theorem}[Discrete case]\label{thm-main-dis}
		Let $T>0$, $L\geq1$, and let $N>0$ denote the BEC density. Choose
		\eqn
			\lambda \, = \,  \frac{\log\log N}{\log N} \, .
		\eeqn 
		Then, under the assumptions stated in Section \ref{sec-assumptions} and with the notations in Section \ref{sec-notation}, there exist constants
		\eqn
			C_0 & = & C_0(\vd,\fd,\vol,T) \, , \\
			N_0 &=& N_0(\vd,\vol,T) \, , 
		\eeqn 
		such that, for all $N\geq N_0$ we have that
		\eqn
		\rate_1 \, := \,  \frac{\log \frac1\lambda}{\log N} >0 \, , \quad \rate_2 \, := \, \frac12-\frac{4\log\frac1\lambda}{\log N} >0  \, , 
		\eeqn 
		and that
		\begin{fleqn}[\parindent]
			\eqn
			\lefteqn{\Big|\Psi_T +\frac{i }{N^{\frac12}\lambda} \int_0^T dS \, \int_{\lattice} dp\, \hat{v}(p)F_S(p)\Big|}\nonumber\\
			&& \hspace{2cm} \leq \, \frac{C_0}{N^{\frac12+\rate_1}\lambda} \, , \label{eq-mainthm-dis-1}\\
			\lefteqn{\Big|\int_\lattice dp \, \big(F_T(p)-F_0(p)\big)J(p)\, - \, \frac1{N}\Big(\int_0^T dS \,  Q_{d;T-S,\lambda}(F_S)[J]}\nonumber\\
			&&+ \,  \int_{[0,T]^2} d\bS_2 \, \mathds{1}_{s_1\geq s_2} \sum_{j=1}^2\lambda^j q^{(j)}_{d,F;\bS_2,\lambda}(F_{S_2})[J]\Big)\Big|\nonumber\\
			&& \hspace{2cm} \leq \, \frac{C_0\|J\|_{\ell^\infty(\lattice)}}{N^{1+\rate_1}} \, , \label{eq-mainthm-dis-2}\\
			\lefteqn{\Big| \int_\lattice dp\, G_T(p)J(p) \, - \, \frac{1}N\Big(\cA_{d;T,\lambda}(F)[J] \, + \, \cQ_{d,G;T,\lambda}(F)[J]\Big)\Big|} \nonumber\\
			&& \hspace{2cm} \leq \, \frac{C_0\Jtn}{N^{1+\rate_2}} \label{eq-mainthm-dis-3}
			\eeqn 
		\end{fleqn}
		for all test functions $J$. 
		The error terms on the right-hand sides of \eqref{eq-mainthm-dis-1}--\eqref{eq-mainthm-dis-3} are subleading in $N$ with respect to the main terms
		appearing on the respective left-hand sides.
	\end{theorem}

	\begin{theorem}[Continuum approximation]\label{thm-main-con}
		Let $T>0$, $\reg>6$, $\veps>0$, and let $N>0$ denote the BEC density. Fix 
		\eqn
			\lambda &=& \Big(\frac{\log \log N}{\log N}\Big)^{\frac{\reg}{(7+\veps)\reg+6}} \, , \\
			L&=& \lambda^{-2-\frac2r-\veps} \, .
		\eeqn 
		Then, under the assumptions stated in Section \ref{sec-assumptions} and with the notations in Section \ref{sec-notation}, there exist constants
		\eqn
		C_0 & = & C_0(\vc,\fc,\reg,\veps,T) \, , \\
		C_1 & =& C_1(\reg,\veps) \, , \\
		N_0 &=& N_0(\vc,\reg,\veps,T) \, ,
		\eeqn 
		such that, for all $N\geq N_0$, and for
		\eqn
		\rate \, := \, \frac{C_1\log\frac1\lambda }{\log N} >0 \, , 
		\eeqn 
		we have that
		\begin{fleqn}[\parindent]
			\eqn
			\lefteqn{\Big|\Psi_T +\frac{i }{(2\pi)^3N^{\frac12}\lambda} \int_0^T \, dS \, \int_{\R^3} dp\, \hat{v}(p)F_S(p) \Big|}\nonumber\\
			&& \hspace{2cm} \leq \, \,\frac{C_0}{N^{\frac12+\rate}\lambda}  \, , \label{eq-mainthm-con-1}\\ 
			\lefteqn{\Big|\int_\lattice dp \, \big(F_T(p)-F_0(p)\big)J(p)-  \frac1{N}\int_0^T \, dS \,Q_{c;T-S,\lambda}(F_S)\Big|} \nonumber\\
			&& \hspace{2cm} \leq \, \frac{C_0\|J\|_{W^{2\floor{\frac{\reg}2}+2,\infty}}}{N^{1+\rate}} \, , \label{eq-mainthm-con-2}\\
			\lefteqn{\Big| \int_\lattice dp \, G_T(p)J(p) \, - \, \frac{1}N\Big(\cA_{c;T,\lambda}(F)[J] \, + \, \cQ_{c,G;T,\lambda}(F)[J]\Big)\Big|} \nonumber\\
			&& \hspace{2cm} \leq \, \frac{C_0\Jcn}{N^{1+ \rate}} \label{eq-mainthm-con-3}
			\eeqn 
		\end{fleqn}
		for all test functions $J$. 
		The error terms on the right-hand sides of \eqref{eq-mainthm-con-1}--\eqref{eq-mainthm-con-2}  are subleading in $N$ with respect to the main terms
		appearing on the respective left-hand sides.
		The main order term in the evolution of $F$ is given by $\frac{T}NQ(f_0)[J]$, where
		\eqn
		\lefteqn{Q(f_0)[J]}\nonumber\\
		&:=& \frac\pi{(2\pi)^6} \int_{\R^9} \, dp_1 \, dp_2 \, dp_3 \,	 \delta\big(E(p_1)+E(p_2)-E(p_3)\big)\delta(p_1+p_2-p_3)\nonumber\\
		&& (\hat{v}(p_1)+\hat{v}(p_2))^2 \big(J(p_1)+J(p_2)-J(p_3)\big)\nonumber\\
		&& \big((1+f_0(p_1))(1+f_0(p_2))f_0(p_3) \, - \, f_0(p_1)f_0(p_2)(1+f_0(p_3))\big)  \label{eq-Q-literature}
		\eeqn 
		denotes the (energy conserving) quantum Boltzmann collision operator.
	\end{theorem}
	
	\begin{remark}
		Theorem \ref{thm-main-con} is {\normalfont not} the continuum limit for the dynamics. Instead, it quantifies how the collision operator can be approximated by its continuous counterpart. In particular, energy conservation cannot hold precisely for finite times, in consistence with the Heisenberg uncertainty principle, see Remark \ref{rem-approx-ec}.
	\end{remark}

	\begin{remark}
		In this work, we are not attempting to analyze $G_T$ in more detail beyond \eqref{eq-mainthm-con-3}. We expect a more detailed analysis to yield 
		\eqn
		\cA_{c;T,\lambda}(F)[J] \, + \, \cQ_{c,G;T,\lambda}(F)[J] \sim  T + T^2 \, ,
		\eeqn 
		based on similar arguments as we present to control the Boltzmann collision term for $F$.
	\end{remark}

	\begin{remark}\label{rem-approx-ec}
		We point out that $Q_{d;T-S,\lambda}$ resp. $Q_{c;T-S,\lambda}$ contain the Bogoliubov dispersion $\Omega$ for sound waves propagating as fluctuations around the BEC. As stated in Theorem \ref{thm-main-con}, the collision operator $Q$ emerges in the limit $\lambda\searrow0$. In the limit $\lambda\searrow0$, we have that $\Omega(p)\to E(p)$, whereby we retrieve the Dirac-$\delta$ on the hypersurface $\{(p_1,p_2)\in\R^6\mid E(p_1)+E(p_2)=E(p_1+p_2)\}$ of energy conservation. Let 
		\eqn
		\delta_\vep(x) \, := \, \frac1{\vep\pi} \frac{\sin^2(x/\vep)}{(x/\vep)^2} \, = \, \frac\vep\pi \frac{\sin^2(x/\vep)}{x^2} \, ,
		\eeqn 
		which defines a Dirac sequence, $\int_\R dx\, \delta_\vep(x)=1$. Observing that
		\eqn
		\frac{\sin\Big(\dOcub\frac{T-S}{\lambda^2}\Big)}{\dOcub} \, = \, -\partial_S\Big((T-S)\delta_{\frac{2\lambda^2}{T-S}}(\dOcub)\Big) \, ,
		\eeqn 
		where $\dOcub(\bp_2)=\Omega(p_1)+\Omega(p_2)-\Omega(p_1+p_2)$, this implies that the mollified quantum Boltzmann collision terms in \eqref{eq-mainthm-dis-2} and \eqref{eq-mainthm-con-2} have the form 
		\eqn
		\lefteqn{\int_0^T dS \, Q_{c/d;T-S,\lambda}(F_S)[J]}\nonumber\\
		&=&\int_0^T dS\, \int_{\cD} d\bp_3 \, \frac{\sin\Big(\dOcub(\bp_2)\frac{T-S}{\lambda^2}\Big)}{\dOcub(\bp_2)} H_S(\bp_3)\nonumber\\
		&=& \pi T\int_{\cD} d\bp_3 \,  \delta_{\frac{2\lambda^2}{T}}(\dOcub(\bp_2)) H_0(\bp_3;J) \nonumber\\
		&& + \, \pi\int_0^T dS\,\int_{\cD} d\bp_3\,  \delta_{\frac{2\lambda^2}{T-S}}(\dOcub(\bp_2)) (T-S) \partial_S H_S(\bp_3;J) \, ,
		\eeqn 
		where $\bp_3=(p_1,p_2,p_3)$ and $\cD=(\lattice)^3$ in the discrete case, and $\cD=\R^9$ for the continuum approximation. Here, 
		\eqn
		\lefteqn{H_S(\bp_3;J)}\nonumber\\
		&:=& |\hat{v}(p_1)+\hat{v}(p_2)|^2\big(J(p_1)+J(p_2)-J(p_3)\big) \delta(p_1+p_2-p_3) \nonumber\\
		&&\big((1+F_S(p_1))(1+F_S(p_2)) F_S(p_3) \, - \, F_S(p_1) F_S(p_2)(1+F_S(p_3))\big) \, ,
		\eeqn 
		where, in the continuum approximation, $H$ contains an additional factor $1/(2\pi)^6$. In particular, this representation makes manifest that the mollification of the quantum Boltzmann collision operator corresponds to approximate energy conservation up to an error of order $O(t^{-1})=O(\lambda^2/T)$, in compliance with the time-energy Heisenberg uncertainty principle.
	\end{remark}

	\begin{remark}
		$Q$ in \eqref{eq-Q-literature} can be evaluated via the Coarea Formula, yielding
		\eqn 
		\lefteqn{Q(f_0)[J]}\nonumber\\
		&=& \frac\pi{(2\pi)^6} \int_{E(p_1)+E(p_2)=E(p_1+p_2)} \, \frac{d\cH^5(\bp_2)}{|\bp_2|} \, (\hat{v}(p_1)+\hat{v}(p_2))^2 \nonumber\\
		&&\big(J(p_1)+J(p_2)-J(p_1+p_2)\big)\nonumber\\
		&& \big((1+f_0(p_1))(1+f_0(p_2))f_0(p_1+p_2) \, - \, f_0(p_1)f_0(p_2)(1+f_0(p_1+p_2)) \big) \, , \label{eq-Q-hausdorff}
		\eeqn 
		where $d\cH^5$ is the induced Hausdorff measure on the hypersurface $\{(p_1,p_2)\in\R^6\mid E(p_1)+E(p_2)=E(p_1+p_2)\}$. 
	\end{remark}

	\begin{remark}
		Theorem \ref{thm-main-con} implies that, for high regularity $\reg\gg6$, for the maximal time scale $t_{max}\sim \lambda^{-2}\sim (\log N/\log\log N)^{\frac27-}$ and length $L\sim \lambda^{-2-}\ll t_{max}$, we obtain that 
		\eqn
		\Psi_T &\sim& \frac{T}{N^{\frac12-}} \, + \, O(N^{-\frac12}) \, ,\\
		F_T \, - \, F_0&\sim& \frac{T}{N} \, + \, O(N^{-1-})
		\eeqn 
		for $N\gg1$, where $F_0\sim1$. We point out that one of the difficulties in extracting a quantum Boltzmann dynamics for $F$ stems from the fact that $F$ is centered with $|\Psi|^2$, which is at least an order $O(N^{0+})$ larger than the Boltzmann collision term for $F$ itself.
		\par There are five characteristic length scales involved in our derivation:
		\begin{enumerate}
			\item a BEC with large density $N$,
			\item thermal fluctuations with density $\sim1$, 
			\item the HFB coupling of size $\lambda$,
			\item the linear system size $L$,
			\item the time scale $t$.
		\end{enumerate}
		A major difficulty that is overcome in this work is to identify a parameter regime which allows the Boltzmann dynamics to dominate over error terms.
	\end{remark}
	
	\begin{remark}
		The collision term for $G$ is a functional only of $F$, due to our choice of initial data with $G_0=0$. Therefore, solving the Boltzmann equation for $F$, and substituting into 
		\eqn 
		\cI_G(F)[J] \,:= \, \frac{1}N\cA_{c;T,\lambda}(F)[J] +\frac{1}{N\lambda^2} \cQ_{c,G;T,\lambda}(F)[J] \, , 
		\eeqn 
		integration time, yields $G_T$. A key reason for which the extraction of the Boltzmann equation for $F$ is a difficult problem, is the fact that we expect $\cI_G(F)[J]$ to be of the same order of magnitude as the collision operator for $F$, 
		\eqn 
		\cI_F(F)[J] \, := \, \frac1N\int_0^T  \, dS \,  Q_{c;T-S,\lambda}(F_S)[J] \, .
		\eeqn 
		In our case, $\cI_G(F)[J]$ does not depend on $G$ because the initial state $\nu_0$ is chosen to conserve the particle number $\nb$. The explicit expression for $\cI_G(F)$ is somewhat lengthy and not sufficiently enlightening to be presented here. We obtain a closed system of equations for $(\Psi,F,G)$ because of the approximate persistence of (restricted) quasifreeness of $\nu_0$ in the scaling regime of this problem. 
		\par If we expand the dynamics of $(\Psi,F,G)$ to lower orders, we expect $\cI_F(\Psi,F,G)$ and $\cI_G(\Psi,F,G)$ to be coupled non-trivially.
	\end{remark}
	
	\begin{remark}
	Our purpose of introducing a condensate of large density $N$ is due to the fact that its subleading order interactions with the fluctuation field are of Boltzmann type (the leading order is determined by the HFB dynamics); the latter are not drowned out by the error term, due to the largeness of $N$.
	On the other hand, a quartic Boltzmann collision term emerges, as expected, from our analysis, but it is a lower order term that is buried in the error terms because it does not couple to the condensate.
		We refer to Proposition \ref{prop-tail-est} for more details. 
	We do expect an analysis to the next subleading order, conjecturally involving appropriate quantum corrections to the BEC and HFB dynamics, to reveal this fourth order collision term of Boltzmann-Uhlenbeck-Uehling type, separated from the error term. 
	\end{remark}
	
	\begin{remark}
		 If it is assumed that the BEC is non-stationary and that both the BEC and the thermal fluctuation field are not translationally invariant, the first term in the Duhamel expansion for $\nu_t(a_p^+a_q)$, expressed with a single time integral, does not vanish, contrary to the situation considered in the paper at hand. This additional term couples nontrivially to both the HFB and the Boltzmann dynamics, and will lead to a different system of PDEs than the one derived here. We expect the rigorous analysis of this system to be considerably more involved, and leave it for future work. 
	\end{remark}

	\subsection{Sketch of the proof}
	The strategy of our proof is to start by calculating one collision, corresponding to calculating the Duhamel expansion to second order in the coupling $\lambda$. The tail in the Duhamel expansion corresponds to recollisions. Using bounds available in the second-quantization formulation, we are able to bound this tail. For that, it is crucial to exploit the fact that the initial condensate density $N$ provides a large perturbation parameter. Accordingly, the resulting leading order cubic Boltzmann collision operator appears with an overall factor $1/N$. This allows us to compare the density of thermal fluctuation particles for positive times with the initial density. We can observe that the superposition of HFB oscillations results in corrections to the leading order $2\leftrightarrow1$ processes coming from the collision of two thermal particles and one of those being absorbed into or emitted from the BEC. In addition, if we consider the continuous approximation, we show that the collision operator can be approximated by its continuous counterpart. 
	\par We start by applying Duhamel's principle to \eqref{LVN}, using one recursion for $\Phi$, and two for $f$ and $g$. The main term is recovered by evaluating the terms involving $\nu_0$. In Section \ref{sec-stat-phys}, in order to control higher-order terms involving the full dynamics $\nu_t$, we establish uniform-in-time bounds on $f[J]$, $g[J] $, $\cHcubt(t)$, and $\cHquart(t)$ with respect to the number operator $\nb$. We use an a-priori bound on the growth of $\nu_t(\nb^k)$ established in \cite{busasc,rosc}. This will yield the restriction $\vol/\lambda=O(\log N/\log\log N)$. Moreover, we show closeness of $\nu_t$ to $\nu_0$ in a suitable sense, which allows us to exploit that $\nu_0$ is approximately quasifree. 
	\par In Section \ref{sec-expansion}, we control the tail term in the Duhamel expansion using the previously established operator bounds and quantify propagation of moments of the number operator, using \cite{busasc,rosc}. The terms in the Duhamel expansion are expressed by way of $a(t)=\cVmf^*(t)a\cVmf(t)$ and $a^+(t)$. $\cHmft$ is quadratic in $a$ and $a^+$, thus $a(t)$ and $a^+(t)$ are linear combinations of $a$ and $a^+$. That enables us to control the proximity of $\nu_t$ to $\nu_0$, which is quasifree, in order to evaluate the main terms in the Duhamel expansion by means of Wick's Theorem. 
	\par In the discrete case, we observe interference phenomena in our scaling regime $L\leq t\sim\lambda^{-2}$. The Boltzmann collision terms are lattice sums in momentum space, and their magnitudes depend on whether terms that vary with time have particular values in $\Q$. This effect becomes negligible for box length $L\sim \lambda^{-2-}\gg t$ and $\lambda=\lambda(N)>0$ chosen small enough. The latter means that we will ultimately set $\lambda = \lambda(N)$, and choose $N$ large enough. 
	\par We prove a discretization Lemma \ref{lem-disc} that allows us to improve the rate of convergence dependent on the regularity of $f_0$ and $\hat{v}$, beyond the trivial bound. As we noticed after completing this work, our approach here appears to be related to the numerical error estimates for the trapezoid rule in numerical mathematics via Poisson summation, see \cite{TW}.
	\par We use the Duhamel expansion and the approximate quasifreeness of $\nu_t$ to relate $f_0$ back to $f_s$. In order to control the large magnitude of condensate terms in the expansion, we need to rewrite the equations for the centered expectations $f_s(p)-|\Phi_s|^2\delta(p)$ and $g_s(p)-\Phi_s^2\delta(p)$. 
	\par In Section \ref{sec-main2}, we collect all results to compute the main order terms in Theorem \ref{thm-main-dis} and \ref{thm-main-con}, and in section \ref{sec-main1}, we compute the effective equations.

	\begin{remark}
		We note that the Boltzmann-type equations in Theorem \ref{thm-main-dis} and \ref{thm-main-con} are presented in their integral and weak forms. No smallness assumption on $f$ is necessary for our result to hold. In this work, we will not further investigate questions regarding the well-posedness of the corresponding Cauchy problem of kinetic equations in the context of nonlinear PDE. Some works in this direction are referenced in Section \ref{sec-wellposed}, 
		see in particular \cite{arno15,arno17,esve,esve2014,esve2015}.

	\end{remark}
	
	\section{Preliminaries}
	
	Some of our estimates will be formulated for finite number subspaces of $\cF$. We introduce the projectors
	\eqn
	P_n \, := \, P_{\cF_n} \, .
	\eeqn
	We will consider $\cF_n$ embedded into $\cF$ and we identify $P_n$ as maps $\cF\to\cF$. By the spectral theorem,
	\eqn
	P_n \, = \, \frac1{2\pi i} \oint_{\partial B_{1/2}(n)} \frac{dz}{\nb-z} \, . \label{eq-pn-spectral}
	\eeqn 
	Observe that $a_x\mid_{F_n}$ defines a map
	\eqn
		a_x \, : \, F_n \to F_{n-1} \label{eq-annihilation-fn}
	\eeqn 
	for all $n\in\N$, with formal adjoint $a_x^+\mid_{F_{n-1}}:F_{n-1}\to F_n$. To study the weak formulation of the effective equations, we introduce
	\eqn\label{def-fjgj}
	f[J]   & := & \int \, dp \, J(p) a_p^+a_p \, ,\nonumber\\
	g[J]  & := & \int \, dp \, J(p) a_{-p}a_p \, , \\
	g^*[J] & := & g[J]^+ \, = \, \int \, dp \, J(p) a_p^+a_{-p}^+ \, . \nonumber
	\eeqn 
	As a convenient notation for iterating Duhamel's formula, let 
	\eqn
	\Delta(t,j):=\{\mathbf{s}_j\in [0,t]^j\mid s_1\geq\ldots\geq s_j\}
	\eeqn 
	be a $j$-simplex, where $t>0$ and $j\in\N$. We also recall that
	\eqn
	\cH_I(t) &=& \cHcubt(t) \, + \, \cHquart(t) \, . \label{def-HI}
	\eeqn 
	Let $(\cVmf(t))_{t\in \R}$ be the solution of
	\begin{equation}\label{def-HFB}
		\begin{cases}
			i\partial_t \cVmf(t) & = \cHmft \cVmf(t) \, , \\
			\cVmf(0) & = \1 \, .
		\end{cases} 
	\end{equation}
	
	We denote the $d-1$-dimensional Hausdorff measure of a smooth, embedded hypersurface in $\R^d$ by $d\cH^{d-1}$.
	\par If needed, we will keep track of the dependence of constants on parameters by adding the respective parameters in a subscript. Whenever the constants have no explicit dependence, they are assumed to be universal.
	\par The following result is a direct consequence from iterating \eqref{LVN}.
	
	\begin{lemma}
		\label{lem-duham}
		Assume that $A\in\alg$ is an observable, and
		that $\nu_t$ is defined as in \eqref{LVN}.
		Then, for any $k\in\N$, we have that
		\eqn\label{eq-Duham-main-tder-2} 
		\nu_t( A )
		&=&\sum_{\ell=0}^{k-1} (-i)^\ell\int_{\Delta[t,\ell]} \, d\bs_\ell \,
		\nu_0(
		[[\cdots[A,\cH_I(s_{1})],\cdots], \cH_I(s_{\ell})]) \nonumber\\
		&&+\, (-i)^{k}\int_{\Delta[t,k]} \, d\bs_k \,
		\nu_{s_k}\big(
		[[\cdots[A,\cH_I(s_{1})],\cdots],\cH_I(s_k)]\big) \, .
		\eeqn
	\end{lemma}
	
	We will be particularly interested in the cases $A\in\{a_0,f[J],g[J] \}$. To study the expansion, we derive the following useful identity.
	
	\begin{lemma}\label{lem-apcV-diag-1}
		Let 
		\eqn
		\Omega(p) := \sqrt{ E(p)(E(p)+2 \lambda\widehat v(p))}
		\eeqn
		denote the dispersion function for acoustic excitations, and let
		\eqn
		a_p^+(t)& :=& \cVmf^*(t) a_p^+ \cVmf(t) \, , \label{eq-apt-exp-1}\\
		a_p(t)& =& \cVmf^*(t) a_p \cVmf(t) \, . \label{eq-apt-exp-2}
		\eeqn
		Moreover, let
		\eqn
		\cM(p) :=
		\begin{pmatrix}-E(p)-\lambda\widehat v(p) & -\lambda \widehat v(p) \\
			 \lambda\widehat v(p) & E(p) +\lambda\widehat v(p) 
		 \end{pmatrix}
		 \,.
		\eeqn
		Then,
		\eqn\label{eq-creannih-t-expl-1}
		\begin{pmatrix}
			a_p^+(t)\\a_{-p}(t)
		\end{pmatrix}
		&=&
		\Big[\cos(t\Omega(p)) \, \1 - i \frac{\sin(t\Omega(p))}{\Omega(p)}\cM(p)\Big] 
		\begin{pmatrix}
			a_p^+\\a_{-p}
		\end{pmatrix}
		\,.
		\eeqn
	\end{lemma}
	
	\prf
	We obtain the following system of ODEs,
	\eqn
	i\partial_t a_p^+(t) &=&\cVmf^*(t) \, [a_p^+,\cHmfz] \, \cVmf(t)
	\nonumber\\
	&=& -(E(p)+\lambda\widehat v(p)) a^+_p(t) - \lambda\widehat v(p) a_{-p}(t)\, , \\
	i\partial_t a_{-p}(t) &=&\cVmf^*(t) \, [a_{-p},\cHmfz] \, \cVmf(t)
	\nonumber\\
	&=& (E(p)+\lambda\widehat v(p)) a_{-p}(t) + \lambda\widehat v(p) a_p^+(t)\, ,
	\eeqn
	where $E(p)=\frac{p^2}{2}$, so that
	\eqn
	i\partial_t 
	\begin{pmatrix}
		a_p^+(t)\\a_{-p}(t)
	\end{pmatrix}
	=
	\cM(p)
	\begin{pmatrix}
		a_p^+(t)\\a_{-p}(t)
	\end{pmatrix}\,
	\eeqn
	with
	\eqn
	\cM(p) := 
	\begin{pmatrix}
		-E(p)-\lambda\widehat v(p) & -\lambda \widehat v(p) \\
		\lambda\widehat v(p) & E(p) +\lambda\widehat v(p) 
	\end{pmatrix} 
	\,.
	\eeqn
	We observe that $\cM(p)$ satisfies
	\eqn
	\cM^2(p) = \Omega^2(p) \, \1 \,,
	\eeqn
	where
	\eqn
	\Omega(p) := \sqrt{ E(p)(E(p)+2 \lambda\widehat v(p))} \label{eq-omega-def-0}
	\eeqn
	denotes the dispersion function  of acoustic excitations; see, for instance, \cite{defrpipi,sei}. In particular, we have $\cM^{-1}(p)=\frac{1}{\Omega^2(p)}\cM(p)$.
	
	Hence,
	\eqn\label{eq-apt-exp-3}
	\begin{pmatrix}
		a_p^+(t)\\a_{-p}(t)
	\end{pmatrix}
	&=&
	\exp\left(-it
	\begin{pmatrix}
		-E(p)-\lambda\widehat v(p) & -\lambda \widehat v(p) \\
		 \lambda\widehat v(p) &  E(p) +\lambda\widehat v(p) 
	 \end{pmatrix}
	\right)
	\begin{pmatrix}
		a_p^+\\a_{-p}
	\end{pmatrix}
	\nonumber\\
	&=&\Big(\cos(t\Omega(p)) \, \1 - i \sin(t\Omega(p)) \frac{\cM(p)}{\Omega(p)}\Big) 
	\begin{pmatrix}
		a_p^+\\a_{-p}
	\end{pmatrix} \,.
	\eeqn
	This finishes the proof.
	\endprf

	\begin{corollary}\label{cor-errorest}
		Assume $q\in[1,\infty]$ and $n\in\{0,1,2,3\}$. 
		Moreover, recall the assumptions $\widehat v\geq0$ with $\widehat v(0)=0$ from Section \ref{sec-assumptions}.
		$V_1,V_2$, defined by
		\eqn\label{eq-aptapprox}
		a_p^+(t)\,=\,\big(e^{i\Omega(p)t}+i\lambda^2\sin(\Omega(p)t)V_1(p)^2\big) a_p^++i\lambda\sin(\Omega(p)t)V_2(p)a_{-p} \, ,
		\eeqn
		are given by 
		\eqn
		V_1(p)^2 & := & \frac{\hat{v}(p)^2}{\Omega(p)(\Omega(p)+E(p)+\lambda\hat{v}(p))} \, , \\
		V_2(p) & := & \frac{\hat{v}(p)}{\Omega(p)} \, .
		\eeqn 
		They satisfy the bounds
		\eqn\label{eq-errorest-linftybd}
		\|D^nV_1^2\|_q  &\leq&  C\vc^{n+2} \, ,\\
		\|D^nV_2\|_q &\leq & C\vc^{n+1} \, .
		\eeqn 
		Similarly,
		\eqn
		u_\lambda(t,p) &:=& e^{i\Omega(p)t}+i\lambda^2\sin(\Omega(p)t)V_1(p)^2 \label{def-chi1}\\
		v_\lambda(t,p) &:=& i\lambda\sin(\Omega(p)t)V_2(p)\label{def-chi2}
		\eeqn 
		satisfy
		\eqn
		\|u_\lambda(t,\cdot)\|_{\ell^\infty(\lattice)} &\leq & C(1+\lambda^2\vd^2) \, ,\\
		 \nd{v_\lambda(t,\cdot)}  &\leq& C \lambda\vd 
		\eeqn
		for all $q\in[1,\infty]$, all $\lambda\in(0,1]$, and all $t>0$.
	\end{corollary}
	
	\prf
	Lemma \ref{lem-apcV-diag-1} implies
	\eqn
		a_p^+(t) \, = \, \cos(t\Omega(p))a_p^+-i\sin(t\Omega(p))\Big(\frac{-E(p)-\lambda\hat{v}(p)}{\Omega(p)}a_p^+-\lambda\frac{\hat{v}(p)}{\Omega(p)}a_{-p}\Big) \, .
	\eeqn 
	Define $a:=E(p)$ and $b:=E(p)+2\lambda\hat{v}(p)$. Setting
	\eqn
	V_1(p)^2 & = & \frac1{\lambda^2}\Big(\frac{a+b}{2\sqrt{ab}}-1\Big) \, ,\\
	V_2(p) & = & \frac1{\lambda}\frac{b-a}{2\sqrt{ab}} \,=\, \frac{\hat{v}(p)}{\Omega(p)} \, , 	
	\eeqn
	we have that
	\eqn
		a_p^+(t)\,=\,\big(e^{i\Omega(p)t}+i\lambda^2\sin(\Omega(p)t)V_1(p)^2\big) a_p^++i\lambda\sin(\Omega(p)t)V_2(p)a_{-p} \, .
	\eeqn 
	We start by establishing pointwise bounds. We have
	\eqn
	\frac{a+b}{2\sqrt{ab}}-1 &=& \frac{(\sqrt{a}-\sqrt{b})^2}{2\sqrt{ab}}=\frac{(b-a)^2}{2\sqrt{ab}(\sqrt{a}+\sqrt{b})^2} \, .
	\eeqn
	In particular, we get 
	\eqn
	V_1(p)^2 & = & \frac{2\hat{v}(p)^2}{\Omega(p)(\sqrt{E(p)}+\sqrt{E(p)+2\lambda\hat{v}(p)})^2} \nonumber\\
	&=& \frac{\hat{v}(p)^2}{\Omega(p)(\Omega(p)+E(p)+\lambda\hat{v}(p))}. 
	\eeqn 
	Using $\hat{v}\geq0$, we have the pointwise bounds 
	\eqn
	|V_1^2| &\leq& 2\frac{\hat{v}^2}{E^2} \, , \label{eq-eta1-pw-0}\\
	|V_2| &\leq& \frac{\hat{v}}E \, .\label{eq-eta2-pw-0}
	\eeqn
	We thus have that
	\eqn
		\nd{V_1^2} &\leq & C\vd^2 \, , \nonumber\\
		\nd{V_2} & \leq & C\vd \, .
	\eeqn 
	\par Finally, set 
	\eqn
	h_1 (x) & :=& \frac{2x^2}{\sqrt{1+2\lambda x}\big(1+\sqrt{1+2\lambda x}\big)^2} \, , \\
	h_2(x) & :=& \frac{x}{\sqrt{1+2\lambda x}} \, .
	\eeqn  
	We have that
	\eqn
		V_1(p)^2 &=& h_1\big(\frac{\hat{v}(p)}{E(p)}\big) \, , \\
		V_2(p) &=& h_2\big(\frac{\hat{v}(p)}{E(p)}\big) \, .
	\eeqn 
	Using Lemma \ref{lem-dnomega} with $n\in\{1,2,3\}$, we obtain 
	\eqn
	\|D^nV_1^2\|_q &\leq & C\|h_1\|_{C^n([0,\vc])}\vc^n \, , \label{eq-Deta1-0}\\
	\|D^nV_2\|_q &\leq & C\|h_2\|_{C^n([0,\vc])}\vc^n \, . \label{eq-Deta2-0}
	\eeqn 
	A straight-forward calculation yields
	\eqn
	\|h_1\|_{C^n([0,\vc])} &\leq & C\vc^2 \, , \label{eq-h1-norm-0}\\
	\|h_2\|_{C^n([0,\vc])} &\leq & C\vc \label{eq-h2-norm-0}\, .
	\eeqn 
	Combining \eqref{eq-Deta1-0}, \eqref{eq-Deta2-0}, \eqref{eq-h1-norm-0}, and \eqref{eq-h2-norm-0} finishes the proof.
	\endprf

	\section{Trace estimates \label{sec-stat-phys}}
	
	In this chapter, we omit the subscript $\lattice$ from $\int_\lattice$ in our notation, as we will only consider $\int\equiv\int_\lattice$ as sums over $\lattice$ in the sense of \eqref{eq-disc-int}.
	
	\subsection{Computation of the partition function}
	
	We establish estimates on the partition function on finite particle number subspaces of the Fock space $\cF$. The estimates derived here are needed to control the error terms coming from the expansion in Lemma \ref{lem-duham}.
	\par For the next lemma, let us introduce some notation. Let us denote 
	\eqn\label{def-opabrev-0}
	a^{(1)}_p \, := \, a^+_p \quad\mbox{and\ }\quad a^{(-1)}_p \, := \, a_p \, .
	\eeqn
	Let 
	\eqn
	\cP[a,a^+]& := & \Big\{\int \, d\bp_k h(\bp_k,\vec{\sigma}_k)\prod_{j=1}^k a_{p_j}^{(\sigma_j)} \mid k\in\N, \, h(\cdot,\vec{\sigma}_k)\in\cS'(\R^3), \nonumber\\
	&&\vec{\sigma}_k=(\sigma_1,\ldots,\sigma_k)\in\{\pm1\}^k \Big\}
	\eeqn 
	denote all monomials in $a,a^+$. Here and henceforth, we define ordered operator products via multiplication from the right, by 
	\eqn
	\prod_{j=1}^{n+1} A_j & := & \Big(\prod_{j=1}^n A_j\Big) A_{n+1} \, ,  \, n\in\N_0 \, , \\
	\prod_{j=1}^0 A_j & :=& \1 \, .
	\eeqn 
	Let us denote 
	\eqn
	\osg(A) \, := \, \sum_{j=1}^k\sigma_j \label{def-osg}
	\eeqn 
	whenever $A=\int \, d\mathbf{p_k} h(\bp_k,\vec{\sigma}_k)\prod_{j=1}^k a_{p_j}^{(\sigma_j)}$. Note that we have
	\eqn\label{eq-com-sign-0}
	[\nb, \prod_{j=1}^k a_{p_j}^{(\sigma_j)}] \, = \, \sum_{j=1}^k\sigma_j\prod_{j=1}^k a_{p_j}^{(\sigma_j)} \, ,
	\eeqn
	which, in turn, gives
	\eqn \label{eq-com-sign-1}
	[\nb,A] \, = \, \osg(A) A
	\eeqn 
	whenever $A=\int \, d\mathbf{p_k} h(\mathbf{p_k},\mathbf{\sigma_k})\prod_{j=1}^k a_{p_j}^{(\sigma_j)}$. Using the spectral decomposition
	\eqn
	P_n \, = \, \frac1{2\pi i} \oint_{\partial B_{\frac12}(n)} \frac{dz}{\nb-z} \, , 
	\eeqn 
	with counter-clockwise contour, it follows that
	\eqn\label{pull-through}
	P_n A \, =\, AP_{n-\osg(A)} 
	\eeqn 
	for all $A\in \cP[a,a^+]$. We will refer to \eqref{pull-through} as the {\it Pull-Through Formula} for projectors, see \cite{bafrsi}. 
	\par Moreover, if $\osg(A)\neq0$, \eqref{eq-com-sign-1} implies that 
	\eqn
		\nu(A) \, = \, \frac1{\osg(A)}\nu([\nb,A]) \, = \, 0 \, . \label{eq-num-con}
	\eeqn

	\begin{lemma}[Moments of the number operator] \label{lem-num-mom}
		We have for all $\ell\in\N_0$ that
		\begin{equation}
			\nu_0((\nb+1)^{\frac\ell2}) \, \leq \, C_{\ell,\fd} \vol^{\frac\ell2} \, .  
		\end{equation}
	\end{lemma}
	
	\prf
	Let $\cK_\mu:=\int_{\lattice}dp \, (K(p)-\mu)a_p^+a_p$.  Then we obtain that
	\eqn
	Z_0(\mu) &:=& \tr(e^{-\cK_\mu}) \nonumber\\
	&=& \prod_{p\in\lattice} \sum_{n=0}^\infty e^{-n(K(p)-\mu)}\nonumber\\
	&=& \prod_{p\in\lattice} \frac1{1-e^{-(K(p)-\mu)}}	\nonumber\\
	&=& e^{-\vol\int_{\lattice} dp \log\big(1-e^{-( K(p)-\mu)}\big)} \label{eq-partition-0}
	\eeqn 
	for any $\mu<\kappa_0$. Let 
	\begin{equation}
		\kappa_n \, := \, -(-\partial_\mu)^n\big|_{\mu=0}\int_{\lattice} dp \log\big(1-e^{-( K(p)-\mu)}\big) \label{def-cumulant-n}
	\end{equation}
	denote the $n^{th}$ cumulant for $n\in\N$. 
	Let ${n \choose \br_\ell}$ denote the multinomial coefficient, and $R(\ell):=\{\br_\ell\in\N_0^\ell\mid \sum_{n=1}^\ell nr_n=\ell\}$. Then the Fa\`a di Bruno formula, see \cite{tbn},  yields
	\begin{align}
		\nu_0(\nb^\ell) \, = & \, \frac{(-\partial_\mu)^\ell\big|_{\mu=0}Z_0(\mu)}{Z_0(0)}\nonumber\\
		=&\, \ell! \sum_{\br_\ell\in R(\ell)}\prod_{n=1}^\ell \frac1{r_n!}\Big(\frac{\vol\kappa_n}{n!}\Big)^{r_n} \, . \label{eq-cum-formula-0}
	\end{align}
	Observe that we have 
	\begin{equation}
		\kappa_1 \, = \, \int_{\lattice}dp \, \frac1{e^{K(p)}-1} \,  = \, \int_{\lattice} dp f_0(p) \, , \label{eq-kappa1}
	\end{equation}
	recalling $f_0$ from \eqref{def-f0}. More generally, Lemma \ref{lem-cumulant} implies
	\begin{equation}
		\kappa_n \, \leq \, C_{n,\fd} \label{eq-cum-est-0}
	\end{equation}
	for all $n\in\N$. As a consequence of \eqref{eq-cum-formula-0} and \eqref{eq-cum-est-0}, we find that
	\begin{equation}
		\nu_0((\nb+1)^\ell) \, \leq \, C_{\ell,\fd} \vol^\ell \, ,
	\end{equation}
	where we used that $\vol\geq1$.	Using 
	\eqn
	\nu_0((\nb+1)^{\frac\ell2}) \, \leq \, \nu_0((\nb+1)^{\ell-1})^{\frac12}\nu_0((\nb+1)^{\ell+1})^\frac12 \label{eq-halfint-int-0}
	\eeqn 
	yields the half-integer case. This concludes the proof.
	\endprf 
	
	\begin{remark}\label{rem-f0-computation}
		We use an argument from \cite{denteneer}. Using the fact that
		\eqn 
		e^{-\cK}a_p^+e^{\cK} \, = \, e^{- K(p)} a_p^+ \, ,
		\eeqn 
		cyclicity of the trace implies
		\eqn
		f_0(p) &=&  \frac{\tr \big( e^{-\cK} a_p^+a_p \big)}{\tr e^{-\cK}\vol} \nonumber\\
		&=& e^{-K(p)}  \frac{\tr \big( a_p^+ e^{-\cK}a_p\big)}{\tr e^{-\cK}\vol} \nonumber\\
		&=& e^{-K(p)}  \frac{\tr \big(e^{-\cK}a_p a_p^+ \big)}{\tr e^{-\cK}\vol} \nonumber\\
		&=& e^{-K(p)}(f_0(p) \, + \, 1) \, ,
		\eeqn 
		Thus, we have
		\eqn
		f_0(p) \, = \, \frac1{e^{K(p)}-1} \, . \label{eq-f0-expression}
		\eeqn 
	\end{remark}

	\begin{lemma}[Operator product bound]\label{lem-prod}
		Let $A_j\in\cP[a,a^+]$ be monomials in $a,a^+$, $\gamma_j>0$, and $k_j\in\N$ be such that
		\eqn
		\|P_m A_j P_{m-\osg(A_j)}\|	\, \leq \, \gamma_j (m+\vol)^{k_j/2} \label{eq-Aj-bd-ass-0}
		\eeqn 
		for all $j\in\{1,\ldots,\ell\}$ and all $m\in\N_0$. Then we have that
		\eqn
		|\nu(\prod_{j=1}^\ell A_j)|  \, \leq \, \Big(\prod_{j=1}^\ell \gamma_j\Big) \nu\Big( \big(\cN+\sum_{m=1}^\ell|\osg(A_m)|+\vol \big)^{\sum_{j=1}^\ell k_j/2}\Big)
		\eeqn 
		for any state $\nu$. 
	\end{lemma}
	
	\prf
	Observe that
	\eqn
	\osg(A) \, = \, \sum_{j=1}^\ell \osg(A_j) \, . \label{eq-sigma-sum-0}
	\eeqn 
	In addition, 
	\eqn
	\sum_{n=0}^\infty P_n & = & \1 \, , \\
	P_nf(\nb) & = & P_nf(n) \label{eq-N-FC}
	\eeqn 
	holds for any measurable function $f$. Hence and by applying \eqref{pull-through}, we have that
	\eqn
	|\nu(A)| &\leq& \sum_{n=0}^{\infty} |\nu(P_nA P_{n-\osg(A)})| \nonumber\\
	&=& \sum_{n=0}^\infty |\nu\Big(P_n\big(\prod_{j=1}^\ell P_{n-\sum_{m=1}^{j-1}\osg(A_m)}A_j P_{n-\sum_{m=1}^j\osg(A_m)}\big)P_{n-\osg(A)}\Big)| \, .
	\eeqn 
	Denoting $B_n:=\prod_{j=1}^\ell P_{n-\sum_{m=1}^{j-1}\osg(A_m)}A_j P_{n-\sum_{m=1}^j\osg(A_m)}$, we can apply Cauchy-Schwarz to $\nu$ to obtain the upper bound
	\eqn
	|\nu(A)| &\leq& \sum_{n=0}^\infty \nu(P_n)^\frac12 \nu(P_{n-\osg(A)}B_n^+B_nP_{n-\osg(A)})^\frac12 \nonumber\\
	&\leq & \sum_{n=0}^\infty \|B_n\| \nu(P_n)^\frac12 \nu(P_{n-\osg(A)})^\frac12\nonumber\\
	&\leq & \Big(\sum_{n=0}^\infty \|B_n\| \nu(P_n)\Big)^\frac12  \Big(\sum_{n=0}^\infty \|B_n\| \nu(P_{n-\osg(A)})\Big)^\frac12 \nonumber\\
	&=& \Big(\sum_{n=0}^\infty \|B_n\| \nu(P_n)\Big)^\frac12 \Big(\sum_{n=\osg(A)_-}^\infty \|B_{n+\osg(A)}\| \nu(P_n)\Big)^\frac12\, . \label{eq-prod-bd-0}
	\eeqn 
	Here, we used Cauchy-Schwarz to $\sum_{n=0}^\infty$, followed by the fact that $P_{n-\osg(A)}\equiv0$ for all $n<\osg(A)_+$, as $\nb\geq0$ in the sense of quadratic forms.  Notice that we have
	\eqn
	\|B_n\| &\leq & \prod_{j=1}^\ell \|P_{n-\sum_{m=1}^{j-1}\osg(A_m)}A_j P_{n-\sum_{m=1}^j\osg(A_m)}\| \nonumber\\
	&\leq & \prod_{j=1}^\ell \gamma_j \Big(\big(n-\sum_{m=1}^{j-1}\osg(A_m)\big)_++\vol\Big)^{k_j/2} \nonumber\\
	&\leq & \Big(\prod_{j=1}^\ell \gamma_j\Big) \Big(n+\sum_{m=1}^\ell|\osg(A_m)|+\vol \Big)^{\sum_{j=1}^\ell k_j/2}\label{eq-prod-bd-1}
	\eeqn 
	by assumption \eqref{eq-Aj-bd-ass-0}. Similarly, we have that
	\eqn
	\|B_{n+\osg(A)}\| &\leq& \prod_{j=1}^\ell \gamma_j \Big(\big(n+\osg(A)-\sum_{m=1}^{j-1}\osg(A_m)\big)_++\vol\Big)^{k_j/2}\nonumber\\
	&=& \prod_{j=1}^\ell \gamma_j \Big(\big(n+\sum_{m=j}^{\ell}\osg(A_m)\big)_++\vol\Big)^{k_j/2}\nonumber\\
	&\leq & \Big(\prod_{j=1}^\ell \gamma_j\Big) \Big(n+\sum_{m=1}^\ell|\osg(A_m)|+\vol \Big)^{\sum_{j=1}^\ell k_j/2}\label{eq-prod-bd-2}
	\eeqn 
	due to \eqref{eq-sigma-sum-0}. Collecting \eqref{eq-prod-bd-0}--\eqref{eq-prod-bd-2} and applying \eqref{eq-N-FC} again, we then obtain
	\eqn
	|\nu(A)| &\leq & \Big(\prod_{j=1}^\ell \gamma_j\Big) \sum_{n=0}^\infty \Big(n+\sum_{m=1}^\ell|\osg(A_m)|+\max_{1\leq j\leq \ell} k_j \Big)^{\sum_{j=1}^\ell k_j/2} \nu(P_n)\nonumber\\
	&=& \Big(\prod_{j=1}^\ell \gamma_j\Big) \nu\Big( \big(\cN+\sum_{m=1}^\ell|\osg(A_m)|+\vol \big)^{\sum_{j=1}^\ell k_j/2}\Big)
	\eeqn 
	\endprf

	\subsection{Operator estimates w.r.t. $\nb$}
	
	In our analysis, we will bound various operators relative to powers of the particle number operator $\nb$, in conjunction with previously established estimates on the partition function.

	\begin{lemma}[$\cH_{cub}(t)$ bound]\label{lem-hcub-est}
		We have that 
		\eqn 
		\cHcubt(t) & = & \sum_{k=1}^{16} A_{cub}^{(k)}(t)
		\eeqn 
		with monomials $A_{cub}^{(k)}(t)\in \cP[a,a^+]$ such that
		\eqn
		\|A_{cub}^{(k)}(t) P_M\| \, \leq \, C\vd(1+\lambda\vd)^6 \frac{\lambda(M+\vol)^{\frac32}}{\sqrt{N}} \label{eq-acub-est-0}
		\eeqn 
		for any $M\in\N_0$, and $|\osg(A_{cub}^{(k)}(t))|\leq 3$.
	\end{lemma}
	\prf
	Corollary \ref{cor-errorest} and evenness of $\hat{v}$ imply that
	\eqn
	\lefteqn{\cHcubt(t) }\nonumber\\
	& =& \frac{\lambda}{\sqrt{N}}\int_{(\lattice)^3} d\bp_3 \, \hat{v}(p_2) \delta(p_1+p_2-p_3) a_{p_1}^+(t)a_{p_2}^+(t)a_{p_3}(t) \, + \, h.c.   \nonumber\\
	&=&\frac{\lambda}{\sqrt{N}}\sum_{\vec{\sigma}_3\in\{\pm1\}^3} \int_{(\lattice)^3} d\bp_3 \,  \delta(\sum_{j=1}^3\sigma_jp_j) \hat{v}(p_2)\prod_{j=1}^3(g_j(p_j,\sigma_j)a_{p_j}^{(\sigma_j)}) \, + \, h.c. \, , \label{eq-hcub-dec-1}
	\eeqn 
	where $\vec{\sigma}_3=(\sigma_1,\sigma_2,\sigma_3)$, and 
	\eqn
	g_j(p_j,\sigma_j) &:=& \delta_{\sigma_j,1} u_\lambda(t,p_j) +\delta_{\sigma_j,-1}v_\lambda(t,p_j) \, , \quad  j\in\{1,2\} \, , \\
	g_3(p_3,\sigma_3) &:=& \delta_{\sigma_3,-1} \overline{u_\lambda}(t,p_3) +\delta_{\sigma_3,1}\overline{v_\lambda}(t,p_3) \, .
	\eeqn 
	Abbreviating 
	\eqn
		A_{cub}(\vec{\sigma}_3) \, := \, \int d\bp_3 \,  \delta(\sum_{j=1}^3\sigma_jp_j) \hat{v}(p_2)\prod_{j=1}^3(g_j(p_j,\sigma_j)a_{p_j}^{(\sigma_j)}) \, ,
	\eeqn
	it is sufficient to prove \eqref{eq-acub-est-0} for $A_{cub}(\vec{\sigma}_3)$, since the adjoint satisfies
	\eqn
		\|A_{cub}(\vec{\sigma}_3)^* P_M\| & = & \|P_{M-\sum_{j=1}^3\sigma_j} A_{cub}(\vec{\sigma}_3)^* P_M\|\nonumber\\
		&=&\|A_{cub}(\vec{\sigma}_3) P_{M-\sum_{j=1}^3\sigma_j}\| \label{eq-acub-adj-1}
	\eeqn  	
	due to the Pull-Through formula \eqref{pull-through}.
	\par Using \nameref{wick-thm} in Appendix \ref{app-calc}, we have that
	\eqn
	a^{(\sigma_1)}_{p_1} a^{(\sigma_2)}_{p_2} a^{(\sigma_3)}_{p_3} \delta(\sum_{j=1}^3\sigma_j p_j)&=& :a^{(\sigma_1)}_{p_1} a^{(\sigma_2)}_{p_2} a^{(\sigma_3)}_{p_3}: \delta(\sum_{j=1}^3\sigma_j p_j) \nonumber\\
	&&+ \, \delta(p_1-p_2)\delta(p_3)\delta_{\sigma_1,-1}\delta_{\sigma_2,1}a_0^{(\sigma_3)} \nonumber\\
	&&+ \, \delta(p_1-p_3)\delta(p_2)\delta_{\sigma_1,-1}\delta_{\sigma_3,1}a_0^{(\sigma_2)}\nonumber\\
	&&+ \, \delta(p_2-p_3)\delta(p_1)\delta_{\sigma_2,-1}\delta_{\sigma_3,1}a_0^{(\sigma_1)} \, . \label{eq-acub-wick-ord-0}
	\eeqn 	
	We want to apply Lemma \ref{lem-wick-ord-bd} to each of the terms associated with the terms in \eqref{eq-acub-wick-ord-0}. For that, we need to ensure integrability for each of those terms. The terms involving $a_0^{(\sigma)}$ contain 2 momentum-$\delta$ where the only free momentum comes with a coefficient $v_\lambda$ resp. $\overline{v_\lambda}$. Using
	\eqn
		\|a_0^+ P_M\| \, = \, \|a_0 P_{M+1}\| \, , 
	\eeqn 	
	Lemma \ref{lem-wick-ord-bd} then implies
	\eqn
		\|a_0^{(\sigma)}P_M\| & = & \delta_{\sigma,1}\|a_0^+P_M\|+\delta_{\sigma,-1}\|a_0 P_M\|  \nonumber\\
		& \leq & \sqrt{\vol(M+1)} \, . \label{eq-a0-bd-1}
	\eeqn 
	For the cubic term in \eqref{eq-acub-wick-ord-0} and if not all $\sigma_j=1$ or all $\sigma_j=-1$, Lemma \ref{lem-wick-ord-bd} implies
	\eqn
		\lefteqn{\|:A_{cub}(\vec{\sigma}_3):P_M\|}\nonumber\\
		 & \leq &  \| |\hat{v}(p_2)|^{\frac12}\prod_{j=1}^2|v_\lambda(p_j)|^{\delta_{\sigma_j,-1}} |u_\lambda(p_3)|^{\delta_{\sigma_3,-1}}\delta(\sum_{j=1}^3p_j\sigma_j)^{\frac12}\|_{L^\infty_{\bp_{J_+}}L^2_{\bp_{J_-}}}\nonumber\\
		&& \||\hat{v}(p_2)|^{\frac12}\prod_{j=1}^2|u_\lambda(p_j)|^{\delta_{\sigma_j,1}} |v_\lambda(p_3)|^{\delta_{\sigma_3,1}}\delta(\sum_{j=1}^3p_j\sigma_j)^{\frac12}\|_{L^\infty_{\bp_{J_-}}L^2_{\bp_{J_+}}}\nonumber\\ &&(M+\sum_{j=1}^3\sigma_j)_{|J_+|}^{\frac12}(M)_{|J_-|}^{\frac12} \, , \label{eq-acub-bd-1}
	\eeqn 
	using the notation in Lemma \ref{lem-wick-ord-bd} with $n=3$. Integrability of $\hat{v}(p_2)$ is sufficient to yield that the RHS is, indeed, finite. If $\sigma_j=1$ for all $j$, we have that
	\eqn
		\lefteqn{\|:A_{cub}(1,1,1):P_M\|}\nonumber\\
		&\leq& \big(\| \hat{v}(p_2) u_\lambda(t,p_1)u_\lambda(t,p_2)\overline{v_\lambda}(t,p_3)\|_{L^2_{p_2}L^\infty_{p_1,p_3}}^2(M-2)\nonumber\\
		&&+\|\hat{v}(p_2)u_\lambda(t,p_1)u_\lambda(t,p_2)\overline{v_\lambda}(t,p_1+p_2)\|_{L^2_{\bp_2}}^2 \vol \big)^{\frac12}\sqrt{M(M-1)} \, . \label{eq-acub-bd-2}
	\eeqn 
	Similarly, we find that
	\eqn
		\lefteqn{\|:A_{cub}(-1,-1,-1):P_M\|}\nonumber\\
		&\leq& \big(\| \hat{v}(p_2) v_\lambda(t,p_1)v_\lambda(t,p_2)\overline{u_\lambda}(t,p_3)\|_{L^2_{p_2}L^\infty_{p_1,p_3}}^2(M+1)\nonumber\\
		&&+\|\hat{v}(p_2)v_\lambda(t,p_1)v_\lambda(t,p_2)\overline{u_\lambda}(t,p_1+p_2)\|_{L^2_{p_2}L^\infty_{p_1}} \vol \big)^{\frac12}\sqrt{(M+3)(M+2)} \label{eq-acub-bd-3}
	\eeqn 
	Observe that
	\eqn
		\|H\|_{L^2(\lattice)} & = & \frac{1}{\sqrt{\vol}} \|H\|_{\ell^{2}(\lattice)} \nonumber\\
		 & \leq & \nd{H} \, . \label{eq-L2-emb-1}
	\eeqn 
	Since we are summing over a lattice, there are terms for which momentum requires the summation over $\hat{v}(2p)$ or $v_\lambda(2p)$ resp. $\overline{v_\lambda}(2p)$ . In order to use the upper bound $\vd$, we employ the fact that
	\eqn
		\Big|\sum_{p\in\lattice} H(2p)\Big| \, \leq \, \|H\|_{\ell^1(\lattice)} \, .
	\eeqn 
	Collecting \eqref{eq-acub-adj-1}, \eqref{eq-a0-bd-1}, \eqref{eq-acub-bd-1}, \eqref{eq-acub-bd-2}, \eqref{eq-acub-bd-3}, \eqref{eq-L2-emb-1}, and employing Corollary \ref{cor-errorest}, we have shown that
	\eqn
		\|A_{cub}(\vec{\sigma}_3)P_M\| \, \leq \, C\vd(1+\lambda\vd)^6 (M+\vol)^{\frac32}	\, ,
	\eeqn 
	where we also used the fact that $\vol\geq1$. Together with \eqref{eq-hcub-dec-1}, this finishes the proof.
	\endprf
	
	\begin{lemma}[$\cH_{quart}(t)$ bound]\label{lem-hquart-est}
		We have that 
		\eqn 
		\cHquart(t) & = & \sum_{k=1}^{16} A_{quart}^{(k)}(t)
		\eeqn 
		with monomials $A_{quart}^{(k)}(t)\in \cP[a,a^+]$ such that
		\eqn
		\|A_{quart}^{(k)}(t) P_M\| \, \leq \, C\vd (1+\lambda\vd)^8\frac{\lambda(M+\vol)^2}{N} 
		\eeqn 
		for any $M\in\N_0$, and $|\osg(A_{quart}^{(k)}(t))|\leq 4$. 
	\end{lemma}
	\prf
	We follow the steps of the proof of Lemma \ref{lem-hcub-est}. We have that
	\eqn
	\lefteqn{\cHquart(t)}\nonumber\\
	 & =& \frac{\lambda}{2N}\int d\bp_4 \, \hat{v}(p_1-p_3) \delta(p_1+p_2-p_3-p_4) a_{p_1}^+(t)a_{p_2}^+(t)a_{p_3}(t)a_{p_4}(t) \nonumber\\
	&=&\frac{\lambda}{2N}\sum_{\vec{\sigma}_4\in\{\pm1\}^4} \int d\bp_4 \,  \delta(\sum_{j=1}^4\sigma_jp_j) \hat{v}(\sigma_1p_1+\sigma_3p_3)\prod_{j=1}^4(g_j(p_j,\sigma_j)a_{p_j}^{(\sigma_j)}) \, , \label{eq-hquart-dec-1}
	\eeqn 
	where $\vec{\sigma}_4=(\sigma_1,\sigma_2,\sigma_3,\sigma_4)$, and 
	\eqn
	g_j(p_j,\sigma_j) :=\begin{cases}
		 u_\lambda(t,p_j)^{\delta_{\sigma_j,1}} v_\lambda(t,p_j)^{\delta_{\sigma_j,-1}} \, , \quad  j\in\{1,2\} \, , \\
		 \overline{u_\lambda}(t,p_j)^{\delta_{\sigma_j,-1}} \overline{v_\lambda}(t,p_j)^{\delta_{\sigma_j,1}} \, , \quad  j\in\{3,4\} \, .
	\end{cases} 
	\eeqn 
	Analogously to above, we denote
	\eqn
		A_{quart}(\vec{\sigma}_4) \, := \, \int d\bp_4 \,  \delta(\sum_{j=1}^4\sigma_jp_j) \hat{v}(\sigma_1p_1+\sigma_3p_3)\prod_{j=1}^4(g_j(p_j,\sigma_j)a_{p_j}^{(\sigma_j)}) \, .
	\eeqn 
	Wick-ordering using Lemma \ref{wick-thm} yields
	\eqn
		\lefteqn{\delta(\sum_{j=1}^4\sigma_jp_j) \prod_{j=1}^4a_{p_j}^{(\sigma_j)}}\nonumber\\
		&=&\delta(\sum_{j=1}^4\sigma_jp_j) :\prod_{j=1}^4a_{p_j}^{(\sigma_j)}: \, + \, \sum_{\substack{ j_1<j_2,\\ j_3<j_4}} \delta(p_{j_1}-p_{j_2})\delta_{\sigma_{j_1},-1}\delta_{\sigma_{j_2},1}\nonumber\\
		&&\big(\delta(\sigma_{j_3} p_{j_3}+\sigma_{j_4} p_{j_4}):a_{p_{j_3}}^{(\sigma_{j_3})}a_{p_{j_4}}^{(\sigma_{j_4})}: \, + \, \vol \delta(p_{j_3}-p_{j_4})\delta_{\sigma_{j_3},-1}\delta_{\sigma_{j_4},1}\big) \, ,
	\eeqn 
	where $\{j_1,\ldots,j_4\}=\{1,\ldots,4\}$.
	\par We start with the 0-order terms. These occur whenever two pairs $aa^+$ exist, each with an annihilation operator to the left of a creation operator. This implies an integrable coefficient $v_\lambda$ resp. $\overline{v_\lambda}$ for each of the corresponding momenta. Together with the deltas coming from the contraction, the resulting terms are integrable and are bounded by
	\eqn
		\lefteqn{\vol \Big\|\sum_{\substack{ j_1<j_2,\\ j_3<j_4}} \int d\bp_4 \,   \hat{v}(\sigma_1p_1+\sigma_3p_3)\prod_{j=1}^4g_j(p_j,\sigma_j)}\nonumber\\ &&\delta(p_{j_1}-p_{j_2})\delta(p_{j_3}-p_{j_4})\delta_{\sigma_{j_1},-1}\delta_{\sigma_{j_2},1}\delta_{\sigma_{j_3},-1}\delta_{\sigma_{j_4},1}P_M\Big\|\nonumber\\
		&\leq & C\lambda^2\vol \vd^3 (1+\lambda^2\vd^2)^2 \, , \label{eq-hquart-est-1}
	\eeqn 
	where, as above $\{j_1,\ldots,j_4\}=\{1,\ldots,4\}$.
	\par For terms quadratic in $a$, $a^+$, there again remain two independent momenta in the integration after integrating out the $\delta's$ coming from momentum conservation and the single contraction. Observe that one of these $\delta$'s comes from a contraction with an annihilation operator left to a creation operator. Thus, the corresponding momentum comes with an integrable factor $v_\lambda(t,p_{j_1})$ resp. $\overline{v_\lambda}(t,p_{j_1})$. The other momentum comes with a coefficient $v_\lambda (t,p_{j_3})^{\delta_{\sigma_{j_3},-1}}:a_{p_{j_3}}^{(\sigma_{j_3})}a_{-\sigma_{j_3}\sigma_{j_4}p_{j_3}}^{(\sigma_{j_4})}:$, unless $j_3=3$. In that case, we have a coefficient $v(\sigma_1 p_1+\sigma_3 p_3):a_{p_3}^{(\sigma_3)}a_{-\sigma_3\sigma_4p_3}^{(\sigma_4)}:$. Then, integrating first over $p_{j_2},p_{j_4}$ and then $p_{j_3}$, Lemma \ref{lem-wick-ord-bd} implies
	\eqn
	\lefteqn{\Big\|\sum_{\substack{ j_1<j_2,\\ j_3<j_4}}\int d\bp_4 \,   \hat{v}(\sigma_1p_1+\sigma_3p_3)\prod_{j=1}^4g_j(p_j,\sigma_j) }\nonumber\\
	&&\delta(p_{j_1}-p_{j_2})\delta(\sigma_{j_3} p_{j_3}+\sigma_{j_4} p_{j_4})\delta_{\sigma_{j_1},-1}\delta_{\sigma_{j_2},1}:a_{p_{j_3}}^{(\sigma_{j_3})}a_{p_{j_4}}^{(\sigma_{j_4})}: P_M\Big\|\nonumber\\
	& \leq & C\vd(1+\lambda^2\vd)^4(M+\vol) \, . \label{eq-hquart-est-2}
	\eeqn 	
	Finally, if not all $\sigma_j=1$ or $\sigma_j =-1$, Lemma \ref{lem-wick-ord-bd} implies
	\eqn
		\lefteqn{\|:A_{quart}(\vec{\sigma}_4): P_M\|}\nonumber\\
		&\leq & C\||\hat{v}(\sigma_1p_1+\sigma_3p_3)|^\frac12\prod_{j=1}^2|v_\lambda(t,p_j)|^{\delta_{\sigma_j,-1}}|u_\lambda(t,p_{j+2})|^{\delta_{\sigma_{j+2},-1}}\delta(\sum_{j=1}^4\sigma_jp_j)^{\frac12}\|_{L^\infty_{p_{J_+}L^2_{p_{J_-}}}}\nonumber\\
		&&\||\hat{v}(\sigma_1p_1+\sigma_3p_3)|^\frac12\prod_{j=1}^2|u_\lambda(t,p_j)|^{\delta_{\sigma_j,1}}|v_\lambda(t,p_{j+2})|^{\delta_{\sigma_{j+2},1}}\delta(\sum_{j=1}^4\sigma_jp_j)^{\frac12}\|_{L^\infty_{p_{J_-}L^2_{p_{J_+}}}}\nonumber\\
		&&(M+1)^2 \, . \label{eq-aquart-est-1}
	\eeqn 
	By flipping the roles of $(1,2)$ and $(3,4)$, it is sufficient to show boundedness of only one of the norms. If $1\in J_-$, we can use integrability of $|v_\lambda(t,p_1)|$, if $2\in J_-$, there is a factor $|v_\lambda(t,p_2)|$. Integrability w.r.t. to $p_3$ and $p_4$ is always given due to the factor $|\hat{v}(\sigma_1p_1+\sigma_3p_3)|\delta(\sum_{j=1}^4\sigma_jp_j)$. In case $J_-=\emptyset$, Lemma \ref{lem-wick-ord-bd} implies 
	\eqn
		\lefteqn{\|:A_{quart}(1,1,1,1): P_M\|}\nonumber\\
		 & \leq & (\|\hat{v}(p_1+p_3)v_\lambda(t,p_1) v_\lambda(t,p_2)\overline{u_\lambda}(t,p_3)\overline{u_\lambda}(t,p_4)\|_{L^2_{\bp_2}L^\infty_{p_3,p_4}}^2(M-3)\nonumber\\
		 && + \, \|\hat{v}(p_1+p_3)v_\lambda(t,p_1) v_\lambda(t,p_2)\overline{u_\lambda}(t,p_3)\overline{u_\lambda}(t,p_1+p_2-p_3)\|_{L^2_{\bp_3}}^2\vol)^{\frac12} \nonumber\\
		 && \sqrt{M(M-1)(M-2)} \, . \label{eq-aquart-est-2}
	\eeqn 
	$\|:A_{quart}(-1,-1,-1,-1): P_M\|$ can be similarly bounded. Collecting \eqref{eq-hquart-est-1}, \eqref{eq-hquart-est-2}, \eqref{eq-aquart-est-1} and \eqref{eq-aquart-est-2}, and using $\vol\geq1$, we have shown that
	\eqn
		\|A_{quart}(\vec{\sigma}_4) P_M\| \, \leq\, C\vd (1+\lambda\vd)^8(M+\vol)^2 \, .
	\eeqn 
	Using \eqref{eq-hquart-dec-1}, this concludes the proof.
	\endprf

	\begin{proposition}\label{prop-h2-est}
		Let $A\in\cP[a,a^+]$ be a monomial in $a,a^+$, $\gamma>0$ and $\ell\in\N$ be such that
		\eqn
		\|AP_M\| \, \leq \, \gamma (M+\vol)^{\frac\ell2} \, , \label{ass-h2-est}
		\eeqn 
		and that $|\osg(A)|\leq\ell$. Let $\bs_{j+k}\in\Delta[t,j+k]$ and $t\geq0$. For the moment, we abbreviate by $A\cHcubt^j\cHquart^k(\bs_{j+k})$ all terms that contain one factor $A$, $j$ factors $\cHcubt$, $k$ factors $\cHquart$, all at possibly different times $s_m$. Then we have that
		\eqn
		\lefteqn{|\nu(A\cHcubt^j\cHquart^k(\bs_{j+k}))|}\nonumber\\
		& \leq &\frac{\gamma (C\lambda\vd)^{j+k} (1+\lambda\vd)^{6j+8k}}{N^{\frac{j}2+k}}\nonumber\\
		&& \nu\Big( \big(\nb+3j+4k+\ell+\vol\big)^{\frac{3j+4k+\ell}2}\Big) \, .
		\eeqn
		for any state $\nu$ and all $\lambda>0$ small enough. 
	\end{proposition}
	\prf
	Decomposing $\cHcubt$ and $\cHquart$ into monomials as in Lemmata \ref{lem-hcub-est} and \ref{lem-hquart-est}, we apply Lemmata \ref{lem-prod}, \ref{lem-hcub-est}, \ref{lem-hquart-est} together with assumption \eqref{ass-h2-est} on $A$, we obtain that
	\eqn
	|\nu(AA_{cub}^j A_{quart}^k(\bs_{j+k}))| & \leq & \frac{\gamma (C\lambda\vd)^{j+k}(1+\lambda\vd)^{6j+8k}}{N^{\frac{j}2+k}}\nonumber\\
	&& \nu\Big( \big(\nb+3j+4k+\ell+\vol\big)^{\frac{3j+4k+\ell}2}\Big) \, .
	\eeqn 
	As a consequence, we obtain that
	\eqn
	|\nu(A\cHcubt^j\cHquart^k(\bs_{j+k}))| & \leq & \frac{\gamma (C\lambda\vd)^{j+k} (1+\lambda\vd)^{6j+8k}}{N^{\frac{j}2+k}}\nonumber\\
	&& \nu\Big( \big(\nb+3j+4k+\ell+\vol\big)^{\frac{3j+4k+\ell}2}\Big) \, .
	\eeqn
	This finishes the proof.
	\endprf

	\begin{lemma}\label{lem-fgphi-bound}
		Given a test function $J\in \ell^2(\lattice)\cap \ell^{\infty}(\lattice)$, we recall that
		\eqn
		f[J]  & = & \int_\lattice \, dp \, J(p) a_p^+a_p \, ,\\
		g[J] & = & \int_\lattice \, dp \, J(p) a_{-p}a_p \, .
		\eeqn 
		Then we have 
		\eqn
		\|P_m f[J]  P_n\| & \leq & \delta_{m,n}\Jin m \, , \\
		\|P_m g[J]  P_n\| &\leq & \delta_{n,m+2}\Jtn(m+1+\vol) \, .
		\eeqn 
	\end{lemma}
	\prf 
	Due to $[\cN,f[J] ]=0$, we have that
	\eqn
	P_m f[J]  P_n \, = \, \delta_{m,n} \int d\bp_2 \, J(p_1) \delta(p_1-p_2) a_{p_1}^+a_{p_2} P_n \, .
	\eeqn 
	Then Lemma \ref{lem-wick-ord-bd} in Appendix \ref{app-calc} implies 
	\eqn
	\|P_m f[J]  P_n\| \, \leq \, \delta_{m,n} \|J\|_\infty m
	\eeqn 
	For $g[J] $, we have that
	\eqn
	P_m g[J]  P_n \, = \, \delta_{n,m+2} \int d\bp_2 \, J(p_1) \delta(p_1+p_2) a_{p_1}a_{p_2}
	\eeqn 
	Then Lemma \ref{lem-wick-ord-bd} implies
	\eqn
		\|P_m g[J]  P_n\| & \leq & \delta_{n,m+2} \sqrt{\|J\|_\infty^2 (m+1)+\vol \|J\|_2^2}\sqrt{m+2} \nonumber\\
		&\leq & \delta_{n,m+2}(\|J\|_{L^2(\lattice)}+\|J\|_\infty)(m+1+\vol)
	\eeqn 
	This finishes the proof. 
	\endprf

	\begin{lemma}[Propagation of moments in HFB evolution]\label{lem-mom-prop-HFB}
		The HFB evolution $\cVmf(t)$ satisfies
		\eqn
		\big\|(\nb+\vol)^{\frac\ell2}\cVmf(t)(\nb+\vol)^{-\frac\ell2}\big\| \, \leq \, e^{K_\ell \vd \lambda t}
		\eeqn 
		for all $\ell\in\N$ and some positive constants $K_\ell>0$.
	\end{lemma}
	\prf
	Let $\psi\in\cF$. We have that
	\eqn
	\lefteqn{i\partial_t\scp{\cVmf(t)\psi}{(\nb+\vol)^{\ell}\cVmf(t)\psi} }\nonumber\\
	&=& \scp{\cVmf(t)\psi}{[(\nb+\vol)^{\ell},\cHmfz]\cVmf(t)\psi}\nonumber\\
	&=& \sum_{n=0}^\infty (n+\vol)^{\ell}\scp{\cVmf(t)\psi}{ [P_n,\cHcorz]\cVmf(t)\psi} \, . 
	\eeqn 
	Employing \eqref{pull-through} and recalling \eqref{def-hmf}, we have that
	\eqn
	[P_n,\cHmfz] & = & \lambda (P_ng[\hat{v}]P_{n+2}-P_{n-2}g[\hat{v}]P_n-h.c.) \, .
	\eeqn 
	As a consequence, we have that 
	\eqn
	\lefteqn{\big|\partial_t\scp{\cVmf(t)\psi}{(\nb+\vol)^{\ell}\cVmf(t)\psi} \big|}\nonumber\\
	&=& 2\lambda \Big|\sum_{n=0}^\infty (n+\vol)^{\ell} \Im\scp{\cVmf(t)\psi}{(P_ng[\hat{v}]P_{n+2}-P_{n-2}g[\hat{v}]P_n)\cVmf(t)\psi} \Big| \nonumber\\
	&=& 2\lambda \Big|\sum_{n=0}^\infty [(n+\vol)^{\ell}-(n+2+\vol)^{\ell}] \Im\scp{\cVmf(t)\psi}{P_ng[\hat{v}]P_{n+2}\cVmf(t)\psi} \Big| \, .
	\eeqn
	Using the Mean-Value Theorem and Cauchy-Schwarz, the last inequality implies
	\eqn	
	\lefteqn{\big|\partial_t\scp{\cVmf(t)\psi}{(\nb+\vol)^{\ell}\cVmf(t)\psi} \big|}\nonumber\\
	&\leq& K_\ell \lambda\sum_{n=0}^\infty(n+2+\vol)^{\ell-1}\|P_n\cVmf(t)\psi\|\|P_ng[\hat{v}]P_{n+2}\cVmf(t)\psi\| \, .
	\eeqn 
	Applying Lemma \ref{lem-fgphi-bound} followed by Young's inequality, we find
	\eqn 
	\lefteqn{\big|\partial_t\scp{\cVmf(t)\psi}{(\nb+\vol)^{\ell}\cVmf(t)\psi} \big|}\nonumber\\
	&\leq& K_\ell \lambda\vd \sum_{n=0}^\infty(n+2+\vol)^{\ell-1} (n+1+\vol)\nonumber\\
	&& \big(\scp{\cVmf(t)\psi}{P_n \cVmf(t)\psi}+\scp{\cVmf(t)\psi}{P_{n+2}\cVmf(t)\psi}\big)\nonumber\\
	&\leq & K_\ell \lambda \vd\scp{\cVmf(t)\psi}{(\nb+\vol)^{\ell}\cVmf(t)\psi} \, .
	\eeqn 
	Employing Gronwall's Lemma concludes the proof.
	\endprf 
	\begin{remark}\label{rem-HFB-density}
		Using Corollary \ref{cor-errorest}, we find that
		\eqn
		\cVmf^*(t)\nb\cVmf(t) &=& \int dp \, \big(|u_\lambda(t,p)|^2a_p^+a_p + |v_\lambda(t,p)|^2a_{-p}a_{-p}^+ \nonumber\\
		&& + \, u_\lambda(t,p)\overline{v_\lambda}(t,p)a_p^+a_{-p}^+ + \overline{u_\lambda}(t,p)v_\lambda(t,p)a_pa_{-p} \nonumber\\
		&=& \int dp \, \big((|u_\lambda(t,p)|^2+|v_\lambda(t,p)|^2)a_p^+a_p +\vol |v_\lambda(t,p)|^2 \nonumber\\
		&& + \, u_\lambda(t,p)\overline{v_\lambda}(t,p)a_p^+a_{-p}^+ + \overline{u_\lambda}(t,p)v_\lambda(t,p)a_pa_{-p} \, , 
		\eeqn 
		where we used the CCR together with the fact that $v_\lambda$ is even in $p$. Lemma \ref{lem-fgphi-bound} then implies
		\eqn
			\big\|\cVmf^*(t)(\nb+\vol)^{\frac12}\cVmf(t)(\nb+\vol)^{-\frac12}\big\| \, \leq \, C_{\vd} \, .
		\eeqn 
	\end{remark}
	As an immediate consequence of Lemmata \ref{lem-mom-prop-HFB} and \ref{lem-mom-prop-con}, we obtain the following statement.
	
	\begin{corollary}[Propagation of moments]\label{cor-mom-prop}
		Let $\vol\geq1$. There exist constants $C_{\ell,\mu_0}, K_\ell>0$ such that for any $\ell\in\N$, we have that
		\eqn
		\nu_t\big( (\nb+\vol)^{\frac\ell2}\big) &\leq & C_{\ell,\fd} e^{K_\ell \vd\lambda\vol t} \vol^{\frac\ell2} \, .
		\eeqn 
		
	\end{corollary}
	\prf
	Assume that $\frac{\ell}{2}\in\N$. Define
	\eqn
	\widetilde\cU_N(t)\,:=\, \cW^*[\sqrt{N\vol}\phi_0]e^{-i\cH_Nt}\cW[\sqrt{N\vol}\phi_0] \, ,
	\eeqn 
	and let
	\eqn
	\mu_t(A) \, := \, \nu_0 \big(\widetilde{\cU}_N^*(t)A\widetilde{\cU}_N(t)\big) \, .
	\eeqn 
	There exist $\lambda_k\geq0$, $\sum_{k=1}^\infty\lambda_k=1$, and $\Psi_k\in\cF$ such that
	\eqn
	\mu_t(A) \, = \, \sum_{k=1}^\infty \lambda_k \scp{\Psi_k}{A\Psi_k} \, .
	\eeqn 
	Then we have that 
	\eqn
	\nu_t\big( (\nb+1)^{\frac\ell2}\big)  & =  &\mu_t\Big(\cVmf^*(t)(\nb+1)^{\frac\ell2}\cVmf(t)\Big) \nonumber\\
	&=& \sum_{k=1}^\infty \lambda_k \scp{\Psi_k}{\cVmf^*(t)(\nb+1)^{\frac\ell2}\cVmf(t)\Psi_k}\nonumber\\
	&\leq &  e^{K_\ell\vd\lambda t}\sum_{k=1}^\infty \lambda_k \scp{\Psi_k}{(\nb+1)^{\frac\ell2}\Psi_k}\nonumber\\
	&=&  e^{K_\ell\vd\lambda t} \mu_t\big((\nb+1)^{\frac\ell2}\big) \, . \label{eq-HFB-mom-prop-0}
	\eeqn 
	using Lemma \ref{lem-mom-prop-HFB}. Similarly, we write
	\eqn
	\nu_0(A) = \sum_{k=1}^\infty \alpha_k \scp{\Phi_k}{A\Phi_k} 
	\eeqn 
	for some $\alpha_k\geq0$, $\sum_{k=1}^\infty\alpha_k=1$, $\Phi_k\in\cF$, and we obtain that
	\eqn
	\mu_t\big((\nb+1)^{\frac\ell2}\big) &=& \sum_{k=1}^\infty \alpha_k \scp{\Phi_k}{\widetilde{\cU}_N^*(t)(\nb+1)^{\frac\ell2}\widetilde{\cU}_N(t)\Phi_k}\nonumber\\
	&\leq & C_\ell e^{K_\ell \vd\lambda\vol t}\sum_{k=1}^\infty \mu_k \scp{\Phi_k}{(\nb+1)^{\frac\ell2}\big(1+\frac{\nb}{N\vol}\big)\Phi_k}\nonumber\\
	&=& C_\ell e^{K_\ell \vd\lambda\vol t}\nu_0\Big( (\nb+1)^{\frac\ell2}\big(1+\frac{\nb}{N\vol}\big)\Big) \nonumber\\
	&\leq & C_{\ell,\fd} e^{K_\ell \vd\lambda\vol t} \vol^{\frac\ell2} \, ,\label{eq-con-mom-prop-0}
	\eeqn 
	where we used Lemma \ref{lem-mom-prop-con} followed by Lemma \ref{lem-num-mom}. Collecting \eqref{eq-HFB-mom-prop-0} and \eqref{eq-con-mom-prop-0} and using $\vol\geq1$, we obtain that
	\eqn
	\nu_t\big( (\nb+1)^{\frac\ell2}\big) \, \leq \, C_{\ell,\fd} e^{K_\ell \vd\lambda\vol t} \vol^{\frac\ell2} \, . \label{eq-total-mom-prop-0}
	\eeqn 
	The half-integer case follows analogously to \eqref{eq-halfint-int-0}. This concludes the proof.
	\endprf

	\section{Control of error terms in the expansion \label{sec-expansion}}
	
	Again, we write $\int\equiv\int_\lattice$ for brevity, to account for lattice sums over $\lattice$ in the sense of \eqref{eq-disc-int}.
	
	\begin{proposition}[Tail estimates]\label{prop-tail-est}
		Let $T>0$ and $t\leq T\lambda^{-2}$, $\lambda\in(0,1)$, $\vol\geq1$, and $J\in \ell^1(\lattice)\cap \ell^\infty(\lattice)$. Then the following holds true.
		\begin{enumerate}
			\item
		\eqn
		\Phi_t& = & -i\int_0^t \, ds \,  \frac{\nu_0\big([a_0,\cHcubt(s)]\big)}\vol \, +\,  \Rem_2(t;\Phi)
		\eeqn
		with 
		\eqn
		|\Rem_2(t;\Phi) | &\leq &  C_{\vd,\fd}T^2e^{C\vd\vol/\lambda T}\frac{\vol^{\frac32}}{N\lambda^2}\Big(1+\frac\vol{N}\Big)
		 \, .
		\eeqn
		\item
		\eqn 
		\int dp \, J(p)(f_t(p)-f_0(p)) & = & -i\int_0^t \, ds \,  \frac{\nu_0\big([f[J],\cHquart(s)]\big)}{\vol}\nonumber\\
		&& - \int_{\Delta[t,2]} \, d\mathbf{s}_2 \frac{\nu_0\big([[f[J],\cHcubt(s_1)],\cHcubt(s_2)]\big)}{\vol}\nonumber\\
		&& +\Rem_2(t;f[J]) 
		\eeqn 
		with
		\eqn
		\lefteqn{|\Rem_2(t;f[J])|}\nonumber\\
		 &\leq&  C_{\vd,\fd} \jb{T}^4\Jin \frac{e^{C\vd\vol/\lambda T} \vol^6}{\lambda^4N^2}\Big(1 + \frac{\vol}{N}\Big)^2 \, .
		\eeqn 
		\item 
		\eqn 
		\int dp \, J(p)g_t(p) & = & -i\int_0^t \, ds \,  \frac{\nu_0\big([g[J],\cHquart(s)]\big)}{\vol}\nonumber\\
		&& - \int_{\Delta[t,2]} \, d\mathbf{s}_2 \frac{\nu_0\big([[g[J],\cHcubt(s_1)],\cHcubt(s_2)]\big)}{\vol}\nonumber\\
		&& +\Rem_2(t;g[J]) 
		\eeqn 
		with
		\eqn
		\lefteqn{|\Rem_2(t;g[J])|}\nonumber\\
		&\leq&  C_{\vd,\fd} \jb{T}^4\Jtn \frac{e^{C\vd\vol/\lambda T} \vol^6}{\lambda^4N^2}\Big(1 + \frac{\vol}{N}\Big)^2 \, .
		\eeqn 
		\end{enumerate}
	\end{proposition}
	\prf
	Recall that $\nu_0$ being number conserving implies 
	\eqn
	\nu_0(a_0) \, = \, \nu_0(a_pa_q) \, = \, 0 \, . \label{eq-being number conserving}
	\eeqn
	Let $A\in\cP[a,a^+]$, be a monomial with data such that
	\eqn
	\|P_nAP_{n-\osg(A)}\| \, \leq \, \gamma (n+\vol)^{\frac\ell2} 
	\eeqn 
	with $\ell,|\osg(A)| \leq 2$. Lemma \ref{lem-duham} implies
	\eqn 
	\nu_t(A) & = & \nu_0(A)-i\int_0^t \, ds \,  \nu_0\big([A,\cH_I(s)]\big)\nonumber\\
	&& - \int_{\Delta[t,2]} \, d\mathbf{s}_2 \nu_{s_2}\big([[A,\cH_I(s_1)],\cH_I(s_2)]\big)\label{eq-phi-error-exp-0}\\
	&=& \nu_0(A)-i\int_0^t \, ds \,  \nu_0\big([A,\cH_I(s)]\big)\nonumber\\
	&& - \int_{\Delta[t,2]} \, d\mathbf{s}_2 \nu_0\big([[A,\cH_I(s_1)],\cH_I(s_2)]\big)\nonumber\\
	&& +i\int_{\Delta[t,3]} \, d\mathbf{s}_3 \nu_0\big([[[A,\cH_I(s_1)],\cH_I(s_2)],\cH_I(s_3)]\big) \nonumber\\
	&& + \int_{\Delta[t,4]} \, d\mathbf{s}_4 \nu_{s_4}\big([[[[A,\cH_I(s_1)],\cH_I(s_2)],\cH_I(s_3)],\cH_I(s_4)]\big) \, , \label{eq-fg-error-exp-0}
	\eeqn
	where
	\eqn
	\cH_I(s) \, = \, \cHcubt(s)+\cHquart(s) \, ,
	\eeqn 
	see \eqref{def-HI}. In order to simplify notation, we shall abbreviate by $A\cHcubt^j\cHquart^k(\bs_{j+k})$ all terms that contain one factor $A$, $j$ factors $\cHcubt(s_{\ell})$, $k$ factors $\cHquart(s_m)$, all at possibly different times. Let $\bs_{j+k}\in\Delta[t,j+k]$, $t\leq T\lambda^{-2}$, $\lambda\in(0,1)$. Proposition \ref{prop-h2-est} then implies 
	\eqn
	\lefteqn{|\nu_{s_{j+k}}\big(A\cHcubt^j\cHquart^k(\bs_{j+k})\big)|}\nonumber\\
	&\leq& \frac{C_{j,k}\gamma\vd^{j+k}\lambda^{j+k}(1+\lambda\vd)^{6j+8k}}{N^{\frac{j}2+k}}\nu_{s_{j+k}}\Big( (\nb+\vol)^{\frac{3j+4k+\ell}2}\Big)  \label{eq-r2error-exp-0}
	\eeqn 
	for all $N>0$ large enough. By Corollary \ref{cor-mom-prop}, we obtain
	\eqn
	\lefteqn{|\nu_{s_{j+k}}\big(A\cHcubt^j\cHquart^k(\bs_{j+k})\big)|}\nonumber\\
	&\leq& C_{\vd,\fd,j,k}\gamma e^{K_{j,k}\vd \vol/\lambda T}\frac{\lambda^{j+k}\vol^{\frac{3j+4k+\ell}2}}{N^{\frac{j}2+k}}\, .\label{eq-r2error-exp-1}
	\eeqn
	Similarly to \eqref{eq-r2error-exp-0}, we have that Lemma \ref{lem-num-mom} implies
	\eqn
	\lefteqn{|\nu_0\big(A\cHcubt^j\cHquart^k(\bs_{j+k})\big)|}\nonumber\\
	&\leq& C_{\vd,\fd,j,k}\gamma\frac{\lambda^{j+k}\vol^{\frac{3j+4k+\ell}2}}{N^{\frac{j}2+k}}\, .\label{eq-r2error-exp-2}
	\eeqn 
	\par Due to $\nu_0$ being number conserving, see \eqref{eq-num-con}, we have that
	\eqn
	|\nu_0\big(A\cHcubt^j\cHquart^k(\bs_{j+k})\big)| \, = \, 0 \, ,\label{eq-nu0-odd-0}
	\eeqn 
	whenever the total number of creation and annihilation operators in $A\cHcubt^j\cHquart^k(\bs_{j+k})$ is odd. 
	\par In the case $A=a_0$, due to Corollary \ref{cor-errorest} and using $\hat{v}(0)=0$, we have that
	\eqn	
	a_0 \, = \, a_0(s) 
	\eeqn 
	for all $s\in\R$. With that, an easy computation yields
	\eqn\label{eq-a0-hcub-0}
	[a_0,\cHcubt(s)] &=& [a_0,\cHcubz](s) \nonumber\\
	&=& \frac{\lambda}{\sqrt{N}} \int \, dp \, \hat{v}(p)\big(a_p^+(s)a_p(s)+a_p(s)a_{-p}(s) \big) \, .
	\eeqn 
	Using Corollary \ref{cor-errorest}, we obtain
	\eqn
	[a_0,\cHcubt(s)] &=& f[J_1(s)]+g[J_2(s)]+g^*[J_3(s)]+J_4(s) \, , \label{eq-a0cub-exp}
	\eeqn 
	where, with the notation of  Corollary \ref{cor-errorest},
	\eqn
	J_1(s,p) & :=& \frac{\lambda\hat{v}(p)}{\sqrt{N}}\big(|u_\lambda(s,p)|^2+|v_\lambda(s,p)|^2+2\overline{v_\lambda}(s,p)\overline{u_\lambda}(s,p)\big) \, , \label{def-j1}\\
	J_2(s,p)&:=&\frac{\lambda\hat{v}(p)}{\sqrt{N}}\big(v_\lambda(s,p)+\overline{u_\lambda}(s,p)\big)\overline{u_\lambda}(s,p) \, ,\\
	J_3(s,p) &:=& \frac{\lambda\hat{v}(p)}{\sqrt{N}}\big(u_\lambda(s,p)+\overline{v_\lambda}(s,p)\big)\overline{v_\lambda}(s,p) \, ,\\
	J_4(s) &:=& \frac{\lambda}{\sqrt{N}}\int \, dp \, \hat{v}(p)\big(|v_\lambda(s,p)|^2+\overline{u_\lambda}(s,p)\overline{v_\lambda}(s,p)\big) \, . \label{def-j4}
	\eeqn 
	Note, that in \eqref{eq-a0cub-exp}, we used the notation
	\eqn
	g^*[J] \, = \, \int \, dp \, J(p) a_p^+a_{-p}^+ \, . \label{def-gbar}
	\eeqn
	Using Corollary \ref{cor-errorest} again, we find that 
	\eqn
	\nd{J_i(s)} & \leq & C_{\vd}\frac{\lambda}{\sqrt{N}}\quad\mbox{for\ } i\in\{1,2,3\} \, ,\label{eq-j123-bd}\\
	|J_4(s)| &\leq& C_{\vd}\frac{\lambda}{\sqrt{N}}\label{eq-j4-bd}
	\eeqn
	for all $s\leq T\lambda^{-2}$. 
	\par Similarly, we compute
	\eqn
	[a_0,\cHquart(s)] &=& [a_0,\cHquart](s) \nonumber\\
	&=& \frac{\lambda}{N} \int \, dp \, \hat{v}(p_2)(a_{p_1+p_2}^+a_{p_2}a_{p_1})(s) \nonumber\\
	&=& \frac{1}{\sqrt{N}} \cHcubt^{(2)}(s) \, .\label{eq-a0-h4-0}
	\eeqn 
	Here, we used the notation
	\eqn
	\cHcubt(s) & = & \frac{\lambda}{\sqrt{N}}\int \, dp \, \hat{v}(p_2)\big((a_{p_1}^+a_{p_2}^+a_{p_1+p_2})(s)+(a_{p_1+p_2}^+a_{p_2}a_{p_1})(s)\big) \nonumber\\
	&=:& \cHcubt^{(1)}(s)+\cHcubt^{(2)}(s) \, .  \label{eq-hcub-dec-0}
	\eeqn 
	Notice that bounds of the form \eqref{eq-r2error-exp-0}--\eqref{eq-r2error-exp-2} hold with $\cHcubt$ replaced by $\cHcubt^{(2)}$, due to the proof of Proposition \ref{prop-h2-est}.
	\par Using \eqref{eq-phi-error-exp-0} and \eqref{eq-a0-h4-0}, we have that
	\eqn
	\Phi_t & = & -\frac{i}{\vol}\int_0^t \, ds \,  \nu_0\big([a_0,\cHcubt(s)]\big) \, + \, \Rem_2(t;\Phi) \, ,
	\eeqn 
	where 
	\eqn
	\Rem_2(t;\Phi) & :=& -\frac{i}{2\sqrt{N}}\int_0^t \, ds \, \frac{\nu_0(\cHcubt^{(2)}(s))}{\vol}\nonumber\\
	&& -\int_{\Delta[t,2]} \, d\mathbf{s}_2 \frac{\nu_{s_2}\big([f[J_1(s_1)]+g[J_2(s_1)]+g^*[J_3(s_1)],\cH_I(s_2)]\big)}{\vol} \nonumber\\
	&&+\frac1{2\sqrt{N}}\int_{\Delta[t,2]} \, d\mathbf{s}_2 \, \frac{\nu_{s_2}\big([\cHcubt^{(2)}(s_1),\cH_I(s_2)]\big)}{\vol } \, . \label{eq-a0-tail-exp-0}
	\eeqn 
	Recall that, by definition \eqref{def-gbar},
	\eqn
	|\nu_{s_2}([g^*[J_3(s_1)],\cH_I(s_2)])|  \, = \, |\nu_{s_2}([g[\overline{J_3}(s_1)],\cH_I(s_2)])| \, . 
	\eeqn 
	Then Lemma \ref{lem-fgphi-bound} together with the bounds \eqref{eq-r2error-exp-1} and \eqref{eq-j123-bd} imply 
	\eqn
	\lefteqn{\frac{\big| \nu_{s_2}\big([f[J_1(s_1)],\cH_I(s_2)]\big)\big|}{\vol} \, , \, \frac{\big|\nu_{s_2}\big([g[J_2(s_1)],\cH_I(s_2)]\big)\big|}{\vol} \, ,} \nonumber\\
	&& \frac{\big|\nu_{s_2}\big([g^*[J_3(s_1)],\cH_I(s_2)]\big)\big|}{\vol}\nonumber\\
	& \leq &  \frac{C_{\vd,\fd}e^{C\vd\vol/\lambda T}\lambda^2\vol^{\frac32}}{N}\Big(1+\sqrt{\frac\vol{N}}\Big) \, . \label{eq-a0-tail-1}
	\eeqn 
	\eqref{eq-nu0-odd-0} with $A=\1$ implies
	\eqn
	\frac{|\nu_0(\cHcubt^{(2)}(s))|}{\vol\sqrt{N}} \, = \, 0 \, . \label{eq-a0-tail-2}
	\eeqn 
	\eqref{eq-r2error-exp-1} with $A=\1$ implies
	\eqn
	\lefteqn{\frac{\big|\nu_{s_2}\big([\cHcubt^{(2)}(s_1),\cH_I(s_2)]\big)\big|}{\vol \sqrt{N}}}\nonumber\\
	&\leq & \frac{C_{\vd,\vd}e^{C\vd\vol/\lambda T}\lambda^2\vol^2}{N^{\frac32}}\Big(1+\sqrt{\frac\vol{N}}\Big) \, . \label{eq-a0-tail-3}
	\eeqn 
	Collecting \eqref{eq-a0-tail-exp-0}, \eqref{eq-a0-tail-1}, \eqref{eq-a0-tail-2}, and \eqref{eq-a0-tail-3}, yields the bound
	\eqn
	\lefteqn{|\Rem_2(t;\Phi)|}\nonumber\\
	&\leq & C_{\vd,\fd}e^{C\vd\vol/\lambda T}\frac{\lambda^2t^2 \vol^{\frac32}}{N}\Big(1+\sqrt{\frac\vol{N}}\Big) ^2\nonumber\\
	&\leq & C_{\vd,\fd}T^2e^{C\vd\vol/\lambda T}\frac{\vol^{\frac32}}{N\lambda^2}\Big(1+\frac\vol{N}\Big)
	\eeqn 
	for all $t\leq T\lambda^{-2}$, $\lambda\in(0,1)$.
	\par In the cases $A=f[J] $ and $A=g[J] $, we have that
	\eqn
	\frac{\nu_t(A)}\vol  & = &\frac{\nu_0(f[J] )}\vol \chif-i\int_0^t \, ds \,  \frac{\nu_0\big([A,\cHquart(s)]\big)}{\vol }\nonumber\\
	&& - \int_{\Delta[t,2]} \, d\mathbf{s}_2 \frac{\nu_0\big([[A,\cHcubt(s_1)],\cHcubt(s_2)]\big)}{\vol }\nonumber\\
	&& +\Rem_2(t;A)  \, ,
	\eeqn 	
	where, using \eqref{eq-being number conserving} and \eqref{eq-fg-error-exp-0},
	\eqn
	\Rem_2(t;A) & := & -i\int_0^t \, ds \,  \frac{\nu_0\big([A,\cHcubt(s)]\big)}\vol \nonumber\\
	&&- \int_{\Delta[t,2]} \, d\mathbf{s}_2 \frac{\nu_0\big([[A,\cHcubt(s_1)],\cHquart(s_2)]\big)}\vol \nonumber\\
	&&- \int_{\Delta[t,2]} \, d\mathbf{s}_2 \frac{\nu_0\big([[A,\cHquart(s_1)],\cH_I(s_2)]\big)}\vol \nonumber\\
	&& +i\int_{\Delta[t,3]} \, d\mathbf{s}_3 \frac{\nu_0\big([[[A,\cH_I(s_1)],\cH_I(s_2)],\cH_I(s_3)]\big)}{\vol} \nonumber\\
	&& + \int_{\Delta[t,4]} \, d\mathbf{s}_4 \frac{\nu_{s_4}\big([[[[A,\cH_I(s_1)],\cH_I(s_2)],\cH_I(s_3)],\cH_I(s_4)]\big)}{\vol}  \,.
	\eeqn 
	Employing \eqref{eq-r2error-exp-2}, Lemma \ref{lem-fgphi-bound} yields
	\eqn
	\frac{\big|\nu_0\big(A\cHquart^2(\bs_2)\big) \big|}\vol  & \leq & C_{\vd,\fd}\frac{\An{J}\lambda^2\vol^4}{N^2} \, , \label{eq-r2-a-bd-0}\\
	\frac{\big|\nu_0\big(A\cHcubt^2\cHquart(\bs_3)\big) \big|}\vol &\leq & C_{\vd,\fd}\frac{\An{J}\lambda^3\vol^5}{N^2} \, , \\
	\frac{\big|\nu_0\big(A\cHquart^3(\bs_3)\big) \big|}\vol &\leq & C_{\vd,\fd}\frac{\An{J}\lambda^3\vol^6}{N^3}  \, ,
	\eeqn 
	where we abbreviated
	\eqn
		\An{J} \, := \, \Jin\chif \, + \, \Jtn\chig \, .
	\eeqn 
	Using \eqref{eq-nu0-odd-0}, we have that
	\eqn
	\lefteqn{
		\frac{\big|\nu_0\big([A,\cHcubt(s)]\big)\big|}\vol\, = \, 
		\frac{\big|\nu_0\big(A\cHcubt\cHquart(\bs_2)\big)\big|}\vol}\nonumber\\
	&=&
	\frac{\big|\nu_0\big(A\cHcubt^3(\bs_3)\big) \big|}\vol \,= \, 
	\frac{\big|\nu_0\big(A\cHcubt\cHquart^2(\bs_3)\big)\big|}\vol \, = \, 0 \, .
	\eeqn 
	\eqref{eq-r2error-exp-0} together with Lemma \ref{lem-fgphi-bound} implies 
	\eqn
	\lefteqn{\frac{\big|\nu_{s_2}\big(A\cH_I^4(\bs_4\big)\big|}\vol}\nonumber\\
	& \leq & C_{\vd,\fd}e^{C\vd\vol/\lambda T}\frac{\An{J}\lambda^4\vol^6}{N^2}\Big(1+\sqrt{\frac\vol{N}}\Big) ^4 \, .\label{eq-r2-a-bd-1}
	\eeqn 
	Collecting \eqref{eq-r2-a-bd-0}-\eqref{eq-r2-a-bd-1}, we arrive at
	\eqn
	\lefteqn{|\Rem_2(t;A)|}\nonumber\\
	&\leq& \frac{C_{\vd,\fd}e^{C\vd\vol/\lambda T}\An{J}\lambda^2 t^2 \vol^4}{N^2}\Big[ 1 \, + \, \lambda t \vol\Big(1+\frac{\vol}{N}\Big) \nonumber\\
	&& + \, \lambda^2 t^2\vol^2\Big(1+\sqrt{\frac\vol{N}}\Big) ^4\Big]\nonumber\\                                                                                     
	&\leq& C_{\vd,\fd} \jb{T}^4\An{J} \frac{e^{C\vd\vol/\lambda T} \vol^6}{\lambda^4N^2}\Big(\big(1+\frac{\vol}{N}\big)^2 + \frac{\lambda^2}{\vol^2} \Big) \nonumber\\
	&\leq& C_{\vd,\fd} \jb{T}^4\An{J} \frac{e^{C\vd\vol/\lambda T} \vol^6}{\lambda^4N^2}\Big(1 + \frac{\vol}{N}\Big)^2
	\eeqn 
	for all $t\leq T\lambda^{-2}$, $\lambda\in(0,1)$, $\vol\geq1$. This finishes the proof.
	\endprf

	\par Up until now, all calculations did not further distinguish the cases of fixed $\vol$ and $\vol$ growing with $N$. In the latter situation, we will approximate lattice sums over $\lattice$ with integrals over $\R^3$. It is crucial to note that oscillatory and dispersive properties differ fundamentally in these two cases. 
	
	\subsection{Fixed, $N$-independent lattice $\lattice\cong \Z^3$} 
	
	\subsubsection {Notation  \label{sec-fixed-notation}} 
	For the next result, we define
	\eqn
	\Delta_{cub} H(\mathbf{p_2}) &:= &H(p_1)+H(p_2)-H(p_1+p_2) \, ,\label{def-dhcub}\\
	\hb & := &1+h \, , \label{def-hbar}\\
	p_{12} &:=& p_1+p_2 \, . \label{def-p12}
	\eeqn
	Recall from Corollary \ref{cor-errorest} that 
	\eqn
	V_1(p)^2 & = & \frac{\hat{v}(p)^2}{\Omega(p)(\Omega(p)+E(p)+\lambda\hat{v}(p))} \, , \\
	V_2(p) & = & \frac{\hat{v}(p)}{\Omega(p)} \, .
	\eeqn 
	\par In order to describe the dynamics of $\Phi$, we introduce the condensate operator
	\eqn
	\cCd(h)(T;\lambda) &:=& -i\int_0^T \, dS \, \int_\lattice \, dp\, \hat{v}(p)h_{\frac{S}{\lambda^2}}(p) \, . \label{def-condensate-0} 
	\eeqn 
	For the dynamics of $f$, we define the generalized Boltzmann operators
	\eqn
	\lefteqn{\cBd(h)[J](T;\lambda)}\nonumber\\
	&:=& 
	\frac1{\lambda^2} \int_{\Delta[T,2]} \, d\bS_2 \, \int_{(\lattice)^2} \, d\bp_2 \, \cos\big(\frac{\Delta_{cub}\Omega(\bp_2)(S_1-S_2)}{\lambda^2}\big)\nonumber\\	 && (\hat{v}(p_1)+\hat{v}(p_2))^2\Delta_{cub}J(\bp_2) \big(\hb_{{\frac{S_2}{\lambda^2}}}(p_1)\hb_{\frac{S_2}{\lambda^2}}(p_2) h_{\frac{S_2}{\lambda^2}}(p_1+p_2)\nonumber\\
	&&- h_{\frac{S_2}{\lambda^2}}(p_1)h_{\frac{S_2}{\lambda^2}}(p_2)\hb_{\frac{S_2}{\lambda^2}}(p_1+p_2)\big) \, , \label{def-boltzmann-0}\\
	\lefteqn{ \bbf^{(1)}(h)[J](\bS_2/\lambda^2)}\nonumber\\
	& := & \frac{2}{\lambda^2} \Im  \int_{(\lattice)^2} \, d\bp_2 \, (\hat{v}(p_1)+\hat{v}(p_2))(\hat{v}(p_1)+\hat{v}(p_{12}))\nonumber\\
	&&V_2(p_1) e^{-i(\Omega(p_2)-\Omega(p_{12}))(S_1-S_2)/\lambda^2}\nonumber\\
	&&\Big(\big(-J(-p_1)+J(p_2)-J(p_{12})\big) e^{i\Omega(p_1)S_2/\lambda^2}\sin(\Omega(p_1)S_1/\lambda^2)\nonumber\\
	&&\big(\hb_{\frac{S_2}{\lambda^2}}(-p_1)h_{\frac{S_2}{\lambda^2}}(p_2)\hb_{\frac{S_2}{\lambda^2}}(p_{12})-h_{\frac{S_2}{\lambda^2}}(-p_1)\hb_{\frac{S_2}{\lambda^2}}(p_2)h_{\frac{S_2}{\lambda^2}}(p_{12})\big)\nonumber\\
	&&+ \, \dJcub(\bp_2)e^{i\Omega(p_1)S_1/\lambda^2}\sin(\Omega(p_1)S_2/\lambda^2)\nonumber\\
	&& \big(h_{\frac{S_2}{\lambda^2}}(p_1)h_{\frac{S_2}{\lambda^2}}(p_2)\hb_{\frac{S_2}{\lambda^2}}(p_{12})-\hb_{\frac{S_2}{\lambda^2}}(p_1)\hb_{\frac{S_2}{\lambda^2}}(p_2)h_{\frac{S_2}{\lambda^2}}(p_{12})\big)\Big) \, , \label{def-q1} \\
	\lefteqn{\cBd^{(1)}(h)[J](T;\lambda)}\nonumber\\
	&:=& \int_{\Delta[T,2]} d\bS_2 \,\bbf^{(1)}(h)[J](\bS_2/\lambda^2) \, , \label{def-cBd1} \\
	\lefteqn{\bbf^{(2,1)}(h)[J](\bS_2;\lambda)}\nonumber\\
	& := & \frac1{\lambda^2}\Im \int_{(\lattice)^2} \, d\bp_2 \, (\hat{v}(p_1)+\hat{v}(p_2))^2\dJcub(\bp_2) e^{i\Omega(\bp_2)(S_1-S_2)/\lambda^2}\nonumber\\
	&&\Big(V_1(p_1)^2\big(\sin(\Omega(p_1)S_1/\lambda^2)e^{-i\Omega(p_1)S_1/\lambda^2} -\sin(\Omega(p_1)S_2/\lambda^2)e^{i\Omega(p_1)S_2/\lambda^2}\big)\nonumber\\
	&&+ V_1(p_2)^2\big(\sin(\Omega(p_2)S_1/\lambda^2)e^{-i\Omega(p_2)S_1/\lambda^2} - \sin(\Omega(p_2)S_2/\lambda^2)e^{i\Omega(p_2)S_2/\lambda^2 }\big)\nonumber\\
	&&- V_1(p_{12})^2\big(\sin(\Omega(p_{12})S_1/\lambda^2)e^{i\Omega(p_{12})S_1/\lambda^2} - \sin(\Omega(p_{12})S_2/\lambda^2)e^{i\Omega(p_{12})S_2/\lambda^2}\big)\Big) \nonumber\\ 	&&\big(h_{\frac{S_2}{\lambda^2}}(p_1)h_{\frac{S_2}{\lambda^2}}(p_2)\hb_{\frac{S_2}{\lambda^2}}(p_{12}))-\hb_{\frac{S_2}{\lambda^2}}(p_1)\hb_{\frac{S_2}{\lambda^2}}(p_2)h_{\frac{S_2}{\lambda^2}}(p_{12})\big)\, , \label{def-cBd21}\\
	\lefteqn{\bbf^{(2,2)}(h)[J](\bS_2/\lambda^2)}\nonumber\\
	& := & \frac{2}{\lambda^2} \Re \int_{(\lattice)^2} \, d\bp_2 \, (\hat{v}(p_1)+\hat{v}(p_2))\nonumber\\
	&&\Big[\sin(\Omega(p_1)S_1/\lambda^2)\sin(\Omega(p_{12})S_1/\lambda^2) \nonumber\\
	&&e^{i\Omega(p_2)(S_1-S_2)/\lambda^2+i(\Omega(p_1)-\Omega(p_{12}))S_2/\lambda^2}\nonumber\\
	&&V_2(p_1)V_2(p_{12})(\hat{v}(p_2)+\hat{v}(p_{12})) \nonumber\\
	&&(-J(-p_1)+J(p_2)+J(-p_{12}))\nonumber\\
	&&\big(h_{\frac{S_2}{\lambda^2}}(-p_1)\hb_{\frac{S_2}{\lambda^2}}(p_2)\hb_{\frac{S_2}{\lambda^2}}(-p_{12})-\hb_{\frac{S_2}{\lambda^2}}(-p_1)h_{\frac{S_2}{\lambda^2}}(p_2)f(-p_{12})\big)\nonumber\\
	&&+ \, \sin(\Omega(p_1)S_1/\lambda^2)e^{i\Omega(p_2)(S_1-S_2)/\lambda^2-i\Omega(p_{12})S_1/\lambda^2} V_2(p_1) \nonumber\\
	&&(-J(-p_1)+J(p_2)-J(p_{12}))\nonumber\\
	&&\Big((\hat{v}(p_1)+ \hat{v}(p_2))V_2(p_1)\sin(\Omega(p_1)S_2/\lambda^2)e^{i\Omega(p_{12})S_2/\lambda^2}\nonumber\\
	&&+ \, (\hat{v}(p_2)+\hat{v}(p_{12}))V_2(p_{12}) \sin(\Omega(p_{12})S_2/\lambda^2)e^{-i\Omega(p_1)S_2/\lambda^2}\Big)\nonumber\\
	&&\big(h_{\frac{S_2}{\lambda^2}}(-p_1)\hb_{\frac{S_2}{\lambda^2}}(p_2)h_{\frac{S_2}{\lambda^2}}(p_{12})-\hb_{\frac{S_2}{\lambda^2}}(-p_1)h_{\frac{S_2}{\lambda^2}}(p_2)\hb_{\frac{S_2}{\lambda^2}}(p_{12})\big)\nonumber\\
	&&+ \Big(\frac12e^{-i(\Omega(p_1)+\Omega(p_2))S_2/\lambda^2}\sin(\Omega(p_{12})S_2/\lambda^2)V_2(p_{12})(\hat{v}(p_1)+\hat{v}(p_2))\nonumber\\
	&&+ \, e^{-i(\Omega(p_1)+\Omega(p_{12}))s_2}\sin(\Omega(p_2)S_2/\lambda^2)V_2(p_2)(\hat{v}(p_1)+\hat{v}(p_{12}))\Big)\nonumber\\
	&&\big(\hb_{\frac{S_2}{\lambda^2}}(p_1)\hb_{\frac{S_2}{\lambda^2}}(p_2)\hb_{\frac{S_2}{\lambda^2}}(-p_{12})-h_{\frac{S_2}{\lambda^2}}(p_1)h_{\frac{S_2}{\lambda^2}}(p_2)h_{\frac{S_2}{\lambda^2}}(-p_{12})\big) \Big] \, , \label{def-cBd22}\\
	\lefteqn{\bbf^{(2)}(h)[J](\bS_2/\lambda^2)}\nonumber\\
	& := & \bbf^{(2,1)}(h)[J](\bS_2;\lambda) + \bbf^{(2,2)}(h)[J](\bS_2/\lambda^2) \, , \label{def-q2} \\
	\lefteqn{\cBd^{(2)}(h)[J](T;\lambda)}\nonumber\\
	&:=& \int_{\Delta[T,2]} d\bS_2 \, \bbf^{(2)}(h)[J](\bS_2/\lambda^2) \, . \label{def-cBd2}
	\eeqn 
	
	\subsubsection{Results}
	
	\begin{proposition}\label{prop-bog-disc}
		Let $T>0$, $\lambda\in(0,1)$, $\vol\geq1$, and $J\in L^{\infty}(\R^3;\R)$. Then the following holds. 
		\begin{enumerate}
			\item \eqn
			-i\int_0^{T\lambda^{-2}} \, ds \,  \frac{\nu_0\big([a_0,\cHcubt(s)]\big)}{\vol} &=&  \frac{1}{N^{\frac12}\lambda}\cCd(f_0)(T;\lambda) \, + \, \rrb_{1,d}(\frac{T}{\lambda^2};\Phi) \, , \label{eq-rrb1d-def}
			\eeqn 
			where 
			\eqn
			|\rrb_{1,d}(\frac{T}{\lambda^2};\Phi)| \, \leq \, \frac{C_{\vd,\fd}T}{N^{\frac12}} \, ,
			\eeqn 
			
			\item \eqn
			\nu_0([f[J] ,\cHquart(s)]) \, = \, 0 \, ,
			\eeqn 
			
			\item \eqn
			&&-\int_{\Delta[T\lambda^{-2},2]} \, d\mathbf{s}_2 \frac{\nu_0\big([[f[J] ,\cHcubt(s_1)],\cHcubt(s_2)]\big)}{\vol }\nonumber\\
			&=&\frac1N\big(\cBd(f_0)[J](T;\lambda) \, + \, \lambda\cBd^{(1)}(f_0)[J](T;\lambda) \, + \, \lambda^2\cBd^{(2)}(f_0)[J](T;\lambda)\big)\nonumber\\
			&&+ \, J(0)|\Phi_{\frac{T}{\lambda^2}}|^2 \, + \, \rrbogb_{2,d}(\frac{T}{\lambda^2};f[J] ) \, + \, J(0)\rrbogc_{2,d}(\frac{T}{\lambda^2};f)\, , \label{eq-f-main-0}
			\eeqn 	
			where 
			\eqn
			|\rrbogb_{2,d}(\frac{T}{\lambda^2};f[J] )| &\leq&  C_{\vd,\fd}\|J\|_{\ell^\infty(\lattice)} T^2 \frac{\lambda}{N} \, , \\
			|\rrbogc_{2,d}(t;f)| &\leq &C_{\vd,\fd}e^{C\vd\vol/\lambda T}\nonumber\\
			&&\frac{\jb{T}^4\vol^{\frac32}}{\lambda^3N^{\frac32}}\Big(1+\frac\vol{N}\Big)^2\Big(1+\frac{\vol^{\frac32}}{\lambda N^{\frac12}}\Big)\, ,
			\eeqn 
			
			\item
			\eqn
			-i\int_0^{T\lambda^{-2}} \, ds\, \frac{\nu_0(g[J] ,\cHquart(s))}{\vol} \, = \, \frac{1}N\int_0^T \, dS \, \acqd(f_0)[J](S/\lambda^2) \, , \label{def-acquart}
			\eeqn 
			The absorption operator $\acqd(f_0)[J](S/\lambda^2)$ consists of terms of the form
			\eqn
			\lefteqn{(-i)^{\ell} \lambda^{j} \int \, dp \, dk \, e^{-i(m_1\Omega(p)+m_2\Omega(k))S/\lambda^2}(1+f(p)+f(-p))J(p)}\nonumber\\
			&&V_1(p)^{2\alpha_1}V_2(p)^{\alpha_2}\hat{v}(p- k) V_1(k)^{2\alpha_3}V_2(k)^{\alpha_4} (f_0(k)+\iota) \, ,
			\eeqn 
			with $\ell\in\N_0$, $\ell\leq 3$, $j\in\N_0$, $j \leq 7$, $m_1,m_2\in\{0,\pm2\}$, $\alpha_j\in\{0,1,2\}$ and $\iota\in\{0,1\}$. The integrand contains a factor $f_0(k)$ or a factor $V_2(k)$.
			
			\item \eqn
			\lefteqn{-\int_{\Delta[T\lambda^{-2},2]} \, d\bs_2 \frac{\nu_0\big([[g[J] ,\cHcubt(s_1)],\cHcubt(s_2)]\big)}\vol }\nonumber\\
			&=& \frac{1}{N\lambda^2}\int_{\Delta[T,2]} \, d\bS_2 \, \big( \bbgd(f_0)[J](\bS_2/\lambda^2) + \acd(f_0)[J](\bS_2/\lambda^2)\big)\nonumber\\
			&& + \, J(0) \Big(\Phi_{T\lambda^{-2}}\Big)^2 \, + \, J(0)\rrbogc_2(\frac{T}{\lambda^2};g) \, . \label{def-bbgd}
			\eeqn 
			Here, the collision operator $\bbgd(f_0)[J](\bs_2)$ consists of terms of the form
			\eqn
			\lefteqn{(-i)^{\ell_0}\lambda^{j_0}\int_{(\lattice)^3} \, d\bp_3 \, e^{-i\sum_{\ell=1}^2s_\ell\sum_{k=1}^3\sigma_{k,\ell}\Omega(p_k)}\delta(p_1+p_2-p_3)}\nonumber\\
			&&J(p_{j_1})\hat{v}(p_2)\hat{v}(p_{j_2})\prod_{k=1}^3V_1(p_k)^{2\alpha_{\ell_k}}V_2(p_k)^{\beta_{\ell_k}}\nonumber\\
			&&\Big(\prod_{k=1}^3\big(f_0(\tau_{k,1}p_k)+\frac{1+\tau_{k,2}}2\big) \, - \, \prod_{k=1}^3\big(f_0(-\tau_{k,1}p_k)+\frac{1-\tau_{k,2}}2\big) \Big) \, , \label{eq-bbgd-reg-0}
			\eeqn 
			where $\ell_0\in\N_0$, $\ell_0\leq 3$, $j_0\in\N_0$,$j_0\leq 12$, $\sigma_{k,\ell},\tau_{k,\ell}\in\{\pm 1\}$, $j_1,j_2\in\{1,2,3\}$ and $\alpha_{\ell_k},\beta_{\ell_k}\in\{0,1,2\}$. Any term contains a product of at least two of the functions $\hat{v}$, $f_0$, $V_1$, and $V_2$ depending on at least two of the momenta $\{p_1,p_2,p_3\}$, which implies that the integrand in \eqref{eq-bbgd-reg-0} exhibits the necessary regularity properties used in Propositions \ref{prop-bog-cont} and  \ref{prop-eff} below. The absorption operator $\acd(f_0)[J](\bs_2)$ consists of terms of the form
			\eqn
			\lefteqn{(-i)^{\ell} \lambda^{j}\int_{(\lattice)^2} \, dp\, dk \, e^{-is_1m_1\Omega(p)-is_2m_2\Omega(k)}J(p)(1+f_0(p)+f_0(-p))}\nonumber\\
			&&\hat{v}(p)V_1(p)^{2\alpha_1}V_2(p)^{\alpha_2} \hat{v}(k) V_1(k)^{2\alpha_3}V_2(k)^{\alpha_4}(f_0(k)+\iota) \, , \label{def-accub}
			\eeqn 
			where $\ell\in\N_0$, $\ell \leq 3$, $j\in\N_0$, $j\leq 12$, $m_1,m_2\in\{0,\pm2\}$, $\alpha_j\in\{0,1,2\}$ and $\iota\in\{0,1\}$. $|\rrbogc_2(\frac{T}{\lambda^2};g)|$ satisfies the same bound as $|\rrbogc_2(\frac{T}{\lambda^2};f)|$.
		\end{enumerate}
	\end{proposition}
	\prf
	Set $t=T\lambda^{-2}$. Since here all integrals range over $\lattice$ in the sense of \eqref{eq-disc-int}, we will omit the domain of integration in this proof. Using that $\nu_0(Q)=0$ for $[Q,\nb]\neq0$, \eqref{eq-a0cub-exp}, and recalling $J_j$ from \eqref{def-j1} and \eqref{def-j4}, we start by observing that
	\eqn
	-i\int_0^t \, ds \,  \frac{\nu_0\big([a_0,\cHcubt(s)]\big)}\vol \, = \, -i\int_0^t \, ds \, \big(\int \, dp \, f_0(p)J_1(s,p) \, + \, \frac{J_4(s)}\vol\big) \, .\label{eq-phiqf-exp-0}
	\eeqn 
	Here, we also used the fact that translation-invariance implies
	\eqn 
	\wick{
		\c a_p^{(\sigma)} \c a_q^{(-\sigma)} \, = \, \delta(p-q) (f_0(p)+\delta_{\sigma,-1}) \, .          
	} \label{eq-2con-0}
	\eeqn 
	Recall that, by Corollary \ref{cor-errorest},
	\eqn
	J_1(s,p) & =& \frac{\lambda\hat{v}(p)}{\sqrt{N}}\big(|u_\lambda(s,p)|^2 \, + \, |v_\lambda(s,p)|^2 \, + \, 2\overline{v_\lambda}(s,p)\overline{u_\lambda}(s,p)\big)\nonumber\\ 
	&=& \frac{\lambda\hat{v}(p)}{\sqrt{N}}\Big(1 \, + \, 2\lambda^2\Re\big(e^{-i\Omega(p)s}i\sin(\Omega(p)s)V_1(p)\big)\nonumber\\
	&&+ \, \lambda^4\sin^2(\Omega(p)s)V_1(p)^4 \, + \, |v_\lambda(s,p)|^2\nonumber\\
	&& + \, 2\overline{v_\lambda}(s,p)\overline{u_\lambda}(s,p) \Big)\, ,\\	 	
	J_4(s) &=& \frac{\lambda}{\sqrt{N}}\int \, dp \, \hat{v}(p)\big(|v_\lambda(s,p)|^2+\overline{u_\lambda}(s,p)\overline{v_\lambda}(s,p)\big) \, .
	\eeqn 
	Thus, again by Corollary \ref{cor-errorest}, we obtain
	\eqn
	\|J_1(s)+\frac{\lambda\hat{v}}{\sqrt{N}}\|_\infty &\leq& C\frac{\lambda\|\hat{v}\|_\infty}{\sqrt{N}}\Big(\lambda^2\|V_1\|_\infty^2 \,+ \, \lambda^4\|V_1\|_\infty^4 \, + \, \|v_\lambda(s,\cdot)\|_\infty^2\nonumber\\
	&& + \, \|v_\lambda(s,\cdot)\|_\infty\|u_\lambda(t,\cdot)\|_\infty \Big)\nonumber\\
	&\leq& C\frac{\lambda^2\vd^2(1+\lambda\vd)^3}{\sqrt{N}} \, ,\label{eq-J1-bd-0}\\
	|J_4(s)| &\leq & C\frac{\lambda\|\hat{v}\|_1}{\sqrt{N}}\big(\|v_\lambda(s,\cdot)\|_\infty^2 \, + \, \|v_\lambda(s,\cdot)\|_\infty\|u_\lambda(t,\cdot)\|_\infty\big)\nonumber\\
	&\leq& C\frac{\lambda^2\vd^2(1+\lambda\vd)^3}{\sqrt{N}} \label{eq-J4-bd-0}
	\eeqn 
	for all $s\geq0$. Then \eqref{eq-phiqf-exp-0}, Lemma \ref{lem-fgphi-bound}, and the definition \eqref{eq-rrb1d-def} of $\rrb_{1,d}(t;\Phi)$ yield
	\eqn
	|\rrb_{1,d}(t;\Phi)| & \leq & C\frac{\lambda^2t\vd^2(1+\lambda\vd)^3(\vol^{-1}+\|f_0\|_1)}{N^{\frac12}}\nonumber\\
	&\leq& \frac{C_{\vd,\fd}T}{N^{\frac12}} \, . \label{eq-phi-bog-err-1}
	\eeqn 
	Collecting \eqref{eq-phiqf-exp-0}, \eqref{eq-J1-bd-0}, \eqref{eq-J4-bd-0}, and \eqref{eq-phi-bog-err-1}, we have proved \eqref{eq-rrb1d-def}.
	\par Next, we compute the dynamics of $f$. We use the fact that $\nu_0$ is quasifree to obtain that
	\eqn
	\nu_0([a_p^+a_p,\cHquart(s)])&=&
	\wick{
		[\c1 a_p^+ \c2 a_p,  \settowidth{\wdth}{$\cHquart$}\hspace{0.4\wdth}\c2{\vphantom{\cHquart}} \hspace{-0.4\wdth}\underbracket{\c1\cHquart} ( s)]}
	\wick{+[\c1 a_p^+\c1 a_p,\underbracket{\cHquart}(s)]
	}\nonumber\\
	&&\propto f_0(p)\fbar(p)-\fbar(p)f_0(p) + 0\,=\,0 \, ,
	\eeqn 
	since, due to Corollary \eqref{cor-errorest}, $\cHquart(s)$ is a quartic polynomial.
	\par We observe that for self-adjoint operators $A$, $B$, $C$ and any state $\nu$ we have that
	\eqn
	\nu([[A,B],C]) = \nu\big(([[A,B],C])^+\big) = \Re \big(\nu([[A,B],C])\big) \, .
	\eeqn 
	Similarly to the previous case, we have
	\eqn\label{eq-zero-contr-0}
	[[\wick{\c1 f[\c2 J] ,\c1 \cHcubt(\c2 s_1)}],\cHcubt(s_2)]\,&=&\,[[\wick{\c1 f[\c2 J] ,\cHcubt(s_1)],\c1 \cHcubt(\c2 s_2)}]\nonumber\\
	&=&\,[[\wick{\c1 f[\c1 J] },\cHcubt(s_1)],\cHcubt(s_2)]
	\nonumber\\
	&=&\,0\,.
	\eeqn 
	This leaves us with exactly two types of possibles contractions. These are
	\eqn
	(-i)^2\wick{
		[[\c1 f[\c2 J] , \c1 \cHcubt(\c3 s_1)],  \c2 \cHcubt(\c3 s_2)] \label{eq-f-bol-con-0}
	}
	\eeqn 
	and, using translation invariance and employing Lemma \ref{lem-TI},
	\eqn
	\lefteqn{(-i)^2\wick{
			[[\c1 f[\c2 J] ,\underbracket{\c1 \cHcubt}(s_1)],\underbracket{\c2 \cHcubt}(s_2)]
	}}\nonumber\\
	& = & -\frac{2J(0)}\vol \Re\Big(
	\wick{
		[\c1 a_0,\underbracket{\c1 \cHcubt}(s_1)]
		[\c2 a_0^+,\underbracket{\c2 \cHcubt}(s_2)] 
	}
	\Big)\, .\label{eq-f-condensate-0}
	\eeqn 
	\eqref{eq-f-bol-con-0} corresponds to a scattering or Boltzmann term, while \eqref{eq-f-condensate-0} describes corrections to the evolution of the condensate. 
	
	\par We start by analyzing the condensate term. We have that 
	\eqn
	\lefteqn{\frac{-2J(0)}{\vol^2} \Re\int_{\Delta[t,2]} \, d\bs_2\, \wick{
			[\c1 a_0,\underbracket{\c1 \cHcubt}(s_1)]
			[\c2 a_0^+,\underbracket{\c2 \cHcubt}(s_2)]
	}} \nonumber\\
	&=& \frac{-J(0)}{\vol^2}\int_{\Delta[t,2]} \, d\bs_2\, \Big( 
	\wick{
		[\c1 a_0,\underbracket{\c1 \cHcubt}(s_1)]
		[\c2 a_0^+,\underbracket{\c2 \cHcubt}(s_2)]
	} 
	\nonumber\\
	&& + \, \wick{
		[\c1 a_0,\underbracket{\c1 \cHcubt}(s_2)]
		[\c2 a_0^+,\underbracket{\c2 \cHcubt}(s_1)] \Big)	
	}\nonumber\\
	&=& -\frac{J(0)}{2\vol^2} \int \, d\bs_2\, \Big( 
	\wick{
		[\c1 a_0,\underbracket{\c1 \cHcubt}(s_1)]
		[\c2 a_0^+,\underbracket{\c2 \cHcubt}(s_2)]
	} 
	\nonumber\\
	&& + \, \wick{
		[\c1 a_0,\underbracket{\c1 \cHcubt}(s_2)]
		[\c2 a_0^+,\underbracket{\c2 \cHcubt}(s_1)] \Big)	
	}\nonumber\\
	&=& J(0) \Big|-i\int_0^t\, \frac{ds}\vol\,\wick{
		[\c1 a_0,\underbracket{\c1 \cHcubt}(s)]
	}\Big|^2 \, .
	\eeqn 
	Here, we used the fact that
	\eqn
	\overline{
		\wick{
			[\c1 a_0,\underbracket{\c1 \cHcubt}(s)]
		}
	} \, = \, -\wick{
		[\c1 a_0^+,\underbracket{\c1 \cHcubt}(s)] \, .
	}
	\eeqn 
	Next, we apply quasifreeness of $\nu_0$, followed by Proposition \ref{prop-tail-est}, to get that
	\eqn 
	-i\int_0^t\, ds\,\frac{
		\wick{
			[\c1 a_0,\underbracket{\c1 \cHcubt}(s)]
		}
	}\vol &= & -i\int_0^t\, ds\, \frac{\nu_0([ a_0,\cHcubt(s)])}\vol \nonumber\\
	&=& \Phi_t \, - \, \Rem_2(t;\Phi) \, . \label{eq-phi-f-dep-0}
	\eeqn 
	In particular, we have that the condensate term is given by
	\eqn 
	J(0)|\Phi_t|^2+J(0)\rrbogc_{2,d}(t;f) \, ,
	\eeqn 
	where 
	\eqn
	\rrbogc_{2,d}(t;f) & = & -2\Re\Big((-i)\int_0^t\, ds\, \frac{\nu_0([ a_0,\cHcubt(s)])}\vol\overline{\Rem_2(t;\Phi)}\Big)\nonumber\\
	&&-\,|\Rem_2(t;\Phi)|^2 \, . \label{eq-f-con-err-0}
	\eeqn 
	Using \eqref{eq-phiqf-exp-0}, \eqref{eq-J1-bd-0}, and \eqref{eq-J4-bd-0}, we have that
	\eqn
	\Big|\int_0^t\, ds\, \frac{\nu_0([ a_0,\cHcubt(s)])}\vol\Big| &\leq& C_{\vd}\frac{\lambda t (1+\fd)}{\sqrt{N}} \nonumber\\
	&\leq& \frac{C_{\vd,\fd}T}{N^{\frac12}\lambda}	\label{eq-con-main-est-0}
	\eeqn 
	for all $\lambda\in(0,1)$. Thus, Proposition \ref{prop-tail-est} and \eqref{eq-con-main-est-0} yield 
	\eqn
	\lefteqn{|\rrbogc_{2,d}(t;f)|}\nonumber\\
	&\leq & C_{\vd,\fd}\Big[\frac{T}{N^{\frac12}\lambda} T^2e^{C\vd\vol/\lambda T}\frac{\vol^{\frac32}}{N\lambda^2}\Big(1+\frac\vol{N}\Big)\nonumber\\
	&&+\,  T^4e^{C\vd\vol/\lambda T}\frac{\vol^3}{N^2\lambda^4}\Big(1+\frac\vol{N}\Big)^2\Big] \nonumber\\
	&\leq & C_{\vd,\fd}e^{C\vd\vol/\lambda T}\frac{\jb{T}^4\vol^{\frac32}}{\lambda^3N^{\frac32}}\Big(1+\frac\vol{N}\Big)^2\Big(1+\frac{\vol^{\frac32}}{\lambda N^{\frac12}}\Big) \, .\label{eq-f-condensate-1}
	\eeqn 
	\par Next, we compute the Boltzmann term for $f$. Observe that with the decomposition \eqref{eq-hcub-dec-0}, we have that
	\eqn
	\cHcubt(t) \, = \, \cHcubt^{(1)}(t) \, + \, \cHcubt^{(2)}(t) \, .
	\eeqn 
	Notice that $\cHcubt^{(1)}(t)$ and $\cHcubt^{(2)}(t)$ are formal adjoints. In particular, we have that
	\eqn
	\lefteqn{\nu_0\big([[f[J] ,\cHcubt(s_1)],\cHcubt(s_2)]\big)}\nonumber\\
	&=& 2\Re \nu_0\big([[f[J] ,\cHcubt^{(1)}(s_1)],\cHcubt^{(1)}(s_2)]\big) \nonumber\\
	&&+ \, 2\Re \nu_0\big([[f[J] ,\cHcubt^{(1)}(s_1)],\cHcubt^{(2)}(s_2)]\big) \label{eq-f-disc-exp-0}
	\eeqn 
	With that, we sort the Boltzmann contractions of $\nu_0\big([[f[J] ,\cHcubt(s_1)],\cHcubt(s_2)]\big)$ by powers of $\lambda$, i.e., 
	\eqn
	- \frac1\vol 
	\wick{
		[[\c1 f[\c2 J], \c1 \cHcubt(\c3 s_1)],\c2 \cHcubt(\c3 s_2)]
	}&
	= & \frac{\lambda^2}N \big(\bbf^{(0)}(f_0)[J](s_1,s_2) \nonumber\\
	&& + \, \lambda\bbf^{(1)}(f_0)[J](s_1,s_2)\nonumber\\
	&&+ \, \lambda^2 \bbf^{(2)}(f_0)[J](s_1,s_2) \nonumber\\
	&&+ \, \lambda^3 \bbf^{(3)}(f_0)[J](s_1,s_2)\big) \, , \label{eq-f-bol-dec-0}
	\eeqn
	Notice that the CCR imply 
	\eqn
	[f[J] ,a_{\sigma_1 p_1}^{(\sigma_1)}a_{\sigma_2 p_2}^{(\sigma_2)}a_{\sigma_3 p_{12}}^{(-\sigma_3)}] \, = \, \sum_{j=1}^3\sigma_j(-1)^{\delta_{j,3}} J(p_j) a_{\sigma_1 p_1}^{(\sigma_1)}a_{\sigma_2 p_2}^{(\sigma_2)}a_{\sigma_3p_{12}}^{(-\sigma_3)} \, . \label{eq-fj-com-0}
	\eeqn 
	We have that
	\eqn
	\bbf^{(0)}(f_0)[J](\bs_2) & = & -2\Re\int \, d\bp_2  d\bk_2 \, e^{i\Delta_{cub}\Omega(\bp_2)s_1-i \Delta_{cub}\Omega(\bk_2) s_2} \nonumber\\
	&&\hat{v}(p_2)\hat{v}(k_2) \Delta_{cub}J(\bp_2)\nonumber\\
	&&\frac1\vol\Big( 
	\wick{
		[\c1 a_{p_1}^+\c2 a_{p_2}^+ \c3 a_{p_{12}},\c3 a_{k_{12}}^+\c2 a_{k_2} \c1 a_{k_1}]
	} \, + \,
	\wick{
		[\c1 a_{p_1}^+\c2 a_{p_2}^+ \c3 a_{p_{12}},\c3 a_{k_{12}}^+\c1 a_{k_2} \c2 a_{k_1}]
	}\nonumber\\
	& = & -2\Re\int d\bp_2 e^{i\Delta_{cub}\Omega(\bp_2)(s_1-s_2)}\hat{v}(p_2)\nonumber\\
	&&(\hat{v}(p_1)+\hat{v}(p_2)) \Delta_{cub}J(\bp_2)\nonumber\\
	&&\big(f_0(p_1)f_0(p_2)\fbar(p_{12})-\fbar(p_1)\fbar(p_2)f_0(p_{12})\big)\nonumber\\
	& = & \Re\int d\bp_2 e^{i\Delta_{cub}\Omega(\bp_2)(s_1-s_2)}(\hat{v}(p_1)+\hat{v}(p_2))^2 \Delta_{cub}J(\bp_2)\nonumber\\	  
	&&\big(\fbar(p_1)\fbar(p_2)f_0(p_{12}) - f_0(p_1)f_0(p_2)\fbar(p_{12})\big) \, . \label{eq-f-bol-0}
	\eeqn  
	Here, we used \eqref{eq-2con-0} followed by symmetry $p_1\leftrightarrow p_2$. 
	\par Recalling Corollary \ref{cor-errorest}, $\bbf^{(1)}(f_0)[J](s_1,s_2)$ is given by the Boltzmann contractions of
	\eqn
	\lefteqn{-\frac{2}{\vol} \Re \, i \int \, d\bp_2 d\bk_2 \, \hat{v}(p_2)\hat{v}(k_2) e^{i\dOcub(\bp_2)s_1+i\dOcub(\bk_2)s_2}} \nonumber\\
	&&\Big( e^{-i\Omega(p_1)s_1}\sin(\Omega(p_1)s_1)V_2(p_1) \nu_0\big([[f[J] ,a_{-p_1}a_{p_2}^+a_{p_{12}}],a_{k_1}^+a_{k_2}^+a_{k_{12}}]\big) \nonumber\\
	&&+ \, e^{-i\Omega(p_2)s_1}\sin(\Omega(p_2)s_1)V_2(p_2) \nu_0\big([[f[J] ,a_{p_1}^+a_{-p_2}a_{p_{12}}],a_{k_1}^+a_{k_2}^+a_{k_{12}}]\big) \nonumber\\
	&&+ \, e^{-i\Omega(k_1)s_2}\sin(\Omega(k_1)s_2)V_2(k_1) \nu_0\big([[f[J] ,a_{p_1}^+a_{p_2}^+a_{p_{12}}],a_{-k_1}a_{k_2}^+a_{k_{12}}]\big) \nonumber\\
	&& + \, e^{-i\Omega(k_2)s_2}\sin(\Omega(k_2)s_2)V_2(k_2) \nu_0\big([[f[J] ,a_{p_1}^+a_{p_2}^+a_{p_{12}}],a_{k_1}^+a_{-k_2}a_{k_{12}}]\big)\Big) 
	\eeqn 
	Using \eqref{eq-fj-com-0}, obtain, similarly to above,
	\eqn
	\lefteqn{2 \Im \int \, d\bp_2 \, \hat{v}(p_2)
		\Big[\big(-J(-p_1)+J(p_2)-J(p_{12})\big)}\nonumber\\ &&(\hat{v}(p_1)+\hat{v}(p_{12}))		e^{i(\Omega(p_2)-\Omega(p_{12}))(s_1-s_2)+i\Omega(p_1)s_2}\sin(\Omega(p_1)s_1)V_2(p_1)\nonumber\\
	&&\big(\fbar(-p_1)f_0(p_2)\fbar(p_{12})-f_0(-p_1)\fbar(p_2)f_0(p_{12})\big)\nonumber\\
	&&+ \, \big(J(p_1)-J(-p_2)-J(p_{12})\big) (\hat{v}(p_2)+\hat{v}(p_{12}))\nonumber\\
	&&e^{i(\Omega(p_1)-\Omega(p_{12}))(s_1-s_2)+i\Omega(p_2)s_2}\sin(\Omega(p_2)s_1)V_2(p_2)\nonumber\\
	&&\big(f_0(p_1)\fbar(-p_2)\fbar(p_{12})-\fbar(p_1)f_0(-p_2)f_0(p_{12})\big)\nonumber\\
	&&+ \, \Big(e^{i(\Omega(p_1)-\Omega(p_{12}))(s_1-s_2)+i\Omega(p_2)s_1}\sin(\Omega(p_2)s_2)V_2(p_2)(\hat{v}(p_2)+\hat{v}(p_{12}))\nonumber\\
	&&+\, e^{i(\Omega(p_2)-\Omega(p_{12}))(s_1-s_2)+i\Omega(p_1)s_1}\sin(\Omega(p_1)s_2)V_2(p_1)(\hat{v}(p_1)+\hat{v}(p_{12}))\Big)\nonumber\\
	&&\dJcub(\bp_2)\big(f_0(p_1)f_0(p_2)\fbar(p_{12})-\fbar(p_1)\fbar(p_2)f_0(p_{12})\big)\Big] \, .
	\eeqn 	
	Here, we used the fact that $\hat{v}$, and thus $\Omega$, are even functions. Using symmetry in $p_1\leftrightarrow p_2$, we can further simplify this expression to
	\eqn 
	\lefteqn{\bbf^{(1)}(f_0)[J](\bs_2)}\nonumber\\
	& :=&  
	2\Im \int \, d\bp_2 \, (\hat{v}(p_1)+\hat{v}(p_2))(\hat{v}(p_1)+\hat{v}(p_{12}))\nonumber\\
	&&V_2(p_1) e^{i(\Omega(p_2)-\Omega(p_{12}))(s_1-s_2)}\nonumber\\
	&&\Big(\big(-J(-p_1)+J(p_2)-J(p_{12})\big) e^{i\Omega(p_1)s_2}\sin(\Omega(p_1)s_1)\nonumber\\
	&&\big(\fbar(-p_1)f_0(p_2)\fbar(p_{12})-f_0(-p_1)\fbar(p_2)f_0(p_{12})\big)\nonumber\\
	&&+ \, \dJcub(\bp_2)e^{i\Omega(p_1)s_1}\sin(\Omega(p_1)s_2)\nonumber\\
	&& \big(f_0(p_1)f_0(p_2)\fbar(p_{12})-\fbar(p_1)\fbar(p_2)f_0(p_{12})\big)\Big) \, . \label{eq-f-bol-cor-1}
	\eeqn 
	Recalling \eqref{def-cBd1}, we have that
	\eqn
		\int_{\Delta[T\lambda^{-2},2]} d\bs_2 \, \bbf^{(1)}(f_0)[J](\bs_2) \, = \, \cBd^{(1)}(f_0)[J](T;\lambda) \, .
	\eeqn 
	Next, we compute the corrections with coefficient $\frac{\lambda^4}N$. These come either from a correction to the same momentum, $a_p^+\to i\lambda^2V_1(p)^2\sin(\Omega(p)t)a_p^+$, or from two momentum flips, $a_p^+\to i\lambda V_2(p) \sin(\Omega(p)t)a_{-p}$. In the first case, computations analogous to \eqref{eq-f-bol-0} yield
	\eqn
	\lefteqn{-2\Re \, i \int \, d\bp_2 \, \hat{v}(p_2) (\hat{v}(p_1)+\hat{v}(p_2))\dJcub(\bp_2) e^{i\dOcub(\bp_2)(s_1-s_2)}}\nonumber\\
	&&\Big(V_1(p_1)^2\big(\sin(\Omega(p_1)s_1)e^{-i\Omega(p_1)s_1} -\sin(\Omega(p_1)s_2)e^{i\Omega(p_1)s_2}\big)\nonumber\\
	&&+\, V_1(p_2)^2\big(\sin(\Omega(p_2)s_1)e^{-i\Omega(p_2)s_1} - \sin(\Omega(p_2)s_2)e^{i\Omega(p_2)s_2 }\big)\nonumber\\
	&& - \, V_1(p_{12})^2\big(\sin(\Omega(p_{12})s_1)e^{i\Omega(p_{12})s_1} - \sin(\Omega(p_{12})s_2)e^{i\Omega(p_{12})s_2}\big)\Big) \nonumber\\
	&&\big(f_0(p_1)f_0(p_2)\fbar(p_{12})-\fbar(p_1)\fbar(p_2)f_0(p_{12})\big) \, .
	\eeqn   
	Using symmetry in $p_1\leftrightarrow p_2$ again, we obtain
	\eqn
	\lefteqn{\bbf^{(2)}_1(f_0)[J](\bs_2)}\nonumber\\
	& := & 
	\Im \int \, d\bp_2 \, (\hat{v}(p_1)+\hat{v}(p_2))^2\dJcub(\bp_2) e^{i\dOcub(\bp_2)(s_1-s_2)}\nonumber\\
	&&\Big(V_1(p_1)^2\big(\sin(\Omega(p_1)s_1)e^{-i\Omega(p_1)s_1} -\sin(\Omega(p_1)s_2)e^{i\Omega(p_1)s_2}\big)\nonumber\\
	&&+\, V_1(p_2)^2\big(\sin(\Omega(p_2)s_1)e^{-i\Omega(p_2)s_1} - \sin(\Omega(p_2)s_2)e^{i\Omega(p_2)s_2 }\big)\nonumber\\
	&& - \, V_1(p_{12})^2\big(\sin(\Omega(p_{12})s_1)e^{i\Omega(p_{12})s_1} - \sin(\Omega(p_{12})s_2)e^{i\Omega(p_{12})s_2}\big)\Big) \nonumber\\
	&&\big(f_0(p_1)f_0(p_2)\fbar(p_{12})-\fbar(p_1)\fbar(p_2)f_0(p_{12})\big) \, . \label{eq-f-bol-cor-22}	
	\eeqn 
	In the case of two momentum flips, the correction terms are given by the Boltzmann contractions of
	\eqn
	-\frac{2}{\vol} \Re \int \, d\bp_2 d\bk_2 \, \hat{v}(p_2)\hat{v}(k_2) e^{i\dOcub(\bp_2)s_1-i\dOcub(\bk_2)s_2} \,\nonumber\\
	\Big(e^{-i(\Omega(p_1)-\Omega(p_{12}))s_1}i\sin(\Omega(p_1)s_1)(-i)\sin(\Omega(p_{12})s_1) V_2(p_1)V_2(p_{12})\nonumber\\
	(-J(-p_1)+J(p_2)+J(-p_{12}))\nu_0([a_{-p_1}a_{p_2}^+a_{-p_{12}}^+,a_{k_{12}}^+a_{k_2}a_{k_1}])\nonumber\\
	+ \, e^{-i(\Omega(p_2)-\Omega(p_{12}))s_1}i\sin(\Omega(p_2)s_1)(-i)\sin(\Omega(p_{12})s_1) V_2(p_2)V_2(p_{12})\nonumber\\
	(J(p_1)-J(-p_2)+J(-p_{12}))\nu_0([a_{p_1}^+a_{-p_2}a_{-p_{12}}^+,a_{k_{12}}^+a_{k_2}a_{k_1}])\nonumber\\
	+ \, e^{-i\Omega(p_1)s_1+i\Omega(k_2)s_2}i\sin(\Omega(p_1)s_1)(-i)\sin(\Omega(k_2)s_2) V_2(p_1)V_2(k_2)\nonumber\\
	(-J(-p_1)+J(p_2)-J(p_{12}))\nu_0([a_{-p_1}a_{p_2}^+a_{p_{12}},a_{k_{12}}^+a_{-k_2}^+a_{k_1}])\nonumber\\
	+ \, e^{-i\Omega(p_2)s_1+i\Omega(k_2)s_2}i\sin(\Omega(p_2)s_1)(-i)\sin(\Omega(k_2)s_2) V_2(p_2)V_2(k_2)\nonumber\\
	(J(p_1)-J(-p_2)-J(p_{12}))\nu_0([a_{p_1}^+a_{-p_2}a_{p_{12}},a_{k_{12}}^+a_{-k_2}^+a_{k_1}])\nonumber\\
	+ \, e^{-i\Omega(p_1)s_1+i\Omega(k_1)s_2}i\sin(\Omega(p_1)s_1)(-i)\sin(\Omega(k_1)s_2) V_2(p_1)V_2(k_1)\nonumber\\
	(-J(-p_1)+J(p_2)-J(p_{12}))\nu_0([a_{-p_1}a_{p_2}^+a_{p_{12}},a_{k_{12}}^+a_{k_2}a_{-k_1}^+]) \nonumber\\
	+ \, e^{-i\Omega(p_2)s_1+i\Omega(k_1)s_2}i\sin(\Omega(p_2)s_1)(-i)\sin(\Omega(k_1)s_2) V_2(p_2)V_2(k_1)\nonumber\\
	(J(p_1)-J(-p_2)-J(p_{12}))\nu_0([a_{p_1}^+a_{-p_2}a_{p_{12}},a_{k_{12}}^+a_{k_2}a_{-k_1}^+])\nonumber\\
	+ \, e^{i\Omega(p_{12})s_1-i\Omega(k_{12})s_2} (-i)\sin(\Omega(p_{12})s_1) i\sin(\Omega(k_{12})s_2)V_2(p_{12}) V_2(k_{12})\nonumber\\
	(J(p_1)+J(p_2)+J(-p_{12}))\nu_0([a_{p_1}^+a_{p_2}^+a_{-p_{12}}^+,a_{-k_{12}}a_{k_2}a_{k_1}])\Big) \, .
	\eeqn 
	This expression equals
	\eqn
	\lefteqn{-2 \Re\int \, d\bp_2 \, \hat{v}(p_2)
		\Big[\sin(\Omega(p_1)s_1)\sin(\Omega(p_{12})s_1) e^{i\Omega(p_2)(s_1-s_2)+i(\Omega(p_1)-\Omega(p_{12}))s_2}} \nonumber\\
	&&V_2(p_1)V_2(p_{12})(\hat{v}(p_2)+\hat{v}(p_{12})) (-J(-p_1)+J(p_2)+J(-p_{12}))\nonumber\\
	&&\big(\fbar(-p_1)f_0(p_2)f(-p_{12})-f_0(-p_1)\fbar(p_2)\fbar(-p_{12})\big)\nonumber\\
	&&+ \, \sin(\Omega(p_2)s_1)\sin(\Omega(p_{12})s_1) e^{i\Omega(p_1)(s_1-s_2)+i(\Omega(p_2)-\Omega(p_{12}))s_2} \nonumber\\
	&&V_2(p_2)V_2(p_{12})(\hat{v}(p_1)+\hat{v}(p_{12})) (J(p_1)-J(-p_2)+J(-p_{12}))\nonumber\\
	&&\big(f_0(p_1)\fbar(-p_2)f(-p_{12})-\fbar(p_1) f_0(-p_2)\fbar(-p_{12})\big)\nonumber\\
	&&+ \, \sin(\Omega(p_1)s_1)e^{i\Omega(p_2)(s_1-s_2)-i\Omega(p_{12})s_1} V_2(p_1) (-J(-p_1)+J(p_2)-J(p_{12}))\nonumber\\
	&&\big(\hat{v}(p_1)V_2(p_1)\sin(\Omega(p_1)s_2)e^{i\Omega(p_{12})s_2}+ \hat{v}(p_{12})V_2(p_{12}) \sin(\Omega(p_{12})s_2)e^{-i\Omega(p_1)s_2}\big)\nonumber\\
	&&\big(\fbar(-p_1)f_0(p_2)\fbar(p_{12})-f_0(-p_1)\fbar(p_2)f_0(p_{12})\big)\nonumber\\
	&&+ \, \sin(\Omega(p_2)s_1)e^{i\Omega(p_1)(s_1-s_2)-i\Omega(p_{12})s_1} V_2(p_2) (J(p_1)-J(-p_2)-J(p_{12}))\nonumber\\
	&&\big(\hat{v}(p_2)V_2(p_2)\sin(\Omega(p_2)s_2)e^{i\Omega(p_{12})s_2}+ \hat{v}(p_{12})V_2(p_{12}) \sin(\Omega(p_{12})s_2)e^{-i\Omega(p_2)s_2}\big)\nonumber\\
	&&\big(f_0(p_1)\fbar(-p_2)\fbar(p_{12})-\fbar(p_1) f_0(-p_2)f_0(p_{12})\big)\nonumber\\
	&&+ \, \sin(\Omega(p_1)s_1)e^{i\Omega(p_2)(s_1-s_2)-i\Omega(p_{12})s_1} V_2(p_1) \hat{v}(p_2)(-J(-p_1)+J(p_2)-J(p_{12}))\nonumber\\
	&&\big(V_2(p_1)\sin(\Omega(p_1)s_2)e^{i\Omega(p_{12})s_2}+ V_2(p_{12}) \sin(\Omega(p_{12})s_2)e^{i\Omega(p_1)s_2}\big)\nonumber\\
	&&\big(\fbar(-p_1)f_0(p_2)\fbar(p_{12})-f_0(-p_1)\fbar(p_2)f_0(p_{12})\big)\nonumber\\
	&&+ \, \sin(\Omega(p_2)s_1)e^{i\Omega(p_1)(s_1-s_2)-i\Omega(p_{12})s_1} V_2(p_2) \hat{v}(p_1)(J(p_1)-J(-p_2)-J(p_{12}))\nonumber\\
	&&\big(V_2(p_2)\sin(\Omega(p_2)s_2)e^{i\Omega(p_{12})s_2}+ V_2(p_{12}) \sin(\Omega(p_{12})s_2)e^{i\Omega(p_2)s_2}\big)\nonumber\\
	&&\big(f_0(p_1) \fbar(-p_2)\fbar(p_{12})-\fbar(p_1) f_0(-p_2) f_0(p_{12})\big)\nonumber\\
	&&+V_2(p_2)e^{i(\Omega(p_1)+\Omega(p_2))s_1}\sin(\Omega(p_{12})s_1)\big(J(p_1)+J(p_2)+J(-p_{12})\big)\nonumber\\
	&&+ \, \Big(e^{-i(\Omega(p_1)+\Omega(p_2))s_2}\sin(\Omega(p_{12})s_2)V_2(p_{12})(\hat{v}(p_1)+\hat{v}(p_2))\nonumber\\
	&&+ \, e^{-i(\Omega(p_1)+\Omega(p_{12}))s_2}\sin(\Omega(p_2)s_2)V_2(p_2)(\hat{v}(p_1)+\hat{v}(p_{12}))\nonumber\\
	&&+ \, \sin(\Omega(p_1)s_2) e^{-i(\Omega(p_2)+\Omega(p_{12}))s_2}V_2(p_1)(\hat{v}(p_2)+\hat{v}(p_{12}))\Big)\nonumber\\
	&&\big(f_0(p_1)f_0(p_2)f_0(-p_{12})-\fbar(p_1)\fbar(p_2)\fbar(-p_{12})\big) \Big]\, .
	\eeqn 
	Using symmetry in $p_1\leftrightarrow p_2$, this can be reduced to
	\eqn
	\lefteqn{-2 \Re\int \, d\bp_2 \, (\hat{v}(p_1)+\hat{v}(p_2))}\nonumber\\
	&&\Big[\sin(\Omega(p_1)s_1)\sin(\Omega(p_{12})s_1) e^{i\Omega(p_2)(s_1-s_2)+i(\Omega(p_1)-\Omega(p_{12}))s_2} \nonumber\\
	&&V_2(p_1)V_2(p_{12})(\hat{v}(p_2)+\hat{v}(p_{12})) (-J(-p_1)+J(p_2)+J(-p_{12}))\nonumber\\
	&&\big(\fbar(-p_1)f_0(p_2)f(-p_{12})-f_0(-p_1)\fbar(p_2)\fbar(-p_{12})\big)\nonumber\\
	&&+ \, \sin(\Omega(p_1)s_1)e^{i\Omega(p_2)(s_1-s_2)-i\Omega(p_{12})s_1} V_2(p_1) (-J(-p_1)+J(p_2)-J(p_{12}))\nonumber\\
	&&\big(\hat{v}(p_1)V_2(p_1)\sin(\Omega(p_1)s_2)e^{i\Omega(p_{12})s_2}+ \hat{v}(p_{12})V_2(p_{12}) \sin(\Omega(p_{12})s_2)e^{-i\Omega(p_1)s_2}\big)\nonumber\\
	&&\big(\fbar(-p_1)f_0(p_2)\fbar(p_{12})-f_0(-p_1)\fbar(p_2)f_0(p_{12})\big)\nonumber\\
	&&+ \, \sin(\Omega(p_1)s_1)e^{i\Omega(p_2)(s_1-s_2)-i\Omega(p_{12})s_1} V_2(p_1) \hat{v}(p_2)(-J(-p_1)+J(p_2)-J(p_{12}))\nonumber\\
	&&\big(V_2(p_1)\sin(\Omega(p_1)s_2)e^{i\Omega(p_{12})s_2}+ V_2(p_{12}) \sin(\Omega(p_{12})s_2)e^{i\Omega(p_1)s_2}\big)\nonumber\\
	&&\big(\fbar(-p_1)f_0(p_2)\fbar(p_{12})-f_0(-p_1)\fbar(p_2)f_0(p_{12})\big)\nonumber\\
	&&\Big(\frac12e^{-i(\Omega(p_1)-\Omega(p_2))s_2}\sin(\Omega(p_{12})s_2)V_2(p_{12})(\hat{v}(p_1)+\hat{v}(p_2))\nonumber\\
	&&+ \, e^{-i(\Omega(p_1)-\Omega(p_{12}))s_2}\sin(\Omega(p_2)s_2)V_2(p_2)(\hat{v}(p_1)+\hat{v}(p_{12}))\Big)\nonumber\\
	&&\big(f_0(p_1)f_0(p_2)f_0(-p_{12})-\fbar(p_1)\fbar(p_2)\fbar(-p_{12})\big) \Big]\, .
	\eeqn 
	This can be further be simplified to
	\eqn
	\lefteqn{\bbf^{(2)}_2(f_0)[J](\bs_2)}\nonumber\\
	& := & 2\Re\int \, d\bp_2 \, (\hat{v}(p_1)+\hat{v}(p_2))\nonumber\\
	&&\Big[\sin(\Omega(p_1)s_1)\sin(\Omega(p_{12})s_1) e^{i\Omega(p_2)(s_1-s_2)+i(\Omega(p_1)-\Omega(p_{12}))s_2}\nonumber\\
	&&V_2(p_1)V_2(p_{12})(\hat{v}(p_2)+\hat{v}(p_{12})) (-J(-p_1)+J(p_2)+J(-p_{12}))\nonumber\\
	&&\big(f_0(-p_1)\fbar(p_2)\fbar(-p_{12})-\fbar(-p_1)f_0(p_2)f(-p_{12})\big)\nonumber\\
	&&+ \, \sin(\Omega(p_1)s_1)e^{i\Omega(p_2)(s_1-s_2)-i\Omega(p_{12})s_1} V_2(p_1) \nonumber\\
	&&(-J(-p_1)+J(p_2)-J(p_{12}))\nonumber\\
	&&\Big((\hat{v}(p_1)+ \hat{v}(p_2))V_2(p_1)\sin(\Omega(p_1)s_2)e^{-i\Omega(p_{12})s_2}\nonumber\\
	&&+ \, (\hat{v}(p_2)+\hat{v}(p_{12}))V_2(p_{12}) \sin(\Omega(p_{12})s_2)e^{-i\Omega(p_1)s_2}\Big)\nonumber\\
	&&\big(f_0(-p_1)\fbar(p_2)f_0(p_{12})-\fbar(-p_1)f_0(p_2)\fbar(p_{12})\big)\nonumber\\
	&&\Big(\frac12e^{-i(\Omega(p_1)+\Omega(p_2))s_2}\sin(\Omega(p_{12})s_2)V_2(p_{12})(\hat{v}(p_1)+\hat{v}(p_2))\nonumber\\
	&&+ \, e^{-i(\Omega(p_1)+\Omega(p_{12}))s_2}\sin(\Omega(p_2)s_2)V_2(p_2)(\hat{v}(p_1)+\hat{v}(p_{12}))\Big)\nonumber\\
	&&\big(\fbar(p_1)\fbar(p_2)\fbar(-p_{12})-f_0(p_1)f_0(p_2)f_0(-p_{12})\big) \Big] \, . \label{eq-f-bol-cor-21}
	\eeqn 
	In addition, let
	\eqn
	\bbf^{(2)}(f_0)[J](\bs_2) \, := \, \bbf^{(2)}_1(f_0)[J](\bs_2) \, + \, \bbf^{(2)}_2(f_0)[J](\bs_2) \, . \label{eq-f-bol-cor-2}
	\eeqn 
	Recalling definitions \eqref{def-cBd21}, \eqref{def-cBd22}, and \eqref{def-cBd2}, we obtain that
	\eqn
		\int_{\Delta[T\lambda^{-2},2]} d\bs_2 \, \bbf^{(2)}(f_0)[J](\bs_2) \, = \, \cBd^{(2)}(f_0)[J](T;\lambda) \, .
	\eeqn 
	In order to bound $\bbf^{(3)}(f_0)[J](\bs_2)$, we need to, first, look at integrability of the terms. Notice that, for Boltzmann contractions, the order of the creation/annihilation operators within a single argument of a commutator do not matter. In particular, we are interested in evaluating expressions involving the Boltzmann contractions of
	\eqn
		[a_{\sigma_1 p_1}^{(\sigma_1)}a_{\sigma_2 p_2}^{(\sigma_2)}a_{\sigma_3p_{12}}^{(-\sigma_3)},a_{\tau_1k_1}^{(\tau_1)}a_{\tau_2k_2}^{(\tau_2)}a_{\tau_3k_{12}}^{(-\tau_3)}] \, ,
	\eeqn 
	which contain a factor
	\eqn
		\lefteqn{\big(f_0(\sigma_1p_1)+\frac{1-\sigma_1}2\big)\big(f_0(\sigma_2p_2)+\frac{1-\sigma_2}2\big)\big(f_0(\sigma_3p_{12})+\frac{1+\sigma_3}2\big)}\nonumber\\
		&&-\big(f_0(\sigma_1p_1)+\frac{1+\sigma_1}2\big)\big(f_0(\sigma_2p_2)+\frac{1+\sigma_2}2\big)\big(f_0(\sigma_3p_{12})+\frac{1-\sigma_3}2\big)\nonumber\\
		&=& \sigma_3f_0(\sigma_1p_1)f_0(\sigma_2p_2) \, - \, \sigma_1 f_0(\sigma_2p_2)f_0(\sigma_3p_{12}) \, - \, \sigma_2 f_0(\sigma_1p_1)f_0(\sigma_3p_{12}) \nonumber\\
		&&+ \, (\sigma_3-\sigma_2) f_0(\sigma_1p_1) \, + \, (\sigma_3-\sigma_1)f_0(\sigma_2p_2) \, - \, (\sigma_1+\sigma_2)f_0(\sigma_3p_{12}) \, . \label{eq-fbol-0}
	\eeqn 
	Observe that there is a global coefficient $\hat{v}(p_2)$. After evaluating all $\delta$ coming from contractions between $p$ and $k$ momenta, we want to verify integrability w.r.t. $dp_1 \,dp_2$. We are thus left with verifying integrability for the terms involving $(\sigma_3-\sigma_1)f_0(\sigma_2 p_2)$. This term occurs only if $\sigma_3=-\sigma_1$. Another global factor then is
	\eqn
		V_2(p_1)^{\frac{1-\sigma_1}2}V_2(p_{12})^{\frac{1-\sigma_3}2}=	V_2(p_1)^{\frac{1-\sigma_1}2}V_2(p_{12})^{\frac{1+\sigma_1}2} \, ,
	\eeqn 
	where we recall from \eqref{eq-f-disc-exp-0} that we only need to consider $\cHcubt^{(1)}(s_1)$ in the first argument of the commutator. In particular, there is an integrable factor w.r.t. $dp_1$.	With that, we have the estimate
	\eqn
	|\bbf^{(3)}(f_0)[J](\bs_2)| \, \leq \, C_{\vd,\fd}\|J\|_{\ell^\infty(\lattice)} \, . \label{eq-f-bol-cor-3}
	\eeqn 
	As a consequence, we have that
	\eqn
	|\rrbogb_{2,d}(t;f[J] )| &=& \Big|\frac{\lambda^5}N \int_{\Delta[t,2]} \, d\bs_2 \, \bbf^{(3)}(f_0)[J] (\bs_2)\Big|\nonumber\\
	& \leq & C_{\vd,\fd}\|J\|_{\ell^\infty(\lattice)} \frac{\lambda^5 t^2 }{N} \nonumber\\
	&\leq & C_{\vd,\fd}\|J\|_{\ell^\infty(\lattice)} T^2 \frac{\lambda}{N} \, . 
	\eeqn
	This concludes the proof of \eqref{eq-f-main-0}.
	\par Finally, we compute the dynamics of $g$. We write
	\eqn
	-i\int_0^t \, \frac{ds}{\vol} \wick{
		[ \c g[\c2 J] , \underbracket{\settowidth{\wdth}{$\cHquart$}\hspace{0.4\wdth}\c2{\vphantom{\cHquart}} \hspace{-0.4\wdth} \c \cHquart}(s)]
	} &= & \frac{\lambda^2}N\int_0^t \, ds \, \acqd(f_0)[J](s) \, , \label{eq-gquartabs-0}
	\eeqn 
	where $\acqd(f_0)[J](s)$ consists of terms of the form
	\eqn
	\lefteqn{(-i)^{\ell} \lambda^{j} \int \, dp \, dk \, e^{-is(m_1\Omega(p)+m_2\Omega(k))}J(p)}\nonumber\\
	&& [(1+f_0(p))(1+f_0(-p))-f_0(p)f_0(-p)] \nonumber\\
	&&V_1(p)^{2\alpha_1}V_2(p)^{\alpha_2}\hat{v}(p\pm k) V_1(k)^{2\alpha_3}V_2(k)^{\alpha_4} (f_0(k)+\iota) \, ,\label{eq-gabsquart-0}
	\eeqn 
	where $\ell\in\N_0$, $\ell\leq 3$, $j\in\N_0$, $j \leq 7$, $m_1,m_2\in\{0,\pm2\}$, $\alpha_j\in\{0,1,2\}$ and $\iota\in\{0,1\}$. Here, we already employed the fact that $\hat{v}(0)=0$. Using symmetry of the integrand w.r.t. $p\leftrightarrow -p$, we can further simplify the expression \eqref{eq-gabsquart-0} to 
	\eqn
	\lefteqn{(-i)^{\ell} \lambda^{j} \int \, dp \, dk \, e^{-is(m_1\Omega(p)+m_2\Omega(k))}(1+f_0(p)+f_0(-p))J(p)}\nonumber\\
	&&V_1(p)^{2\alpha_1}V_2(p)^{\alpha_2}\hat{v}(p-k) V_1(k)^{2\alpha_3}V_2(k)^{\alpha_4} (f_0(k)+\iota) \, .\label{eq-gabsquart-1}
	\eeqn 	
	Observe that the only terms contributing to \eqref{eq-gquartabs-0} are of the form $[aa,(a^+)^3a]$ with any permutation of $(a^+)^3a$, which is why the terms contain at least one momentum flip $a_p\to i\lambda V_2(p)\sin(\Omega(p)s)a_{-p}$. This justifies the extra factor $\lambda$ on the RHS of \eqref{eq-gquartabs-0}. In the case $\iota=1$, we need to have at least one annihilation operator left of a creation operator in the second argument of the commutator. This yields an additional factor $V_2(k)$. In particular, the integrand in \eqref{eq-gabsquart-0} contains $\hat{v}(p\pm k)f_0(k)$ or $\hat{v}(p\pm k)V_2(k)$. 
	\par Next, we compute
	\eqn
	\lefteqn{\frac{(-i)^2}\vol \int_{\Delta[t,2]} \, d\bs_2 \, \nu_0([[g[J] ,\cHcubt(s_1)],\cHcubt(s_2)])} \nonumber\\
	&=& -\frac{1}\vol \int_{\Delta[t,2]} \, d\bs_2 \, \Big([[\wick{ \c1 g(\c2 J),\underbracket{\c1\cHcubt}(s_1)],\underbracket{\c2\cHcubt}(s_2)]
	} \label{eq-g-con-con-0} \\
	&& + \, [[\wick{ \c1 g[\c2 J],\c1\cHcubt\c1(\c3 s_1)],\c2\cHcubt\c2(\c3 s_2)]
	} \label{eq-g-bol-con-0} \\
	&& + \, [[\wick{ \c1 g[\c2 J],\settowidth{\wdth}{$\cHcubt$}\hspace{0.4\wdth}\c2{\vphantom{\cHcubt}}\hspace{0.2\wdth}\c3{\vphantom{\cHcubt}} \hspace{-0.6\wdth} \c1\cHcubt( \c2 s_1)],\underbracket{\c3 \cHcubt}(\c2 s_2)]
	} \label{eq-g-pa-con-0}
	\eeqn
	Similarly to above, we will refer to \eqref{eq-g-con-con-0} as condensate contraction and to \eqref{eq-g-bol-con-0} as Boltzmann contraction. In addition, we call \eqref{eq-g-pa-con-0} pair absorption contractions. 
	\par We start again with the condensate term. We obtain that
	\eqn
	\lefteqn{-\frac{1}\vol \int_{\Delta[t,2]} \, d\bs_2 \, [[\wick{ \c1 g(\c2 J),\underbracket{\c1\cHcubt}(s_1)],\underbracket{\c2\cHcubt}(s_2)]
	}}\nonumber\\
	& = & -2J(0)\int_{\Delta[t,2]} \, \frac{d\bs_2}{\vol} \,\Big(
	\wick{
		[\c1 a_0,\underbracket{\c1 \cHcubt}(s_1)]
		}
	\wick{
		[\c1 a_0,\underbracket{\c1 \cHcubt}(s_2)] 
	}
	\Big)\nonumber\\
	&=& J(0) \Phi_t^2+J(0)\rrbogc_{2,d}(t;g) \, .\label{eq-g-condensate-0}
	\eeqn 
	where, analogously to \eqref{eq-f-con-err-0}, 
	\eqn
	\rrbogc_{2,d}(t;g) & := & -2\Big(-i\int_0^t\, ds\, \frac{\nu_0([ a_0,\cHcubt(s)])}\vol \Rem_2(t;\Phi)\Big)\nonumber\\
	&&-\,\big(\Rem_2(t;\Phi)\big)^2 \, .
	\eeqn 
	Using analogous estimates as for \eqref{eq-f-condensate-1}, we find that $|\rrbogc_{2,d}(t;g)|$ satisfies the same bounds as $|\rrbogc_{2,d}(t;f)|$. 
	\par For the Boltzmann contraction, we write 
	\eqn 
	\lefteqn{-\int_{\Delta[t,2]} \, \frac{d\bs_2}\vol \, [[\wick{ \c1 g[\c2 J],\c1\cHcubt\c1(\c3 s_1)],\c2\cHcubt\c2(\c3 s_2)]
	}}\nonumber\\
	& = &             \frac{\lambda^2}N\int_{\Delta[t,2]} \, d\bs_2 \, \bbgd(f_0)[J](\bs_2) \, . \label{eq-gboltzmann-0}
	\eeqn 
	The expressions in $\bbgd(f_0)[J](\bs_2)$ are of the form
	\eqn
	\lefteqn{(-i)^{\ell_0}\lambda^{j_0}\int \, d\bp_3 \, e^{-i\sum_{\ell=1}^2s_\ell\sum_{k=1}^3\sigma_{k,\ell}\Omega(p_k)}\delta(p_1+p_2-p_3)}\nonumber\\
	&&J(p_{j_1})\hat{v}(p_2)\hat{v}(p_{j_2})\prod_{k=1}^3V_1(p_k)^{2\alpha_{\ell_k}}V_2(p_k)^{\beta_{\ell_k}}\nonumber\\
	&&\Big(\prod_{k=1}^3\big(f_0(\tau_{k,1}p_k)+\frac{1+\tau_{k,2}}2\big) \, - \, \prod_{k=1}^3\big(f_0(-\tau_{k,1}p_k)+\frac{1-\tau_{k,2}}2\big) \Big) \, , \label{eq-gboltzmann-1}
	\eeqn 
	where $\ell_0\in\N_0^{\leq 3}$, $j_0\in\N_0$, $j_0\leq 12$, $\sigma_{k,\ell},\tau_{k,\ell}\in\{\pm 1\}$, $j_1,j_2\in\{1,2,3\}$ and $\alpha_{\ell_k},\beta_{\ell_k}\in\{0,1,2\}$. We need to ensure integrability of each of these terms. More precisely, we will show that any term contains a product of at least two of the functions $\hat{v}$, $f_0$, $V_1$, and $V_2$ depending on at least two of the momenta $p_1$, $p_2$, or $p_3$. We have that 
	\eqn
	\int \, dp \, J(p)a_pa_{-p}  \, = \, \int \, dp \, J(p) a_{-p}a_p \, = \, \int \, dp \, J(-p) a_p a_{-p} \, , 
	\eeqn 
	where we used the CCR followed by substitution. In particular, we may assume without loss of generality that $J$ is even. Then the CCR imply
	\eqn
	[g[J] , a_{\sigma_1p_1}^{(\sigma_1)}a_{\sigma_2p_2}^{(\sigma_2)}a_{\sigma_3p_3}^{(-\sigma_3)}] &=&  \delta_{\sigma_1,1}J(p_1)a_{-p_1}a_{\sigma_2p_2}^{(\sigma_2)}a_{\sigma_3p_3}^{(-\sigma_3)}\nonumber\\
	&& + \, \delta_{\sigma_2,1}J(p_2)a_{\sigma_1p_1}^{(\sigma_1)} a_{-p_2} a_{\sigma_3p_3}^{(-\sigma_3)}\nonumber\\
	&& + \, \delta_{\sigma_3,-1}J(p_3)a_{\sigma_1p_1}^{(\sigma_1)} a_{\sigma_2p_2}^{(\sigma_2)} a_{p_3} \, . \label{eq-gj-com}
	\eeqn 
	We will discuss the expressions related to one these three terms in detail; the remaining follow with analogous computations. Consider the Boltzmann contractions of 
	\eqn
	\nu_0\big( [a_{-p_1}a_{\sigma_2p_2}^{(\sigma_2)}a_{\sigma_3p_3}^{(-\sigma_3)},a_{\tau_1k_1}^{(\tau_1)}a_{\tau_2k_2}^{(\tau_2)}a_{\tau_3k_3}^{(-\tau_3)}]\big) \, , \label{eq-gboltzcon-0}
	\eeqn 
	which is why, again, the order of the operators $a$ and $a^+$ does not matter.	Observe that it is sufficient to have a factor $\hat{v}$, $f_0$, $V_1$, or $V_2$ with momentum $p_1$ or $p_{12}$, since, due to \eqref{eq-gboltzmann-1}, we always have a coefficient $\hat{v}(p_2)$. The Boltzmann contractions in \eqref{eq-gboltzcon-0} yield a factor
	\eqn
	\lefteqn{\big(f_0(-p_1)+1\big)\big(f_0(\sigma_2p_2)+\frac{1-\sigma_2}2\big)\big(f_0(\sigma_3p_{12})+\frac{1+\sigma_3}2\big)}\nonumber\\
	&&-f_0(-p_1)\big(f_0(\sigma_2p_2)+\frac{1+\sigma_2}2\big)\big(f_0(\sigma_3p_{12})+\frac{1-\sigma_3}2\big)\nonumber\\
	&=& \sigma_3f_0(-p_1)f_0(\sigma_2p_2) \, + \,  f_0(\sigma_2p_2)f_0(\sigma_3p_{12}) \, - \, \sigma_2 f_0(-p_1)f_0(\sigma_3p_{12}) \nonumber\\
	&&+ \, (\sigma_3-\sigma_2) f_0(-p_1) \, + \, (1+\sigma_3)f_0(\sigma_2p_2) \, + \, (1-\sigma_2)f_0(\sigma_3p_{12}) \, . \label{eq-gbol-0}
	\eeqn 
	The only term in \eqref{eq-gbol-0} that does not already involve a factor depending on a momentum other than $p_2$, see also \eqref{eq-gboltzmann-1}, is $(1+\sigma_3)f_0(\sigma_2p_2)$. This term only appears if $\sigma_3=1$, which we now want to consider.
	\par Due to momentum flips in $p_j$, the corresponding term associated with $\cHcubt^{(2)}(s_1)$ in the first argument of the commutator has a coefficient $V_2(p_1)V_2(p_3)$, yielding the remaining integrability w.r.t. $dp_1$.
	\par Thus, let us consider the terms associated with $\cHcubt^{(2)}(s_1)$. If we contract $a_{\tau_2k_2}^{(\tau_2)}$ in with $a_{-\sigma_1p_1}^{(-\sigma_1)}$ or $a_{\sigma_3p_3}^{(-\sigma_3)}$, we obtain a factor $\hat{v}(p_1)$ or $\hat{v}(p_{12})$, yielding integrability w.r.t. $dp_1$.
	\par So, it remains to consider the case when $a_{\tau_2k_2}^{(\tau_2)}$ is contracted with $a_{\sigma_2p_2}^{(\sigma_2)}$. The remaining contractions yield either $(\tau_1,-\tau_3)=(1,1)$ or $(-\tau_3,\tau_1)=(1,1)$. In those cases, we obtain an additional factor $V_2(p_{12})$ or $V_2(p_1)$. This concludes the argument.
	\par We are left with evaluating the pair absorption terms
	\eqn
	-\frac1\vol\int_{\Delta[t,2]} \, d\bs_2 \,[[\wick{ \c1 g[\c2 J],\settowidth{\wdth}{$\cHcubt$}\hspace{0.4\wdth}\c2{\vphantom{\cHcubt}}\hspace{0.2\wdth}\c3{\vphantom{\cHcubt}} \hspace{-0.6\wdth} \c1\cHcubt( \c2 s_1)],\underbracket{\c3 \cHcubt}(\c2 s_2)]
	}\nonumber\\
	= \frac{\lambda^2}{N}\int_{\Delta[t,2]} \, d\bs_2 \, \acd(f_0)[J](\bs_2) \, . \label{eq-g-pair-ann-cub2-0} 
	\eeqn 
	$\acd(f_0)[J](\bs_2)$ consists of terms of the form
	\eqn
	\lefteqn{(-i)^{\ell} \lambda^{j}\int \, dp\, dk \, e^{-is_1m_1\Omega(p)-is_2m_2\Omega(k)}J(p)(1+f_0(p)+f_0(-p))\hat{v}(p)}\nonumber\\
	&&V_1(p)^{2\alpha_1}V_2(p)^{\alpha_2} \hat{v}(k) V_1(k)^{2\alpha_3}V_2(k)^{\alpha_4}(f_0(k)+\iota) \, ,
	\eeqn 
	where $\ell\in\N_0$, $\ell\leq 3$, $j\in\N_0$, $j \leq 12$, $m_1,m_2\in\{0,\pm2\}$, $\alpha_j\in\{0,1,2\}$ and $\iota\in\{0,1\}$. Here again, we take into account that terms involving $\hat{v}(0)=0$ vanish. This concludes the proof.
	\endprf
	
	\begin{lemma}
		\label{lem-bog-expansion}
		We have the following expansions.
		\begin{enumerate}
			\item Identifying the RHS with its continuous extension, we have that
			\eqn
			\int_{\Delta[t,2]} \, d\bs_2 \, e^{i(\omega_1s_1+ \omega_2s_2)} & = & \frac2{\omega_2}\Big(\frac{\sin^2\big((\omega_1+ \omega_2)t/2\big)}{\omega_1+ \omega_2} \, - \, \frac{\sin^2\big(\omega_1t/2\big)}{\omega_1}\Big)\nonumber\\
			&& - \, \frac{i}{\omega_2}\Big(\frac{\sin\big((\omega_1+ \omega_2)t\big)}{\omega_1+ \omega_2} \, - \, \frac{\sin(\omega_1t)}{\omega_1}\Big)
			\eeqn 
			for all $\omega_1,\omega_2\in\R$. 
			\item The Bogoliubov dispersion $\Omega$ in Lemma \ref{lem-apcV-diag-1} satisfies 
			\eqn
			\Omega &=& E \, + \, \lambda\hat{v}_{Bog} \nonumber\\
			&=& E \, + \, \lambda \hat{v} \, - \, \lambda^2 \frac{\hat{v}^2}{2E} \, + \, \lambda^3\eeb  \, , 
			\eeqn  
			where $\hat{v}_{Bog} := \frac{2\hat{v}}{1+\sqrt{1+2\lambda\frac{\hat{v}}E}}$ and $\eeb$ satisfies
			\eqn
			\|\eeb\|_\infty \, \leq \, C\vd^3
			\eeqn 
			for all $\lambda>0$.
		\end{enumerate}
	\end{lemma}
	
	\prf
	For the first part, let $\omega_1,\omega_2,\omega_1+\omega_2\neq0$. Then
	\eqn
	\int_{\Delta[t,2]} \, d\bs_2 \, e^{i(\omega_1s_1+ \omega_2s_2)} & = & \int_0^t \, ds \, \frac{e^{i\omega_1s}}{i\omega_2}(e^{i\omega_2 s} \, - \, 1 )\nonumber\\
	&=& -\frac1{\omega_2}\Big(\frac{e^{i(\omega_1+\omega_2)t}-1}{\omega_1+\omega_2} \, - \, \frac{e^{i\omega_1t}-1}{\omega_1}\Big) \, .
	\eeqn 
	Using $1-\cos(x)=2\sin(x/2)$, we have shown the first statement.
	\par For the second part, we expand $\Omega$ using
	\eqn
	\Omega &=& E \, + \, \frac{\Omega^2-E^2}{E+\Omega} \nonumber\\
	&=& E \, + \, \frac{2\lambda \hat{v}}{1+\sqrt{1+2\lambda\frac{\hat{v}}E}} \, .
	\eeqn 
	We Taylor expand
	\eqn
	\frac1{1+\sqrt{1+x}} \, = \, \frac12 \, - \, \frac{x}8 \, + \, R(x)
	\eeqn 
	for some $R(x)$. An easy computation yields
	\eqn
	\frac{|R(x)|}{x^2} \, \leq \, C 
	\eeqn 
	for all $x>0$, and in the limit $x\searrow0$. This concludes the proof.
	\endprf

	\begin{remark}[Talbot effect]\label{rem-talbot}
		Lemma \ref{lem-bog-expansion} shows that the leading term in $\cB_d(f_0)[J]$ is a sum of the form 
		\eqn
		\frac{T}{\lambda^2} \int_{(\lattice)^2} \, d\bp_2 \, \frac{\sin^2\Big(\frac{T(\dEcub \, + \, \lambda\Delta_{cub} \hat{v}_{Bog})}{2\lambda^2}\Big)}{\Big(\frac{T(\dEcub \, + \, \lambda\Delta_{cub} \hat{v}_{Bog})}{2\lambda^2}\Big)^2} \dJcub(\bp_2) H(\bp_2)
		\eeqn 
		for some $H\sim1$ with $\hat{v}_{Bog}$ as defined in Lemma \ref{lem-bog-expansion}. Thus, as a function of $T$, the modulus of this sum oscillates between $0$ and $O(\lambda^{-2})$. The size of these oscillations is dependent on the the interaction profile $\hat{v}$, the lattice $\Lambda$, and the chosen sequence $\lambda=\lambda(N)$. A similar phenomenon occurs for $\cBd^{(j)}$, which thereby also oscillate in modulus between $0$ and $O(\lambda^{j-2})$, depending on $T$, and we also observe it for $\bbgd$, $\acqd$, and $\acd$. Therefore, $\cB_d^{(j)}$ can dominate $\cB_d$, depending on whether $\frac{T}{4\pi\lambda^2}\dOcub(\bp_2)$ lies in a small vicinity of $\frac14+\Z$ for some $\bp_2\in(\lattice)^2$. This is reminiscent of the Talbot effect, see \cite{ertz,karo}. As heuristically explained above, $\cBd^{(j)}$ are negligible in the case of large system length $L\sim \lambda^{-2-}$. Next, we will present a proof of this fact.
	\end{remark}

	\subsection{Continuum approximation $\lattice\to\R^3$}\label{sec-cont} 
	
	In this section, we will write out the summation over the lattice $\lattice$ explicitly. In contrast, all integrals will be understood as integrals over $\R^d$ in the respective dimension $d$. 
	
	\begin{lemma}\label{lem-disc}
		Let $F_1(\bp_2),F_2(\bp_2)\in\{\pm\Omega(p_1)\pm\Omega(p_2)\pm\Omega(p_1+p_2),0\}$, $H\in C^\infty_c(\R^6)$, $\chi\in C^\infty_c(\R^3)$, and $\tau_1,\tau_2\in\R$. Let $\jb{x}:=(1+x^2)^{\frac12}$. Abbreviate
		\eqn
		\eed^{(1)} &:=& \lambda^2\int_0^{T\lambda^{-2}} \, ds \, \Big(\frac{1}{\vol^2}\sum_{\bp_2\in(\lattice)^2} e^{i(\tau_1\Omega(p_1)+\tau_2\Omega(p_2))s}H(\bp_2) \nonumber\\
		&&- \, \frac1{(2\pi)^6}\int \, d\bp_2 \, e^{i(\tau_1\Omega(p_1)+\tau_2\Omega(p_2))s}H(\bp_2) \Big) \, , \nonumber\\
		\eed^{(2)} &:=& \lambda^2  \int_{\Delta[T\lambda^{-2},2]} \, d\bs_2 \, \Big(\frac{1}{\vol^2}\sum_{\bp_2\in(\lattice)^2} e^{iF_1(\bp_2)s_1+iF_2(\bp_2)s_2}H(\bp_2) \nonumber\\
		&&- \, \frac1{(2\pi)^6}\int \, d\bp_2 \, e^{iF_1(\bp_2)s_1+iF_2(\bp_2)s_2}H(\bp_2) \Big) \, .
		\eeqn 
		Then we have for any $\reg>6$ and $\lambda>0$ the following. 
		\begin{enumerate}
			\item \eqn
			\Big|\frac{1}{\vol}\sum_{p\in\lattice} \chi(p) \, - \, \frac1{(2\pi)^3}\int_{\R^3} \, dp \, \chi(p)\Big| \, \leq \, 
			C_r
			\frac{\||\nabla|^{\frac{\reg}2}\chi\|_1}{\vol^{\frac{\reg}6}} \, ,
			\eeqn 
			\item\eqn
			|\eed^{(1)}| \, \leq \, C_{\vc,r} \frac{\jb{T}^{\reg+1}}{\lambda^{2\reg}\vol^{\frac{r}3}} \sum_{n=0}^{2(\floor{\frac\reg2}+1)}\Big\|\jb{|\bp_2|}^nD^{2(\floor{\frac\reg2}+1)-n} H\Big\|_1 \, ,
			\eeqn 
			\item if $F_1=F_2\equiv 0$, 
			\eqn
			|\eed^{(2)}| &\leq & C_{\vc,r} \frac{T^2}{\lambda^2\vol^{\frac\reg3}} \||\nabla|^\reg H\|_1 \, ,
			\eeqn
			\item if $(F_1,F_2)\not\equiv(0,0)$, 
			\eqn
			|\eed^{(2)}| \, \leq \, C_{\vc,r} \frac{\jb{T}^{\reg+2}}{\lambda^{2\reg+2}\vol^{\frac{r}3}} \sum_{n=0}^{2(\floor{\frac\reg2}+1)}\Big\|\jb{|\bp_2|}^nD^{2(\floor{\frac\reg2}+1)-n} H\Big\|_1 \, .
			\eeqn 
		\end{enumerate}
		
	\end{lemma}
	
	\prf
	Let
	\eqn
	\ups \, :=  \, \frac{2\pi}{\vol^{\frac13}} \, , \label{def-nudef}
	\eeqn 
	so that
	\eqn
	\int_\lattice \, dp \, \tilde{H}(p) \, = \, \frac{\ups^3}{(2\pi)^3} \sum_{p\in\lattice} \tilde{H}(p) \, .
	\eeqn 
	Denote
	\eqn
	G_\lambda(\bp_2) \, := \, \lambda^2  \int_{\Delta[T\lambda^{-2},2]} \, d\bs_2 \, e^{iF_1(\bp_2)s_1+iF_2(\bp_2)s_2} \, .
	\eeqn 
	Poisson summation implies
	\eqn
	\ups^6 \sum_{\bp_2\in(\lattice)^2} G_\lambda(\bp_2) H(\bp_2)  &=& \ups^6 \sum_{\bP_2\in\Z^6} G_\lambda(\tau\bP_2) H(\ups\bP_2) \nonumber\\
	&=& \sum_{\bX_2\in\Z^6} \ups^6 \int_{\R^6} \, d\bP_2 \, e^{2\pi i \bP_2\cdot \bX_2}G_\lambda(\tau\bP_2) H(\ups\bP_2) \nonumber\\
	& = & \sum_{\bX_2\in\Z^6} \int_{\R^6} \, d\bp_2 \, e^{2\pi i \bp_2\cdot \bX_2/\ups} G_\lambda(\bp_2)H(\bp_2)  \, . \label{eq-poisson-0}
	\eeqn  
	As a consequence of \eqref{eq-poisson-0}, we obtain that
	\eqn
	\eed^{(2)} \, = \, \frac1{(2\pi)^6}\sum_{\bX_2\in\Z^6\setminus\{0\}} \int_{\R^6} \, d\bp_2 \, e^{2\pi i \bp_2\cdot \bX_2/\ups} G_\lambda(\bp_2)H(\bp_2)  \, . \label{eq-dis-cont-diff-0}
	\eeqn 
	Next, we have that
	\eqn
	|\bX_2|^\reg \int_{\R^6} \, d\bp_2 \, e^{2\pi i \bp_2\cdot \bX_2/\ups} G_\lambda(\bp_2)H(\bp_2)\nonumber\\
	= \, \frac{\ups^\reg}{(2\pi)^\reg}\int_{\R^6} \, d\bp_2 \,  e^{2\pi i\bp_2\cdot\bX_2/\ups} |\nabla_{\bp_2}|^\reg \big(G_\lambda(\bp_2)H(\bp_2)\big) \, . \label{eq-ehrenfest-0}
	\eeqn 
	In particular, we have that
	\eqn
	|\eed^{(2)}| &\leq & C_{\reg}\ups^\reg \sum_{\bX_2\in\Z^6\setminus\{0\}} \frac{1}{|\bX_2|^\reg} \||\nabla|^\reg \big(G_\lambda H\big)\|_1 \nonumber\\
	&\leq & \frac{C_{\reg}}{\vol^\frac\reg3} \||\nabla|^\reg \big(G_\lambda H\big)\|_1  \label{eq-discerror-1}
	\eeqn 
	due to $\reg>6$. With analogous steps, we obtain
	\eqn
	\Big|\ups^3\sum_{p\in\lattice} \chi(p) \, - \, \int_{\R^3} \, dp \, \chi(p)\Big| &= & \frac{\ups^{\frac{\reg}2}}{(2\pi)^{\frac{\reg}2}}\Big| \sum_{X\in\Z^3\setminus\{0\}} \frac1{|X|^{\frac{\reg}2}} \int_{\R^3} \, dp \, e^{2\pi i p\cdot X/\ups}|\nabla|^{\frac{\reg}2}\chi(p)\Big|\nonumber\\
	&\leq& \frac{C_{\reg}}{\vol^\frac\reg6} \| |\nabla|^{\frac{\reg}2}\chi\|_1 \, .
	\eeqn 	
	Interpolation implies
	\eqn
	\||\nabla|^\reg \big(G_\lambda H\big)\|_1  \, \leq\, C \|(-\Delta)^{\floor{\frac\reg2}+1}\big(G_\lambda H\big)\|_1^{\frac{\frac\reg2}{\floor{\frac\reg2}+1}}\|G_\lambda H\|_1^{\frac{1+\floor{\frac\reg2}-\frac\reg2}{\floor{\frac\reg2}+1}}\big) \, , \label{eq-discerror-2}
	\eeqn  
	where $t_0>0$ will be fixed below. The Leibniz rule implies for $k\in\N_0$ that
	\eqn
	|(-\Delta)^k G_\lambda H| &
	\leq & C_k \sum_{n=0}^{2k} |D^n G_\lambda| |D^{2k-n} H| \, .\label{eq-de-leib-1}
	\eeqn
	\par We start with the case $F_1=F_2\equiv 0$. Lemma \ref{lem-bog-expansion} yields $G_\lambda=C\frac{T^2}2\lambda^{-2}$, which, employing \eqref{eq-discerror-1}, implies
	\eqn
	|\eed^{(2)}| &\leq & C_{\reg} T^2\lambda^{-2}\ups^\reg \||\nabla|^\reg H\|_1 \nonumber\\
	&\leq & C_{\reg} \frac{T^2}{\lambda^2\vol^{\frac\reg3}} \||\nabla|^\reg H\|_1 \, .
	\eeqn  
	\par Next, let $F_1\equiv 0\wedge F_2\not\equiv 0$, $F_1\not\equiv 0\wedge F_2\equiv 0$, or $F_1= -F_2\not\equiv0$. Let $F_j\not\equiv0$ for some $j\in\{1,2,3\}$. Then, by Lemma \ref{lem-bog-expansion}, we obtain that 
	\eqn
	G_\lambda(\bp_2) \, = \, \frac{T^2}{\lambda^2} G(F_j(\bp_2)T/\lambda^2)
	\eeqn 
	for some continuous function $G\in C^\infty_b(\R)$. Moreover, the Fa\`a di Bruno formula, see \cite{tbn}, implies
	\eqn
	|D^n G(F_jT/\lambda^2)| \, \leq \, C_n \sum_{\substack{\br_{n}\in\\ R(n)}} |G^{(S(\br_n))}|(F_jT/\lambda^2) \prod_{\ell=1}^{n}\Big[ \frac{T}{\lambda^2}|D^\ell F_j |\Big]^{r_\ell} \, , \label{eq-g-1zero-fdb-0}
	\eeqn 
	where 
	\eqn
	R(n) &:=& \{\br_n\in\N_0^n \mid \sum_{\ell=1}^n\ell r_\ell=n\} \, , \label{def-Rn}\\
	S(\br_n) &:=& \sum_{\ell=1}^n r_\ell \, . \label{def-srn}
	\eeqn 
	Observe that due to Lemma \ref{lem-bog-expansion}, we have that
	\eqn
	|\nabla F_j| & \leq & C_{\vc} \jb{|\bp_2|} \, , \label{eq-fj-bd-1} \\
	|D^\ell F_j| &\leq & C_{\vc} \label{eq-fj-bd-2}
	\eeqn 
	for all $\lambda\in(0,1)$, and all $\ell\geq2$. Then \eqref{eq-g-1zero-fdb-0} together with \eqref{eq-fj-bd-1}, \eqref{eq-fj-bd-2}, and the fact that $1\leq S(\br_n)\leq n$ imply
	\eqn
	|D^n G(F_jT/\lambda^2)| & \leq & C_n \frac{T\jb{T}^{n-1}\jb{|\nabla F_j|}^n}{\lambda^{2n}}\Big(\sum_{\ell=1}^n|G^{(\ell)}|\Big)(F_jT/\lambda^2) \nonumber\\
	& \leq & C_{\vc,n} \frac{T\jb{T}^{n-1}\jb{|\bp_2|}^n}{\lambda^{2n}}\Big(\sum_{\ell=1}^n|G^{(\ell)}|\Big)(F_jT/\lambda^2) \label{eq-g-1zero-fdb-1}
	\eeqn 
	for all $\lambda\in(0,1)$. Collecting \eqref{eq-de-leib-1}, \ref{eq-g-1zero-fdb-1}, we have that
	\eqn
	\lefteqn{\|(-\Delta)^k \big(G_\lambda H\big)\|_1}\nonumber\\
	& \leq & C_{\vc,r} \sum_{n=0}^{2k}\frac{T^3\jb{T}^{n-1}}{\lambda^{2n+2}}\nonumber \Big\|\Big(\sum_{\ell=1}^n|G^{(\ell)}|\Big)(F_jT/\lambda^2) \jb{|\bp_2|}^nD^{2k-n} H\Big\|_1 \nonumber\\
	&\leq & C_{\vc,r} \frac{\jb{T}^{2k+2}}{\lambda^{4k+2}} \Big\| \Big(\sum_{\ell=0}^{2k}|G^{(\ell)}|\Big)\Big\|_\infty \sum_{n=0}^{2k}\Big\|\jb{|\bp_2|}^nD^{2k-n} H\Big\|_1  \label{eq-g-1zero-1}
	\eeqn 
	for all $k\in\{0,1,2,\ldots,\floor{\frac\reg2}+1\}$. Then \eqref{eq-discerror-2} and \eqref{eq-g-1zero-1} yield
	\eqn
	\lefteqn{\||\nabla|^\reg \big(G_\lambda H\big)\|_1}\nonumber\\
	& \leq&  C_{\vc,r} \frac{\jb{T}^{\reg+2}}{\lambda^{2\reg+2}}\sum_{n=0}^{2(\floor{\frac\reg2}+1)}\Big\|\jb{|\bp_2|}^nD^{2(\floor{\frac\reg2}+1)-n} H\Big\|_1 \, , \label{eq-g-1zero-4}
	\eeqn 
	where we also used the fact that $\|H\|_1 \leq \|\jb{\bp_2}^{2(\floor{\frac\reg2}+1)}H\|_1$. \eqref{eq-discerror-1} and \eqref{eq-g-1zero-4} yield
	\eqn
	|\eed^{(2)}| \, \leq \, C_{\vc,r} \frac{\jb{T}^{\reg+2}}{\lambda^{2\reg+2}\vol^{\frac\reg3}} \sum_{n=0}^{2(\floor{\frac\reg2}+1)}\Big\|\jb{|\bp_2|}^nD^{2(\floor{\frac\reg2}+1)-n} H\Big\|_1  \, .  \label{eq-eed2-fj0}
	\eeqn 
	\par Now, let $F_j\not\equiv0$ for all $j\in\{1,2,3\}$. In this case, Lemma \ref{lem-bog-expansion} implies
	\eqn
	G_\lambda(\bp_2) \, = \, \frac{T^2}{\lambda^2}G\big(\frac{T}{\lambda^2}(F_1(\bp_2),F_2(\bp_2)) \big) 
	\eeqn 
	for some smooth function $G\in C^{\infty}_b(\R^2)$. The Fa\`a di Bruno formula implies
	\eqn
	\big|D^n G\big(\frac{T}{\lambda^2}(F_1(\bp_2),F_2(\bp_2)) \big) \big| & \leq & C_n \sum_{\substack{\br_{n}\in\\ R(n)}} |D^{S(\br_n)}G|\big(\frac{T}{\lambda^2}(F_1(\bp_2),F_2(\bp_2)) \big) \nonumber\\
	&&\prod_{\ell=1}^{n}\Big[ \frac{T}{\lambda^2}\sum_{j=1}^2|D^\ell F_j |\Big]^{r_\ell} \nonumber\\
	&\leq & C_n \frac{T\jb{T}^{n-1}\jb{\sum_{j=1}^2|\nabla F_j|}^n}{\lambda^{2n}}\nonumber\\
	&&\Big(\sum_{\ell=1}^n|D^\ell G|\Big)\big(\frac{T}{\lambda^2}(F_1(\bp_2),F_2(\bp_2)) \big) \, . \label{eq-g-nonzero-fdb-0}
	\eeqn 
	Collecting \eqref{eq-de-leib-1}, \eqref{eq-fj-bd-1}, \eqref{eq-fj-bd-2}, and \ref{eq-g-nonzero-fdb-0}, we have that
	\eqn
	\lefteqn{\|(-\Delta)^k \big(G_\lambda H\big)\|_1}\nonumber\\
	& \leq & C_{\vc,r} \sum_{n=0}^{2k}\frac{T^3\jb{T}^{n-1}}{\lambda^{2n+2}}\nonumber\\
	&& \Big\|\Big(\sum_{\ell=1}^n|D^\ell G|\Big)\big(\frac{T}{\lambda^2}(F_1,F_2) \big) \jb{\sum_{j=1}^2|\nabla F_j|}^nD^{2k-n} H\Big\|_1 \nonumber\\
	&\leq & C_{\vc,r} \frac{\jb{T}^{2k+2}}{\lambda^{4k+2}}
	\Big\|\Big(\sum_{\ell=0}^{2k}|D^\ell G|\Big)\Big\|_\infty \sum_{n=0}^{2k}\Big\|\jb{|\bp_2|}^nD^{2k-n} H\Big\|_1  \label{eq-g-1zero-2}
	\eeqn 
	for all $k\in\{0,1,2,\ldots,\floor{\frac\reg2}+1\}$. As a consequence of \eqref{eq-discerror-2}, we thus obtain
	\eqn
	\||\nabla|^\reg \big(G_\lambda H)\|_1 & \leq& C_{\vc,r} \frac{\jb{T}^{\reg+2}}{\lambda^{2\reg+2}}\sum_{n=0}^{2(\floor{\frac\reg2}+1)}\Big\|\jb{|\bp_2|}^nD^{2(\floor{\frac\reg2}+1)-n} H\Big\|_1 \, . \label{eq-g-nonzero-3}
	\eeqn 
	\eqref{eq-discerror-1} and \eqref{eq-g-nonzero-3} imply
	\eqn
	|\eed^{(2)}| \, \leq\, C_{\vc,r} \frac{\jb{T}^{\reg+2}}{\lambda^{2\reg+2}\vol^{\frac{r}3}} \sum_{n=0}^{2(\floor{\frac\reg2}+1)}\Big\|\jb{|\bp_2|}^nD^{2(\floor{\frac\reg2}+1)-n} H\Big\|_1 \, .
	\eeqn 
	\par Finally, let 
	\eqn
	\mathring{G}_\lambda(\bp_2) \, := \, \lambda^2\int_0^{T\lambda^{-2}} \, ds \, e^{i(\tau_1 \Omega(p_1)+\tau_2 \Omega(p_2))s} \, .
	\eeqn 
	A simple computation shows
	\eqn
	\mathring{G}_\lambda(\bp_2) \, = \, T \mathring{G}\big((\tau_1\Omega(p_1)+\tau_2\Omega(p_2))\frac{T}{\lambda^2}\big) 
	\eeqn 	
	for some smooth $\mathring{G}\in C^\infty_b(\R)$. If $\tau_1=\tau_2=0$, we have $\mathring{G}\equiv C$. If $(\tau_1,\tau_2)\neq(0,0)$, a computation analogous to \eqref{eq-g-1zero-fdb-1} yields
	\eqn
	\lefteqn{|D^n \mathring{G}\big((\tau_1\Omega(p_1)+\tau_2\Omega(p_2))T/\lambda^2\big) |}\nonumber\\
	& \leq & C_{\vc,n} \frac{T\jb{T}^{n-1}\jb{|\bp_2|}^n}{\lambda^{2n}}\Big(\sum_{\ell=1}^n|\mathring{G}^{(\ell)}|\Big)((\tau_1\Omega(p_1)+\tau_2\Omega(p_2))T/\lambda^2) 
	\eeqn 
	for all $n\in\{0,1,2,\ldots,2(\floor{\frac\reg2}+1)\}$. In particular, with analogous steps that led to \eqref{eq-eed2-fj0}, we obtain
	\eqn
	|\eed^{(1)}| \, \leq \, C_{\vc,r} \frac{\jb{T}^{\reg+1}}{\lambda^{2\reg}\vol^{\frac{r}3}} \sum_{n=0}^{2(\floor{\frac\reg2}+1)}\Big\|\jb{|\bp_2|}^nD^{2(\floor{\frac\reg2}+1)-n} H\Big\|_1 \, . 
	\eeqn 
	This concludes the proof. 
	\endprf
	
	\begin{remark} \label{rem-dis-to-con}
		Observe that Lemma \ref{lem-disc} implies that 
		\eqn
		\int_\lattice \, dp \, f_0(p) & \leq& \|f_0\|_1+C_{\reg}\frac{\||\nabla|^{\frac\reg2}f_0\|_1}{\vol^{\frac{\reg}6}} \nonumber\\
		&\leq & C_{\reg}\fc
		\eeqn 
		for any $r>6$ as in Theorem \ref{thm-main-con}, where $C_{\reg}$ is independent of $\vol\geq1$. In particular, we have that
		\eqn
		\fd \, \leq \, C_{\reg} \fc \, .
		\eeqn 
	    This is allows us to use the previous estimates to prove Theorem \ref{thm-main-con} as well. Analogously, we have that
		\eqn
			\vd \, \leq \, C_{\reg}\vc \, .
		\eeqn 
		in the assumption of Theorem \ref{thm-main-dis}, where, again $C_{\reg}$ is independent of $\vol\geq1$.
		\par Likewise, we have that
		\eqn
			\Jtn \, \leq \, C_{\reg} \Jcn \, .
		\eeqn 
	\end{remark}
	
	For the next statement, define the continuous analogues $\cCc$, $\cBc$, etc. of $\cCd$, $\cBd$, etc. in \eqref{def-condensate-0}--\eqref{def-cBd2}, and \eqref{def-acquart}, \eqref{def-bbgd} by replacing the lattice sums $\int_\lattice$ over $\lattice$ in the sense of \eqref{eq-disc-int} by Lebesgue integrals $\frac1{(2\pi)^3}\int_{\R^3}$ over $\R^3$. 
	
	\begin{proposition}\label{prop-bog-cont}
		Let $T>0$, $J\in L^{\infty}(\R^3;\R)$ and $\reg>6$. Then the following holds for all $\lambda>0$ small enough, dependent on $\vc$.
		\begin{enumerate}
			\item \eqn
			-i\int_0^{T\lambda^{-2}} \, ds \,  \frac{\nu_0\big([a_0,\cHcubt(s)]\big)}{\vol} &=&  \frac1{N^{\frac12}\lambda}\cCc(f_0)(T;\lambda) \, + \, \rrb_{1,c}(\frac{T}{\lambda^2};\Phi) \, , \label{eq-rrb1c-def}
			\eeqn 
			where 
			\eqn
			|\rrb_{1,c}(\frac{T}{\lambda^2};\Phi)| \, \leq \, \frac{C_{\vc,\fc,\reg}T}{N^{\frac12}\lambda}\Big(\lambda \, + \, \frac1{\vol^{\frac{r}6}}\Big) \, ,
			\eeqn 
			
			\item \eqn
			&&-\int_{\Delta[t,2]} \, d\mathbf{s}_2 \frac{\nu_0\big([[f[J] ,\cHcubt(s_1)],\cHcubt(s_2)]\big)}{\vol }\nonumber\\
			&=&\frac1N\cBc(f_0)[J](T;\lambda) \, + \, J(0)|\Phi_{\frac{T}{\lambda^2}}|^2 \nonumber\\
			&& + \, \rrbogb_{2,c}(\frac{T}{\lambda^2};f[J] ) \, + \, J(0)\rrbogc_{2,c}(\frac{T}{\lambda^2};f)\, ,
			\eeqn 	
			where 
			\eqn
			|\rrbogb_{2,c}(\frac{T}{\lambda^2};f[J] )| &\leq& \frac{C_{\vc,\fc,\reg}\jb{T}^{\reg+2}\|J\|_{W^{2\floor{\frac{\reg}2}+2,\infty}}}N\nonumber\\
			&& \Big(\frac{1}{\lambda^{2(\reg+1)}\vol^{\frac{\reg}3}} +\lambda|\log(\lambda)|\Big) \, , \nonumber\\
			|\rrbogc_{2,c}(\frac{T}{\lambda^2};f[J] )| &\leq& C_{\vc,\fc,\reg}e^{C_{\reg}\vc\vol/\lambda T}\nonumber\\
			&&\frac{\jb{T}^4\vol^{\frac32}}{\lambda^3N^{\frac32}}\Big(1+\frac\vol{N}\Big)^2\Big(1+\frac{\vol^{\frac32}}{\lambda N^{\frac12}}\Big) \, ,
			\eeqn 
			\item 
			\eqn
			-i\int_0^{T\lambda^{-2}} \, ds\, \frac{\nu_0(g[J] ,\cHquart(s))}{\vol} \nonumber\\
			= \, \frac{1}N\int_0^T \, dS \, \acqc(f_0)[J](S/\lambda^2) \, + \, \rrd_{1,c}(\frac{T}{\lambda^2};g[J])
			\eeqn 
			with
			\eqn
			|\rrd_{1,c}(\frac{T}{\lambda^2};g[J])| & \leq & \frac{C_{\vc,\fc,\reg}\jb{T}^{\reg+1} \|J\|_{W^{2\floor{\frac{\reg}2}+2,\infty}}}{N\lambda^{2\reg}\vol^{\frac{\reg}3}} \, ,
			\eeqn
			\item \eqn
			-\int_{\Delta[T\lambda^{-2},2]} \, d\bs_2 \frac{\nu_0\big([[g[J] ,\cHcubt(s_1)],\cHcubt(s_2)]\big)}\vol\nonumber\\
			= \frac{1}{N\lambda^2}\int_{\Delta[T,2]} \, d\bS_2 \, \big( \bbgc(f_0)[J](\bS_2/\lambda^2) + \acc(f_0)[J](\bS_2/\lambda^2)\big)\nonumber\\
			+ \, \rrd_{2,c}(\frac{T}{\lambda^2};g[J]) \, + \, J(0) \Big(\Phi_{T\lambda^{-2}}\Big)^2 \, + \, J(0)\rrbogc_{2,c}(\frac{T}{\lambda^2};g) \, .
			\eeqn 
			where
			\eqn
			|\rrd_{2,c}(\frac{T}{\lambda^2};g[J])|	& \leq & \frac{C_{\vc,\fc,\reg}\jb{T}^{\reg+2}\|J\|_{W^{2\floor{\frac{\reg}2}+2,\infty}}}{N\lambda^{2(\reg+1)}\vol^{\frac\reg3}} 
			\eeqn 
			and $\rrbogc_{2,c}(\frac{T}{\lambda^2};g)$ satisfies the same bound as $\rrbogc_{2,c}(\frac{T}{\lambda^2};f)$.
		\end{enumerate}
		
	\end{proposition}
	
	\prf	
	\par Recall from Proposition \ref{prop-bog-disc} that
	\eqn
	\lefteqn{-i\int_0^{T\lambda^{-2}} \, ds \,  \frac{\nu_0\big([a_0,\cHcubt(s)]\big)}{\vol}}\nonumber\\
	 & = &   \frac1{N^{\frac12}\lambda}\cCd(f_0)(T;\lambda) \, + \, \rrb_{1,d}(\frac{T}{\lambda^2};\Phi) 
	\eeqn 
	with
	\eqn
	|\rrb_{1,d}(\frac{T}{\lambda^2};\Phi)| & \leq & \frac{C_{\vd,\fd}T}{N^{\frac12}} \nonumber\\
	& \leq & \frac{C_{\vc,\fc,\reg}T}{N^{\frac12}} \label{eq-rrb1-phi-1}
	\eeqn 
	due to Remark \ref{rem-dis-to-con}.
	Let 
	\eqn 
	\rrd_{1,c}(\frac{T}{\lambda^2};\Phi) \, := \, \frac1{N^{\frac12}\lambda}\Big(\cCd(f_0)(T;\lambda)-\cCc(f_0)(T;\lambda)\Big) \, .
	\eeqn 
	Then Lemma \ref{lem-disc} implies  
	\eqn
	|\rrd_{1,c}(\frac{T}{\lambda^2};\Phi)| &\leq & \frac{T}{N^{\frac12}\lambda} \Big|  \frac{(2\pi)^3}\vol\sum_{p\in\lattice} \hat{v}(p)f_0(p) \, - \, \int_{\R^3} \, dp\, \hat{v}(p)f_0(p)\Big| \nonumber\\
	&\leq & C_{\reg}\frac{T}{N^{\frac12}\lambda\vol^{\frac{r}6}} \| |\nabla|^{\frac{r}2}\hat{v}f_0\|_1\nonumber\\
	&\leq&  \frac{C_{\vc,\fc,\reg} T}{N^{\frac12}\lambda \vol^{\frac{r}6}} \, . \label{eq-r1-phi-1}
	\eeqn 
	Here, we used interpolation as in \eqref{eq-discerror-2}, together with the Leibniz rule. Applying \eqref{eq-rrb1-phi-1}, \eqref{eq-r1-phi-1} implies
	\eqn
	|\rrb_{1,c}(\frac{T}{\lambda^2};\Phi)| &\leq &|\rrb_{1,d}(\frac{T}{\lambda^2};\Phi)| \, + \,  |\rrd_{1,c}(\frac{T}{\lambda^2};\Phi)| \nonumber\\
	& \leq & \frac{C_{\vc,\fc,\reg}T}{N^{\frac12}\lambda}\Big(\lambda \, + \, \frac1{\vol^{\frac{r}6}}\Big) \, .
	\eeqn 
	\par Next, Proposition \ref{prop-bog-disc} together with Remark \ref{rem-dis-to-con} yield that
	\eqn
	&&-\int_{\Delta[T\lambda^{-2},2]} \, d\mathbf{s}_2 \frac{\nu_0\big([[f[J] ,\cHcubt(s_1)],\cHcubt(s_2)]\big)}{\vol }\nonumber\\
	&=&\frac1N\big(\cBd(f_0)[J](T;\lambda) \, + \, \lambda \cBd^{(1)}(f_0)[J](T;\lambda) \, + \, \lambda^2 \cBd^{(2)}(f_0)[J](T;\lambda)\big)\nonumber\\
	&&+ \, J(0)|\Phi_{\frac{T}{\lambda^2}}|^2 \, + \, \rrbogb_{2,d}(\frac{T}{\lambda^2};f[J] ) \, + \, J(0)\rrbogc_{2,d}(\frac{T}{\lambda^2};f)\, ,
	\eeqn 	
	where 
	\eqn
	|\rrbogb_{2,d}(\frac{T}{\lambda^2};f[J] )| &\leq& C_{\vc,\fc,\reg}T^2  \|J\|_{\ell^\infty(\lattice)} \frac{\lambda}{N} \, , \label{eq-rrbogbf-0}\\
	|\rrbogc_{2,d}(t;f)| &\leq &C_{\vc,\fc,\reg}e^{C_{\reg}\vc\vol/\lambda T}\nonumber\\
	&&\frac{\jb{T}^4\vol^{\frac32}}{\lambda^3N^{\frac32}}\Big(1+\frac\vol{N}\Big)^2\Big(1+\frac{\vol^{\frac32}}{\lambda N^{\frac12}}\Big) \, .
	\eeqn 
	Let 
	\eqn
	\rrd_{2,0}(\frac{T}{\lambda^2};f[J] ) & := & \frac1N\Big(\cBd(f_0)[J](T;\lambda) \, - \, \cBc(f_0)[J](T;\lambda)\Big) \, ,\\
	\rrd_{2,1}(\frac{T}{\lambda^2};f[J] ) & := & \frac{\lambda}N\Big(\cBd^{(1)}(f_0)[J](T;\lambda) \, - \, \cBc^{(1)}(f_0)[J](T;\lambda)\Big) \, , \\
	\rrd_{2,2}(\frac{T}{\lambda^2};f[J] ) & := & \frac{\lambda^2}N\Big(\cBd^{(2)}(f_0)[J](T;\lambda) \, - \, \cBc^{(2)}(f_0)[J](T;\lambda)\Big) \, .
	\eeqn 
	Recalling \eqref{def-boltzmann-0}, \eqref{def-cBd1}, \eqref{def-cBd2}, \eqref{def-cBd21}, and \eqref{def-cBd22}, Lemma \ref{lem-disc} then implies 
	\eqn
	\rrd_{2,j}(\frac{T}{\lambda^2};f[J] ) & \leq & \frac{\lambda^j}{N\lambda^{2\reg+2}\vol^{\frac{\reg}3}}   \nonumber\\
	&\leq & \frac{C_{\vc,\fc,\reg}\jb{T}^{\reg+2}\|J\|_{W^{2\floor{\frac{\reg}2}+2,\infty}} }{N \lambda^{2\reg+2} \vol^{\frac{\reg}3}}  \label{eq-f-dis-0}
	\eeqn 
	for $j\in\{0,1,2\}$. Lemma \ref{lem-bog-bol} yields
	\eqn
	\lefteqn{\frac{\lambda^j}{N}|\cBc^{(j)}(f_0)[J](T;\lambda)|}\nonumber\\
	& \leq & C_{\vc,\fc,\reg}T(1+\log(T))\|J\|_{C^1}\frac{\lambda|\log(\lambda)| }{N}  \label{eq-cBcj-bd-0}
	\eeqn 
	for $j\in\{1,2\}$, and all $\lambda>0$ small enough, dependent on $\vc$ Let 
	\eqn
	\lefteqn{\frac1N\cBc(f_0)[J](T;\lambda) \, + \, \rrbogb_{2,c}(\frac{T}{\lambda^2};f[J] )}\nonumber\\
	&:=& \frac1N\big(\cBd(f_0)[J](T;\lambda) \, + \, \lambda\cBd^{(1)}(f_0)[J](T;\lambda) \, \nonumber\\
	&& + \, \lambda^2\cBd^{(2)}(f_0)[J](T;\lambda)\big) \, + \, \rrbogb_{2,d}(\frac{T}{\lambda^2};f[J] ) \, .
	\eeqn 
	Collecting \eqref{eq-rrbogbf-0}, \eqref{eq-f-dis-0}, and \eqref{eq-cBcj-bd-0}, we thus proved that
	\eqn
	\lefteqn{|\rrbogb_{2,c}(\frac{T}{\lambda^2};f[J]  )|}\nonumber\\
	&\leq & C_{\vc,\fc,\reg}\Big(\frac{ T^2 \|J\|_{\ell^\infty(\lattice)}\lambda}{N} \nonumber\\
	&& + \, \frac{\jb{T}^{\reg+2}\|J\|_{W^{2\floor{\frac{\reg}2}+2,\infty}} }{N \lambda^{2(\reg+1)}\vol^{\frac{\reg}3}}  \nonumber\\
	&& + \, \frac{\lambda |\log(\lambda)|T(1+\log(T))\|J\|_{W^{1,\infty}}}{N}\Big) \nonumber\\
	&\leq & \frac{C_{\vc,\fc,\reg}\jb{T}^{\reg+2}\|J\|_{W^{2\floor{\frac{\reg}2}+2,\infty}}}N\nonumber\\
	&& \Big(\frac{1}{\lambda^{2(\reg+1)}\vol^{\frac{\reg}3}} +\lambda|\log(\lambda)|\Big) \, .
	\eeqn 
	\par Finally, we compute the discretization error in the dynamics of $g$. Let 
	\eqn
	\lefteqn{\rrd_{1,c}(\frac{T}{\lambda^2};g[J])}\nonumber\\
	& := & \frac1{N}\int_0^T \, dS\, \big(\acqd(f_0)[J](S/\lambda^2) \, - \, \acqc(f_0)[J](S/\lambda^2)\big) \, .
	\eeqn 
	Lemma \ref{lem-disc} together with Proposition \ref{prop-bog-disc} implies 
	\eqn
	|\rrd_{1,c}(\frac{T}{\lambda^2};g[J])| & \leq & \frac{C_{\vc,\fc,\reg}\jb{T}^{\reg+1} \|J\|_{W^{2\floor{\frac{\reg}2}+2,\infty}}}{N\lambda^{2\reg}\vol^{\frac{\reg}3}}  \, .
	\eeqn
	Moreover, let 
	\eqn
	\lefteqn{\rrd_{2,c}(\frac{T}{\lambda^2};g[J])}\nonumber\\
	&:=& \frac1{N\lambda^2} \int_{\Delta[T,2]} \, d\bS_2 \, \big(\bbgd(f_0)[J](\bS_2/\lambda^2)) \, - \, \bbgc(f_0)[J](\bS_2/\lambda^2)) 
	\nonumber\\
	&& + \, \acd(f_0)[J](\bS_2/\lambda^2)) \, - \, \acc(f_0)[J](\bS_2/\lambda^2))\big)
	\eeqn 
	Again, applying Lemma \ref{lem-disc} together with Proposition \ref{prop-bog-disc}, yields
	\eqn
	|\rrd_{2,c}(\frac{T}{\lambda^2};g[J])| &\leq& \frac{C_{\vc,\fc,\reg}\jb{T}^{\reg+2}\|J\|_{W^{2\floor{\frac{\reg}2}+2,\infty}}}{N\lambda^{2\reg+2}\vol^{\frac\reg3}}
	\eeqn 
	for all $\lambda>0$ small enough, dependent on $\vc$. This concludes the proof. 
	\endprf
	
	For the next result, observe that
	\eqn
	\dEcub \, = \,E(p_1)+E(p_2)-E(p_1+p_2) \, =\, -p_1\cdot p_2 \, .\label{eq-dEcub-exp-0}
	\eeqn 
	Let
	\eqn
	\lefteqn{\cBfec(f_0)[J](T)}\nonumber\\
	&:=& \frac{ \pi T}{(2\pi)^6} \int_{p_1\perp p_2} \, d\cH^5(\bp_2) \,  \dJcub(\bp_2)\frac{(\hat{v}(p_1)+\hat{v}(p_2))^2}{|\bp_2|} \nonumber\\
	&& \big(\fbar(p_1)\fbar(p_2)f_0(p_1+p_2) \, - \, f_0(p_1)f_0(p_2)\fbar(p_1+p_2)\big)  \label{def-Bfec} 
	\eeqn 
	be the Boltzmann operator with energy conserving collision kernel for the free energy dispersion $E(p)=\frac{|p|^2}2$.
	\begin{proposition}\label{prop-f-order}
		We have that 
		\eqn
		\lefteqn{\cB_c(f_0)[J](T;\lambda)}\nonumber\\
		&= &  \frac{\pi T}{(2\pi)^6}\int \, d\bp_2 \, \delta_{\frac{2\lambda^2}{T}}(\dOcub(\bp_2))\dJcub(\bp_2)(\hat{v}(p_1)+\hat{v}(p_2))^2\nonumber\\
		&& \big(\fbar(p_1)\fbar(p_2)f_0(p_1+p_2) \, - \, f_0(p_1)f_0(p_2)\fbar(p_1+p_2)\big) \, ,
		\eeqn
		where $\delta_\vep(x) :=\frac{\vep\sin^2(x/\vep)}{\pi x^2}$.	Moreover, approximating $\cB(f_0)[J](T\lambda^{-2})$ with its continuum counterpart, i.e., replacing $\frac{1}{\vol} \sum_{\lattice}$ by the (Lebesgue-) integral $\frac1{(2\pi)^3}\int_{\R^3} $, we find that
		\eqn
		\frac1N\cBc(f_0)[J](T;\lambda) &=& \frac1N\cBfec(f_0)[J](T) \, + \, \rrfec_{2,c}(\frac{T}{\lambda^2};f[J] ) \, ,
		\eeqn 
		where
		\eqn
		|\rrfec_{2,c}(\frac{T}{\lambda^2};f[J] )| \, \leq \, C_{\vc,\fc,\reg}\jb{T} \|J\|_{W^{2,\infty}}\frac{\lambda}{N}
		\eeqn 
		for all $\lambda>0$ small enough, dependent on $\vc$.
	\end{proposition}
	\prf
	Observe that $\dEcub=0$ is equivalent to $p_1\perp p_2$, see \eqref{eq-dEcub-exp-0}, and that
	\eqn
	|\nabla\dEcub(\bp_2)| \, = \, \Big|\begin{pmatrix}
		-p_2\\ -p_1
	\end{pmatrix}\Big| \, = \, |\bp_2| \, .
	\eeqn 
	Moreover, notice that for any $\omega\in\R\setminus\{0\}$, we have
	\eqn
	\Re\int_{\Delta[t,2]} d\mathbf{s}_2\, e^{-i\omega(s_1-s_2)} & = & \frac{1-\cos(\omega t)}{\omega^2}\nonumber\\
	&=& 2\frac{\sin^2(\frac{\omega t}2)}{\omega^2} \, .
	\eeqn 
	Observe that $\delta_a(x) =\frac{a\sin^2(x/a)}{\pi x^2}$ defines an approximate identity in the sense that we have
	\eqn
	\delta_a(x) &\geq& 0  \quad \forall a>0 \, ,\label{eq-approxid-pos}\\
	\int \, dx \, \delta_a(x) &= & 1 \quad \forall a>0 \, ,\label{eq-approxid-norm}\\
	\int_{|x|>\rho} \, dx \, \delta_a(x) &\leq & \frac{Ca}{\rho} \quad\forall a,\rho>0 \, .\label{eq-approxid-supp}
	\eeqn  
	With that, we obtain
	\eqn
	\lefteqn{\BBF(f_0)[J](T;\lambda)}\nonumber\\
	&= & \frac{\pi T}{(2\pi)^6} \int \, d\bp_2 \, \delta_{\frac{2\lambda^2}{T}}(\dOcub(\bp_2))\dJcub(\bp_2)(\hat{v}(p_1)+\hat{v}(p_2))^2\nonumber\\
	&&\big(f_0(p_1+p_2)+f_0(p_1)f_0(p_1+p_2)+f_0(p_1+p_2)f_0(p_2)\nonumber\\
	&&-f_0(p_1)f_0(p_2) \big) \label{eq-bol-order-0} 
	\eeqn
	This proves the first part of the statement. 
	\par For the second part, assume $\bp_2$ ranges over $\R^6$. Emphasizing the dependence of $\cE(\lambda,\bp_2) := \dOcub_\lambda (\bp_2)$ on $\lambda>0$, we abbreviate \eqref{eq-bol-order-0} as
	\eqn
	\lambda^2 \BBF(f_0)[J](T\lambda^{-2}) &=:& \int \, d\bp_2 \, \delta_{\frac{2\lambda^2}{T}}(\cE(\lambda,\bp_2)) H(\bp_2)\, , \label{eq-bol-order-1} 
	\eeqn 
	where we emphasize. Observe that $\cE(0,\bp_2)=\dEcub(\bp_2)$. Using the Fundamental Theorem of Calculus by the Coarea Formula, we obtain that
	\eqn
	N \rrf_{2}(\frac{T}{\lambda^2};f[J] )&:=&\int \, d\bp_2 \, \Big[\delta_{\frac{2\lambda^2}{T}}(\cE(\lambda,\bp_2)) \, - \, \delta_{\frac{2\lambda^2}{T}}(\cE(0,\bp_2))\Big] H(\bp_2) \nonumber\\
	&=& \int_0^\lambda \, d\tau \,  \delta_{\frac{2\lambda^2}{T}}'(\cE(\tau,\bp_2)) \partial_\tau \cE(\tau,\bp_2) H(\bp_2)\nonumber\\
	&=& \int_0^\lambda \, d\tau \, \int \,  d\omega\, \delta_{\frac{2\lambda^2}{T}}'(\omega) \nonumber\\
	&&\int_{ \cE(\tau,\bp_2)=\omega} \, d\cH^5\, \frac{\partial_\tau \cE(\tau,\bp_2) H(\bp_2)}{|\nabla_{\bp_2}\cE(\tau,\bp_2)|} \, . \label{eq-rrfd-1}
	\eeqn 
	We prove in Lemma \ref{lem-r2fd} that  
	\eqn
	|\rrf_{2}(t;f[J] )| &\leq& C_{\vc,\fc,\reg}T\|J\|_{W^{2,\infty}}\frac{\lambda}{N}
	\eeqn 
	for all $\lambda>0$ small enough, dependent on $\vc$. 
	\par Next, using \eqref{eq-approxid-norm} together with the Coarea Formula, we have that
	\eqn
	N\rrec_2(\frac{T}{\lambda^2};f[J] ) &:=& \int \, d\bp_2 \, \delta_{\frac{2\lambda^2}{T}}(\dEcub(\bp_2)) H(\bp_2) \nonumber\\
	&& - \, \int_{\dEcub=0} \, d\cH^5\, \frac{H}{|\nabla\dEcub|}\nonumber\\
	&=& \int \, d\omega\, \delta_{\frac{2\lambda^2}{T}}(\omega)\Big(\int_{\dEcub=\omega} \, d\cH^5\, \frac{H}{|\nabla\dEcub|}\nonumber\\
	&&+\, \int_{\dEcub=0} \, d\cH^5\, \frac{H}{|\nabla\dEcub|} \Big) \, .\label{eq-rr2ec-def-0}
	\eeqn 
	We prove in Lemma \ref{lem-rr2ec} that
	\eqn
	|\rrec_{2}(\frac{T}{\lambda^2};f[J] )| \, \leq \, C_{\vc,\fc,\reg}\sqrt{T} \|J\|_{W^{2,\infty}}\frac{\lambda}N   
	\eeqn
	for all $\lambda>0$ small enough, dependent on $\vc$. In particular, we have that  	
	\eqn
	\rrfec_{2,c}(\frac{T}{\lambda^2};f[J] ) \, = \, \rrf_2(\frac{T}{\lambda^2};f[J] ) \, + \, \rrec_2(\frac{T}{\lambda^2};f[J] )
	\eeqn 
	satisfies		
	\eqn
	|\rrfec_{2,c}(\frac{T}{\lambda^2};f[J] )| \, \leq \, C_{\vc,\fc,\reg}\jb{T} \|J\|_{W^{2,\infty}} \frac{\lambda }{ N}  \, .
	\eeqn 	
	This concludes the proof.
	\endprf

	\subsection{Centered expectations}
	In order to resolve the fluctuations around the HFB dynamics, we have to consider the dynamics relative to the condensate term, i.e.,
	\eqn
	f_t^{(\Phi)}(p) &:=& f_t(p) \, - \, |\Phi_t|^2 \delta(p) \, . \label{def-f-rel-dyn}
	\eeqn 
	
	\begin{proposition}\label{prop-eff}
		Let $T>0$, $\lambda\in(0,1)$ and $\vol\geq1$. Let $J\in L^{\infty}(\R^3;\R)$. Then the following holds. 
		\begin{enumerate}
			\item
			\eqn
			\frac1{N^{\frac12}\lambda}\cCd(f_0)(T;\lambda) \, = \, \frac1{N^{\frac12}\lambda} \cCd(f^{(\Phi)})(T;\lambda) \, + \, \rreff_{1,d}(\frac{T}{\lambda^2};\Phi) \, ,
			\eeqn   
			where
			\eqn
			\lefteqn{|\rreff_{1,d}(\frac{T}{\lambda^2};\Phi)|} \nonumber\\
			& \leq &
			C_{\vd,\fd}\jb{T}^3\frac{\vol^{\frac32} e^{C\vd \vol/\lambda T}}{N\lambda^2}  \nonumber\\
			&&\Big(1+\frac{\vol}{N}\Big)\Big(1 \, + \, \frac{1}{N^{\frac12}\lambda\vol^{\frac32}}\Big) \, ,
			\eeqn 
			\item  
			\begin{enumerate}
				\item
				\eqn
				\lefteqn{\frac1N\big(\cBd(f_0)[J](T;\lambda) \, + \, \lambda \cBd^{(1)}(f_0)[J](T;\lambda) \, + \, \lambda^2 \cBd^{(2)}(f_0)[J](T;\lambda)\big)}\nonumber\\
				& = &  \frac1N\big(\cBd(f^{(\Phi)})[J](T;\lambda) \, + \, \lambda \cBd^{(1)}(f^{(\Phi)})[J](T;\lambda) \nonumber\\
				&&+ \, \lambda^2 \cBd^{(2)}(f^{(\Phi)})[J](T;\lambda) \big) \, + \,  \rreff_{2,d}(\frac{T}{\lambda^2};f[J] ) \, ,
				\eeqn
				with
				\eqn
				\lefteqn{|\rreff_{2,d}(\frac{T}{\lambda^2};f[J] )|}\nonumber\\
				& \leq & C_{\vd,\fd} \|J\|_{\ell^\infty(\lattice)} \jb{T}^4\frac{\vol^{\frac32} e^{C\vd \vol/\lambda T}}{N^{\frac32}\lambda^3}  \nonumber\\
				&&\Big(1+\frac{\vol}{N}\Big)\Big(1 \, + \, \frac{1}{N^{\frac12}\lambda\vol^{\frac32}}\Big) \nonumber\\
				&& \Big[1 \, + \, \jb{T}^2\frac{\vol^{\frac32} e^{C\vd \vol/\lambda T}}{N^{\frac12}\lambda} \Big(1+\frac{\vol}{N}\Big)\Big(1 \, + \, \frac{1}{N^{\frac12}\lambda\vol^{\frac32}}\Big)\Big] \, ,
				\eeqn 
				\item 
				\eqn
				\lefteqn{\frac1N\cBc(f_0)[J](T;\lambda)}\nonumber\\
				&= &  \frac1N\cBc(f^{(\Phi)})[J](T;\lambda) \, + \,  \rreff_{2,c}(\frac{T}{\lambda^2};f[J] ) \, ,
				\eeqn 
				where $\rreff_{2,c}(\frac{T}{\lambda^2};f[J] )$ satisfies the same bound as $\rreff_{2,d}(\frac{T}{\lambda^2};f[J] )$,
			\end{enumerate}
			\item 
			\begin{enumerate}
				\item 
				\eqn
				\lefteqn{\frac{1}N\int_0^T \, dS \, \acqd(f_0)[J](S/\lambda^2)} \nonumber\\
				& = & \frac{1}N\int_0^T \, dS \, \acqd(f^{(\Phi)})[J](S/\lambda^2) \, +\, \rrf_{1,d}(\frac{T}{\lambda^2};g[J])
				\eeqn 
				with 
				\eqn
				\lefteqn{|\rreff_{1,d}(\frac{T}{\lambda^2};g[J] )| }\nonumber\\
				& \leq & C_{\vd,\fd}\|J\|_{\ell^\infty(\lattice)} \jb{T}^3\frac{\vol^{\frac32} e^{C\vd \vol/\lambda T}}{N^{\frac32}\lambda}  \nonumber\\
				&&\Big(1+\frac{\vol}{N}\Big)\Big(1 \, + \, \frac{1}{N^{\frac12}\lambda\vol^{\frac32}}\Big)\nonumber\\
				&& \Big[1 \, + \, \jb{T}^2\frac{\vol^{\frac32} e^{C\vd \vol/\lambda T}}{N^{\frac12}\lambda} \Big(1+\frac{\vol}{N}\Big)\Big(1 \, + \, \frac{1}{N^{\frac12}\lambda\vol^{\frac32}}\Big)\Big] \, , 
				\eeqn 
				\item 
				\eqn
				\lefteqn{\frac{1}{N\lambda^2}\int_{\Delta[T,2]} \, d\bS_2 \, \big( \bbgd(f_0)[J](\bS_2/\lambda^2) + \acd(f_0)[J](\bS_2/\lambda^2)\big)} \nonumber\\
				&=& \frac{1}{N\lambda^2}\int_{\Delta[T,2]} \, d\bS_2 \, \big( \bbgd(f^{(\Phi)})[J](\bS_2/\lambda^2)\nonumber\\
				&& + \acd(f^{(\Phi)})[J](\bS_2/\lambda^2)\big) \, + \, \rreff_{2,d}(\frac{T}{\lambda^2};g[J])
				\eeqn 
				with
				\eqn
				\lefteqn{|\rreff_{2,d}(\frac{T}{\lambda^2};g[J] )|}\nonumber\\
				& \leq & C_{\vd,\fd}\|J\|_{\ell^\infty(\lattice)} \jb{T}^3 \frac{\vol^{\frac32} e^{C\vd \vol/\lambda T}}{N^{\frac32}\lambda^3}  \nonumber\\
				&&\Big(1+\frac{\vol}{N}\Big)\Big(1 \, + \, \frac{1}{N^{\frac12}\lambda\vol^{\frac32}}\Big) \nonumber\\
				&& \Big[1 \, + \, \jb{T}^2\frac{\vol^{\frac32} e^{C\vd \vol/\lambda T}}{N^{\frac12}\lambda} \Big(1+\frac{\vol}{N}\Big)\Big(1 \, + \, \frac{1}{N^{\frac12}\lambda\vol^{\frac32}}\Big)\Big] \, .
				\eeqn 
			\end{enumerate}
		\end{enumerate}
	The analogous statements hold true for the continuum approximation if one replaces $(\vd,\fd)$ on the RHS of the inequalities by $(C_{\reg}\vc,C_{\reg}\fc)$.
	\end{proposition} 
	\prf 
	In the notation of the proof, we will focus on the discrete case. The bounds for the continuum approximation follow by replacing $(\vd,\fd)$ by 
	\eqn
		(C_{\reg}\vc,C_{\reg}\fc) \, , 
	\eeqn 
	see Remark \ref{rem-dis-to-con}.
	\par Let $t=T\lambda^{-2}$. Next, Lemma \ref{lem-duham} and the definition \eqref{def-f-rel-dyn} of $f^{(\Phi)}$ imply that
	\eqn\label{eq-f0-ft-bd-0}
	\int \, dp \, J(p)\big(f_0(p)-f_t^{(\Phi)}(p)\big) & = &  \int \, dp \, J(p)\big(f_0(p)-f_t(p)\big) \, + \, J(0)|\Phi_t|^2\nonumber\\
	&=& -i\int_0^t \,ds \, \frac{\nu_s([f[J] ,\cH_I(s)])}{\vol } \, + \, J(0)|\Phi_t|^2\, . 
	\eeqn
	Using \eqref{eq-r2error-exp-1}, we have the estimate
	\eqn
	\lefteqn{\Big|\frac{\nu_s([f[J] ,\cH_I(s)])}\vol\Big|} \nonumber\\
	 & \leq &  C_{\vd,\fd}e^{C\fd \vol/\lambda T}\|J\|_{\ell^\infty(\lattice)} \frac{\lambda \vol^{\frac32}}{N^{\frac12}}\Big(1 \, + \, \frac{\vol^{\frac12}}{N^{\frac12}}\Big) \, . \label{eq-nmhi-bd-0}
	\eeqn 
	We have that
	\eqn
	\Phi_t & = & -i\int_0^t \, ds \,  \frac{\nu_s\big([a_0,\cH_I(s)]\big)}\vol \, . \label{eq-phi-diff-1}
	\eeqn
	Recall from \eqref{eq-a0cub-exp} and \eqref{eq-a0-h4-0} that
	\eqn
		[a_0,\cHcubt(s)] & = & f[J_1(s)]+g[J_2(s)]+g^*[J_3(s)]+J_4(s) \, , \\
		[a_0,\cHquart(s)] &=& \frac1{N^{\frac12}} \cHcubt^{(1)}(s) \, .
	\eeqn 
	Collecting \eqref{eq-j123-bd}, \eqref{eq-j4-bd}, and using Lemma \ref{lem-hcub-est}, \eqref{eq-r2error-exp-1} yields
	\eqn
	|\Phi_t| &\leq & C_{\vd,\fd}T e^{C\fd \vol/\lambda T}\frac{1}{N^{\frac12}\lambda}  \Big(1 \, + \, \frac{\vol^{\frac12}}{N^{\frac12}}\Big) \, . \label{eq-phi-est-0}
	\eeqn 
	Then \eqref{eq-f0-ft-bd-0}, \eqref{eq-nmhi-bd-0}, and \eqref{eq-phi-est-0} yield that
	\eqn
	\lefteqn{\Big| \int \, dp \, J(p)\big(f_0(p)-f_t^{(\Phi)}(p)\big)\Big|}\nonumber\\
	&\leq& C_{\vd,\fd}\jb{T}^2\|J\|_{\ell^\infty(\lattice)}\frac{\vol^{\frac32} e^{C\vd \vol/\lambda T}}{N^{\frac12}\lambda}  \nonumber\\
	&&\Big(1+\frac{\vol}{N}\Big)\Big(1 \, + \, \frac{1}{N^{\frac12}\lambda\vol^{\frac32}}\Big) \, . \label{eq-f0-ft-bd-1}
	\eeqn 
	\par As a consequence of \eqref{eq-f0-ft-bd-1} and recalling definition \eqref{def-condensate-0} of $\con(f)$, we find that
	\eqn
	\lefteqn{|\rreff_1(\frac{T}{\lambda^2};\Phi)|}\nonumber\\
		& \leq &
	C_{\vd,\fd}\jb{T}^3\frac{\vol^{\frac32} e^{C\vd \vol/\lambda T}}{N\lambda^2}  \nonumber\\
	&&\Big(1+\frac{\vol}{N}\Big)\Big(1 \, + \, \frac{1}{N^{\frac12}\lambda\vol^{\frac32}}\Big) \, . \label{eq-r1c-bd-0}
	\eeqn 
	\par After substitution and using the notation in the proof of Proposition \ref{prop-tail-est}, the terms in $\bbf^{(j)}(f_0)[J](t)-\bbf^{(j)}(f^{(\Phi)})[J](t)$ and $\bbg^{(k)}(f_0)[J](t)-\bbg^{(k)}(f^{(\Phi)})[J](t)$  are of the form
	\eqn
	\lefteqn{\int \, d\bp_2 \, H_1(p_1)H_2(p_2)H_3(p_1\pm p_2) f_{s_2}^{(\Phi)}(p_1)(f_0(p_2)-f_{s_2}^{(\Phi)}(p_2)) \, , } \label{eq-f0-fs-1}\\
	&&\int \, d\bp_2 \, H_1(p_1)H_2(p_2)H_3(p_1\pm p_2) f_0(p_1)(f_0(p_2)-f_{s_2}^{(\Phi)}(p_2)) \, ,\label{eq-f0-fs-2}\\
	&&\int \, d\bp_2 \, H_1(p_1)H_2(p_2)H_3(p_{12})(f_0(p_{12})-f_{s_2}^{(\Phi)}(p_{12})) \, ,\label{eq-f0-fs-3}
	\eeqn 
	where $H_j\in L^\infty_{\bs_2}(\R^2;L^\infty_p(\R^3))$ can differ in every line and in the last line, we require $H_1\in L^\infty_{\bs_2}(\R^2; L^1_p(\R^3))$. We have that $s_2\leq T\lambda^{-2}$ with $\lambda\in(0,1)$.
	\par Observe that
	\eqn
	\int \, dp \, |H_1(p)| f_0(p) & = & \frac{\nu_0(f(|H_1|))}{\vol } \nonumber\\
	& \leq & \frac{\|H_1\|_\infty\nu_0(\nb)}{\vol } \nonumber\\
	& \leq & \fd\|H_1\|_\infty\label{eq-f0-test-0}
	\eeqn 
	due to Lemma \ref{lem-fgphi-bound} and 
	\eqn
	\frac{\nu_0(\nb)}\vol \, = \, \int \, dp\, f_0(p) \, \leq \, \fd \, .
	\eeqn
	\eqref{eq-f0-ft-bd-1} and \eqref{eq-f0-test-0} imply that the terms of the form \eqref{eq-f0-fs-2} satisfy
	\eqn
	\lefteqn{|\int \, d\bp_2 \, H_1(p_1)H_2(p_2)H_3(p_1\pm p_2) f_0(p_1)(f_0(p_2)-f_{s_2}^{(\Phi)}(p_2))|} \nonumber\\
	&\leq & \int \, dp_1 \, |H_1(p_1)|f_0(p_1)\Big|\int \, dp_2 \, H_2(p_2)H_3(p_1\pm p_2)(f_0(p_2)-f_{s_2}^{(\Phi)}(p_2))\Big| \nonumber\\
	&\leq &  C_{\vd,\fd}\|H_1\|_\infty\|H_2\|_\infty\|H_3\|_\infty\jb{T}^2\frac{\vol^{\frac32} e^{C\vd \vol/\lambda T}}{N^{\frac12}\lambda}  \nonumber\\
	&&\Big(1+\frac{\vol}{N}\Big)\Big(1 \, + \, \frac{1}{N^{\frac12}\lambda\vol^{\frac32}}\Big) \, . \label{eq-f0-fs-2-bd}
	\eeqn 	
	For \eqref{eq-f0-fs-1}, \eqref{eq-f0-ft-bd-1} and \eqref{eq-f0-fs-2-bd} yield that
	\eqn \label{eq-f0-fs-sample-comp}
	\lefteqn{|\int \, d\bp_2 \, H_1(p_1)H_2(p_2)H_3(p_1\pm p_2) f_{s_2}^{(\Phi)}(p_1)(f_0(p_2)-f_{s_2}^{(\Phi)}(p_2))|} \nonumber\\
	&\leq&|\int \, d\bp_2 \, H_1(p_1)H_2(p_2)H_3(p_1\pm p_2) f_0(p_1)(f_0(p_2)-f_{s_2}^{(\Phi)}(p_2))| \nonumber\\
	&&+   |\int \, d\bp_2 \, H_1(p_1)H_2(p_2)H_3(p_1\pm p_2) \big(f_{s_2}^{(\Phi)}(p_1)-f_0(p_1)\big)(f_0(p_2)-f_{s_2}^{(\Phi)}(p_2))|\nonumber\\
	&\leq& C_{\vd,\fd} \|H_1\|_\infty \jb{T}^2\frac{\vol^{\frac32} e^{C\vd \vol/\lambda T}}{N^{\frac12}\lambda} \nonumber\\
	&&\Big(1+\frac{\vol}{N}\Big)\Big(1 \, + \, \frac{1}{N^{\frac12}\lambda\vol^{\frac32}}\Big) \nonumber\\
	&&\Big(\|H_2\|_\infty\|H_3\|_\infty \, + \, \sup_{p_1}\Big|\int \, dp_2 \, H_2(p_2)H_3(p_1\pm p_2)(f_0(p_2)-f_{s_2}^{(\Phi)}(p_2))\Big|\Big) \nonumber\\
	&\leq& C_{\vd,\fd} \|H_1\|_\infty\|H_2\|_\infty\|H_3\|_\infty \jb{T}^2\frac{\vol^{\frac32} e^{C\vd \vol/\lambda T}}{N^{\frac12}\lambda}  \nonumber\\
	&&\Big(1+\frac{\vol}{N}\Big)\Big(1 \, + \, \frac{1}{N^{\frac12}\lambda\vol^{\frac32}}\Big) \nonumber\\
	&& \Big[1 \, + \, \jb{T}^2\frac{\vol^{\frac32} e^{C\vd \vol/\lambda T}}{N^{\frac12}\lambda} \Big(1+\frac{\vol}{N}\Big)\Big(1 \, + \, \frac{1}{N^{\frac12}\lambda\vol^{\frac32}}\Big)\Big] \, .
	\eeqn 
	Employing \eqref{eq-f0-ft-bd-1} again, we estimate \eqref{eq-f0-fs-3} by
	\eqn\label{eq-f0-fs-3-bd}
	\lefteqn{|\int \, d\bp_2 \, H_1(p_1)H_2(p_2)H_3(p_1+ p_2)(f_0(p_1+p_2)-f_{s_2}^{(\Phi)}(p_1+p_2))|} \nonumber\\
	&\leq& C_{\vd,\fd}\|H_1\|_1\|H_2\|_\infty\|H_3\|_\infty \jb{T}^2\frac{\vol^{\frac32} e^{C\vd \vol/\lambda T}}{N^{\frac12}\lambda}  \nonumber\\
	&&\Big(1+\frac{\vol}{N}\Big)\Big(1 \, + \, \frac{1}{N^{\frac12}\lambda\vol^{\frac32}}\Big) \, .
	\eeqn 
	Collecting \eqref{eq-f0-fs-2-bd}--\eqref{eq-f0-fs-3-bd}, recalling definitions \eqref{def-boltzmann-0}--\eqref{def-cBd2}, and denoting $\cB^{(0)}:=\cB$, we find the upper bound
	\eqn 
	\lefteqn{\frac{\lambda^j}N|\cB^{(j)}(f_0)[J](T;\lambda)-\cB^{(j)}(f^{(\Phi)})[J](T;\lambda)|}\nonumber\\
	&\leq & C_{\vd,\fd}\frac{\lambda^{2+j}t^2\|J\|_{\ell^\infty(\lattice)}}{N} \jb{T}^2\frac{\vol^{\frac32} e^{C\vd \vol/\lambda T}}{N^{\frac12}\lambda}  \nonumber\\
	&&\Big(1+\frac{\vol}{N}\Big)\Big(1 \, + \, \frac{1}{N^{\frac12}\lambda\vol^{\frac32}}\Big) \nonumber\\
	&& \Big[1 \, + \, \jb{T}^2\frac{\vol^{\frac32} e^{C\vd \vol/\lambda T}}{N^{\frac12}\lambda} \Big(1+\frac{\vol}{N}\Big)\Big(1 \, + \, \frac{1}{N^{\frac12}\lambda\vol^{\frac32}}\Big)\Big] \, . \label{eq-bol-f-bd-0}
	\eeqn 
	In particular, we have that
	\eqn\label{eq-r2b-bd-0}
	\lefteqn{|\rreff_{2,d}(t;f[J] )| \, , \, | \rreff_{2,c}(t;f[J] )|}\nonumber\\
	& \leq & C_{\vd,\fd} \|J\|_{\ell^\infty(\lattice)} \jb{T}^4\frac{\vol^{\frac32} e^{C\vd \vol/\lambda T}}{N^{\frac32}\lambda^3}  \nonumber\\
	&&\Big(1+\frac{\vol}{N}\Big)\Big(1 \, + \, \frac{1}{N^{\frac12}\lambda\vol^{\frac32}}\Big) \nonumber\\
	&& \Big[1 \, + \, \jb{T}^2\frac{\vol^{\frac32} e^{C\vd \vol/\lambda T}}{N^{\frac12}\lambda} \Big(1+\frac{\vol}{N}\Big)\Big(1 \, + \, \frac{1}{N^{\frac12}\lambda\vol^{\frac32}}\Big)\Big] \, .
	\eeqn 
	\par Finally, it remains to estimate $\rreff_{j}(\frac{T}{\lambda^2};g[J] )$, $j\in\{1,2\}$. Following analogous steps that lead to \eqref{eq-r2b-bd-0}, Proposition \ref{prop-bog-disc} implies
	\eqn
	\lefteqn{|\rreff_{1}(\frac{T}{\lambda^2};g[J] )| }\nonumber\\
	& \leq & C_{\vd,\fd}\|J\|_{\ell^\infty(\lattice)} \jb{T}^3\frac{\vol^{\frac32} e^{C\vd \vol/\lambda T}}{N^{\frac32}\lambda}  \nonumber\\
	&&\Big(1+\frac{\vol}{N}\Big)\Big(1 \, + \, \frac{1}{N^{\frac12}\lambda\vol^{\frac32}}\Big)\nonumber\\
	&& \Big[1 \, + \, \jb{T}^2\frac{\vol^{\frac32} e^{C\vd \vol/\lambda T}}{N^{\frac12}\lambda} \Big(1+\frac{\vol}{N}\Big)\Big(1 \, + \, \frac{1}{N^{\frac12}\lambda\vol^{\frac32}}\Big)\Big] \, , \\
	\lefteqn{|\rreff_{2}(\frac{T}{\lambda^2};g[J] )|}\nonumber\\
	& \leq & C_{\vd,\fd}\|J\|_{\ell^\infty(\lattice)} \jb{T}^3 \frac{\vol^{\frac32} e^{C\vd \vol/\lambda T}}{N^{\frac32}\lambda^3}  \nonumber\\
	&&\Big(1+\frac{\vol}{N}\Big)\Big(1 \, + \, \frac{1}{N^{\frac12}\lambda\vol^{\frac32}}\Big) \nonumber\\
	&& \Big[1 \, + \, \jb{T}^2\frac{\vol^{\frac32} e^{C\vd \vol/\lambda T}}{N^{\frac12}\lambda} \Big(1+\frac{\vol}{N}\Big)\Big(1 \, + \, \frac{1}{N^{\frac12}\lambda\vol^{\frac32}}\Big)\Big] \, . 
	\eeqn
	This concludes the proof. 
	\endprf
	
	\newpage
	
	\section{Main order terms in the evolution of $(\Phi,f,g)$\label{sec-main2}}
	\label{sec-coh-mean-exact}
	
	In this section, we will prove Theorem \ref{thm-main-con}. We will always assume that $\lambda\in(0,1)$ is small enough to comply with the estimates proven in all preceding steps.
	
	\subsection{Discrete case}
	Fix $L\geq1$. Recall that we refer to the case of fixed $\Lambda$ as the 'discrete case'.  
	\subsubsection{Main order term of $\Phi$ \label{sec-phi-main-dis}}
	
	Using Propositions \ref{prop-tail-est}, and \ref{prop-bog-disc}, we have that
	\eqn
	\Phi_{\frac{T}{\lambda^2}} & = & \frac1{N^{\frac12}\lambda}\cCd(f_0)(T;\lambda) \, + \Rem_2(\frac{T}{\lambda^2};\Phi) \, + \, \rrb_{1,d}(\frac{T}{\lambda^2};\Phi)\label{eq-phi-expansion-dis-0}
	\eeqn
	with
	\eqn
	|\Rem_2(\frac{T}{\lambda^2};\Phi) | &\leq & C_{\vd,\fd}T^2e^{C\vd\vol/\lambda T}\frac{\vol^{\frac32}}{N\lambda^2} \, , \\
	|\rrb_{1,d}(\frac{T}{\lambda^2};\Phi)| & \leq & \frac{C_{\vd,\fd}T}{N^{\frac12}} \, .
	\eeqn 
	In particular, the main term is given by
	\eqn
	\frac1{N^{\frac12}\lambda}\cCd(f_0)(T;\lambda) \, = \, -\frac{i T}{N^{\frac12}\lambda} \int_\lattice \, dp\, \hat{v}(p)f_0(p) \, , \label{eq-phi-main-0}
	\eeqn 
	see \eqref{def-condensate-0}, and thus it is of size $N^{-\frac12}\lambda^{-1}$. In order to suppress $\Rem_2(\frac{T}{\lambda^2};\Phi)$, we choose $\lambda=\log\log N/\log N$.

	\subsubsection{Main order term of $f$ \label{sec-f-main-dis}}
	
	For $f$, we apply Propositions \ref{prop-tail-est}, and \ref{prop-bog-disc} to obtain that 
	\eqn
	\lefteqn{\int_\lattice dp \, \big(f_{\frac{T}{\lambda^2}}^{(\Phi)} - f_0^{(\Phi)}\big)J(p)}\nonumber\\
	&=& \frac1N\Big(\cBd(f_0)[J](T;\lambda) \, + \, \lambda \cBd^{(1)}(f_0)[J](T;\lambda) \, + \, \lambda^2 \cBd^{(2)}(f_0)[J](T;\lambda)\Big) \nonumber\\
	&& + \, \Rem_2(\frac{T}{\lambda^2};f[J] ) \, + \, \rrbogb_{2,d}({\frac{T}{\lambda^2}};f[J] ) \, + \, J(0)\rrbogc_2({\frac{T}{\lambda^2}};f) \, , \label{eq-f-expansion-dis-0}
	\eeqn 
	where
	\eqn
	|\Rem_2({\frac{T}{\lambda^2}};f[J] )| &\leq&C_{\vd,\fd} \jb{T}^4\Jin \frac{e^{C\vd\vol/\lambda T} \vol^6}{\lambda^4N^2}  \, , \\
	|\rrbogb_{2,d}({\frac{T}{\lambda^2}};f[J] )| &\leq& C_{\vd,\fd}\|J\|_{\ell^\infty(\lattice)} T^2 \frac{\lambda}{N} \, , \\
	|\rrbogc_{2,d}(t;f)| &\leq &C_{\vd,\fd}e^{C\vd\vol/\lambda T}\frac{\jb{T}^4\vol^{\frac32}}{\lambda^3N^{\frac32}}\Big(1+\frac{\vol^{\frac32}}{\lambda N^{\frac12}}\Big) \, .
	\eeqn
	Due to Lemma \ref{lem-bog-expansion}, we can expand the oscillatory factors in $\cBd^{(j)}$ in powers of $\lambda$. Then the main order term depends on $\hat{v}$, $\lambda(N)$, and $T$. We are interested in terms up to $O(N^{-1})$ which, as we will see below, is the size of the main order term in the continuum approximation. For the same reasons as above, we choose $\lambda=\log\log N/\log N$. 
	
	\subsubsection{Main order term of $g$\label{sec-main-g-dis}}
	Similar to the case of $f$, we need to subtract the dynamics of the condensate in order to resolve the fluctuations defined by $g$. For that, we introduce
	\eqn
	g_t^{(\Phi)}(p) \, := \, g_t(p) \, - \, \Phi_t^2\delta(p) \, .\label{def-g-rel}
	\eeqn  
	Propositions \ref{prop-tail-est} and \ref{prop-bog-disc} then yield
	\eqn
	\lefteqn{\int_\lattice dp \, g_{\frac{T}{\lambda^2}}^{(\Phi)}(p)J(p)}\nonumber\\
	&=& \frac{1}N\int_0^T \, dS \, \acqd(f_0)[J](S/\lambda^2) \nonumber\\
	&&+ \, \frac{1}{N\lambda^2}\int_{\Delta[T,2]} \, d\bS_2 \, \big( \bbgd(f_0)[J](\bS_2/\lambda^2) + \acd(f_0)[J](\bS_2/\lambda^2)\big) \nonumber\\
	&& + \, \Rem_2(\frac{T}{\lambda^2};g[J] ) \, + \, J(0)\rrbogc_2(\frac{T}{\lambda^2};g) \label{eq-g-expansion-dis-0}
	\eeqn 
	with
	\eqn
	|\Rem_2(\frac{T}{\lambda^2};g[J] )| &\leq& C_{\vd,\fd} \jb{T}^4\Jtn \frac{e^{C\vd\vol/\lambda T} \vol^6}{\lambda^4N^2} \, , \\
	|\rrbogc_{2,d}(\frac{T}{\lambda^2};g)| &\leq& 
	C_{\vd,\fd}e^{C\vd\vol/\lambda T}\frac{\jb{T}^4\vol^{\frac32}}{\lambda^3N^{\frac32}}\Big(1+\frac{\vol^{\frac32}}{\lambda N^{\frac12}}\Big) \, . 
	\eeqn 
	As in the case of $f$, we again observe a  phenomenon similar to the Talbot effect on $\acqd$, $\bbgd$, and $\acd$. We choose $N^{-1}$ as reference order. 
	
	\subsubsection{Conclusion for $L\sim 1$} \label{sec-main-dis}
	As described above, we choose $\lambda=\log\log(N)/\log(N)$. Let 
	\eqn
	\rate_{\Psi,d}
	&=& \min\Big\{\frac{\log \frac1\lambda}{\log N},\frac12-\frac{\log\frac1\lambda}{\log N}-\frac{C\vd\vol T}{\log \log N} \Big\} \, , \label{def-rate-psi-d}\\
	\rate_{F,d}
	&=& \min\Big\{\frac{\log\frac1\lambda}{\log N},\frac12-\frac{3\log\frac1\lambda}{\log N}-\frac{C\vd\vol T}{\log \log N}\Big\} \, , \label{def-rate-f-d}\\
	\rate_{G,d} &=& \frac12-\frac{3\log\frac1\lambda}{\log N}-\frac{C\vd\vol T}{\log \log N} \, .  \label{def-rate-g-d}
	\eeqn 
	Then we obtain 
	\eqn
	\lefteqn{\Big|\Phi_{\frac{T}{\lambda^2}} \, - \, \frac1{N^{\frac12}\lambda}\cCd(f_0)(T;\lambda)\Big|}\nonumber\\
	&& \hspace{2cm} \leq \, \frac{C_{\vd,\fd,\vol,T}}{N^{\frac12+\rate_{\Psi,d}}\lambda}\, , \label{eq-phi-main-dis-1} \\
	\lefteqn{\Big|\int_\lattice dp \, \big(f_{\frac{T}{\lambda^2}}^{(\Phi)}(p)- f_0^{(\Phi)}(p)\big)J(p)} \nonumber\\
	&& - \, \frac1N\Big(\cBd(f_0)[J](T;\lambda) \, + \, \sum_{j=1}^2 \lambda^j \cBd^{(j)}(f_0)[J](T;\lambda) \Big)\Big| \nonumber\\
	&& \hspace{2cm} \leq \, \frac{C_{\vd,\fd,\vol,T}\|J\|_{\ell^\infty(\lattice)}}{N^{1+\rate_{F,d}}} \, , \label{eq-f-main-dis-1} \\
	\lefteqn{\Big|\int_\lattice dp \, g_{\frac{T}{\lambda^2}}^{(\Phi)}(p)J(p) \, - \,  \frac{1}N\int_0^T \, dS \, \acqd(f_0)[J](S/\lambda^2)} \nonumber\\
	&&- \, \frac{1}{N\lambda^2}\int_{\Delta[T,2]} \, d\bS_2 \, \big( \bbgd(f_0)[J](\bS_2/\lambda^2) + \acd(f_0)[J](\bS_2/\lambda^2)\big)\Big|\nonumber\\
	&& \hspace{2cm} \leq \, \frac{C_{\vd,\fd,\vol,T}\Jtn}{N^{1+\rate_{G,d}}} \label{eq-g-main-dis-1}
	\eeqn 
	for all $N$ larger than a universal constant. Notice that for all $N$ large enough, we have that $\delta_{j,d}=\delta_{j,d}(N,\vol,T)>0$ for $j\in\{\Psi,F,G\}$.

	\subsection{Continuum approximation}
	
	We have that errors coming from the tail in the Duhamel expansion grow like $\exp(C_{\reg}\vc T\vol /\lambda)$. As a consequence, we require
	\eqn
	\frac\vol\lambda=O(\frac{\log(N)}{\log\log(N)})
	\eeqn 
	as $N\to\infty$. In order to use bounds established for general values of $\vol$ to the limit $\vol\to\infty$, for specific expressions of interest, we employ Remark \ref{rem-dis-to-con}. 
	
	\subsubsection{Main order term of $\Phi$ \label{sec-phi-main-con}}
	
	Using Propositions \ref{prop-tail-est} and \ref{prop-bog-cont}, we have that
	\eqn
	\Phi_{\frac{T}{\lambda^2}} & = & \frac1{N^{\frac12}\lambda}\cCc(f_0)(T;\lambda) \, + \Rem_2(\frac{T}{\lambda^2};\Phi) \, + \, \rrb_{1,c}(\frac{T}{\lambda^2};\Phi) \label{eq-phi-expansion-con-0}
	\eeqn
	with
	\eqn
	|\Rem_2(\frac{T}{\lambda^2};\Phi) | &\leq & C_{\vc,\fc,\reg}T^2e^{C_{\reg}\vc\vol/\lambda T}\frac{\vol^{\frac32}}{N\lambda^2}\nonumber\\
	&& \Big(1+\frac\vol{N}\Big) \, , \nonumber \\
	|\rrb_{1,c}(t;\Phi)| & \leq & \frac{C_{\vc,\fc,\reg}T}{N^{\frac12}\lambda}\Big(\lambda \, + \, \frac1{\vol^{\frac{r}6}}\Big)
	\eeqn 
	for some $\reg>6$. The main order term is given by
	\eqn
		\frac1{N^{\frac12}\lambda}\cCc(f_0)(T;\lambda) \, = \, -\frac{i T}{(2\pi)^3N^{\frac12}\lambda} \int_{\R^3} \, dp\, \hat{v}(p)f_0(p) \, , \label{eq-phi-con-main-0}
	\eeqn 
	and it is of size $N^{\frac12}\lambda^{-1}$.

	\subsubsection{Main order term of $f$ \label{sec-f-main-cont}}
	
	For $f$, we apply Propositions \ref{prop-tail-est}, \ref{prop-bog-cont}, and \ref{prop-f-order} to obtain that 
	\eqn
	\lefteqn{\int_\lattice dp \, \big(f_{\frac{T}{\lambda^2}}^{(\Phi)} (p)- f_0^{(\Phi)}(p)\big)J(p)}\nonumber\\
	&=& \frac1N\cBfec(f_0)[J](T) \nonumber\\
	&& + \, \Rem_2(\frac{T}{\lambda^2};f[J] ) \, + \, \rrbogb_{2,c}({\frac{T}{\lambda^2}};f[J] ) \nonumber\\
	&& + \, J(0)\rrbogc_{2,c}({\frac{T}{\lambda^2}};f) \, + \, \rrfec_{2,c}(T\lambda^{-2};f[J] ) \, , \label{eq-f-expansion-con-0}
	\eeqn 
	where
	\eqn
	|\Rem_2({\frac{T}{\lambda^2}};f[J] )| &\leq&C_{\vc,\fc,\reg} \jb{T}^4\|J\|_\infty \nonumber\\
	&&\frac{e^{C_{\reg}\vc\vol/\lambda T} \vol^6}{\lambda^4N^2}\Big(1 + \frac{\vol}{N}\Big)^2  \, , \\
	|\rrbogb_{2,c}(\frac{T}{\lambda^2};f[J] )| &\leq& \frac{C_{\vc,\fc,\reg}\jb{T}^{\reg+2}\|J\|_{W^{2\floor{\frac{\reg}2}+2,\infty}}}N\nonumber\\
	&& \Big(\frac{1}{\lambda^{2(\reg+1)}\vol^{\frac{\reg}3}} +\lambda|\log(\lambda)|\Big) \, , \\
	|\rrbogc_{2,c}(\frac{T}{\lambda^2};f[J] )| &\leq& C_{\vc,\fc,\reg}e^{C_{\reg}\vc\vol/\lambda T}\nonumber\\
	&&\frac{\jb{T}^4\vol^{\frac32}}{\lambda^3N^{\frac32}}\Big(1+\frac\vol{N}\Big)^2\Big(1+\frac{\vol^{\frac32}}{\lambda N^{\frac12}}\Big) \, , \\
	|\rrfec_{2,c}(\frac{T}{\lambda^2};f[J] )| & \leq & C_{\vc,\fc,\reg}\jb{T} \|J\|_{W^{2,\infty}}\frac{\lambda}{N} 
	\eeqn
	for all $\lambda>0$ small enough, dependent on $\vc$. Recalling definition \eqref{def-Bfec} of $\cBfec$, the main term is given by 
	\eqn
	\lefteqn{\frac{\pi T}{(2\pi)^6N} \int_{p_1\perp p_2} \, d\cH^5(\bp_2) \,  \dJcub(\bp_2)\frac{(\hat{v}(p_1)+\hat{v}(p_2))^2}{|\bp_2|}} \nonumber\\
	&& \big(\fbar(p_1)\fbar(p_2)f_0(p_1+p_2) \, - \, f_0(p_1)f_0(p_2)\fbar(p_1+p_2)\big) \, ,
	\eeqn 
	and it is of size $N^{-1}$. In order to have that the discretization error is negligible, we need to impose $\frac1\lambda=o(\vol^{\frac{\reg}{6(\reg+1)}})$ as $\vol\to\infty$.  In particular, we may choose 
	\eqn
		L \, = \, \lambda^{-2-\frac2\reg-\veps} 
	\eeqn 
	for any arbitrary $\vep>0$. Recall that we required
	\eqn
	\frac{\vol}\lambda=O\big(\frac{\log N}{\log\log N}\big)
	\eeqn 
	as $N\to\infty$ to suppress the tail in the Duhamel expansion. We may thus choose
	\eqn
	 \lambda &=& \Big(\frac{\log \log N}{\log N}\Big)^{\frac{\reg}{(7+\veps)\reg+6}} \, , \\
	 L &=& \Big(\frac{\log N}{\log \log N}\Big)^{\frac{(2+\veps)\reg+2}{(7+\veps)\reg+6}} \, .
	\eeqn

	\subsubsection{Main order term of $g$\label{sec-main-g-cont}}
	
	Propositions \ref{prop-tail-est} and \ref{prop-bog-cont} imply
	\eqn
	\lefteqn{\int_\lattice dp \, g_{\frac{T}{\lambda^2}}^{(\Phi)}(p)J(p) }\nonumber\\
	&=& \frac{1}N\int_0^T \, dS \, \acqc(f_0)[J](S/\lambda^2) \nonumber\\
	&&+ \, \frac{1}{N\lambda^2}\int_{\Delta[T,2]} \, d\bS_2 \, \big( \bbgc(f_0)[J](\bS_2/\lambda^2) + \acc(f_0)[J](\bS_2/\lambda^2)\big) \nonumber\\
	&& + \, \Rem_2(\frac{T}{\lambda^2};g[J] ) \, +\, \rrd_{1,c}(\frac{T}{\lambda^2};g[J]) \nonumber\\
	&& + \, \rrd_{2,c}(\frac{T}{\lambda^2};g[J]) \, + \, J(0)\rrbogc_{2,c}(\frac{T}{\lambda^2};g) \label{eq-g-expansion-con-0}
	\eeqn 
	with
	\eqn
	|\Rem_2(\frac{T}{\lambda^2};g[J] )| &\leq& C_{\vc,\fc,\reg} \jb{T}^4\Jcn \nonumber\\
	&&\frac{e^{C_{\reg}\vc\vol/\lambda T} \vol^6}{\lambda^4N^2}\Big(1 + \frac{\vol}{N}\Big)^2 \, , \\
	|\rrd_{1,c}(\frac{T}{\lambda^2};g[J])| & \leq & \frac{C_{\vc,\fc,\reg}\jb{T}^{\reg+1} \|J\|_{W^{2\floor{\frac{\reg}2}+2,\infty}}}{N\lambda^{2\reg}\vol^{\frac{\reg}3}} \, , \\
	|\rrd_{2,c}(\frac{T}{\lambda^2};g[J])|	& \leq & \frac{C_{\vc,\fc,\reg}\jb{T}^{\reg+2}\|J\|_{W^{2\floor{\frac{\reg}2}+2,\infty}}}{N\lambda^{2(\reg+1)}\vol^{\frac\reg3}}   \, ,\\
	|\rrbogc_{2,c}(\frac{T}{\lambda^2};g[J] )| &\leq& 
	C_{\vc,\fc,\reg}e^{C_{\reg}\vc\vol/\lambda T}\nonumber\\
	&&\frac{\jb{T}^4\vol^{\frac32}}{\lambda^3N^{\frac32}}\Big(1+\frac\vol{N}\Big)^2\Big(1+\frac{\vol^{\frac32}}{\lambda N^{\frac12}}\Big) 
	\eeqn 
	for all $\lambda>0$ small enough, dependent on $\vc$. All the errors are suppressed for the choices of $\vol$ and $\lambda$ as in the case of $f$ above.

	\subsubsection{Conclusion for $L\sim \lambda^{-2-}$}
	
	Recall that we impose
	\eqn
	\lambda &=& \Big(\frac{\log \log N}{\log N}\Big)^{\frac{\reg}{(7+\veps)\reg+6}} \, , \\
	L &=& \lambda^{-2-\frac2\reg-\veps} \, = \, \Big(\frac{\log N}{\log \log N}\Big)^{\frac{(2+\veps)\reg+2}{(7+\veps)\reg+6}} \, .
	\eeqn 
	Let 
	\eqn
	\rate_{\Psi,c} &:=& \min\Big\{\frac12-C_{\reg,\veps}\frac{\log\frac1\lambda }{\log N}-\frac{C_\reg \vc  T}{\log \log N},\frac{\log\frac1\lambda }{\log N}\Big\}  \, , \label{def-rate-psi-c} \\
	\rate_{F,c} &:=& \min\Big\{\frac12-C_{\reg,\veps}\frac{\log\frac1\lambda }{\log N}-\frac{C_\reg\vc T}{\log \log N},\frac{\veps\reg\log\frac1\lambda }{\log N},\nonumber\\
	&&\frac{\log\frac1\lambda-\log\log\frac1\lambda}{\log N}\Big\} \, , \label{def-rate-f-c}\\
	\rate_{G,c} &:=&\min\Big\{\frac12-C_{\reg,\veps}\frac{\log\frac1\lambda }{\log N}-\frac{C_\reg\vc T}{\log \log N}, \frac{\veps\reg\log\frac1\lambda }{\log N}\Big\} \, . \label{def-rate-g-c}
	\eeqn 
	Observe that we have
	\eqn
	\|J\|_{W^{2,\infty}(\R^3)} \, \leq \, \Jcn \, .
	\eeqn  
	Then we have proved that, for some $N_0=N_0(\vc)$,
	\eqn
	\lefteqn{\Big|\Phi_{\frac{T}{\lambda^2}} \, - \, \frac1{N^{\frac12}\lambda}\cCc(f_0)(T;\lambda)\Big|} \nonumber\\
	&& \hspace{2cm} \leq \,\frac{C_{\vc,\fc,\reg,\veps,T}}{N^{\frac12+\rate_{\Psi,c}}\lambda} \, , \label{eq-phi-main-con-1} \\
	\lefteqn{\Big|  \int_\lattice dp \, \big(f_{\frac{T}{\lambda^2}}^{(\Phi)}(p) - f_0^{(\Phi)}(p)\big)J(p)\, - \, \frac1N\cBfec(f_0)[J](T) \Big|} \nonumber\\
	&& \hspace{2cm} \leq \,\frac{C_{\vc,\fc,\reg,\veps,T}\|J\|_{W^{2\floor{\frac{\reg}2}+2,\infty}}}{N^{1+\rate_{F,c}}} \, , \label{eq-f-main-con-1} \\ 
	\lefteqn{\Big| \int_\lattice dp\, g_{\frac{T}{\lambda^2}}^{(\Phi)}(p)J(p) \, - \,   \frac{1}N\int_0^T \, dS \, \acqc(f_0)[J](S/\lambda^2)} \nonumber\\
	&&- \, \frac{1}{N\lambda^2}\int_{\Delta[T,2]} \, d\bS_2 \, \big( \bbgc(f_0)[J](\bS_2/\lambda^2)\nonumber\\
	&&+ \acc(f_0)[J](\bS_2/\lambda^2)\big)\Big| \nonumber\\
	&& \hspace{2cm} \leq \,\frac{C_{\vc,\fc,\reg,\veps,T}\Jcn}{N^{1+ \rate_{G,c}}} \label{eq-g-main-con-1}
	\eeqn 
	for all $N\geq N_0(\vc)$. Observe that for all $N$ large enough, we have that $\rate_{j,c}=\rate_{j,c}(N,T)>0$ for $j\in\{\Psi,F,G\}$.

	\section{Effective equations \label{sec-main1}}
	
	After establishing the sizes of the leading order terms, we can derive the effective equations. This will prove Theorem \ref{thm-main-dis}. 
	
	\subsection{Discrete case}
	
	Let $L\geq1$ be fixed. Choose $\lambda=\log\log(N)/\log(N)$ as explained in section \ref{sec-main-dis}. 
	
	\subsubsection{Evolution of $\Phi$}
	
	We have that
	\eqn\label{eq-phi-evo-0}
	\Phi_{\frac{T}{\lambda^2}} & = & \frac1{N^{\frac12}\lambda}\cCd(f^{(\Phi)})(T;\lambda) \nonumber\\
	&& + \Rem_2(\frac{T}{\lambda^2};\Phi) \, + \, \rrb_{1,d}(\frac{T}{\lambda^2};\Phi) \, + \, \rreff_{1,d}(\frac{T}{\lambda^2};\Phi),
	\eeqn
	see \eqref{eq-phi-expansion-dis-0}. Proposition \ref{prop-eff} implies
	\eqn
	|\rreff_{1,d}(\frac{T}{\lambda^2};\Phi)|&\leq & C_{\vd,\fd}\jb{T}^3\frac{\vol^{\frac32} e^{C\vd \vol/\lambda T}}{N\lambda^2}  \nonumber\\
	&&\Big(1+\frac{\vol}{N}\Big)\Big(1 \, + \, \frac{1}{N^{\frac12}\lambda\vol^{\frac32}}\Big)\nonumber\\
	&\leq & \frac{C_{\vd,\fc,\vol,T}}{N^{\frac12+\rate_{\Psi,d}}\lambda} \, . \label{eq-phi-dis-eff-1}
	\eeqn 
	for all $N$ large enough. Observe that
	\eqn
	\frac1{N^{\frac12}\lambda}\cCd(f^{(\Phi)})(T;\lambda) & =& -\frac{i }{N^{\frac12}\lambda} \int_0^T \, dS \, \int_\lattice \, dp\, \hat{v}(p)f^{(\Phi)}_{\frac{S}{\lambda^2}}(p) \, .
	\eeqn 
	Then, analogously to \eqref{eq-phi-main-dis-1} and employing \eqref{eq-phi-dis-eff-1}, we obtain that
	\eqn
	\Big|\Phi_{\frac{T}{\lambda^2}} \,  - \, \frac1{N^{\frac12}\lambda}\cCd(f^{(\Phi)})(T;\lambda)\Big| \, \leq \, \frac{C_{\vd,\fd,\vol,T}}{N^{\frac12+\rate_{\Psi,d}}\lambda} \, . \label{eq-phi-evo-1}
	\eeqn 
	
	\subsubsection{Evolution of $f$}
	
	\par \eqref{eq-f-expansion-dis-0} and Proposition \ref{prop-eff} imply 
	\eqn
	\lefteqn{\int_\lattice dp \, \big(f_{\frac{T}{\lambda^2}}^{(\Phi)}(p) - f_0^{(\Phi)}(p)\big)J(p)}\nonumber\\
	&=& \frac1N\Big(\cBd(f^{(\Phi)})[J](T;\lambda) \, + \, \sum_{j=1}^2\lambda^j \cBd^{(1)}(f^{(\Phi)})[J](T;\lambda) \Big) \nonumber\\
	&& + \, \Rem_2(\frac{T}{\lambda^2};f[J] ) \, + \, \rrbogb_{2,d}({\frac{T}{\lambda^2}};f[J] )  \nonumber\\
	&& + \, J(0)\rrbogc_{2,d}({\frac{T}{\lambda^2}};f) \, + \, \rreff_{2,d}(\frac{T}{\lambda^2};f[J] ) \, .
	\eeqn 
	Recall that, due to Proposition \ref{prop-eff},  
	\eqn
	|\rreff_{2,d}(\frac{T}{\lambda^2};f[J] )|  &\leq & C_{\vd,\fd} \|J\|_{\ell^\infty(\lattice)} \jb{T}^4\frac{\vol^{\frac32} e^{C\vd \vol/\lambda T}}{N^{\frac32}\lambda^3}  \nonumber\\
	&&\Big(1+\frac{\vol}{N}\Big)\Big(1 \, + \, \frac{1}{N^{\frac12}\lambda\vol^{\frac32}}\Big) \nonumber\\
	&& \Big[1 \, + \, \jb{T}^2\frac{\vol^{\frac32} e^{C\vd \vol/\lambda T}}{N^{\frac12}\lambda} \Big(1+\frac{\vol}{N}\Big)\Big(1 \, + \, \frac{1}{N^{\frac12}\lambda\vol^{\frac32}}\Big)\Big] \nonumber\\
	&\leq & \frac{C_{\vd,\fd,\vol,T} \|J\|_{\ell^\infty(\lattice)}}{N^{1+\rate_{F,d}(N)}}\, , \label{eq-f-rreff-0}
	\eeqn 
	where we possibly enlarge the constant in the definition of $\rate_{F,d}$. In particular, similar to \eqref{eq-f-main-dis-1} we obtain that
	\begin{fleqn}[\parindent]
		\eqn
		\lefteqn{\Big|\int_\lattice dp \, \big(f_{\frac{T}{\lambda^2}}^{(\Phi)}(p)- f_0^{(\Phi)}(p)\big)J(p) \, - \, \frac1N\Big(\cBd(f^{(\Phi)})[J](T;\lambda)} \nonumber\\ 
		&& + \,\sum_{j=1}^2 \lambda^j \cBd^{(j)}(f^{(\Phi)})[J](T;\lambda)\Big)\Big|\nonumber\\
		& \leq & \frac{C_{\vd,\fd,\vol,T}\|J\|_{\ell^\infty(\lattice)}}{N^{1+ \rate_{F,d}}} \, .
		\eeqn 
	\end{fleqn}

	\subsubsection{Evolution of $g$} \eqref{eq-g-expansion-dis-0} yields
	\eqn
	\lefteqn{\int_\lattice dp \, g_{\frac{T}{\lambda^2}}^{(\Phi)}(p)J(p)}\nonumber\\
	&=& \frac{1}N\int_0^T \, dS \, \acqd(f^{(\Phi)})[J](S/\lambda^2) \nonumber\\
	&&+ \, \frac{1}{N\lambda^2}\int_{\Delta[T,2]} \, d\bS_2 \, \big( \bbgd(f^{(\Phi)})[J](\bS_2/\lambda^2) + \acd(f^{(\Phi)})[J](\bS_2/\lambda^2)\big) \nonumber\\
	&& + \, \Rem_2(\frac{T}{\lambda^2};g[J] ) \, + \, J(0)\rrbogc_{2,d}(\frac{T}{\lambda^2};g) \nonumber\\
	&& + \,  \rreff_{1}(\frac{T}{\lambda^2};g[J] ) \, + \,  \rreff_{2}(\frac{T}{\lambda^2};g[J] ) \, ,
	\eeqn 
	where, due to Proposition \ref{prop-eff} and analogously to \eqref{eq-f-rreff-0},
	\eqn
	\lefteqn{|\rreff_{1}(\frac{T}{\lambda^2};g[J] )| \, , \, |\rreff_{2}(\frac{T}{\lambda^2};g[J] )|}\nonumber\\
	& \leq & \frac{C_{\vd,\fd,\vol,T}\|J\|_{\ell^\infty(\lattice)}}{N^{1+ \rate_{G,d}}} \, ,
	\eeqn 
	where we possibly enlarge the constant in the definition of $\rate_{G,d}$. As a consequence of \eqref{eq-g-main-dis-1}, we thus obtain that
	\eqn
	\lefteqn{\Big| \int_\lattice dp \, g_{\frac{T}{\lambda^2}}^{(\Phi)}(p)J(p) \, - \,  \frac{1}N\int_0^T \, dS \, \acqd(f^{(\Phi)})[J](S/\lambda^2)} \nonumber\\
	&&- \, \frac{1}{N\lambda^2}\int_{\Delta[T,2]} \, d\bS_2 \, \big( \bbgd(f^{(\Phi)})[J](\bS_2/\lambda^2) + \acd(f^{(\Phi)})[J](\bS_2/\lambda^2)\big)\Big| \nonumber\\
	& \leq& \frac{C_{\vd,\fd,\vol,T}\Jtn}{N^{1+ \rate_{G,d}(N)}} \, .
	\eeqn 	
	
	\subsubsection{Conclusion for $L\sim1$}
	
	Recall that $\lambda=\log\log(N)/\log(N)$. We consider the effective quantities
	\eqn
	\Psi_T &=& \Phi_{\frac{T}{\lambda^2}} \, , \\
	F_T &=& f_{\frac{T}{\lambda^2}}^{(\Phi)} \, , \\
	G_T &=& g_{\frac{T}{\lambda^2}}^{(\Phi)} \, .
	\eeqn 
	We have that
	\eqn
	\lambda^2\Re \int_{[\Delta[T\lambda^{-2},2]} \, d\bs_2 \, e^{-i\omega(s_1-s_2)}h(s_2) &=& \lambda^2\int_0^{\frac{T}{\lambda^2}} \, ds\, \frac{\sin(\omega(\frac{T}{\lambda^2} -s))}{\omega} h(s) \nonumber\\
	&=&  \int_0^T \, dS\, \frac{\sin\Big(\frac{\omega}{\lambda^2}(T-S)\Big)}{\omega} h(\frac{S}{\lambda^2}) \, . \label{eq-phase-int-0}
	\eeqn  
	For a function $H_t(p)$ and $j\in\{1,2\}$, denote
	\eqn
	Q_{d;T-S,\lambda}(H_S)[J] &=& \int_{(\lattice)^2} d\bp_2 \frac{\sin\Big(\frac{\Omega(p_1)+\Omega(p_2)-\Omega(p_1+p_2)}{\lambda^2}(T-S)\Big)}{\Omega(p_1)+\Omega(p_2)-\Omega(p_1+p_2)}\nonumber\\
	&&(\hat{v}(p_1)+\hat{v}(p_2))^2\big(J(p_1)+J(p_2)-J(p_1+p_2)\big) \nonumber\\
	&&\big(\widetilde{H_S}(p_1)\widetilde{H_S}(p_2) H_S(p_1+p_2) \nonumber\\
	&& - \, H_S(p_1) H_S(p_2)\widetilde{H_S}(p_1+p_2)\big) \, , \\
	q^{(j)}_{d,F;\bS_2,\lambda}(H_{S_2})[J] &=& \bbf^{(j)}(H_{\cdot \lambda^2})[J](\bS_2/\lambda^2) \, , \\
	\cQ_{d,G;T,\lambda}(H)[J]
	&=& \, \frac1{\lambda^2}\int_{\Delta[T,2]} \, d\bS_2 \, \bbgd(H_{\cdot\lambda^2})[J](\bS_2/\lambda^2) \, , \\
	\cA_{d;T,\lambda}(H)[J]
	&=& \int_0^T \, dS \, \acqd(H_{\cdot\lambda^2})[J](S/\lambda^2) \\ 
	&&+ \, \frac{1}{\lambda^2}\int_{\Delta[T,2]} \, d\bS_2 \acd(H_{\cdot\lambda^2})[J](\bS_2/\lambda^2) \, , \nonumber
	\eeqn 
	where we used Proposition \ref{prop-bog-disc}. $\cQ_{d,G;T,\lambda}$ denotes a generalized collision operator and $\cA_{d;T,\lambda}$ a generalized absorption operator. We have proved that
	\begin{fleqn}[\parindent]
		\eqn
		\lefteqn{\Big|\Psi_T +\frac{i }{N^{\frac12}\lambda} \int_0^T \, dS \, \int_{\lattice} \, dp\, \hat{v}(p)F_S(p)\Big|}\nonumber\\
		&& \hspace{2cm} \leq \, \frac{C_{\vd,\fd,\vol,T}}{N^{\frac12+\rate_{\Psi,d}}\lambda} \, , \\
		\lefteqn{\Big|\int_\lattice dp \, \big(F_T(p)-F_0(p)\big)J(p)\, - \, \frac1{N}\Big(\int_0^T \, dS \,  Q_{d;T-S,\lambda}(F_S)[J]}\nonumber\\
		&&+ \,  \int_{\Delta[T,2]} \, d\bS_2 \, \sum_{j=1}^2\lambda^j q^{(j)}_{d,F;\bS_2,\lambda}(F_{S_2})[J]\Big)\Big|\nonumber\\
		&& \hspace{2cm} \leq \, \frac{C_{\vd,\fd,\vol,T}\|J\|_{\ell^\infty(\lattice)}}{N^{1+\rate_{F,d}}} \, , \\
		\lefteqn{\Big| \int_\lattice dp\, G_T(p)J(p) \, - \, \frac{1}N\Big(\cA_{d;T,\lambda}(F)[J] \, + \, \cQ_{d,G;T,\lambda}(F)[J]\Big)\Big|} \nonumber\\
		&& \hspace{2cm} \leq \, \frac{C_{\vd,\fd,\vol,T}\Jtn}{N^{1+\rate_{G,d}}}
		\eeqn 
	\end{fleqn}
	for all $N>0$ larger than a universal constant.
	
	\subsection{Continuum approximation}
	
	Recall that 
	\eqn
	\lambda &=& \Big(\frac{\log \log N}{\log N}\Big)^{\frac{\reg}{(7+\veps)\reg+6}} \, , \\
	L &=& \lambda^{-2-\frac2\reg-\veps} \, = \, \Big(\frac{\log N}{\log \log N}\Big)^{\frac{(2+\veps)\reg+2}{(7+\veps)\reg+6}} \, .
	\eeqn

	\subsubsection{Evolution of $\Phi$}
	
	\eqref{eq-phi-expansion-con-0} and Proposition \ref{prop-eff} imply that
	\eqn
	\Phi_{\frac{T}{\lambda^2}}& = & \frac1{N^{\frac12}\lambda}\cCc(f^{(\Phi)})(T;\lambda) \nonumber\\
	&& + \Rem_2(\frac{T}{\lambda^2};\Phi) \, + \, \rrb_{1,c}(\frac{T}{\lambda^2};\Phi) \, + \, \rreff_1(t;\Phi) \, .
	\eeqn
	We have that
	\eqn
	|\rreff_1(t;\Phi)|&\leq & C_{\vc,\fc,\reg}\jb{T}^3\frac{\vol^{\frac32} e^{C_\reg\vc \vol/\lambda T}}{N\lambda^2}  \nonumber\\
	&&\Big(1+\frac{\vol}{N}\Big)\Big(1 \, + \, \frac{1}{N^{\frac12}\lambda\vol^{\frac32}}\Big) \nonumber\\
	&\leq & \frac{C_{\vc,\fc,\reg,T}}{N^{\frac12+\rate_{\Psi,c}}\lambda}   
	\eeqn 
	by possibly enlarging the constant $C$ in the definition \eqref{def-rate-psi-c} of $\rate_{\Psi,c}$. Combining this inequality with \eqref{eq-phi-main-con-1}, we find that, 
	\eqn
	\Big|\Phi_{\frac{T}{\lambda^2}} \, - \, \frac1{N^{\frac12}\lambda}\cCc(f^{(\Phi)})(T;\lambda)\Big| \, \leq \, \frac{C_{\vc,\fc,\reg,\veps,T}}{N^{\frac12+\rate_{\Psi,c}}\lambda}  \, .\label{eq-phi-evo-con-1}
	\eeqn

	\subsubsection{Evolution of $f$}
	
	As a consequence of \eqref{eq-f-expansion-con-0}, and Proposition \ref{prop-f-order}, \ref{prop-eff}, we have that
	\eqn
	\lefteqn{\int_\lattice dp \, \big(f_{\frac{T}{\lambda^2}}^{(\Phi)}(p) - f_0^{(\Phi)}(p)\big)J(p)}\nonumber\\
	&=& \frac1N\cBc(f^{(\Phi)})[J](T;\lambda) \nonumber\\
	&& + \, \Rem_2(\frac{T}{\lambda^2};f[J] ) \, + \, \rrbogb_{2,c}({\frac{T}{\lambda^2}};f[J] ) \nonumber\\
	&& + \, J(0)\rrbogc_2({\frac{T}{\lambda^2}};f)\, + \, \rreff_{2,c}(\frac{T}{\lambda^2};f[J] ) \, . 
	\eeqn 
	Observe that
	\eqn
	|\rreff_{2,c}(\frac{T}{\lambda^2};f[J] )|  &\leq & C_{\vc,\fc,\reg} \|J\|_{\infty} \jb{T}^4\frac{\vol^{\frac32} e^{C_\reg\vc \vol/\lambda T}}{N^{\frac32}\lambda^3}  \nonumber\\
	&&\Big(1+\frac{\vol}{N}\Big)\Big(1 \, + \, \frac{1}{N^{\frac12}\lambda\vol^{\frac32}}\Big) \nonumber\\
	&& \Big[1 \, + \, \jb{T}^2\frac{\vol^{\frac32} e^{C_\reg\vc \vol/\lambda T}}{N^{\frac12}\lambda} \Big(1+\frac{\vol}{N}\Big)\Big(1 \, + \, \frac{1}{N^{\frac12}\lambda\vol^{\frac32}}\Big)\Big]  \nonumber\\
	&\leq &  \frac{C_{\vc,\fc,\reg,T}\|J\|_{\infty}}{N^{1+\rate_{F,c}}} \label{eq-rreff-f-evo-0}
	\eeqn 
	by possibly enlarging the constants in the definition \eqref{def-rate-f-c} of $\rate_{F,c} (N,\lambda)$. Then, \eqref{eq-f-main-con-1} yields 
	\eqn
	\Big|  \int_\lattice dp \, \big(f_{\frac{T}{\lambda^2}}^{(\Phi)} (p)- f_0^{(\Phi)}(p)\big)J(p) \, - \, \frac1N\cBc(f^{(\Phi)})[J](T;\lambda)\Big| \nonumber\\
	\, \leq \, \frac{C_{\vc,\fc,\reg,\veps,T}\|J\|_{W^{2\floor{\frac{\reg}2}+2,\infty}}}{N^{1+\rate_{F,c}}} \, . \label{eq-f-evo-con-1}
	\eeqn

	\subsubsection{Evolution of $g$} 
	
	\eqref{eq-g-expansion-con-0} and Proposition \ref{prop-eff} imply
	\eqn
	\lefteqn{\int_\lattice dp \, g_{\frac{T}{\lambda^2}}^{(\Phi)} (p)J(p)}\nonumber\\
	&=& \frac{1}N\int_0^T \, dS \, \acqc(f^{(\Phi)})[J](S/\lambda^2) \nonumber\\
	&&+ \, \frac{1}{N\lambda^2}\int_{\Delta[T,2]} \, d\bS_2 \, \big( \bbgc(f^{(\Phi)})[J](\bS_2/\lambda^2) + \acc(f^{(\Phi)})[J](\bS_2/\lambda^2)\big) \nonumber\\
	&& + \, \Rem_2(\frac{T}{\lambda^2};g[J] ) \, +\, \rrd_1(\frac{T}{\lambda^2};g[J]) \, + \, \rrd_{2}(\frac{T}{\lambda^2};g[J])  \nonumber\\
	&&   + \, J(0)\rrbogc_2(\frac{T}{\lambda^2};g)  \, + \,  \rreff_{1}(\frac{T}{\lambda^2};g[J] ) \, + \,  \rreff_{2}(\frac{T}{\lambda^2};g[J] ) \, .
	\eeqn 
	Analogously to \eqref{eq-rreff-f-evo-0}, we have that
	\eqn
	\lefteqn{|\rreff_{1}(\frac{T}{\lambda^2};g[J] )| \, , \, |\rreff_{2}(\frac{T}{\lambda^2};g[J] )|}\nonumber\\
	& \leq & \frac{C_{\vc,\fc,\reg,\veps,T}\|J\|_{\infty}}{N^{1+\rate_{G,c}}}  \, ,
	\eeqn 
	by again possibly enlarging the constants in the definition \eqref{def-rate-g-c} of $\rate_{G,c} $. Applying \eqref{eq-g-main-con-1}, we obtain that
	\eqn
	\lefteqn{\Big| \int_\lattice dp \, g_{\frac{T}{\lambda^2}}^{(\Phi)} (p)J(p) \, - \,   \frac{1}N\int_0^T \, dS \, \acqc(f^{(\Phi)})[J](S/\lambda^2)} \nonumber\\
	&&- \, \frac{1}{N\lambda^2}\int_{\Delta[T,2]} \, d\bS_2 \, \big( \bbgc(f^{(\Phi)})[J](\bS_2/\lambda^2)\nonumber\\
	&&+ \acc(f^{(\Phi)})[J](\bS_2/\lambda^2)\big)\Big| \nonumber\\
	&\leq & \frac{C_{\vc,\fc,\reg,\veps,T}\Jcn}{N^{1+\rate_{G,c}}} \, . \label{eq-absorption-rough-con-1}
	\eeqn

	\subsubsection{Conclusion for $L\sim \lambda^{-2-}$}
	
	Recall that we impose
	\eqn
	\lambda &=& \Big(\frac{\log \log N}{\log N}\Big)^{\frac{\reg}{(7+\veps)\reg+6}} \, , \\
	L &=& \lambda^{-2-\frac2\reg-\veps} \, = \, \Big(\frac{\log N}{\log \log N}\Big)^{\frac{(2+\veps)\reg+2}{(7+\veps)\reg+6}} \, .
	\eeqn 
	Again, we consider the effective quantities
	\eqn
	\Psi_T &=& \Phi_{\frac{T}{\lambda^2}} \, , \\
	F_T &=& f_{\frac{T}{\lambda^2}}^{(\Phi)} \, , \\
	G_T &=& g_{\frac{T}{\lambda^2}}^{(\Phi)} \, .
	\eeqn 
	Let $Q_{c;T-S,\lambda}$, $Q_{c,G;T,\lambda}$ and $\cA_{c;T,\lambda}$ be defined analogously to \eqref{def-Qd}, \eqref{def-qdG} respectively \eqref{def-cBgdf} with sums over $\lattice$ replaced by integrals $\frac1{(2\pi)^3}\int_{\R^3}$ over $\R^3$. Collecting  \eqref{eq-phi-evo-con-1}, \eqref{eq-f-evo-con-1}, and \eqref{eq-absorption-rough-con-1}, we have proved that, for some possibly larger constant $N_0=N_0(\vc)$,
	\begin{fleqn}[\parindent]
		\eqn
		\lefteqn{\Big|\Psi_T +\frac{i }{(2\pi)^3N^{\frac12}\lambda} \int_0^T \, dS \, \int_{\R^3} \, dp\, \hat{v}(p)F_S(p) \Big|}\nonumber\\
		&& \hspace{2cm} \leq \, \,\frac{C_{\vc,\fc,\reg,\veps,T}}{N^{\frac12+\rate_{\Psi,c}}\lambda}  \, , \\ 
		\lefteqn{\Big|\int_\lattice dp \, \big(F_T(p)-F_0(p)\big)J(p)-  \frac1{N}\int_0^T \, dS \,Q_{c;T-S,\lambda}(F_S)\Big|} \nonumber\\
		&& \hspace{2cm} \leq \, \frac{C_{\vc,\fc,\reg,\veps,T}\|J\|_{W^{2\floor{\frac{\reg}2}+2,\infty}}}{N^{1+\rate_{F,c}}} \, , \\
		\lefteqn{\Big| \int_\lattice dp \, G_T(p)J(p) \, - \, \frac{1}N\Big(\cA_{c;T,\lambda}(F)[J] \, + \, \cQ_{c,G;T,\lambda}(F)[J]\Big)\Big|} \nonumber\\
		&& \hspace{2cm} \leq \, \frac{C_{\vc,\fc,\reg,\veps,T}\Jcn}{N^{1+ \rate_{G,c}}}
		\eeqn 
	\end{fleqn}
	for all $N\geq N_0$.

	\appendix 
	
	\section{Calculus for creation and annihilation operators \label{app-calc}}

	\begin{lemma}\label{lem-TI}
		Let $\nu$ be a translation invariant state, i.e.,
		\eqn
		\nu(A) \, = \, \nu(e^{ix\cdot \cP}Ae^{-ix\cdot \cP})
		\eeqn 
		for all $x\in\R^3$ and all observables $A$. Then we have
		\eqn
		\nu(\prod_{i=1}^ma_{p_i}^{(\sigma_i)}) \, = \, \frac{\delta(\sum_{i=1}^m\sigma_i p_i)}{\vol }\nu(\prod_{i=1}^ma_{p_i}^{(\sigma_i)}) \, .
		\eeqn 
	\end{lemma}
	\prf
	By translation invariance, we have that 
	\eqn
	\nu(\prod_{i=1}^ma_{p_i}^{(\sigma_i)}) &=& \nu(e^{ix\cdot \cP}\prod_{i=1}^ma_{p_i}^{(\sigma_i)}e^{-ix\cdot \cP})\nonumber\\
	&=& e^{ix\cdot(\sum_{i=1}^m\sigma_i p_i)}\nu(\prod_{i=1}^ma_{p_i}^{(\sigma_i)})
	\eeqn 
	for all $x\in\R^3$. Integrating both sides $\int_{\Lambda} dx$, we obtain
	\eqn
	\vol  \nu(\prod_{i=1}^ma_{p_i}^{(\sigma_i)}) \, = \, \delta(\sum_{i=1}^m\sigma_i p_i))\nu(\prod_{i=1}^ma_{p_i}^{(\sigma_i)}) \, ,
	\eeqn  
	which yields the statement.
	\endprf

	\begin{lemma}[Cumulant Formula] \label{lem-cumulant}
		Recall from \eqref{def-cumulant-n} that
		\eqn
			\kappa_n \, = \, (-\partial_\mu)^n \big|_{\mu=0}\int_{\lattice} dp \log\big(1-e^{-( K(p)-\mu)}\big) \, .
		\eeqn 
		Then there are constants $a_{n,k}\in\R$ such that
		\eqn
			\kappa_n \, = \, \sum_{k=1}^{n-1}a_{n,k} \int_\lattice dp \, f_0(p)^k(1+f_0(p))^{n-k} \, + \, \delta_{n,1} \int_\lattice dp \, f_0(p) \, ,
		\eeqn 
		where $f_0(p)=(e^{K(p)}-1)^{-1}$.
	\end{lemma}
	\prf
		We will show, in more generality that
		\eqn
			\kappa_n(\mu) &:=&  (-\partial_\mu)^n \int_{\lattice} dp \log\big(1-e^{-( K(p)-\mu)}\big)\\
			&=& \sum_{k=1}^{n-1}a_{n,k} \int_\lattice dp \, f_\mu(p)^k(1+f_\mu(p))^{n-k} \, + \, \delta_{n,1} \int_\lattice dp \, f_\mu(p) \label{eq-kappa-IH-0}
		\eeqn 
		for all $\mu\leq0$, where $f_\mu(p)=(e^{K(p)-\mu}-1)^{-1}$. A straightforward calculation yields
		\eqn
			\kappa_1(\mu) & = & \int_\lattice dp \, f_\mu(p) \, , \\
			\kappa_2(\mu) & = & \int_\lattice dp \, f_\mu(p)(1+f_\mu(p)) \, .
		\eeqn 
		Observe that
		\eqn
			-\partial_\mu f_\mu \, = \, f_\mu(1+f_\mu) \, . \label{eq-df-1}			
		\eeqn 
		Now assume that \eqref{eq-kappa-IH-0} for some fixed $n\in\N$, $n\geq2$. By definition, we have that
		\eqn
			\kappa_{n+1}(\mu) &=& (-\partial_\mu)\kappa_n(\mu) \nonumber\\
			&=& \int_\lattice dp \, \sum_{k=1}^{n-1}a_{n,k} \big(k f_\mu(p)^k(1+f_\mu(p))^{n+1-k} \nonumber\\
			&& + \, (n-k) f_\mu(p)^{k+1}(1+f_\mu(p))^{n-k} \big) \, .
		\eeqn
		After an index shift, we can further simplify this to
		\eqn
			\kappa_{n+1}(\mu) &=& a_{n,1}\int_\lattice dp \, f_\mu(p)(1+f_\mu(p))^n \nonumber\\
			&& + \, a_{n,n-1} \int_\lattice dp \, f_\mu(p)^n(1+f_\mu(p)) \nonumber\\
			&& + \, \sum_{k=2}^{n-1} (ka_{n,k}+(n+1-k)a_{n,k-1}) \nonumber\\
			&&\int_\lattice dp \, f_\mu(p)^k(1+f_\mu(p))^{n+1-k} \nonumber\\
			&=& \sum_{k=1}^n a_{n+1,k}\int_\lattice dp \, f_\mu(p)^k(1+f_\mu(p))^{n+1-k} \, ,
		\eeqn  
		for some $a_{n+1,k}\in\R$. This finishes the proof.
	\endprf 
	
	For the following standard result, we need to introduce some notation. For a proof of the statement, we refer, e.g., to \cite{bachfrsi,bafrsi}. Given a finite ordered subset $J=\{j_1<j_2<\ldots<j_r\}\subset \N$ and $\sigma_{j_k}\in\{\pm1\}$, we define the ordered product
	\eqn
		\prod_{j\in J} a_{p_j}^{(\sigma_j)} \, : = \, a_{p_{j_1}}^{(\sigma_{j_1})}\ldots a_{p_{j_r}}^{(\sigma_{j_r})} \, .
	\eeqn
	In addition, we abbreviate
	\eqn
		\bp_J \, := \, (p_{j_k})_{k=1}^r \, ,
	\eeqn 
	as well as 
	\eqn
	a^{(\sigma)}(\bp_J) \, := \, \prod_{j\in J} a^{(\sigma)}_{p_j} \, .
	\eeqn 
	Furthermore, we define the sets
	\eqn
		J_{\pm} \, := \, \{j\in J\mid \sigma_j=\pm 1\} 
	\eeqn 
	and the Wick-ordered product
	\eqn
		:\prod_{j\in J}a_{p_j}^{(\sigma_j)}: \, := \, a^+(\bp_{J_+})a(\bp_{J_-})
	\eeqn 	
	with all creation operators to the left, and all annihilation operators to the right. 	
	\par Finally, in order to keep track of the correct scaling, it is useful to work with the rescaled $\ell^2(\lattice)$-norm
	\eqn
		\|H\|_{L^2(\lattice)} \, = \, \frac{1}{\sqrt{\vol}} \|H\|_{\ell^2(\lattice)} \, ,
	\eeqn 
	see \eqref{def-rescaled}. More generally, we also define
	\eqn
		\|H\|_{L^\infty_{\bp_m} L^2_{\bk_n}((\lattice)^{m+n})} \, := \, \sup_{\bp_m\in (\lattice)^m}\Big( \int_{(\lattice)^n} d\bk_n |H(\bp_m,\bk_n)|^2\Big)^{\frac12} \, , \\
		\|H\|_{ L^2_{\bk_n}L^\infty_{\bp_m}((\lattice)^{m+n})} \, := \, \Big( \int_{(\lattice)^n} d\bk_n \sup_{\bp_m\in (\lattice)^m}|H(\bp_m,\bk_n)|^2\Big)^{\frac12} \, ,
	\eeqn 
	where in the case $n=0$, this norm reduces to $\|H\|_{L^\infty_{\bp_m}((\lattice)^m)}$, and in the case $m=0$, to $\|H\|_{L^2_{\bk_n}((\lattice)^n)}$.
		
	\begin{lemma}[Wick's Theorem] \label{wick-thm}
		Let $\sigma_j\in\{\pm1\}$, $p_j\in\R^3$ for all $j\in\{1,\ldots,n\}$, $n\in\N$. Then we have that
		\eqn
		\prod_{j=1}^n a_{p_j}^{(\sigma_j)} \, = \, \sum_{\substack{J\subseteq\\\{1,\ldots,n\}}} \scp{\vac}{\prod_{\substack{j\in\\\{1,\ldots,n\}\setminus J}}a_{p_j}^{(\sigma_j)}\vac} :\prod_{j\in J}a_{p_j}^{(\sigma_j)}: \, .
		\eeqn 
	\end{lemma}

	\begin{lemma}[Wick-ordered operator bound] \label{lem-wick-ord-bd}
		Let $M\in\N_0$, $n\in\N$, $J:=\{1,\ldots,n\}$, $\sigma_j\in\{\pm1\}$ for all $j\in J$. Let $H:(\lattice)^n\to \C$, and $g_j:\lattice\to\C$ be given functions. Then the following holds true
		\begin{enumerate}
		\item If $J_\pm\neq\emptyset$, we have that
		 \begin{fleqn}[\parindent]
			\eqn
			\lefteqn{\Big\|\int_{(\lattice)^n} d\bp_n H(\bp_n)\delta(\sum_{j=1}^np_j\sigma_j) \prod_{j=1}^ng_j(p_j):\prod_{j=1}^na_{p_j}^{(\sigma_j)} :P_M \Big\|}\nonumber\\
			&\leq & \| |H|^{\frac12}  \prod_{j\in J_-} g_j(p_j) \delta(\sum_{j=1}^n\sigma_j p_j)^{\frac12}\|_{L^\infty_{\bp_{J_+}}L^2_{\bp_{J_-}}}\| |H|^{\frac12}  \prod_{j\in J_+} g_j(p_j) \delta(\sum_{j=1}^n\sigma_j p_j)^{\frac12}\|_{L^\infty_{\bp_{J_-}}L^2_{\bp_{J_+}}}\nonumber\\
			&&(M+\sum_{j=1}^n\sigma_j)_{|J_+|}^{\frac12}(M)_{|J_-|}^{\frac12}\mathds{1}_{M\geq |J_-|} \, , \nonumber
			\eeqn 
			where
			\eqn
			(x)_m \, := \, \prod_{k=0}^{m-1}(x-k) \label{def-falling-fac}
			\eeqn 
			denotes the falling factorial. 
			\item If $J_+=\emptyset$ and $n\geq2$, we find that 
			\eqn
			\lefteqn{\Big\|\int_{(\lattice)^n} d\bp_n H(\bp_n)\delta(\sum_{j=1}^np_j) \prod_{j=1}^na_{p_j} P_M \Big\|}\nonumber\\
			&\leq &  \big(\|H\|_{L^2_{\bp_{n-2}L^\infty_{p_{n-1},p_n}}}^2(M-n+1)+\vol \|\delta(\sum_{j=1}^np_j)^{\frac12} H\|_{L^2_{\bp_n}}^2\big)^{\frac12} (M)_{n-1}^{\frac12} \, .
			\eeqn 
		\end{fleqn}
	Similarly, in the case $J_-=\emptyset$, we have that 
	\eqn
	\lefteqn{\Big\|\int_{(\lattice)^n} d\bp_n \, H(\bp_n) \delta(\sum_{j=1}^np_j) a^+(\bp_n) P_M\Big\|}\\
	& \leq & \big(\|H\|_{L^2_{\bp_{n-2}}L^\infty_{p_{n-1},p_n}}^2(M+1)+\vol \|\delta(\sum_{j=1}^np_j)^{\frac12} H\|_{L^2_{\bp_n}}^2\big)^{\frac12} (M+n)_{n-1}^{\frac12} \, .
	\eeqn 
	\item If $n=1$, we obtain
	\eqn
	\|a_0 P_M\| \, = \, \sqrt{M\vol} \, .
	\eeqn 
	\end{enumerate}
	\end{lemma}
	\prf
	Let $\Phi_+\in\cF_{M+\sum_{j=1}^n\sigma_j}$ and $\Phi_-\in \cF_M$ be two normalized test functions. 
	\par In the case $J_{\pm}\neq\emptyset$, we have that
	\eqn
	\lefteqn{\Big|\Bra\Phi_+,\int_{(\lattice)^n} d\bp_n H(\bp_n)\delta(\sum_{j=1}^np_j\sigma_j) \prod_{j=1}^ng_j(p_j):\prod_{j=1}^na_{p_j}^{(\sigma_j)} :\Phi_-\Ket\Big|}\nonumber\\
	&\leq & \int_{(\lattice)^{n}} d\bp_n |H(\bp_n)|\delta(\sum_{j=1}^np_j\sigma_j) \prod_{j=1}^n|g_j(p_j)| \|a(\bp_{J_+})\Phi_+\| \|a(\bp_{J_-})\Phi_-\| \, , \label{eq-wick-ord-0}
	\eeqn 
	where we applied Cauchy-Schwarz w.r.t. the inner product on $\cF$. Let $\alpha\in[0,1]$ be arbitrary. Then Cauchy-Schwarz w.r.t. $d\bp_n$ implies that we can estimate the last expression by
	\eqn
	\lefteqn{\Big(\int_{(\lattice)^{n}} d\bp_n \delta(\sum_{j=1}^np_j\sigma_j)|H(\bp_n)|\prod_{j\in J_-}|g_j(p_j)|^2 \|a(\bp_{J_+})\Phi_+\|^2 \Big)^{\frac12}}\nonumber\\
	&&\Big(\int_{(\lattice)^{n}} d\bp_n \delta(\sum_{j=1}^np_j\sigma_j)|H(\bp_n)|\prod_{j\in J_+}|g_j(p_j)|^2 \|a(\bp_{J_-})\Phi_-\|^2 \Big)^{\frac12} \, . \label{eq-wick-ord-1}
	\eeqn 
	Observe that the Pull-Through Formula implies that
	\eqn
	\int d\bp_m a^+(\bp_m)a(\bp_m) &=& \int d\bp_{m-1} a^+(\bp_{m-1}) a(\bp_{m-1})(\nb-m+1)_+\nonumber\\
	&=& (\nb)_m \, , \label{eq-nb-ff}
	\eeqn 
	where we define $(\nb)_m \big|_{\cF_j}:=0$ for all $j\in\{0,1,\ldots,m-1\}$, in accordance with \eqref{eq-annihilation-fn}. In particular, we have that $(\nb)_m\geq0$ as a quadratic form on $\cF$.
	\par Then, integrating the first factor first w.r.t. the momenta $p_{J_-}$ and then $p_{J_+}$, and opposite for the second factor, \eqref{eq-wick-ord-1} has the upper bound
	\eqn
		\lefteqn{\| |H|^{\frac12} \prod_{j\in J_-} g_j(p_j) \delta(\sum_{j=1}^n\sigma_j p_j)^{\frac12}\|_{L^\infty_{\bp_{J_+}}L^2_{\bp_{J_-}}} \|(\nb )_{|J_+|}^{\frac12}\Phi_+\|}	\nonumber\\
		&&\| |H|^{\frac12} \prod_{j\in J_+} g_j(p_j) \delta(\sum_{j=1}^n\sigma_j p_j)^{\frac12}\|_{L^\infty_{\bp_{J_-}}L^2_{\bp_{J_+}}} \|(\nb )_{|J_-|}^{\frac12}\Phi_-\| \, . \label{eq-wick-ord-2}
	\eeqn 
	Using the fact that $\Phi_+\in\cF_{M+\sum_{j=1}^n \sigma_j}$, $\Phi_-\in\cF_M$ and that both are normalized with norm 1, we have that
	\eqn
	\|(\nb)_{|J_+|}^{\frac12}\Phi_+\| &\leq & (M+\sum_{j=1}^n\sigma_j)_{|J_+|}^{\frac12} \mathds{1}_{M+\sum_{j=1}^n\sigma_j\geq |J_+|} \, , \label{eq-wick-ord-4}\\
	\|(\nb)_{|J_-|}^{\frac12}\Phi_-\| &\leq & (M)_{|J_-|}^{\frac12}\mathds{1}_{M\geq |J_-|} \, . \label{eq-wick-ord-5}
	\eeqn 
	Observe that $\sum_{j=1}^n\sigma_j=|J_+|-|J_-|$. Collecting \eqref{eq-wick-ord-0}--\eqref{eq-wick-ord-5}, we have proved that
	\eqn
		\lefteqn{\Big|\Bra\Phi_+,\int_{(\lattice)^n} d\bp_n H(\bp_n)\delta(\sum_{j=1}^np_j\sigma_j) \prod_{j=1}^ng_j(p_j):\prod_{j=1}^na_{p_j}^{(\sigma_j)} :\Phi_-\Ket\Big|}\nonumber\\
		&\leq & \| |H|^{\frac12} \prod_{j\in J_-} g_j(p_j) \delta(\sum_{j=1}^n\sigma_j p_j)^{\frac12}\|_{L^\infty_{\bp_{J_+}}L^2_{\bp_{J_-}}}\| |H|^{\frac12} \prod_{j\in J_+} g_j(p_j) \delta(\sum_{j=1}^n\sigma_j p_j)^{\frac12}\|_{L^\infty_{\bp_{J_-}}L^2_{\bp_{J_+}}}\nonumber\\
		&&(M+\sum_{j=1}^n\sigma_j)_{|J_+|}^{\frac12}(M)_{|J_-|}^{\frac12}\mathds{1}_{M\geq |J_-|} \, .
	\eeqn 
	\par In the case $J_-=\emptyset$, we have that 
	\eqn
	\lefteqn{\Big\|\int_{(\lattice)^n} d\bp_n \, H(\bp_n) \delta(\sum_{j=1}^np_j) a^+(\bp_n) P_M\Big\|}\\
	& = & \Big\|\int_{(\lattice)^n} d\bp_n \, \overline{H}(\bp_n) \delta(\sum_{j=1}^np_j) a(\bp_n) P_{M+n}\Big\| \, ,
	\eeqn 
	which reduces to the case $J_+=\emptyset$. 
	\par In the case $J_+=\emptyset$ and $n\geq2$, we find that
	\eqn
		\lefteqn{\Big\|\int_{(\lattice)^n} d\bp_n \, H(\bp_n) \delta(\sum_{j=1}^np_j)a(\bp_n)\Phi_-\Big\|^2}\nonumber\\ 
		&=& \int_{(\lattice)^{2n}} d\bp_n d\bq_n\, \overline{H}(\bp_n)H(\bq_n) \delta(\sum_{j=1}^np_j) \delta(\sum_{j=1}^nq_j)\scp{\Phi_-}{a^+(\bp_n)a(\bq_n)\Phi_-}\nonumber\\
		&=& \int_{(\lattice)^{2n}} d\bp_n d\bq_n\, \overline{H}(\bp_n)H(\bq_n) \delta(\sum_{j=1}^np_j) \delta(\sum_{j=1}^nq_j)\nonumber\\
		&& \scp{\Phi_-}{a^+(\bp_{n-1})a_{q_n}a^+_{p_n}a(\bq_{n-1})\Phi_-}\nonumber\\
		&&-\int_{(\lattice)} dp_n \, \|A_{p_n}\Phi_-\|^2 \, ,
	\eeqn 
	where
	\eqn
		A_{p_n}  := \int_{(\lattice)^{n-1}} d\bp_{n-1} \, H(\bp_n) \delta(\sum_{j=1}^np_j) a(\bp_{n-1}) \, .
	\eeqn 
	Using Cauchy-Schwarz first on the inner product on $\cF$, and then w.r.t. $d\bp_n d\bq_n$, we find the upper bound
	\eqn
			\lefteqn{\Big\|\int_{(\lattice)^n} d\bp_n \, H(\bp_n) \delta(\sum_{j=1}^np_j)a(\bp_n)\Phi_-\Big\|^2}\nonumber\\ 
			&\leq& \int_{(\lattice)^{2n}} d\bp_n d\bq_n\, |H(\bp_n)||H(\bq_n)| \delta(\sum_{j=1}^np_j) \delta(\sum_{j=1}^nq_j)\nonumber\\
			&&\|a_{q_n}^+a(\bp_{n-1})\Phi_-\|\|a_{p_n}^+a(\bq_{n-1})\Phi_-\|\nonumber\\
			&\leq & \Big( \int_{(\lattice)^{2n}} d\bp_n d\bq_n\, \delta(\sum_{j=1}^np_j) \delta(\sum_{j=1}^nq_j)|H(\bq_n)|^2 \|a_{q_n}^+a(\bp_{n-1})\Phi_-\|^2\Big)^{\frac12}\nonumber\\
			&&\Big( \int_{(\lattice)^{2n}} d\bp_n d\bq_n\, \delta(\sum_{j=1}^np_j) \delta(\sum_{j=1}^nq_j)|H(\bp_n)|^2 \|a_{p_n}^+a(\bq_{n-1})\Phi_-\|^2\Big)^{\frac12}\nonumber\\
			& = &  \int_{(\lattice)^{2n-1}} d\bp_{n-1}d\bq_n \, \delta(\sum_{j=1}^nq_j)|H(\bq_n)|^2 \|a_{q_n}a(\bp_{n-1})\Phi_-\|^2\nonumber\\
			&& + \, \vol \int_{(\lattice)^n}d\bq_n \, \delta(\sum_{j=1}^nq_j)|H(\bq_n)|^2\|(\nb)_{n-1}^{\frac12}\Phi_-\|^2 \, ,
	\eeqn 
	where we used that $[a_{q_n},a_{q_n}^+]=\vol$ together with \eqref{eq-nb-ff}. Using \eqref{eq-nb-ff} again, we conclude
	\eqn
		\lefteqn{\Big\|\int_{(\lattice)^n} d\bp_n \, H(\bp_n) \delta(\sum_{j=1}^np_j)a(\bp_n)\Phi_-\Big\|^2}\nonumber\\ 
		&\leq& \|H\|_{L^2_{\bp_{n-2}}L^\infty_{p_{n-1},p_n}}^2\|(\nb)_n^{\frac12}\Phi_-\|^2+\vol \|\delta(\sum_{j=1}^np_j)^{\frac12} H\|_{L^2_{\bp_n}}^2\|(\nb)_{n-1}^{\frac12}\Phi_-\|^2\nonumber\\
		&\leq & \big(\|H\|_{L^2_{\bp_{n-2}}L^\infty_{p_{n-1},p_n}}^2(M-n+1)+\vol \|\delta(\sum_{j=1}^np_j)^{\frac12} H\|_{L^2_{\bp_n}}^2\big) (M)_{n-1} \, .
	\eeqn 
	Finally, we have
	\eqn
		\|a_0 P_M\|^2 &=& \sup_{\|\Phi\|=1} \scp{P_M\Phi}{a_0^+a_0P_M\Phi} \nonumber\\
		&=& \scp{P_M\Phi}{a_0a_0^+P_M\Phi} - \vol\|P_M\| \nonumber\\
		&=& \|a_0^+P_M\|^2 \, - \, \vol \, , \label{eq-a0-1}
	\eeqn 
	where we used that $\|P_M\|=1$. In addition, we have that
	\eqn
		\|a_0^+P_M\| \, = \, \|P_{M+1}a_0^+\| \, = \, \|a_0 P_{M+1}\| \, . \label{eq-a0-2}
	\eeqn 
	\eqref{eq-a0-1} and \eqref{eq-a0-2} imply
	\eqn
		\|a_0P_M\|^2 \, = \, M\vol \, .
	\eeqn 
	This finishes the proof.
	\endprf
	
	\section{Propagation of approximate restricted quasifreeness}
	
	\begin{lemma}\label{lem-FD-1}
		Let
		\eqn
		\cU_N(t) \; := \; \cVmf^*(t)\cW^*[\sqrt{N\vol}\phi_0]e^{-it\cH_N} \cW[\sqrt{N\vol}\phi_0] \, ,  
		\eeqn
		where for the definition of $\cVmf$, we refer to \eqref{def-HFB}. Then fluctuation dynamics $\cU_N$ obeys
		\begin{equation}\label{eq-fluct}
			\begin{cases}
				i \partial_t \cU_N(t)\,&=\,(\cHcubt(t)+\cHquart(t))\cU_N(t),\\
				\cU_N(0)\,&=\,\1 \, , 
			\end{cases}
		\end{equation}
		where $\cHcubt(t)$ is defined in \eqref{def-hcubt}, $\cHquart(t)$ in \eqref{def-hquart}.
	\end{lemma}
	
	\prf
	We start by defining the auxiliary dynamics
	\eqn
	\widetilde\cU_N(t)\,:=\, \cW^*[\sqrt{N\vol}\phi_0]e^{-i\cH_Nt}\cW[\sqrt{N\vol}\phi_0].
	\eeqn 
	We have that
	\eqn
	i\partial_t\widetilde\cU_N(t)\,&=&\,\cW^*[\sqrt{N\vol}\phi_0] \cH_N\cW[\sqrt{N\vol}\phi_0]\widetilde\cU_N(t) \, .
	\eeqn 
	Using $\phi_0=\vol^{-1/2}$, the explicit expressions for the terms on the right hand side are given by
	\eqn
	\lefteqn{
		\cW^*[\sqrt{N\vol}\phi_0] \, \cH_N \, \cW[\sqrt{N\vol}\phi_0]
	}
	\nonumber\\
	&=&
	\frac{N\vol\lambda}2 \int_\Lambda dx \, v(x) \, + \, \frac12\int_\Lambda dx \, a_x^+(-\Delta_x)a_x
	\nonumber\\
	&&  
	+  \lambda \sqrt{N}
	\int_{\Lambda^2} dx \, dy \,  v(x-y) ( a^+_y  + a_y ) \, + \, \lambda
	\int_{\Lambda^2} dx \, dy \,  v(x-y)     a^+_y a_y 
	\nonumber\\
	&&  
	+ \lambda
	\int_{\Lambda^2} dx \, dy \, v(x-y)  \,  \big( a^+_x a_y \, + \, \frac12(a^+_x a^+_y \, + \, a_y a_x ) \big) 
	\nonumber\\
	&&  
	+ \frac{\lambda}{\sqrt N}
	\int_{\Lambda^2} dx \, dy \, v(x-y)  a_x^+(a_y  + a^+_y) a_y 
	\nonumber\\
	&&  
	+ \frac{\lambda}{2N }
	\int_{\Lambda^2}  dx \, dy \, v(x-y)  a_x^+ a_y^+ a_y a_x  \, . 
	\eeqn
	Using the fact that $\hat{v}(0)=0$, and recalling definitions \eqref{def-hmf}--\eqref{def-hcubz}, we thus obtain
	\begin{equation}\label{eq-FD-lem-aux}
	\begin{cases}
	i\partial_t\widetilde\cU_N(t) &=(\cHmft+\cHcubt+\cHquart)\widetilde\cU_N(t),\\
	\widetilde\cU_N(0)&= \1 \, .
	\end{cases}
	\end{equation}
	In particular, using
	\eqn
	\cU_N(t)\,=\,\cVmf^*(t)\widetilde\cU_N(t) \, ,     
	\eeqn 
	a straight-forward calculation yields that $\cU_N$ satisfies \eqref{eq-fluct}. This concludes the proof.
	\endprf
	Next, we adjust Proposition 3.1 in \cite{busasc} to our present context.
	\begin{lemma}\label{lem-mom-prop-con}
		Assume $\vol\geq1$. Let $\widetilde{\cU}_N$ be defined as in \eqref{eq-FD-lem-aux}. Then there are $C_\ell, K_\ell>0$ such that
		\eqn
		\Big\|(\nb+\vol)^{\frac\ell2} \widetilde{\cU}_N(t)(\nb+\vol)^{-\frac\ell2}\big(1+\frac{\nb}{N\vol}\big)^{-\frac12}\Big\| \, \leq \, C_\ell e^{K_\ell\vd\lambda\vol t}
		\eeqn 
		for all $\ell\in\N$.
	\end{lemma}
	\prf
	We follow the steps of the proof in \cite{busasc} and point out the differences. We show the statement by induction. Let $\psi\in\cF$ be arbitrary. 
	\paragraph{Step 1:} \eqref{eq-FD-lem-aux} implies that
	\eqn
	\partial_t \scp{\widetilde{\cU}_N(t)\psi}{(\nb+\vol)\widetilde{\cU}_N(t)\psi}&=& -i\scp{\widetilde{\cU}_N(t)\psi}{[\nb,\cHmft+\cHcubt+\cHquart]\widetilde{\cU}_N(t)\psi} \nonumber\\
	&=&-i\scp{\widetilde{\cU}_N(t)\psi}{[\nb,\cHmft+\cHcubt]\widetilde{\cU}_N(t)\psi} \, . \label{eq-dt-nbt-0}
	\eeqn 
	Recalling \eqref{def-hmf} and \eqref{def-hcubz}, we have that 
	\eqn 
		[\nb,\cHmft] &=& \lambda \int d\bp_2 \, \hat{v}(p_1) \delta(p_1-p_2) (a_{p_1}^+a_{-p_2}^+ - a_{p_1}a_{-p_2}) \, , \label{eq-hcor-nb-com}\\
		[\nb,\cHcubt] &=& \frac{\lambda}{\sqrt{N}} \int d\bp_3 \, \hat{v}(p_2)\delta(p_1+p_2-p_3)(a_{p_1}^+a_{p_2}^+a_{p_3} \, - \, a_{p_3}^+a_{p_2}a_{p_1})  \, . \label{eq-hcub-nb-com}
	\eeqn 
	Employing Lemma \ref{lem-wick-ord-bd}, \eqref{eq-dt-nbt-0} thus yields
	\eqn
	\lefteqn{|\partial_t\scp{\widetilde{\cU}_N(t)\psi}{(\nb+\vol)\widetilde{\cU}_N(t)\psi}|}\nonumber\\
	&\leq& C\vd\lambda \Big(\scp{\widetilde{\cU}_N(t)\psi}{(\nb+\vol)\widetilde{\cU}_N(t)\psi} \nonumber\\
	&& + \, \frac1{\sqrt{N}}\scp{\widetilde{\cU}_N(t)\psi}{\nb^{\frac32}\widetilde{\cU}_N(t)\psi}\Big)\nonumber\\
	&\leq & C\vd\lambda \Big(\scp{\widetilde{\cU}_N(t)\psi}{(\nb+\vol)\widetilde{\cU}_N(t)\psi}\nonumber\\
	&&+\frac1N\scp{\widetilde{\cU}_N(t)\psi}{\nb^2\widetilde{\cU}_N(t)\psi}\Big) \, . \label{eq-nb-gron-ib-0}
	\eeqn 
	Using
	\eqn
	\cW[\sqrt{N\vol}\phi_0]a_p \cW^*[\sqrt{N\vol}\phi_0] \, = \, a_p \, - \, \sqrt{N}\delta(p) \, , \label{eq-weyl-0}
	\eeqn  
	we derive that
	\eqn
	[\nb,\cW^*[\sqrt{N\vol}\phi_0]]    &=&  -\sqrt{N}\cW^*[\sqrt{N\vol}\phi_0](a_0+a_0^+)+N\vol \cW^*[\sqrt{N\vol}\phi_0] \nonumber\\
	&=& -\big(\sqrt{N}(a_0+a_0^+)+N\vol \big)\cW^*[\sqrt{N\vol}\phi_0] \, , \label{eq-nbw-hc-com}\\
	[\nb,\cW[\sqrt{N\vol}\phi_0]] &=& \cW[\sqrt{N\vol}\phi_0]\big(\sqrt{N}(a_0+a_0^+)+N\vol\big) \, .\label{eq-nbw-com}
	\eeqn 
	From these identities and using $[\nb,\cH_N]=0$, we obtain that
	\eqn
	[\nb,\widetilde{\cU}_N(t)] &=& [\nb,\cW^*[\sqrt{N\vol}\phi_0]]e^{-it\cH_N}\cW[\sqrt{N\vol}\phi_0] \nonumber\\
	&&+ \, \cW^*[\sqrt{N\vol}\phi_0]e^{-it\cH_N}[\nb,\cW[\sqrt{N\vol}\phi_0]] \nonumber\\
	&=& -\sqrt{N}(a_0+a_0^+)\widetilde{\cU}_N(t) \,  + \, \sqrt{N}\widetilde{\cU}_N(t)(a_0+a_0^+) \, . \label{eq-nb-tUn-com}
	\eeqn
	As a consequence, we have that
	\eqn
		\lefteqn{\scp{\widetilde{\cU}_N(t)\psi}{\nb^2\widetilde{\cU}_N(t)\psi}}\nonumber\\
		&=& \scp{\nb\widetilde{\cU}_N(t)\psi}{\widetilde{\cU}_N(t)\nb\psi} \, - \, \sqrt{N}\scp{\nb\widetilde{\cU}_N(t)\psi}{(a_0+a_0^+)\widetilde{\cU}_N(t)\psi} \,\nonumber\\
		&& + \, \sqrt{N}\scp{\nb\widetilde{\cU}_N(t)\psi}{\widetilde{\cU}_N(t)(a_0+a_0^+)\psi} \, .
	\eeqn 
	Using Cauchy-Schwarz, we thus obtain
	\eqn
		\lefteqn{\scp{\widetilde{\cU}_N(t)\psi}{\nb^2\widetilde{\cU}_N(t)\psi}}\nonumber\\
		&\leq& \|\nb\widetilde{\cU}_N(t)\psi\|\Big(\|\widetilde{\cU}_N(t)\nb\psi\|+\sqrt{N}\big(\|(a_0+a_0^+)\widetilde{\cU}_N(t)\psi\| \, + \, \|\widetilde{\cU}_N(t)(a_0+a_0^+)\psi\|\big)\Big) \nonumber\\
		& \leq & \|\nb\widetilde{\cU}_N(t)\psi\|\Big(\|\nb\psi\|+\sqrt{N}\big(\|a_0\widetilde{\cU}_N(t)\psi\| \, + \, \|a_0^+\widetilde{\cU}_N(t)\psi\| \nonumber\\
		&& + \, \|a_0\psi\| \, + \, \|a_0^+\psi\|\big)\Big) \, , \label{eq-nb2-bd-1}
	\eeqn 
	where we also applied that $\widetilde{\cU}_N(t):\cF\to\cF$ is a unitary transformation. Lemma \ref{lem-wick-ord-bd} implies
	\eqn
		\|a_0 \psi\|^2 & =& \sum_{M=0}\| P_Ma_0 \psi\|^2 \nonumber\\
		&=& \sum_{M=0} \|a_0 P_{M-1}\psi\|^2\nonumber\\
		&\leq & \sum_{M=0} (M-1)\vol \|P_{M-1}\psi\|^2 \nonumber\\
		&=& \vol\|\sqrt{\nb}\psi\|^2 \, . \label{eq-a0-nb-bd-1}
	\eeqn 
	Similarly, we have that
	\eqn
		\|a_0^+ \psi\| \, \leq \, \sqrt{\vol}\|\sqrt{\nb+1}\psi\|  \label{eq-a0+-nb-bd-1}
	\eeqn 	
	Employing \eqref{eq-a0-nb-bd-1} and \eqref{eq-a0+-nb-bd-1}, \eqref{eq-nb2-bd-1} implies
	\eqn
		\lefteqn{\scp{\widetilde{\cU}_N(t)\psi}{\nb^2\widetilde{\cU}_N(t)\psi}}\nonumber\\
		& \leq & \|\nb\widetilde{\cU}_N(t)\psi\|\big(\|\nb\psi\|+2\sqrt{N\vol}(\|\sqrt{\nb+1}\widetilde{\cU}_N(t)\psi\| \, + \, \|\sqrt{\nb+1}\psi\|)\big) \, .
	\eeqn 
	Using Young's inequality implies
	\eqn
		\lefteqn{\scp{\widetilde{\cU}_N(t)\psi}{\nb^2\widetilde{\cU}_N(t)\psi}}\nonumber\\
		& \leq & \frac12\scp{\widetilde{\cU}_N(t)\psi}{\nb^2\widetilde{\cU}_N(t)\psi} + C\Big(\scp{\psi}{\big(\nb^2+N\vol(\nb+1)\big)\psi}\nonumber\\
		&&N\vol \scp{\widetilde{\cU}_N(t)\psi}{(\nb+1)\widetilde{\cU}_N(t)\psi}\Big)
	\eeqn 
	As a consequence, we find that
	\eqn
	\lefteqn{\frac1{N}\scp{\widetilde{\cU}_N(t)\psi}{\nb^2\widetilde{\cU}_N(t)\psi}}\nonumber\\
	&\leq & C\vol\Big(\scp{\widetilde{\cU}_N(t)\psi}{(\nb+1)\widetilde{\cU}_N(t)\psi}+\scp{\psi}{\big(\nb +1+\frac{\nb^2}{N\vol}\big)\psi}\Big) \, . \label{eq-nb2/N-0}
	\eeqn
	Plugging this into \eqref{eq-nb-gron-ib-0} and using $\vol\geq1$, we obtain that
	\eqn 
	\lefteqn{|\partial_t\scp{\widetilde{\cU}_N(t)\psi}{(\nb+\vol)\widetilde{\cU}_N(t)\psi}|}\nonumber\\
	&\leq & C\vd\lambda \vol \Big(\scp{\widetilde{\cU}_N(t)\psi}{(\nb+\vol)\widetilde{\cU}_N(t)\psi} \, + \, \scp{\psi}{\big(\nb+1 +\frac{\nb^2}{N\vol}\big)\psi}\Big) \, .
	\eeqn 
	Gronwall's Lemma then implies 
	\eqn
	\scp{\widetilde{\cU}_N(t)\psi}{(\nb+\vol)\widetilde{\cU}_N(t)\psi} \, \leq \, e^{C\vd\lambda\vol t}\scp{\psi}{\big(\nb +\vol+\frac{\nb^2}{N\vol}\big)\psi} \, .
	\eeqn 
	\paragraph{Step 2:} Assume that 
	\eqn
	\lefteqn{\scp{\widetilde{\cU}_N(t)\psi}{(\nb+\vol)^j\widetilde{\cU}_N(t)\psi}}\nonumber\\
	& \leq  &C_j	e^{K_j\vd\lambda\vol t}\scp{\psi}{(\nb+\vol)^j\big(1+\frac{\nb}{N\vol}\big)\psi} \label{eq-nb-IH-1}
	\eeqn 
	for all $1\leq j\leq \ell$ and some constants $C_j, K_j$, and any $\psi\in\cF$. We compute
	\eqn
	\lefteqn{i\partial_t\scp{\widetilde{\cU}_N(t)\psi}{(\nb+\vol)^{\ell+1}\widetilde{\cU}_N(t)\psi}}\nonumber\\
	&=& \scp{\widetilde{\cU}_N(t)\psi}{\big[(\nb+\vol)^{\ell+1},\cHcorz+\cHcubz]\widetilde{\cU}_N(t)\psi}\nonumber\\
	&=&\sum_{j=1}^{\ell+1}\scp{\widetilde{\cU}_N(t)\psi}{(\nb+\vol)^{j-1}\big[\nb,\cHcorz+\cHcubz\big](\nb+\vol)^{\ell+1-j}\widetilde{\cU}_N(t)\psi} \, . \label{eq-nbt-ind-1}
	\eeqn 
	Let
	\eqn
		A_{cub}[\hat{v}] \, : = \, \int d\bp_3 \, \hat{v}(p_2)\delta(p_1+p_2-p_3)a_{p_1}^+a_{p_2}^+a_{p_3} \, .
	\eeqn 
	Applying \eqref{eq-hcor-nb-com} and \eqref{eq-hcub-nb-com}, \eqref{eq-nbt-ind-1} yields
	\eqn
	\lefteqn{i\partial_t\scp{\widetilde{\cU}_N(t)\psi}{(\nb+\vol)^{\ell+1}\widetilde{\cU}_N(t)\psi}}\nonumber\\
	&=& 2\sum_{j=1}^{\ell+1} \Im\scp{\widetilde{\cU}_N(t)\psi}{(\nb+\vol)^{j-1}(g[\hat{v}]-A_{cub}[\hat{v}])(\nb+\vol)^{\ell+1-j}\widetilde{\cU}_N(t)\psi}\nonumber\\
	&=& 2\sum_{j=1}^{\ell+1}\Im\sum_{m,n=0}^\infty (m+\vol)^{j-1}(n+\vol)^{\ell+1-j}\nonumber\\ &&\scp{\widetilde{\cU}_N(t)\psi}{P_m(\lambda g[\hat{v}]-\frac{\lambda}{\sqrt{N}}A_{cub}[\hat{v}])P_n\widetilde{\cU}_N(t)\psi} \, . \label{eq-nb-IH-0}
	\eeqn 
	Observe that we have
	\eqn
		P_mg[\hat{v}]P_n &=& P_m g[\hat{v}] P_{m+2} \delta_{n,m+2} \, , \\
		P_mA_{cub}[\hat{v}]P_n &=& P_mA_{cub}[\hat{v}] P_{m-1} \delta_{n,m-1} \, .
	\eeqn 
	Lemma \ref{lem-wick-ord-bd} and \eqref{eq-nb-IH-0} then imply
	\eqn
	\lefteqn{|\partial_t\scp{\widetilde{\cU}_N(t)\psi}{(\nb+\vol)^{\ell+1}\widetilde{\cU}_N(t)\psi}|}\nonumber\\
	&\leq & C\lambda\sum_{j=1}^{\ell+1} \sum_{m,n=0}^\infty (m+\vol)^{j-1}(n+\vol)^{\ell+1-j}\|P_m\widetilde{\cU}_N(t)\psi\|\|P_n\widetilde{\cU}_N(t)\psi\| \nonumber\\
	&&\big( \|g[\hat{v}] P_{m+2}\| \delta_{n,m+2} \, + \, \frac{\|A_{cub}[\hat{v}]P_{m-1}\|}{\sqrt{N}}  \delta_{n,m-1} \big) \nonumber\\
	&\leq & C\vd \lambda\sum_{j=1}^{\ell+1} \sum_{m,n=0}^\infty (m+\vol)^{j-1}(n+\vol)^{\ell+1-j}\nonumber\\
	&& \big(\|P_m\widetilde{\cU}_N(t)\psi\|^2+\|P_n\widetilde{\cU}_N(t)\psi\|^2\big)  \big( (m+2+\vol)\delta_{n,m+2} \, + \, \frac{m^{\frac32}}{\sqrt{N}}  \delta_{n,m-1}\big) \nonumber\\
	&\leq& C\vd \lambda\sum_{j=1}^{\ell+1}\sum_{m=0}^\infty (m+\vol)^{j-1}(m+2+\vol)^{\ell+1-j}\|P_m\widetilde{\cU}_N(t)\psi\|^2\nonumber\\
	&& \big(m+2+\vol + \frac{(m+1)^{\frac32}}{\sqrt{N}}\big)
	\, .
	\eeqn 
	We can further estimate this by
	\eqn
	\lefteqn{|\partial_t\scp{\widetilde{\cU}_N(t)\psi}{(\nb+\vol)^{\ell+1}\widetilde{\cU}_N(t)\psi}|}\nonumber\\
	&\leq& C(\ell+1) \vd \lambda \big( \scp{\widetilde{\cU}_N(t)\psi}{(\nb+\vol+2)^{\ell+1} \widetilde{\cU}_N(t)\psi} \nonumber\\
	&& + \, \scp{\widetilde{\cU}_N(t)\psi}{\frac{(\nb+\vol+2)^{\ell+\frac32}}{\sqrt{N}} \widetilde{\cU}_N(t)\psi}\big) \nonumber\\
	&\leq& C_\ell \vd \lambda \big( \scp{\widetilde{\cU}_N(t)\psi}{(\nb+\vol)^{\ell+1} \widetilde{\cU}_N(t)\psi}
	\nonumber\\
	&& + \, \frac1N\scp{\widetilde{\cU}_N(t)\psi}{(\nb+\vol)^{\ell+2} \widetilde{\cU}_N(t)\psi} \label{eq-nb-IH-1.5}
	\eeqn 
	We claim that 
	\eqn
	\lefteqn{\frac1{N\vol}\scp{\widetilde{\cU}_N(t)\psi}{(\nb+\vol)^{j+1}\widetilde{\cU}_N(t)\psi}} \nonumber\\
	&\leq& 	C_j \Big(e^{K_j\vd\lambda\vol t}\scp{\psi}{(\nb+\vol)^j\big(1+\frac{\nb}{N\vol}\big)\psi}\nonumber\\
	&& + \, \scp{\widetilde{\cU}_N(t)\psi}{(\nb+\vol)^j\widetilde{\cU}_N(t)\psi}\Big) \label{eq-nb-IH-2}
	\eeqn 
	for all $1\leq j\leq \ell+1$ and all $\psi\in\cF$. \eqref{eq-nb-IH-2} for $j=\ell+1$ together with \eqref{eq-nb-IH-1.5} implies
	\eqn
	\lefteqn{|\partial_t\scp{\widetilde{\cU}_N(t)\psi}{(\nb+\vol)^{\ell+1}\widetilde{\cU}_N(t)\psi}|}\nonumber\\
	&\leq & C_\ell \vd\lambda \vol \Big( \scp{\widetilde{\cU}_N(t)\psi}{(\nb+\vol)^{\ell+1}\widetilde{\cU}_N(t)\psi}\nonumber\\
	&& + \ e^{K_\ell\vd\lambda\vol t}\scp{\psi}{(\nb+\vol)^{\ell+1}\big(1+\frac{\nb}{N\vol}\big)\psi} \Big) \, . 
	\eeqn 
	Gronwall's Lemma then implies
	\eqn
	\lefteqn{\scp{\widetilde{\cU}_N(t)\psi}{(\nb+\vol)^{\ell+1}\widetilde{\cU}_N(t)\psi}|}\nonumber\\
	&\leq & e^{K_\ell\vd\lambda\vol t}\scp{\psi}{(\nb+\vol)^{\ell+1}\big)\psi}\nonumber\\
	&&+ \, C_\ell \vd\lambda \vol t e^{K_\ell\vd\lambda\vol t}\scp{\psi}{(\nb+\vol)^{\ell+1}\big(1+\frac{\nb}{N\vol}\big)\psi}\nonumber\\
	&\leq & C_\ell e^{K_\ell\vd\lambda\vol t}\scp{\psi}{(\nb+\vol)^{\ell+1}\big(1+\frac{\nb}{N\vol}\big)\psi} \, .
	\eeqn 
	Thus, proving \eqref{eq-nb-IH-2} for $j=\ell+1$ concludes the proof. We have proved \eqref{eq-nb-IH-2} for $j=1$ in Step 1, \eqref{eq-nb2/N-0}. We have that \eqref{eq-nb-IH-2} also holds for $j=0$, observing that 
	\eqn
	\lefteqn{\cW[\sqrt{N\vol}\phi_0]\nb \cW^*[\sqrt{N\vol}\phi_0]}\nonumber\\
	&=&\nb-\sqrt{N}(a_0+a_0^+)+N\vol \nonumber\\
	&\leq& 2(\nb+1+N\vol) \, ,
	\eeqn 
	which then commutes with $e^{-it\cH_N}$. 
	\par Suppose \eqref{eq-nb-IH-2} holds up to some $1\leq j\leq\ell-1$. Applying \eqref{eq-nb-tUn-com}, we have that 
	\eqn
	\lefteqn{\frac1{N\vol}\scp{\widetilde{\cU}_N(t)\psi}{(\nb+\vol)^{j+2}\widetilde{\cU}_N(t)\psi}}\nonumber\\
	&=& \frac1{N\vol}\scp{(\nb+\vol)^{j+1} \widetilde{\cU}_N(t)\psi}{\widetilde{\cU}_N(t)(\nb+\vol)\psi}\nonumber\\
	&& -\frac1{\sqrt{N}\vol}\scp{(\nb+\vol)^{j+1} \widetilde{\cU}_N(t)\psi}{(a_0+a_0^+)\widetilde{\cU}_N(t)\psi} \nonumber\\
	&& + \,\frac1{\sqrt{N}\vol}\scp{(\nb+\vol)^{j+1} \widetilde{\cU}_N(t)\psi}{\widetilde{\cU}_N(t)(a_0+a_0^+)\psi} \, . \label{eq-nb/N-ind-0}
	\eeqn 
	We can bound the second term by
	\eqn
	\lefteqn{\frac1{\sqrt{N}\vol}\big|\scp{(\nb+\vol)^{j+1} \widetilde{\cU}_N(t)\psi}{(a_0+a_0^+)\widetilde{\cU}_N(t)\psi}\big|}\nonumber\\
	&\leq & \alpha \scp{ \widetilde{\cU}_N(t)\psi}{(\nb+\vol)^{j+1}\widetilde{\cU}_N(t)\psi} \nonumber\\
	&& + \, \frac{1}{\alpha N\vol^2}\scp{\widetilde{\cU}_N(t)\psi}{(a_0+a_0^+)(\nb+\vol)^{j+1}(a_0+a_0^+)\widetilde{\cU}_N(t)\psi} \, . \label{eq-nb-U-a0-0}
	\eeqn 	
	Employing \eqref{eq-a0-nb-bd-1} and \eqref{eq-a0+-nb-bd-1}, we find that, for any $\tilde{\psi}\in\cF$,
	\eqn
	\lefteqn{\scp{\tilde{\psi}}{(a_0+a_0)^+(\nb+\vol)^k(a_0+a_0^+)\tilde{\psi}}}\nonumber\\
	 &\leq& (\|(\nb+\vol)^{\frac{k}2}a_0\tilde{\psi}\| + \|(\nb+\vol)^{\frac{k}2}a_0^+\tilde{\psi}\|)^2\nonumber\\
	&=&(\|a_0(\nb+\vol-1)^{\frac{k}2}\tilde{\psi}\| + \|a_0^+(\nb+\vol+1)^{\frac{k}2}\tilde{\psi}\|)^2\nonumber\\
	&\leq & C_k\vol \scp{\tilde{\psi}}{(\nb+\vol)^{k+1}\tilde{\psi}} \, ,
	\eeqn 
	i.e., 
	\eqn
	(a_0+a_0)^+(\nb+\vol)^k(a_0+a_0^+) \, \leq\, C_k\vol (\nb+\vol)^{k+1} \, . \label{eq-a0nba0-bd-0}
	\eeqn 
	Employing \eqref{eq-a0nba0-bd-0} and choosing $\alpha>0$ sufficiently large, \eqref{eq-nb-U-a0-0} implies
	\eqn
	\lefteqn{\frac1{\sqrt{N}\vol}\big|\scp{(\nb+\vol)^{j+1} \widetilde{\cU}_N(t)\psi}{(a_0+a_0^+)\widetilde{\cU}_N(t)\psi}\big|}\nonumber\\
	&\leq &C_j\scp{ \widetilde{\cU}_N(t)\psi}{(\nb+\vol)^{j+1}\widetilde{\cU}_N(t)\psi} \nonumber\\
	&& + \, \frac{1}{4 N\vol}\scp{ \widetilde{\cU}_N(t)\psi}{(\nb+\vol)^{j+2}\widetilde{\cU}_N(t)\psi} \, . \label{eq-un-nb-a0-un}
	\eeqn
	We bound the third term in \eqref{eq-nb/N-ind-0} by 
	\eqn
	\lefteqn{\frac1{\sqrt{N}\vol}\big|\scp{(\nb+\vol)^{j+1} \widetilde{\cU}_N(t)\psi}{\widetilde{\cU}_N(t)(a_0+a_0^+)\psi}\big|}\nonumber\\
	\leq & \frac4\vol\scp{ \widetilde{\cU}_N(t)(a_0+a_0^+)\psi}{(\nb+\vol)^{j}\widetilde{\cU}_N(t)(a_0+a_0^+)\psi}\nonumber\\
	& + \, \frac1{4N\vol}\scp{ \widetilde{\cU}_N(t)\psi}{(\nb+\vol)^{j+2}\widetilde{\cU}_N(t)\psi} \, . 
	\eeqn 
	The induction hypothesis \eqref{eq-nb-IH-1} and \eqref{eq-a0nba0-bd-0} hence imply
	\eqn 
	\lefteqn{\frac1{\sqrt{N}\vol}\big|\scp{(\nb+\vol)^{j+1} \widetilde{\cU}_N(t)\psi}{\widetilde{\cU}_N(t)(a_0+a_0^+)\psi}\big|}\nonumber\\
	&\leq &\frac{C_j}\vol e^{K_j\vd \lambda\vol t}\scp{(a_0+a_0^+)\psi}{(\nb+\vol)^j\big(1+\frac{\nb}{N\vol}\big)(a_0+a_0^+)\psi}\nonumber\\
	&& + \, \frac1{4N\vol}\scp{ \widetilde{\cU}_N(t)\psi}{(\nb+\vol)^{j+2}\widetilde{\cU}_N(t)\psi}\nonumber\\
	&\leq &C_je^{K_j\vd \lambda\vol t}\scp{\psi}{(\nb+\vol)^{j+1}\big(1+\frac{\nb}{N\vol}\big)\psi}\nonumber\\
	&& + \, \frac1{4N\vol}\scp{ \widetilde{\cU}_N(t)\psi}{(\nb+\vol)^{j+2}\widetilde{\cU}_N(t)\psi}\label{eq-un-nb-un-a0}
	\eeqn
	For the first term in \eqref{eq-nb/N-ind-0}, we apply \eqref{eq-nb-tUn-com} to the left and obtain
	
	\eqn
	\lefteqn{\frac1{N\vol}\scp{(\nb+\vol)^{j+1} \widetilde{\cU}_N(t)\psi}{\widetilde{\cU}_N(t)(\nb+\vol)\psi}}\nonumber\\
	&=& \frac1{N\vol}\scp{ \widetilde{\cU}_N(t)(\nb+\vol)\psi}{(\nb+\vol)^j\widetilde{\cU}_N(t)(\nb+\vol)\psi}\nonumber\\
	&& -\frac1{\sqrt{N}\vol}\scp{(\nb+\vol)^j (a_0+a_0^+)\widetilde{\cU}_N(t)\psi}{\widetilde{\cU}_N(t)(\nb+\vol)\psi} \nonumber\\
	&& + \,\frac1{\sqrt{N}\vol}\scp{(\nb+\vol)^j \widetilde{\cU}_N(t)(a_0+a_0^+)\psi}{\widetilde{\cU}_N(t)(\nb+\vol)\psi} \, .\label{eq-nb-un-nb-0}
	\eeqn 	
	For the first term in \eqref{eq-nb-un-nb-0}, we use the induction hypothesis \eqref{eq-nb-IH-2}, and obtain
	
	\eqn
	\lefteqn{\frac1{N\vol}\scp{ \widetilde{\cU}_N(t)(\nb+\vol)\psi}{(\nb+\vol)^j\widetilde{\cU}_N(t)(\nb+\vol)\psi}}\nonumber\\
	&\leq & C_j \Big(e^{K_j\vd\lambda\vol t}\scp{\psi}{(\nb+\vol)^{j+1}\big(1+\frac{\nb}{N\vol}\big)\psi}\nonumber\\
	&& + \, \scp{\widetilde{\cU}_N(t)(\nb+\vol)\psi}{(\nb+\vol)^{j-1}\widetilde{\cU}_N(t)(\nb+\vol)\psi}\Big) \nonumber\\
	&\leq & C_j e^{K_j\vd\lambda\vol t}\scp{\psi}{(\nb+\vol)^{j+1}\big(1+\frac{\nb}{N\vol}\big)\psi} \, , \label{eq-nb-un-nb-un-nb}
	\eeqn 
	where in the last step we employed \eqref{eq-nb-IH-1}. Using Cauchy-Schwarz, followed by Young's inequality, the second term in \eqref{eq-nb-un-nb-0} can be bounded by
	\eqn
	\lefteqn{\frac1{\sqrt{N}\vol}\big|\scp{(\nb+\vol)^j (a_0+a_0^+)\widetilde{\cU}_N(t)\psi}{\widetilde{\cU}_N(t)(\nb+\vol)\psi} \big|}\nonumber\\
	&\leq& \frac1{\sqrt{N}\vol}\|(\nb+\vol)^{\frac{j}2} (a_0+a_0^+)\widetilde{\cU}_N(t)\psi\| \|(\nb+\vol)^{\frac{j}2}\widetilde{\cU}_N(t)(\nb+\vol)\psi\|\nonumber\\
	&\leq & \frac1{\vol}\Big(\scp{\widetilde{\cU}_N(t)\psi}{(a_0+a_0^+)(\nb+\vol)^j(a_0+a_0^+)\widetilde{\cU}_N(t)\psi} \nonumber\\
	&&+ \, \frac1N \scp{ \widetilde{\cU}_N(t)(\nb+\vol)\psi}{(\nb+\vol)^j\widetilde{\cU}_N(t)(\nb+\vol)\psi}\Big) \, . 
	\eeqn
	Employing \eqref{eq-a0nba0-bd-0} and then \eqref{eq-nb-un-nb-un-nb}, we thus obtain
	\eqn 
	\lefteqn{\frac1{\sqrt{N}\vol}\big|\scp{(\nb+\vol)^j (a_0+a_0^+)\widetilde{\cU}_N(t)\psi}{\widetilde{\cU}_N(t)(\nb+\vol)\psi} \big|}\nonumber\\
	&\leq & C_j\scp{ \widetilde{\cU}_N(t)\psi}{(\nb+\vol)^{j+1}\widetilde{\cU}_N(t)\psi} \nonumber\\
	&& + \, \frac1{N\vol} \scp{ \widetilde{\cU}_N(t)(\nb+\vol)\psi}{(\nb+\vol)^j\widetilde{\cU}_N(t)(\nb+\vol)\psi}\nonumber\\
	&\leq &C_j\scp{ \widetilde{\cU}_N(t)\psi}{(\nb+\vol)^{j+1} \widetilde{\cU}_N(t)\psi}\nonumber\\
	&&+\, C_j e^{K_j\vd\lambda\vol t}\scp{\psi}{(\nb+\vol)^{j+1}\big(1+\frac{\nb}{N\vol}\big)\psi} \, . \label{eq-un-a0-nb-un-nb}
	\eeqn 
	Similarly, the third term in \eqref{eq-nb-un-nb-0} can be estimated using
	\eqn
	\lefteqn{\frac1{\sqrt{N}\vol}\big|\scp{(\nb+\vol)^j \widetilde{\cU}_N(t)(a_0+a_0^+)\psi}{\widetilde{\cU}_N(t)(\nb+\vol)\psi}\big|}\nonumber\\
	&\leq& \frac1\vol \scp{ \widetilde{\cU}_N(t)(a_0+a_0^+)\psi}{(\nb+\vol)^j\widetilde{\cU}_N(t)(a_0+a_0^+)\psi}\nonumber\\
	&& + \, \frac1{N\vol} \scp{ \widetilde{\cU}_N(t)(\nb+\vol)\psi}{(\nb+\vol)^j\widetilde{\cU}_N(t)(\nb+\vol)\psi}\nonumber\\
	&\leq & \frac{C_j}{\vol} e^{K_j\vd\lambda\vol t}\scp{\psi}{(a_0+a_0^+)(\nb+\vol)^j\big(1+\frac{\nb}{N\vol}\big)(a_0+a_0^+)\psi} \nonumber\\
	&& + \, C_j e^{K_j\vd\lambda\vol t}\scp{\psi}{(\nb+\vol)^{j+1}\big(1+\frac{\nb}{N\vol}\big)\psi} \nonumber\\
	&\leq & C_j e^{K_j\vd\lambda\vol t}\scp{\psi}{(\nb+\vol)^{j+1}\big(1+\frac{\nb}{N\vol}\big)\psi} \, , \label{eq-a0-un-nb-un-nb}
	\eeqn 
	where we used \eqref{eq-nb-IH-1} and \eqref{eq-nb-un-nb-un-nb}, followed by \eqref{eq-a0nba0-bd-0}. Inserting \eqref{eq-nb-un-nb-un-nb}, \eqref{eq-un-a0-nb-un-nb}, and \eqref{eq-a0-un-nb-un-nb} into \eqref{eq-nb-un-nb-0}, we obtain that
	
	\eqn
	\lefteqn{\frac1{N\vol}\scp{(\nb+\vol)^{j+1} \widetilde{\cU}_N(t)\psi}{\widetilde{\cU}_N(t)(\nb+\vol)\psi}}\nonumber\\
	&\leq & C_j e^{K_j\vd\lambda\vol t}\scp{\psi}{(\nb+\vol)^{j+1}\big(1+\frac{\nb}{N\vol}\big)\psi} \nonumber\\
	&&+ \, C_j\scp{\widetilde{\cU}_N(t)\psi}{(\nb+\vol)^{j+1}\widetilde{\cU}_N(t)\psi} \, . \label{eq-un-nb-un-nb}
	\eeqn	
	
	Plugging \eqref{eq-un-nb-a0-un}, \eqref{eq-un-nb-un-a0}, and \eqref{eq-un-nb-un-nb} into \eqref{eq-nb/N-ind-0}, we find that
	
	\eqn
	\lefteqn{\frac1{N\vol}\scp{\widetilde{\cU}_N(t)\psi}{(\nb+\vol)^{j+2}\widetilde{\cU}_N(t)\psi}}\nonumber\\
	&\leq & C_j e^{K_j\vd\lambda\vol t}\scp{\psi}{(\nb+\vol)^{j+1}\big(1+\frac{\nb}{N\vol}\big)\psi} \nonumber\\
	&&+ \, C_j\scp{\widetilde{\cU}_N(t)\psi}{(\nb+\vol)^{j+1}\widetilde{\cU}_N(t)\psi} \, .
	\eeqn 
	This concludes the proof.
	\endprf

	\section{Proof of convergence to mean field equations}
	\label{sec-coh-mean-lim-1}

	\begin{lemma}\label{lem-levelsets}
		Let $H,F \in C^1(\R^n)$ be such that 
		\eqn
			\Big\| \nabla\cdot\Big(\frac{H\nabla F}{|\nabla F|}\Big) \Big\|_1 \, < \, \infty \, . \label{eq-HF-integrability-0} 
		\eeqn 
		Then the following holds true for all $\omega\in\R$ and $g\in C^1_0(\R)$:
		\begin{enumerate}
			\item $\int_{F=\omega}\frac{d\cH^{n-1}}{|\nabla F|} H =  \int_{F<\omega} dp \, \nabla\cdot \Big(\frac{H\nabla F}{|\nabla F|^2}\Big)$,
			\item $\int d\omega \, g(\omega) \partial_\omega \int_{F=\omega}\frac{d\cH^{n-1}}{|\nabla F|} H  = \int d\omega \, g(\omega) \int_{F=\omega} \frac{d\cH^{n-1}}{|\nabla F|} \nabla\cdot\Big(\frac{H\nabla F}{|\nabla F|^2}\Big)$.	
		\end{enumerate}

	\end{lemma}
	\prf The first statement is a direct consequence of the Divergence Theorem together with the fact that $\nabla F/|\nabla F|$ is the outer normal for $\{F<\omega\}$. For the second statement, we have that the divergence theorem implies that
	\eqn
		\int d\omega \, g(\omega) \partial_\omega \int_{F=\omega}\frac{d\cH^{n-1}}{|\nabla F|} H &=& - \, \int d\omega \, g'(\omega) \int_{F=\omega}\frac{d\cH^{n-1}}{|\nabla F|} H \, ,
	\eeqn 
	where we used the assumptions $g\in C^1_0(\R)$, \eqref{eq-HF-integrability-0}, and the first statement. Employing the Coarea Formula, we have that
	\eqn
		\int d\omega \, g(\omega) \partial_\omega \int_{F=\omega}\frac{d\cH^{n-1}}{|\nabla F|} H &=& - \, \int dp \, g'(F)H \, .
	\eeqn 
	Expanding with the factor $\frac{\nabla F\cdot \nabla F}{|\nabla F|^2}$, the Divergence Theorem, together with $g\in C^1_0(\R)$, \eqref{eq-HF-integrability-0}, then implies
	\eqn
		\int d\omega \, g(\omega) \partial_\omega \int_{F=\omega}\frac{d\cH^{n-1}}{|\nabla F|} H &=& \, \int dp \, g(F) \nabla\cdot\Big(\frac{H\nabla F}{|\nabla F|^2}\Big) \, .
	\eeqn 
	Finally, the Coarea Formula implies
	\eqn
		\int d\omega \, g(\omega) \partial_\omega \int_{F=\omega}\frac{d\cH^{n-1}}{|\nabla F|} H &=& \, \int d\omega \, g(\omega) \int_{F=\omega}\frac{d\cH^{n-1}}{|\nabla F|} \nabla\cdot\Big(\frac{H\nabla F}{|\nabla F|^2}\Big) \, .
	\eeqn 
	This finishes the proof.
	\endprf
	
	\begin{lemma}\label{lem-osc-int-bd}
		Let $H,F_1,F_2\in C^2(\R^6)$ be s.t.  
		\eqn
		H \, , \, \nabla\cdot\frac{\nabla F_2|H|}{|\nabla F_2|^2} \in L^1(\R^6)\, . \label{eq-osc-int-bd-ass-0} 
		\eeqn
		Then, for all $t\geq0$, we have that 
		\eqn
		\lefteqn{\Big|\int \, d\bp_2 \,\int_{\Delta[t,2]} \, d\bs_2\, e^{-i(F_1(\bp_2)s_1+F_2(\bp_2)s_2)}H(\bp_2)\Big|}\nonumber\\
		&\leq & Ct\Big(\log(t+1)\Big\|\nabla\cdot\frac{\nabla F_2|H|}{|\nabla F_2|^2}\Big\|_1+\|H\|_1\Big)\, .
		\eeqn 
	\end{lemma}
	\prf
	Observe that 
	\eqn
	\Big| \int_{\Delta[t,2]} \, d\bs_2\, e^{-i(\omega_1s_1+\omega_2s_2)}\Big| &\leq & \int_0^t \, ds_1 \, \Big|\int_0^{s_1} \, ds_2\, e^{-i\omega_2s_2}\Big| \nonumber\\
	&\leq&  \int_0^t \, ds_1 \, \frac2{|\omega_2|+\frac1{s_1}} \nonumber\\
	&\leq & \frac{2t}{|\omega_2|+\frac1{t}} \label{eq-osc-int-bd-0}
	\eeqn 
	for all $\omega_1,\omega_2\in\R$, $t\geq0$, due to the fact that
	\eqn
	\Big| \int_{x_1}^{x_2} \, dy \, e^{-iay}\Big| \, \leq \, \min\{x_2-x_1;\frac2{|a|}\} \, \leq \, \frac2{|a|+\frac1{x_2-x_1}} 
	\eeqn 	
	for all $x_1,x_2,a\in\R$, $x_2\geq x_1$. 
	\par Then \eqref{eq-osc-int-bd-0} followed by the Coarea Formula implies 
	\eqn
	\lefteqn{\Big|\int \, d\bp_2 \,\int_{\Delta[t,2]} \, d\bs_2\, e^{-i(F_1(\bp_2)s_1+F_2(\bp_2)s_2)}H(\bp_2)\Big|}\nonumber\\
	&\leq& Ct\int \, d\bp_2 \, \frac{|H(\bp_2)|}{|F_2(\bp_2)|+\frac1t} \nonumber\\
	&\leq& Ct \int \, d\omega\, \frac1{|\omega|+\frac1t} \int_{F_2=\omega} \, d\cH^5(\bp_2) \frac{|H(\bp_2)|}{|\nabla F_2(\bp_2)|} \, . \label{eq-osc-int-bd-1}
	\eeqn 
	Lemma \ref{lem-levelsets} yields
	\eqn
	\Big|\int_{F_2=\omega} \, d\cH^5(\bp_2) \frac{|H(\bp_2)|}{|\nabla F_2(\bp_2)|}\Big| &\leq & \int_{F_2>\omega} d\bp_2 \Big|\nabla\cdot\frac{\nabla F_2(\bp_2)|H(p_2)|}{|\nabla F_2(\bp_2)|^2}\Big|\nonumber\\
	&\leq &  \Big\|\nabla\cdot\frac{\nabla F_2|H|}{|\nabla F_2|^2}\Big\|_1\, . \label{eq-div-thm-0}
	\eeqn 
	Then, after splitting the domain of integration w.r.t. $d\omega$ into $(-1,1)\cup(-1,1)^\complement$, \eqref{eq-osc-int-bd-1} can be bounded by
	\eqn
	\lefteqn{\int_{-1}^1 d\omega \frac1{|\omega|+\frac1t} \sup_\omega \Big|\int_{F_2=\omega} \, d\cH^5(\bp_2) \frac{|H(\bp_2)|}{|\nabla F_2(\bp_2)|}\Big|}\nonumber\\
	&& + \, \int d\bp_2 \, \mathds{1}_{|F_2(\bp_2)|\geq 1} |H(\bp_2)| \nonumber\\
	&\leq &Ct\Big(\log(t+1)\Big\|\nabla\cdot\frac{\nabla F_2|H|}{|\nabla F_2|^2}\Big\|_1+\|H\|_1\Big)\, , 
	\eeqn 
	where, again, we used the Coarea Formula.
	\endprf 
	
	\begin{lemma}\label{lem-dnomega}
		Let $h\in C^n_b([0,\vc])$, $n\in\N$, $n\leq 3$. Then the following holds for all $\lambda\in(0,1]$. We have that
		\eqn 
		\Big\|D^n\Big[h\Big(\frac{\lambda\hat{v}}{E}\Big)\Big]\Big\|_\infty &\leq & C_n \lambda \|h\|_{C^n[0,\vc\lambda]} \vc^n \, .
		\eeqn 
		Moreover, $\Omega$ satisfies  
		\eqn
		\nabla\Omega(p) &=& (1+\lambda \dO(p))p \, \\
		\|\dO\|_\infty &\leq & C\vc\, , \\
		\|D^2\Omega-I\|_\infty & \leq  & C \lambda\vc^2 \, ,\nonumber\\
		\|D^3\Omega\|_\infty & \leq & C\lambda\vc^3 \, .
		\eeqn 
	\end{lemma}
	\prf 
	By the Fa\`{a} di Bruno formula, we have that 
	\eqn
	|D^n\Big[h\Big(\frac{\lambda\hat{v}}{E}\Big)\Big]| & \leq & C_n \sum_{\br_n\in R(n)} \Big|h^{(S(\br_n))}\Big(\frac{\lambda\hat{v}}{E}\Big)\Big| \lambda^{S(\br_n)}\prod_{j=1}^n \Big[ \big|D^j\big(\frac{\hat{v}}{E}\big)\big|\Big]^{r_j}\
	\, , \label{eq-fdb-0}
	\eeqn 
	where, $R(n)$ is defined as in \eqref{def-Rn} and 
	\eqn
	S(\br_n) \,:=\, \sum_{k=1}^nr_k \, ,
	\eeqn 
	see \eqref{def-srn}. Notice that $S(\br_n)$ satisfies
	\eqn 
	1\,\leq\,S(\br_n) \, \leq \, n \label{eq-Srn-0}
	\eeqn 
	due to the summation condition
	\eqn
	\sum_{j=1}^njr_j \, =\, n \, . \label{eq-rj-sum-1}
	\eeqn
	\eqref{eq-Srn-0} allows us to extract a factor $\lambda$ in \eqref{eq-fdb-0} since it appears within the sum with power $S(\br_n)$. Then \eqref{eq-fdb-0} implies
	\eqn
	\Big\|D^n\Big[h\Big(\frac{\lambda\hat{v}}{E}\Big)\Big]\Big\|_\infty &\leq & C_n \lambda \|h\|_{C^n[0,\vc\lambda]} \vc^n \, . \label{eq-dn-h-1}
	\eeqn  
	With analogous steps, we have that
	\eqn
	\Big\|\partial_{|p|}^n\Big[h\Big(\frac{\lambda\hat{v}}{E}\Big)\Big]\Big\|_\infty &\leq & C_n \lambda \|h\|_{C^n[0,\vc\lambda]} \vc^n \, . \label{eq-dn-h-radial-1}
	\eeqn 
	\par We have that $\Omega=E\sqrt{1+\frac{2\lambda\hat{v}}{E}}$, see \eqref{eq-omega-def-0}, is radial since $\hat{v}$ is radial. Thus we have that
	\eqn
	\nabla\Omega (p)&=& p \Big[\sqrt{1+\frac{2\lambda\hat{v}(p)}{ E(p)}} \, + \, \frac{|p|}2\Big(\sqrt{1+\frac{2\lambda\hat{v}}{ E}}\Big)'(p)\Big] \, .
	\eeqn	
	Observe that
	\eqn
	\Big|\sqrt{1+\frac{2\lambda\hat{v}}{E}}-1\Big| & \leq &  C\frac{\lambda\vc}{\sqrt{1+\frac{2\lambda\hat{v}}{E}}+1} \nonumber\\
	&\leq & C \lambda\vc \, . \label{eq-id-diff-0}
	\eeqn 
	Then \eqref{eq-dn-h-radial-1}, and \eqref{eq-id-diff-0} imply that
	\eqn
	\nabla\Omega(p) &=& (1+\lambda \dO(p))p \, \\
	\|\dO\|_\infty &\leq & C\vc \, . 
	\eeqn	
	\par Next, using the multivariate Leibniz rule together with \eqref{eq-id-diff-0}, we then find that
	\eqn
	D^2\Omega &=& \sqrt{1+\frac{2\lambda\hat{v}}{ E}}I+2p\vee D\sqrt{1+\frac{2\lambda\hat{v}}{ E}}+ED^2\sqrt{1+\frac{2\lambda\hat{v}}{ E}}\nonumber\\
	&=& I \, + \, \lambda A \label{eq-d2omega-0}
	\eeqn 
	for some bounded matrix $A$. Using \eqref{eq-dn-h-1} with $h(x)=\sqrt{1+2x}$, and \eqref{eq-id-diff-0}, we have that
	\eqn
	\|A\|_\infty \, \leq \, C\vc^2
	\eeqn 
	for all $\lambda\in(0,1]$. Here, we used the fact that $\|h\|_{C^2([0,\vc])}\leq C$.
	\par Finally, the multivariate Leibniz rule implies
	\eqn
	\|D^3\Omega\|_\infty &\leq& C\sum_{k=0}^3 \Big\|D^k(|p|^2)D^{3-k} \sqrt{1+\frac{2\lambda\hat{v}}{ E}}\Big\|_\infty\nonumber\\
	&\leq & C \sum_{k=0}^2 \Big\||p|^{2-k}D^{3-k} \sqrt{1+\frac{2\lambda\hat{v}}{ E}}\Big\|_\infty \, .
	\eeqn 
	Using \eqref{eq-fdb-0} and the fact that
	\eqn
	\||p|^{2-k} D^j\big(\frac{\hat{v}}E\big)\|_\infty \, \leq \, C\vc
	\eeqn 	
	for $j\in\{1,2,3\}$, we find that
	\eqn
	\|D^3\Omega\|_\infty \, \leq \, C \lambda\vc^3 \, .
	\eeqn 
	This concludes the proof.
	\endprf
	
	\begin{lemma}\label{lem-d2f1}
		Let
		\eqn
		F_2(\bp_2) \,: = \, \sigma_1\Omega(p_1)+\sigma_2\Omega(p_2)+\sigma_{12}\Omega(p_{12}) 
		\eeqn 
		and 
		\eqn
		T \, := \, \begin{pmatrix}
			(\sigma_1+\sigma_{12})I & \sigma_{12}I \\
			\sigma_{12} I & (\sigma_2+\sigma_{12})I
		\end{pmatrix}  \, .
		\eeqn 
		$F_2$ satisfies
		\eqn
		\nabla F_2(0) & = & 0 \, , \label{eq-f20-0}\\ 
		\|\det(D^2F_2)-\sigma_{12}(\sigma_{12}\sigma_1\sigma_2+\sigma_1+\sigma_2)^3\|_\infty & \leq & C\lambda \vc^2 \, ,\label{eq-detF1-0} \\
		\|D^2F_2\|_\infty &\leq & C \label{eq-d2f1-0}
		\eeqn 
		For all $\bp_2,\bk_2\in\R^6$, and all $\lambda>0$ small enough, dependent on $\vc$, we have that
		\eqn
		|D^2F_2(\bk_2)\bp_2| \in \Big[\frac{|T\bp_2|}2,\frac{3|T\bp_2|}2\Big] \, .
		\eeqn 
		In particular, in this case, we have
		\eqn
		|D^2F_2(\bk_2)\bp_2| \, \geq \, 
		\begin{cases}
			|\bp_2|/2 \, , \mbox{\ if\ } \sigma_1=\sigma_2=\pm\sigma_{12} \, , \nonumber\\
			|p_1|/2 \, , \mbox{\ if\ } \sigma_1=-\sigma_2=\sigma_{12} \, , \nonumber\\
			|p_2|/2 \, , \mbox{\ if\ } -\sigma_1=\sigma_2=\sigma_{12} \, .
		\end{cases}
		\eeqn 
	\end{lemma}
	\prf 
	Thanks to Lemma \ref{lem-dnomega}, we have that
	\eqn
	\nabla \Omega(p) \, = \, p(1+\lambda \dO(p))
	\eeqn 
	with
	\eqn
	\|\dO\|_\infty \, \leq \, C\vc \, .
	\eeqn 
	This immediately implies \eqref{eq-f20-0}.
	\par Next, again due to Lemma \ref{lem-dnomega}, we have that
	\eqn
	D^2\Omega \, = \, I+\lambda A
	\eeqn 
	for some matrix $A\in \R^{3\times3}$ with 
	\eqn
	\|A\|_\infty \, \leq \, C \vc^2 \label{eq-A-bd-0}
	\eeqn 	
	for all $\lambda\in(0,1)$. Denoting $A_j:=A(p_j)$, we obtain that
	\eqn
	D^2F_2 & =& \begin{pmatrix}
		\sigma_1 (I+\lambda A_1) +\sigma_{12}(I+\lambda A_{12}) & \sigma_{12}(I+\lambda A_{12})\\
		\sigma_{12}(I+\lambda A_{12}) & \sigma_2 (I+\lambda A_2) +\sigma_{12}(I+\lambda A_{12})
	\end{pmatrix}\nonumber\\
	&=& \begin{pmatrix}
		(\sigma_1+\sigma_{12})I & \sigma_{12}I \\
		\sigma_{12} I & (\sigma_2+\sigma_{12})I
	\end{pmatrix} + \lambda 
	\begin{pmatrix}
		\sigma_1A_1+\sigma_{12}A_{12} & \sigma_{12}A_{12}\\
		\sigma_{12}A_{12} & \sigma_2A_2+\sigma_{12}A_{12} 
	\end{pmatrix}\nonumber\\
	&=:& T+\lambda \tilde{B} \, .
	\eeqn 
	This identity together with \eqref{eq-A-bd-0} immediately implies \eqref{eq-d2f1-0}. Notice that due to \eqref{eq-A-bd-0}, we have that 
	\eqn
	\|\tilde{B}\|_\infty \, \leq \, C\vc^2 \, . \label{eq-Btil-bd-0}
	\eeqn 
	Now, observe that
	\eqn
	\det(T) &=& \det \big(\begin{pmatrix}
		0& -(\sigma_{12}\sigma_1\sigma_2+\sigma_1+\sigma_2)I \\
		\sigma_{12} I & (\sigma_2+\sigma_{12})I
	\end{pmatrix} \big)\nonumber\\
	&=& -\det \big(\begin{pmatrix}
		\sigma_{12} I & (\sigma_2+\sigma_{12})I\\
		0& -(\sigma_{12}\sigma_1\sigma_2+\sigma_1+\sigma_2)I 		
	\end{pmatrix} \big) \nonumber\\
	&=& \sigma_{12}(\sigma_{12}\sigma_1\sigma_2+\sigma_1+\sigma_2)^3 \, \neq \, 0 \, . \label{eq-D-inv-0}
	\eeqn 
	Thus we may rewrite $T+\lambda\tilde{B} =: (I+\lambda B)T$ with 
	\eqn
	\|B\|_\infty \, \leq \, C\vc^2 \, . \label{eq-B-bd-0}
	\eeqn 
	due to \eqref{eq-Btil-bd-0} and \eqref{eq-D-inv-0}. Then, we have that
	\eqn
	\|\det(I+\lambda B) -1\|_\infty \, \leq \, C\lambda \|B\|_\infty 
	\eeqn 
	for all $\lambda\in(0,1)$. 
	\par Finally, Gershgorin's Circle Theorem implies that 
	\eqn
	\sigma(|I+\lambda B|^2)  & \subseteq & [1-2\lambda \|B\|_\infty-\lambda^2\|B\|_\infty^2,1+2\lambda \|B\|_\infty+\lambda^2\|B\|_\infty^2 ] \nonumber\\
	& \subseteq & \Big[\frac14,\frac94\Big] \label{eq-d2f2-spectrum-0}
	\eeqn 
	for all $\lambda\in(0,1)$ small enough, dependent on $\vc$. Then we have that
	\eqn 
	|D^2F_2(\bk_2)\bp_2|^2 \, = \,  (T\bp_2)^T|I+\lambda B(\bk_2)|^2T\bp_2 \, ,
	\eeqn 
	which together with \eqref{eq-d2f2-spectrum-0} implies that
	\eqn
	|D^2F_2(\bk_2)\bp_2| \in \Big[\frac{|T\bp_2|}2,\frac{3|T\bp_2|}2\Big] \, .
	\eeqn
	This finishes the proof.
	\endprf
	
	\subsection{Error bounds due to HFB evolution}

	\begin{lemma}\label{lem-bog-bol}
		We have that
		\eqn
		\lefteqn{\frac{\lambda^{j}}{N}|\cBc^{(j)}(f_0)[J](T;\lambda)|}\nonumber\\
		&\leq & C_{\vc,\fc,\reg}T(1+\log(T)) \|J\|_{W^{1,\infty}}\frac{\lambda|\log(\lambda)|}{N} 
		\eeqn 
		for $j\in\{1,2\}$, and all $\lambda>0$ small enough, dependent on $\vc$.
	\end{lemma}
	\prf
	$\frac{\lambda^{j}}{N}|\cBc^{(j)}(f_0)[J](T;\lambda)|$ are of the form
	\eqn
	\frac{\lambda^{j-2}}{N}\int_{\R^6} \, d\bp_2 \, \int_{\Delta[T,2]} \, d\bS_2\, e^{-i(F_1(\bp_2)S_1+F_2(\bp_2)S_2)/\lambda^2} H(\bp_2) \label{eq-osc-int-exp-0}
	\eeqn 
	with
	\eqn
	F_2(\bp_2) \, = \, \sigma_1\Omega(p_1)+\sigma_2\Omega(p_2)+\sigma_{12}\Omega(p_{12}) \label{eq-F1-def-0}
	\eeqn 
	for some $\sigma_j\in\{\pm1\}$ and $H:\R^6\to\R$. Employing the bound on $D^2 F_2$ given in Lemma \ref{lem-d2f1} and the diamagnetic inequality $|\nabla|h||\leq |\nabla h|$, we find the estimate 
	\eqn
	\Big\|\nabla\cdot\frac{\nabla F_2|H|}{|\nabla F_2|^2}\Big\|_1 &\leq & \Big\|\frac{\Delta F_2 H}{|\nabla F_2|^2}\Big\|_1 \, + \, \Big\|\frac{\nabla F_2^TD^2 F_2\nabla F_2 H}{|\nabla F_2|^4}\Big\|_1 \, + \, \Big\|\frac{\nabla |H|}{|\nabla F_2|}\Big\|_1 \nonumber\\
	&\leq & C \Big(\Big\|\frac{H}{|\nabla F_2|^2}\Big\|_1+\Big\|\frac{\nabla H}{|\nabla F_2|}\Big\|_1\Big) \, .\label{eq-nablaf1H-0}
	\eeqn 
	Let $B_{F_2}:=\{|\nabla F_2|\leq 1\}$ denote the unit ball induced by $\nabla F_2$, and let $J_{\nabla F_2}:=|\det(D^2F_2)|$. Notice that for any $m\in[0,4)$ and test function $h$, we have that
	\eqn
	\Big\|\frac{h}{|\nabla F_2|^m}\Big\|_1 &\leq& \Big\|\frac{{J_{\nabla F_2}}^{\frac{12-m}{24-3m}}}{|\nabla F_2|^m}\Big\|_{L^{\frac{24-3m}{12-m}}(B_{F_2})}\Big\|\frac{h}{{J_{\nabla F_2}}^{\frac{12-m}{24-3m}}}\Big\|_{L^{\frac{24-3m}{12-2m}}(B_{F_2})} \nonumber\\
	&&+ \, \Big\|\frac{h}{|\nabla F_2|^m}\Big\|_{L^1(B_{F_2}^\complement)}\nonumber\\
	&\leq & C\big(\|h\|_{\frac{24-3m}{12-2m}}+\|h\|_1\big)\label{eq-nablaf1-inv-int-0}
	\eeqn  
	for all $\lambda>0$ small enough, dependent on $\vc$. Here, we used Lemma \ref{lem-d2f1} to bound $J_{\nabla F_2}\geq C$ for $\lambda\in(0,1)$ small enough, substitution, and the fact that
	\eqn
	\int_{B_1}\, \frac{ d\bx_6}{|\bx_6|^{m\frac{24-3m}{12-m}}} &\leq & C\int_0^1 \frac{d|\bx_6|}{|\bx_6|^{m\frac{24-3m}{12-m}-5}} \, < \, \infty 
	\eeqn 	
	since, due to $m<4$,
	\eqn
	m\frac{24-3m}{12-m}-5 \, =\, -3\frac{(m-5)^2-1}{12-m}+1 \, < \, 1 .
	\eeqn 
	In particular, \eqref{eq-nablaf1H-0} and \eqref{eq-nablaf1-inv-int-0} imply
	\eqn
	\Big\|\nabla\cdot\frac{\nabla F_2|H|}{|\nabla F_2|^2}\Big\|_1 &\leq & C\big(\|H\|_{W^{1,1}} \, + \, \|H\|_{\frac94} \, + \, \|\nabla H\|_{\frac{21}{10}}\big) \nonumber\\
	&\leq & C\big(\|H\|_{W^{1,1}} \, + \, \|H\|_{W^{1,\frac{9}4}}\big) \, ,
	\eeqn 
	where we employed the fact that $m\mapsto \frac{24-3m}{12-2m}$ is an increasing function, and interpolation. With that, we apply Lemma \ref{lem-osc-int-bd} to obtain
	\eqn
	\lefteqn{\Big|\int \, d\bp_2 \, \int_{\Delta[t,2]} \, d\bs_2\, e^{-i(F_1(\bp_2)s_1+F_2(\bp_2)s_2)} H(\bp_2) \Big|} \nonumber\\
	&\leq & Ct\Big(\log(1+t)\Big\|\nabla\cdot\frac{\nabla F_2|H|}{|\nabla F_2|^2}\Big\|_1+\|H\|_1\Big) \nonumber\\
	&\leq & Ct(1+\log(1+t))\big(\|H\|_{W^{1,1}} \, + \, \|H\|_{W^{1,\frac{9}4}}\big) \, . \label{eq-int-osc-bd-suf-0}
	\eeqn 
	As a consequence of \eqref{eq-int-osc-bd-suf-0} and using the chain rule, we obtain that
	\eqn 
	\lefteqn{\frac{\lambda^{j}}{N}|\cBc^{(j)}(f_0)[J](T;\lambda)|}\nonumber\\
	&\leq & C_{\vc,\fc,\reg}T(1+\log(T)) \|J\|_{W^{1,\infty}}\frac{\lambda|\log(\lambda)|}{N} 
	\eeqn 
	for $j\in\{1,2\}$, and all $\lambda>0$ small enough, dependent on $\vc$. This finishes the proof.
	
	\endprf

	\subsection{Mollified energy conservation}

	\begin{lemma}\label{lem-r2fd}
		Let $\rrf_2(t;f[J] )$ be defined as in \eqref{eq-rrfd-1}. Then 
		\eqn
		|\rrf_2(t;f[J] )| &\leq&C_{\vc,\fc,\reg} T \|J\|_{W^{2,\infty}}\frac{ \lambda}{N}
		\eeqn 
		for all $\lambda>0$ small enough, dependent on $\vc$.
	\end{lemma}
	\prf
	We want to apply integration by parts to
	\eqn
	\rrf_2(t;f[J] ) &=& \frac1N \int_0^\lambda \, d\tau \, \int \,  d\omega\, \delta_{\frac{2\lambda^2}{T}}'(\omega) \nonumber\\
	&&\int_{ \cE(\tau,\bp_2)=\omega} \, d\cH^5\, \frac{\partial_\tau \cE(\tau,\bp_2) H(\bp_2)}{|\nabla_{\bp_2}\cE(\tau,\bp_2)|} \, . \label{eq-r2fd-recall}
	\eeqn 
	For that, we need to establish a uniform bound on 
	\eqn
	\int_{ \cE(\tau,\bp_2)=\omega} \, d\cH^5\, \frac{\partial_\tau \cE(\tau,\bp_2) H(\bp_2)}{|\nabla_{\bp_2}\cE(\tau,\bp_2)|} \, .
	\eeqn 
	The divergence theorem implies that
	\eqn
	\Big| \int_{ \cE(\tau,\bp_2)=\omega} \, d\cH^5\, \frac{\partial_\tau \cE(\tau,\bp_2) H(\bp_2)}{|\nabla_{\bp_2}\cE(\tau,\bp_2)|} \Big|&\leq & \Big\| \nabla_{\bp_2}\cdot \Big(\frac{\nabla_{\bp_2} \cE(\tau) \partial_\tau \cE(\tau) H}{|\nabla_{\bp_2}\cE(\tau)|^2} \Big) \Big\|_1 \, . \label{eq-cE-bd-0}
	\eeqn 
	We compute
	\eqn
	\partial_\tau \Omega_\tau\, = \, \frac{2\hat{v}}{\sqrt{1+\frac{2\tau\hat{v}}{E}}} \, . \label{eq-omega-tau-1}
	\eeqn 
	Then $\partial_\tau \Omega_\tau$ satisfies
	\eqn
	\|\partial_\tau \Omega_\tau\|_\infty &\leq & C_{\vc} \, ,\label{eq-omega-tau-2}\\
	\|\nabla_p \partial_\tau \Omega_\tau\|_\infty &\leq & C \Big(\vc \, + \, \|\hat{v}\|_\infty \Big\| D\Big(1+\frac{2\tau\hat{v}}{ E}\Big)^{-\frac12}\Big\|_\infty \Big) \nonumber\\
	&\leq & C_{\vc} \, , \label{eq-omega-tau-3}\\
	\|D_p^2 \partial_\tau \Omega_\tau\|_\infty &\leq & C \Big( \|D^2\hat{v}\|_\infty \, +\, \|D\hat{v}\|_\infty \Big\| D\Big(1+\frac{2\tau\hat{v}}{ E}\Big)^{-\frac12}\Big\|_\infty \nonumber\\
	&&+\, \|\hat{v}\|_\infty \Big\| D^2\Big(1+\frac{2\tau\hat{v}}{E}\Big)^{-\frac12}\Big\|_\infty \Big)\nonumber\\
	&\leq & C_{\vc} \label{eq-omega-tau-4}
	\eeqn 
	for all $\tau\in[0,\lambda]$, $\lambda\in(0,1]$. Here we applied the Leibniz Formula, Lemma \ref{lem-dnomega}. Then \eqref{eq-omega-tau-2}-\eqref{eq-omega-tau-4} imply 
	\eqn 
	\|\partial_\tau \cE(\tau,\bp_2)\|_{W^{2,\infty}} \, \leq\, C_{\vc} \label{eq-omega-tau-bd-0}
	\eeqn 
	for all $\tau\in[0,\lambda]$, $\lambda\in(0,1)$. Following the steps of \eqref{eq-nablaf1H-0} in Lemma \ref{lem-bog-bol}, we find that
	\eqn 
	\Big\| \nabla_{\bp_2}\cdot \Big(\frac{\nabla_{\bp_2} \cE(\tau) \partial_\tau \cE(\tau) H}{|\nabla_{\bp_2}\cE(\tau)|^2} \Big) \Big\|_1 &\leq & C\Big( \Big\| \frac{\partial_\tau \cE(\tau) H}{|\nabla_{\bp_2}\cE(\tau)|^2} \Big\|_1 \nonumber\\
	&& + \, \Big\| \frac{\nabla_{\bp_2}\big(\partial_\tau \cE(\tau) H\big)}{|\nabla_{\bp_2}\cE(\tau)|} \Big\|_1 \Big) \label{eq-cE-bd-1}
	\eeqn 
	due to Lemma \ref{lem-d2f1}. Using the Mean-Value Theorem together with  $\nabla_{\tau,\bp_2}\cE(\tau,0)=0$, see Lemma \ref{lem-d2f1}, we find that, for some $\zeta_{\tau,\bp_2}\in[0,\bp_2]$,                
	\eqn
	\nabla_{\bp_2}\cE(\tau,\bp_2)&=&D_{\bp_2}^2\cE(\tau,\zeta_{\tau,\bp_2})\bp_2 \, .
	\eeqn 
	Using \ref{lem-d2f1} again, we conclude that
	\eqn
	|\nabla_{\bp_2}\cE(\tau,\bp_2)| \, \geq \, \frac12\Big|\begin{pmatrix}
		p_2 \\ p_1
	\end{pmatrix}
	\Big| \, = \, \frac{|\bp_2|}{2}  \label{eq-dE-pw-bd-0}
	\eeqn 
	for all $\lambda>0$ small enough, dependent on $\vc$. Thus, we obtain that
	\eqn
	\lefteqn{\Big|\frac{D^\ell\big[(\hat{v}(p_1)+\hat{v}(p_2))^2f_0(p_{12})\big]}{|\nabla_{\bp_2}\cE(\tau,\bp_2)|^m}\Big|}\nonumber\\
	&\leq & C_{\vc,\fc}\Big(\frac{1}{|\bp_2|^m}\mathds{1}_{B_1}(\bp_2) \nonumber\\
	&&+\,  D^\ell\big[(\hat{v}(p_1)+\hat{v}(p_2))^2f_0(p_{12})\big]\mathds{1}_{B_1^\complement}(\bp_2) \Big)  \, \label{eq-vdE-1} 
	\eeqn 
	for all $\ell,m\in\N_0$, $\ell\leq 2$. Hence,
	\eqn
	\Big\|\frac{D^\ell\big[(\hat{v}(p_1)+\hat{v}(p_2))^2f_0(p_{12})\big]}{|\nabla_{\bp_2}\cE(\tau)|^m}\Big\|_1 &\leq & C_{\vc,\fc} \label{eq-vdE-2}
	\eeqn 
	for all $\ell,m\in\N_0$, $\ell\leq 2$, $m\leq5$, and all $\lambda>0$ small enough, dependent on $\vc$. With analogous arguments, we may replace $f_0(p_{12})$ in \eqref{eq-vdE-2} by $f_0(p_{12})f_0(p_1)$, $f_0(p_{12})f_0(p_2)$, or $f_0(p_1)f_0(p_2)$ and obtain an analogous inequality. Then, \eqref{eq-cE-bd-0}, \eqref{eq-omega-tau-bd-0}, \eqref{eq-cE-bd-1}, \eqref{eq-vdE-2}, and the definition \eqref{eq-bol-order-1} of $H$ imply
	\eqn
	\Big| \int_{ \cE(\tau,\bp_2)=\omega} \, d\cH^5\, \frac{\partial_\tau \cE(\tau,\bp_2) H(\bp_2)}{|\nabla_{\bp_2}\cE(\tau,\bp_2)|} \Big|&\leq & C_{\vc,\fc}T\|J\|_{W^{1,\infty}}\label{eq-cE-bd-2}
	\eeqn 
	for all $\omega\in\R$, $\tau\in[0,\lambda]$, $\lambda$ small enough, dependent on $\vc$. 
	\par Next, by integration by parts w.r.t. $d\omega$, $\delta_{\frac{2\lambda^2}{T}}'(\omega)\to0$ as $|\omega|\to\infty$, \eqref{eq-cE-bd-2}, and employing Lemma \ref{lem-levelsets}, \eqref{eq-r2fd-recall} implies that
	\eqn
	\rrf_2(t;f[J] )&=& -\frac1N\int_0^\lambda \, d\tau \, \int \,  d\omega\, \delta_{\frac{2\lambda^2}{T}}(\omega)\int_{ \cE(\tau,\bp_2)=\omega} \, \frac{d\cH^5}{|\nabla_{\bp_2}\cE(\tau,\bp_2)|}\, \nonumber\\
	&& \nabla_{\bp_2}\cdot \Big(\frac{\nabla_{\bp_2}\cE(\tau,\bp_2)\partial_\tau \cE(\tau,\bp_2) H(\bp_2)}{|\nabla_{\bp_2}\cE(\tau,\bp_2)|^2} \Big) \label{eq-rrfd-2}
	\eeqn 
	Lemma \ref{lem-levelsets} implies
	\eqn
	\lefteqn{\Big|\int_{ \cE(\tau,\bp_2)=\omega} \, \frac{d\cH^5}{|\nabla_{\bp_2}\cE(\tau,\bp_2)|} \, \nabla_{\bp_2}\cdot \Big(\frac{\nabla_{\bp_2}\cE(\tau,\bp_2)\partial_\tau \cE(\tau,\bp_2) H(\bp_2)}{|\nabla_{\bp_2}\cE(\tau,\bp_2)|^2} \Big) \Big|} \nonumber\\
	&\leq & \Big\|\nabla_{\bp_2}\cdot\Big[\frac{\nabla_{\bp_2}\cE(\tau,\bp_2)}{|\nabla_{\bp_2}\cE(\tau,\bp_2)|^2}\nabla_{\bp_2}\cdot \Big(\frac{\nabla_{\bp_2}\cE(\tau,\bp_2)\partial_\tau \cE(\tau,\bp_2) H(\bp_2)}{|\nabla_{\bp_2}\cE(\tau,\bp_2)|^2} \Big)\Big]\Big\|_1 \nonumber\\
	&\leq& C\Big(\|D^2_{\bp_2}\cE(\tau)\|_\infty^2\Big\|\frac{\partial_\tau \cE(\tau)H}{|\nabla_{\bp_2}\cE(\tau,\bp_2)|^4}\Big\|_1 \, + \, \|D^3_{\bp_2}\cE(\tau)\|_\infty\Big\|\frac{\partial_\tau \cE(\tau)H}{|\nabla_{\bp_2}\cE(\tau,\bp_2)|^3}\Big\|_1\nonumber\\
	&&+ \, \|D^2_{\bp_2}\cE(\tau)\|_\infty\Big\|\frac{D_{\bp_2}\big(\partial_\tau \cE(\tau)H\big)}{|\nabla_{\bp_2}\cE(\tau,\bp_2)|^3}\Big\|_1 \, + \, \Big\|\frac{D_{\bp_2}^2\big(\partial_\tau \cE(\tau)H\big)}{|\nabla_{\bp_2}\cE(\tau,\bp_2)|^2}\Big\|_1 \Big)\, . \label{eq-rrfd-3}
	\eeqn 
	Lemma \ref{lem-dnomega} implies 
	\eqn
	\|D^2_{\bp_2}\cE(\tau)\|_\infty &\leq & 1+C\vc^2\tau \, ,\label{eq-d2e-0} \\
	\|D^3_{\bp_2}\cE(\tau)\|_\infty &\leq & C\vc^3\tau \, . \label{eq-d3e-0}
	\eeqn 
	Collecting \eqref{eq-omega-tau-bd-0}, \eqref{eq-vdE-2}, \eqref{eq-rrfd-3}, \eqref{eq-d2e-0}, and \eqref{eq-d3e-0}, we find that
	\eqn
	\Big|\int_{ \cE(\tau,\bp_2)=\omega} \frac{d\cH^5}{|\nabla_{\bp_2}\cE(\tau,\bp_2)|} \, \nabla_{\bp_2}\cdot \Big(\frac{\nabla_{\bp_2}\cE(\tau,\bp_2)\partial_\tau \cE(\tau,\bp_2) H(\bp_2)}{|\nabla_{\bp_2}\cE(\tau,\bp_2)|^2} \Big) \Big|\nonumber\\
	\, \leq \, C_{\vc,\fc} T\|J\|_{W^{2,\infty}} \, . \label{eq-rrfd-4}
	\eeqn
	Employing \eqref{eq-vdE-2}, \eqref{eq-rrfd-2}, and \eqref{eq-rrfd-4}, we have proved that 
	\eqn
	|\rrf_2(t;f[J] )| & \leq & C_{\vc,\fc} \frac{T\|J\|_{W^{2,\infty}}}{N}  \int_0^\lambda \, d\tau \, \int \,  d\omega\, \delta_{\frac{2\lambda^2}{T}}(\omega) \nonumber\\
	&\leq& C_{\vc,\fc}\frac{T\lambda \|J\|_{W^{2,\infty}}}{N} \, , \label{eq-rrfd-bd-0}
	\eeqn 
	where in the last step, we applied the normalization \eqref{eq-approxid-norm} of $\delta_{\frac{2\lambda^2}{T}}$.  This finishes the proof.
	\endprf

	\begin{lemma}\label{lem-rr2ec}
		Recall that, due to \eqref{eq-rr2ec-def-0},
		\eqn
		\rrec_2(t;f[J] ) &=& \frac1N\int \, d\omega\, \delta_{\frac{2\lambda^2}{T}}(\omega)\Big(\int_{\dEcub=\omega} \, d\cH^5\, \frac{H}{|\nabla\dEcub|}\nonumber\\
		&&+\, \int_{\dEcub=0} \, d\cH^5\, \frac{H}{|\nabla\dEcub|} \Big) 
		\eeqn 
		with $H$ as defined in \eqref{eq-bol-order-1}. Then we have that	\eqn
		|\rrec_2(t;f[J] )| \, \leq \, C_{\vc,\fc}\sqrt{T}\|J\|_{W^{2,\infty}}\frac{\lambda }N   
		\eeqn
		for all $\lambda>0$ small enough, dependent on $\vc$.
	\end{lemma}
	\prf
	We start by writing
	\eqn
	\rrec_2(t;f[J] ) &=:& \frac1N\int \, d\omega\, \delta_{\frac{2\lambda^2}{T}}(\omega) \cI(\omega) \, . \label{eq-rr2ec-0}
	\eeqn 
	Analogously to \eqref{eq-cE-bd-2}, we have that
	\eqn
	|\cI(\omega)| & \leq & C_{\vc,\fc}T \|J\|_{W^{1,\infty}}\frac{1}{N} \, . \label{eq-rrec-1}
	\eeqn 
	
	Notice that, due to Lemma \ref{lem-levelsets}, we have that
	\eqn
	\cI(\omega) &=& \int_0^{\omega} \, d\tau \, \int_{\dEcub=\tau} \, \frac{d\cH^5}{|\nabla\dEcub|}\, \nabla\cdot\Big( \frac{\nabla\dEcub H}{|\nabla\dEcub|^2}\Big) \, ,
	\eeqn 
	where the integral respects the orientation of $[0,\omega]$ resp. $[\omega,0]$. Using Lemma \ref{lem-levelsets}, we thus obtain
	\eqn
	|\cI(\omega)| &\leq & |\omega| \Big\|\nabla\cdot\Big[ \frac{\nabla \dEcub}{|\nabla\dEcub|^2}\nabla\cdot\Big( \frac{\nabla\dEcub H}{|\nabla\dEcub|^2}\Big)\Big]\Big\|_1
	\nonumber\\
	&\leq& C|\omega|\Big(\|D^2\dEcub\|_\infty^2\Big\|\frac{H}{|\nabla \dEcub |^4}\Big\|_1 \, + \, \Big\|\frac{|D^3\dEcub |H}{|\nabla\dEcub|^3}\Big\|_1\nonumber\\
	&&+ \, \|D^2 \dEcub\|_\infty\Big\|\frac{DH}{|\nabla\dEcub|^3}\Big\|_1 \, + \, \Big\|\frac{D^2H}{|\nabla \dEcub|^2}\Big\|_1 \Big) \, , \label{eq-rrec-2}
	\eeqn 
	for all $\lambda>0$ small enough, dependent on $\vc$, analogously to \eqref{eq-rrfd-3}. Notice that $D^3\dEcub=0$ and $\|D^2\dEcub\|_\infty\leq C$. Analogously to \eqref{eq-rrfd-bd-0}, we thus obtain the upper bound
	\eqn
	|\cI(\omega)| \, \leq \, C_{\vc,\fc}|\omega|T\|J\|_{W^{2,\infty}} \, . \label{eq-rrec-3}
	\eeqn 
	\par We split the integral in \eqref{eq-rr2ec-0} into the regions $(-\omega_0,\omega_0)\cup (-\omega_0,\omega_0)^\complement$ with $\omega_0$ to be determined below. Then \eqref{eq-rrec-1} and \eqref{eq-rrec-3} yield
	\eqn
	\lefteqn{|\rrec_2(t;f[J] )|}\nonumber\\
	&\leq & \frac1N\int_{(-\omega_0,\omega_0)} \, d\omega\,  \delta_{\frac{2\lambda^2}{T}}(\omega) |\cI(\omega)| \, + \, \int_{(-\omega_0,\omega_0)^\complement} \, d\omega\, \delta_{\frac{2\lambda^2}{T}}(\omega) |\cI(\omega)|\nonumber\\
	&\leq &  \frac{C_{\vc,\fc}T }N  \Big(\|J\|_{W^{2,\infty}}\int_{(-\omega_0,\omega_0)} \, d\omega\,  \delta_{\frac{2\lambda^2}{T}}(\omega) |\omega|\nonumber\\
	&&+ \, \|J\|_{W^{1,\infty}}  \int_{(-\omega_0,\omega_0)^\complement} \, d\omega\, \delta_{\frac{2\lambda^2}{T}}(\omega) \Big) \, . \label{eq-rrec-4}
	\eeqn     
	Employing the normalization condition \eqref{eq-approxid-norm} and the decay condition \eqref{eq-approxid-supp} on $\delta_{\frac{2\lambda^2}{T}}$, we thus obtain
	\eqn
	|\rrec_2(t;f[J] )| \, \leq \, \frac{C_{\vc,\fc}T \|J\|_{W^{2,\infty}}}N \Big(\omega_0 \, +\, \frac{\lambda^2}{\omega_0 T}\Big) \, .
	\eeqn
	By now choosing $\omega_0=\frac{\lambda}{\sqrt{T}}$, we have hence proved that\
	\eqn
	|\rrec_2(t;f[J] )| \, \leq \, \frac{C_{\vc,\fc}\sqrt{T}\|J\|_{W^{2,\infty}}\lambda}N  \, .
	\eeqn
	
	\endprf

	\noindent
	{\bf Acknowledgments:} 
	T.C. gratefully acknowledges support by the NSF through grants DMS-1151414 (CAREER), DMS-1716198, DMS-2009800, and the RTG Grant DMS-1840314 {\em Analysis of PDE}. M.H. was supported by a University of Texas at Austin Provost Graduate Excellence Fellowship, and by NSF grants DMS-1716198 and DMS-2009800 through T.C. M.H. would like to thank Darren King and Daniel Weser for helpful discussions and comments on the use of Geometric Measure Theory for some calculations. We also thank Esteban C\'ardenas, Jacky Chong, Ryan Denlinger, J\"urg Fr\"ohlich, Irene Gamba, Laurent Lafleche, Nata\v{s}a Pavlovi\'c, Israel Michael Sigal, and Avy Soffer for inspiring discussions, and some references. We thank the anonymous referees for valuable comments. 
	\\

	\noindent
	{\bf Conflict of Interest Statement:} 
	No conflicts of interests are connected to this article.
	\\
	
	\noindent
	{\bf Data availability:} 
	Data sharing not applicable to this article as no datasets were generated or analysed during the current study.

\end{document}